\documentclass[11pt]{article}
\usepackage{mymacros}

\title{Finite-cutoff geometry in double-scaled SYK:\\SYK deformations, branes and CMC-Krylov complexity
}
\author[a]{Sergio E. Aguilar-Gutierrez\orcidlink{0000-0003-0308-0061}}
\affiliation[a]{Qubits and Spacetime Unit, Okinawa Institute of Science and Technology Graduate University, 
\footnote{\begin{CJK}{UTF8}{min}沖縄科学技術大学院大学\end{CJK}}
1919-1 Tancha, Onna, Okinawa 904 0495, Japan}
\emailAdd{sergio.ernesto.aguilar@gmail.com}
\abstract{
We develop a finite-cutoff holographic description of the double-scaled SYK (DSSYK) model. We construct microscopic deformations of Kourkoulou-Maldacena and thermofield-double SYK states whose double-scaling limits reproduce a lower-dimensional analogue of $\TT$ and $\TT+\Lambda_2$ deformations, which we denote $T^2$ and $T^2+\Lambda_1$, respectively. The ensemble-averaged theory describes wormhole states in finite-cutoff sine dilaton gravity, and finite-cutoff anti-de Sitter and de Sitter geometries in the low- and high-energy regimes, respectively. The analysis includes both perturbative and non-perturbative solutions to the energy spectrum flow equation, and the regions behind and beyond the respective horizons. Only the perturbative branch admits physical quantum states. We study how to incorporate  end-of-the-world (ETW) branes in the finite cutoff holographic dictionary. The ETW brane leads to an additional boundary theory deformation, which leads to black hole or eternal traversable wormhole geometries depending on the existence of normalizable bound states. We extend the finite cutoff length evaluations for constant mean curvature (CMC) slices, which we interpret as complexity in the boundary theory. We discuss possible extensions with matter. Our results provide a microscopic framework for studying the finite-cutoff boundaries, CMC slices and branes in the holographic dictionary of DSSYK.}

\begin{document}
\maketitle

\section{Introduction}
Finite-cutoff holography is a concrete framework to investigate the relation between quantum field theories with irrelevant deformations and gravitational dynamics away from the asymptotic boundary. The paradigmatic example is the $\TT$ deformation (see \cite{Jiang:2019epa,He:2025ppz,Guica:2025jkq} for modern reviews), which modifies the ultraviolet structure of a two-dimensional quantum field theory while admitting a geometric interpretation in terms of a finite radial cutoff in the holographic dual \cite{McGough:2016lol}. Since the original formulation of $\TT$ deformations \cite{Smirnov:2016lqw,Cavaglia:2016oda} (see also \cite{Zamolodchikov:2004ce}), a broad class of generalizations and applications has been developed \cite{Dubovsky:2012wk,Dubovsky:2013ira, Tsolakidis:2024wut,Dubovsky:2017cnj,Dubovsky:2018bmo, Cardy:2018sdv,Aharony:2018bad,Conti:2018jho,Conti:2018tca,Tolley:2019nmm,Jafari:2019qns,Zhang:2025dgm,Barbon:2020amo, Apolo:2023ckr,Apolo:2023vnm}, including higher-dimensional \cite{Hartman:2018tkw,Taylor:2018xcy,Morone:2024ffm,Cardy:2018sdv,Bonelli:2018kik,Ferko:2024yua} and lower-dimensional analogues \cite{Gross:2019ach,Gross:2019uxi,Iliesiu:2020zld,Johnson:2020lns,Rosso:2020wir,Griguolo:2021wgy,Aguilar-Gutierrez:2024nst,Aguilar-Gutierrez:2024oea,Ebert:2022ehb,Ebert:2022xfh,Griguolo:2025kpi}, mixed boundary conditions \cite{Guica:2019nzm}, and more general multitrace deformations \cite{Witten:2001ua,Parvizi:2025shq,Parvizi:2025wsg,Taghiloo:2025oeu,Coleman:2020jte,Allameh:2025gsa,Sheikh-Jabbari:2025kjd,Ran:2025xas,Sheikh-Jabbari:2026hkh,Sheikh-Jabbari:2026uol}. In the gravitational language, these deformations provide a way of replacing the asymptotic boundary by a finite-radius boundary and thereby retaining access to regions of the bulk that are normally hidden behind the usual holographic cutoff. In two-dimensional gravity this question is especially tractable, since the dynamics of Jackiw-Teitelboim (JT) gravity \cite{JACKIW1985343,TEITELBOIM198341} (reviewed in \cite{Mertens:2022irh}) and its finite-cutoff generalizations \cite{Gross:2019ach,Gross:2019uxi,Iliesiu:2020zld,Johnson:2020lns,Rosso:2020wir,Griguolo:2021wgy,Aguilar-Gutierrez:2024nst,Aguilar-Gutierrez:2024oea,Ebert:2022ehb,Ebert:2022xfh,Griguolo:2025kpi,AliAhmad:2025kki,Griguolo:2026eoi} can often be reduced to solvable quantum mechanical systems.

The double-scaled SYK (DSSYK) model \cite{Cotler:2016fpe,Erdos:2014zgc} offers a complementary boundary realization of this program, which describes ``wormhole states'' \cite{Harlow:2018tqv} in the bulk, denoting Einstein-Rosen bridges connecting the boundaries in the bulk. The model combines a large degree of analytic control through chord-diagram techniques \cite{Berkooz:2018jqr,Berkooz:2018qkz} in terms of an auxiliary quantum mechanical description (denoted chord theory from now on) containing thermal information about the SYK system \cite{kitaevTalks1,kitaevTalks2,kitaevTalks3,Maldacena:2016hyu} and a nontrivial gravitational interpretation: its low-energy sector is governed by the Schwarzian theory associated with JT gravity, while its exact double-scaling description retains information about the full spectrum and nonperturbative effects \cite{Cotler:2016fpe,Erdos:2014zgc,Berkooz:2018jqr,Berkooz:2018qkz,Berkooz:2020fvm,Berkooz:2020uly,Berkooz:2020xne,Berkooz:2022fso,Berkooz:2022mfk,Berkooz:2024evs,Berkooz:2024ofm,Berkooz:2024lgq,Okuyama:2025fhi,Beccaria:2026ndg}. The relation between DSSYK and finite-cutoff gravitational systems has recently been developed from the chord description \cite{Aguilar-Gutierrez:2024oea,Aguilar-Gutierrez:2026ogo} and in relation to finite-cutoff JT gravity in a canonical formulation \cite{Griguolo:2025kpi,Griguolo:2026eoi}. At the same time, the connection between DSSYK and sine dilaton gravity has provided a useful framework for understanding both the anti-de Sitter (AdS) and de Sitter (dS) regimes of the model \cite{Blommaert:2023opb,Blommaert:2024whf,Blommaert:2024ymv,Blommaert:2025avl,Bossi:2024ffa,Arundine:2025mcu,Heller:2025ddj,Heller:2024ldz,Fu:2025kkh}, complex Liouville string \cite{Collier:2024kmo,Collier:2025pbm,Blommaert:2026ofx,Blommaert:2025eps}, dS$_3$ space \cite{Narovlansky:2023lfz,Verlinde:2024znh,Verlinde:2024zrh,Narovlansky:2025tpb,Blommaert:2025eps,Tietto:2025oxn,Aguilar-Gutierrez:2024nau,Marini:2026zjk,Blommaert:2026ofx,Goto:2026ipq} among others \cite{Susskind:2021esx,Susskind:2022bia,Susskind:2023hnj,Lin:2022nss,Rahman:2022jsf,Rahman:2023pgt,Rahman:2024iiu,Rahman:2024vyg,Sekino:2025bsc,Miyashita:2025rpt,Aguilar-Gutierrez:2025otq,Aguilar-Gutierrez:2026ogo,Milekhin:2023bjv,Milekhin:2024vbb,Narovlansky:2025tpb,Okuyama:2025hsd,Yuan:2024utc,Gubankova:2025gbx,Ahn:2025exp,Miyashita:2026zyl,Dong:2026gaa}. These developments suggest that DSSYK can be used not only as a model for asymptotic JT or sine dilaton gravity, but also as a laboratory for studying finite-cutoff gravitational physics directly from a boundary perspective.

Another motivation for this work comes from holographic complexity \cite{Susskind:2014rva,Chapman:2021jbh}. Krylov space methods (reviewed in \cite{Rabinovici:2025otw,Nandy:2024evd,Baiguera:2025dkc,Jeong:2026gdc}) have emerged as a particularly useful language for relating quantum dynamics to bulk geometry. In DSSYK, the Lanczos algorithm provides an explicit construction of a Krylov basis, and the associated Krylov complexity \cite{Parker:2018yvk,Balasubramanian:2022tpr} has been related to bulk geodesic lengths in JT gravity \cite{Rabinovici:2023yex}. This relation has subsequently been explored in a variety of extensions of DSSYK and in other gravitational settings \cite{Ambrosini:2024sre,Ambrosini:2025hvo,Heller:2024ldz,Heller:2025ddj,Xu:2024gfm,Bhattacharjee:2022ave,Fu:2025kkh,Aguilar-Gutierrez:2024nau,Aguilar-Gutierrez:2025hty,Aguilar-Gutierrez:2025mxf,Aguilar-Gutierrez:2025pqp,Aguilar-Gutierrez:2025sqh,Aguilar-Gutierrez:2026jjv,Forste:2025gng,Aguilar-Gutierrez:2026ogo,Miyaji:2025ucp,Alfinito:2026cky,Aguilar-Gutierrez:2026nmd,Chakraborty:2026ssu,Alfinito:2026opj,Pedraza:2026zji,Bhattacharya:2026ybq,Balasubramanian:2024lqk,Nandy:2024zcd,Qu:2026dmv,Bhattacharya:2026yfm}\footnote{See also \cite{Bhattacharyya:2025gvd,Chattopadhyay:2024pdj,Takahashi:2025iol} for discussion about Krylov complexity in other $\TT$-deformed theories.} These results naturally raise a broader question: is Krylov complexity the only boundary complexity measure with a concrete geometric dual interpretation? Conversely, \emph{are there other geometric observables besides geodesic lengths in the two-dimensional gravity context which can serve as a concrete dual complexity measure?}
Several complexity measures have been associated with a geometric dual interpretation (e.g. \cite{Erdmenger:2021wzc,Erdmenger:2022lov,Aguilar-Gutierrez:2024nau,Aguilar-Gutierrez:2023pnn}). More recently, it was found in \cite{Blommaert:2026lvp} a very simple functional relation between the geodesic and constant mean curvature (CMC) slice length in finite causal diamonds. In dS$_2$ JT gravity, an analogous relation follows after the appropriate continuation associated with the different sign of the cosmological constant, with spacelike surfaces in AdS becoming timelike surfaces in dS. This observation is particularly useful because CMC slices probe regions of the geometry that are not captured by the usual geodesic observable and can exhibit qualitatively different late-time behavior \cite{Jorstad:2023kmq,Aguilar-Gutierrez:2023pnn,Belin:2022xmt}.

\paragraph{Questions in this work}
Despite recent progress, there are important questions that remain open. What is the precise boundary dual theory and its microscopic realization (i.e.~before ensemble averaging)? What is the physical meaning of the non-perturbative solutions to the corresponding energy flow equation describing $T^2$ and $T^2+\Lambda_1$ deformed theories? What is the boundary interpretation of other finite geometric structures beyond geodesic lengths, such as CMC slice foliations of finite causal diamonds in JT gravity \cite{Blommaert:2026lvp}? What is the boundary theory deformation describing end-of-the-world (ETW) branes in the bulk with a finite cutoff Dirichlet boundary? What is its effect in AdS eternal traversable wormholes \cite{Maldacena:2018lmt}? We address these questions, and provide clues towards related questions, from three angles:  SYK deformations, the chord Hilbert space \cite{Berkooz:2018jqr}, and finite-cutoff JT and sine dilaton gravity, as explained below.

\paragraph{Answers and summary of results}
To establish the microscopic holographic dual of the finite-cutoff holographic setup, we construct new deformations of SYK before ensemble averaging.\footnote{See \cite{Berkooz:2024evs,Berkooz:2024ofm,Aguilar-Gutierrez:2026nmd,Bhattacharya:2026yfm} for other approaches to deforming the DSSYK model.} We consider both Kourkoulou-Maldacena \cite{Kourkoulou:2017zaj} (KM) states and thermofield-double (TFD) states, and introduce deformation operators adapted to the corresponding states. In the KM construction, the Hamiltonian is built from composite SYK operators that act naturally on the KM states, while in the TFD construction we employ a generalization of the Maldacena-Qi \cite{Maldacena:2018lmt} (MQ) interaction compatible with the double-scaling limit. The deformation also incorporates operators related to the size, or equivalently to the chord number, of DSSYK \cite{Maldacena:2018lmt,Roberts:2018mnp,Lin:2022rbf}, and takes the form (see \eqref{eq:def H SYK PT TFD})
\begin{equation}\label{eq:H0 intro}
\hH_{0'}=\frac{1}{2}\qty(\text{b}(\hH_{\rm MQ})\hH_{\rm def'}^\dagger+\hH_{\rm def'} \text{b}(\hH_{\rm MQ}))~,
\end{equation}
where $\text{b}(\hH_{\rm MQ})$ is a Hermitian operator which depends on an MQ interaction term, $\hH_{\rm def'}$ is a microscopic deformation which, in the double-scaling limit, raises or lowers the index $n$ in the corresponding Krylov basis, $\ket{n}$. In this way, the finite-cutoff structure is not introduced only at the level of an effective chord Hamiltonian: it has an explicit microscopic origin in the underlying SYK theory. Taking the double-scaling limit then produces the desired deformed DSSYK with a chord Hamiltonian (see \eqref{eq:chord PT H})
\begin{eqnarray}\label{eq:intro H PT}
    \hH_{\rm PT}\equiv -\frac{1}{\sqrt{1-q}}\qty(\sqrt{1-\frac{\xi^2}{\cosh^2(\lambda\hat{n}/2)}}\rme^{-\rmi \hat{p}_n}+\rme^{\rmi \hat{p}_n}\sqrt{1-\frac{\xi^2}{\cosh^2(\lambda\hat{n}/2)}})~,
\end{eqnarray}
where $\hat{n}$ and $\hat{p}_n$ are operators acting on boundary states describing the length of Einstein-Rosen bridges in the bulk dual, $\lambda$ is a parameter characterizing the classical limit where $\lambda\rightarrow0$, $q\equiv \rme^{-\lambda}$, and $\xi$ a boundary parameter that reproduces the dilaton evaluated at the cutoff boundary. \eqref{eq:intro H PT} can be further deformed with an additional term (see \eqref{eq:PT improved H}) 
\begin{equation}\label{eq:intro PT improved H}
    \begin{aligned} H_{\rm PT'}&=\hH_{\rm PT}+\frac{1}{\sqrt{1-q}}\qty(\mathcal K_0(\hat{n},\hat{p}_n)-\mathcal K_{\infty}(\hat{n},\hat{p}_n)) \\ 
    \mathcal K_0&=\frac12\left( \sqrt{\text{csch}(\lambda\hat{n})}\sin\hat{p}_n\sqrt{\sinh (\lambda\hat{n})} \sin\hat{p}_n+\sin\hat{p}_n \sqrt{\sinh (\lambda\hat{n})} \sin\hat{p}_n \sqrt{\text{csch}(\lambda\hat{n})} \right)~,\\
    \mathcal K_\infty &=\frac12\left( \rme^{-\lambda\hat{n}/2}\sin\hat{p}_n\rme^{\lambda\hat{n}/2} \sin\hat{p}_n+\sin\hat{p}_n\rme^{\lambda\hat{n}/2} \sin\hat{p}_n\rme^{-\lambda\hat{n}/2} \right)~,
    \end{aligned} 
\end{equation}
representing a choice of operator ordering in finite cutoff sine dilaton gravity (Sec.~\ref{ssec:improvement JT}) which encodes a quantum correction term in finite cutoff JT gravity \cite{Griguolo:2026eoi}.

We define the \emph{low-energy limit} of the chord Hamiltonian \eqref{eq:intro H PT} as (see \eqref{eq:low energy limit})
\begin{equation}\label{eq:low energy limit0}
    \xi\rightarrow\lambda\Phi_b~,\quad \hat{p}_n\rightarrow\lambda\hat{P}~,\quad \lambda\hat{n}\rightarrow\hat{L}~,
\end{equation}
where $\Phi_b\sim\mathcal{O}(\lambda^0)$ as $\lambda\rightarrow0$ is a fixed parameter (the dilaton $\Phi$ in finite cutoff JT gravity at the boundary), and $\hat{L}$ and $\hat{P}$ are canonical variables whose spectrum in the low-energy limit is $\mathcal{O}(\lambda^0)$ as $\lambda\rightarrow0$. The first limit in \eqref{eq:low energy limit0} is in agreement with sine dilaton gravity \cite{Blommaert:2023opb,Blommaert:2024ymv}, which reproduces JT gravity when its dilaton is $\mathcal{O}(\lambda)$. The second limit in \eqref{eq:low energy limit0} indicates that we study the low energy spectrum \cite{Blommaert:2023opb,Blommaert:2024ymv,Blommaert:2024whf}, while the third one means that there is no regularization for the bulk geodesic length dual \cite{Lin:2022rbf}, which is indeed an expected property in finite cutoff holography \cite{Griguolo:2026eoi}. We find that under the proposed low-energy limit \eqref{eq:low energy limit0}, \eqref{eq:intro H PT} (as well as its extension \eqref{eq:intro PT improved H}) reproduces the Pöschl--Teller \cite{Poschl:1933zz} (PT) Hamiltonians that arise from finite-cutoff JT gravity \cite{Griguolo:2025kpi,Griguolo:2026eoi} as well as sine dilaton gravity at finite cutoff, which we discuss in Sec.~\ref{ssec:sine dilaton}; see also \cite{GPRST}. 

We emphasize that in contrast to the \emph{asymptotic boundary limit}, defined by (see \eqref{eq:ABL})
\begin{equation}
\Phi_b\rightarrow\infty~,\quad \hat{L}\rightarrow\infty~,\quad \Phi_b~\rme^{-\hat{L}/2}~:~~\text{fixed}~,
\end{equation}
bulk geodesic lengths at finite cutoff are absolute distances instead of regularized ones, which requires an additional choice of boundary conditions when the distance vanishes \cite{Griguolo:2026eoi}. The quantum correction term encoded in the extension of the chord Hamiltonian in \eqref{eq:intro PT improved H} allows one to identify a natural choice, which is not present without it (Sec.~\ref{ssec:improvement JT}); which is realized by restricting $\ket{n\geq1}$ in the chord basis, a condition that approaches the Friedrichs extension only logarithmically in $\lambda$ (App.~\ref{app:log corrections}). Since $\xi\in[0,1]$ (below \eqref{eq:case 1 b}), the limit $\xi\rightarrow\infty$ at fixed $\lambda$ would leave the domain where \eqref{eq:intro H PT} is Hermitian; with $\xi=\lambda\Phi_b$, the asymptotic boundary limit is $\Phi_b\rightarrow\infty$ after $\lambda\rightarrow0$, as in \eqref{eq:ABL}.

\begin{figure}
    \centering
    \includegraphics[width=0.9\linewidth]{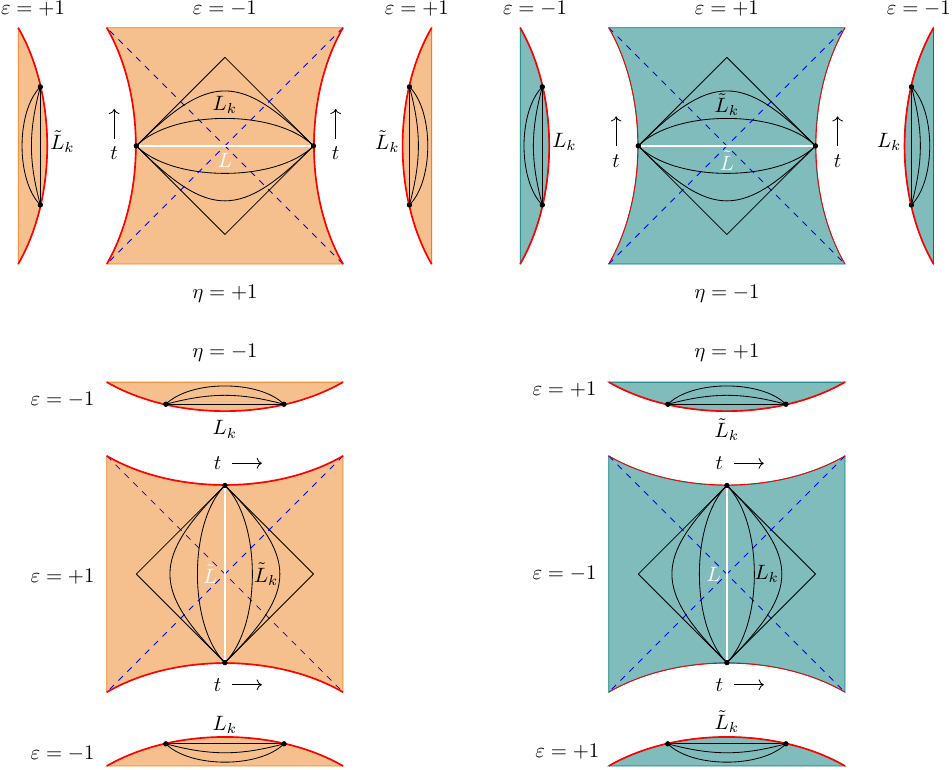}\vspace{0.2cm}
    \caption{CMC slices (black curves) between Dirichlet boundaries (red curves) in an AdS$_2$ black hole (light orange) dS$_2$ space (light blue). $\eta=+1$ denotes the finite regions described by $T^2$ deformations in the boundary dual, and $\eta=-1$ for $T^2+\Lambda_1$. Further, $\varepsilon=-1$ corresponds to regions where $0\leq \Phi\leq \Phi_b$, with $\Phi=\Phi_b$ being the dilaton at the boundary; while $\varepsilon=+1$ corresponds to those where $\Phi\geq \Phi_b$, associated to the different branches in the energy spectrum. $L_k$ denotes the length of CMC slices when $\varepsilon=-1$, $L\equiv L_{k=0}$; and $\tilde{L}_k$ for $\varepsilon=+1$, while $\tilde{L}=\tilde{L}_{k=0}$.}
    \label{fig:CMC-Krylov_AdS_BH}
\end{figure}

At the classical level, the comparison also clarifies the geometric interpretation of the different branches of the deformed energy spectrum of the boundary theory (Sec.~\ref{ssec:bulk interpretation in (dS) JT at finite cutoff}) which connects the perturbative and non-perturbative solutions of the energy spectrum flow equation (see \eqref{eq:flow eq}) to distinct regions of the bulk geometry (illustrated in Fig.~\ref{fig:CMC-Krylov_AdS_BH}). 
In particular, the finite cutoff permits a unified description of different portions of AdS$_2$ black-hole and dS$_2$ geometries. The low-energy regime is associated with the JT/AdS$_2$ description, while the appropriate high-energy regime provides access to the dS$_2$ static-patch description.

Later, to learn about the perturbative and non-perturbative branches in more detail, we studied the evolution of CMC slice lengths, $L_k$ with $k$ the mean curvature, in (A)dS$_2$ spacetimes for the different regions displayed in Fig.~\ref{fig:CMC-Krylov_AdS_BH}. 
A summary of the different bulk lengths evaluated in this work is shown in Tab.~\ref{tab:lengths} (Sec.~\ref{sec:CMC-Krylov complexity}). 

In particular, one finds an Arnowitt-Deser-Misner (ADM) \cite{arnowitt1959dynamical} Hamiltonian for the non-perturbative branches of the form (see \eqref{eq:ADM non-perturbative})
\begin{equation}\label{eq:ADM H non}
\hat{H}_{\rm ADM}=-\hat{P}^2+\frac{\Phi_b^2}{\cos^2(\hat{L}/2)}+\frac{\csc^2\hat{L}}{4}~.
\end{equation}
Due to the negative kinetic term, the canonically quantized theory has energy states that are unbounded from below and does not allow a sensible quantum mechanical interpretation (App.~\ref{sapp:bound states}) in contrast to the perturbative branch investigated in \cite{Griguolo:2026eoi}.

We also arrived at a boundary interpretation of $L_k$ for the perturbative branches, which we identify in terms of a complexity measure in Krylov space, which we denote CMC-Krylov complexity (see \eqref{eq:def CMC-Krylov cmplexity}). 
\begin{eqnarray}\label{eq:intro def CMC-Krylov cmplexity}
    C_{\mfa}(t)\equiv\bra{K_0} \hat{C}_{\mfa}(t)\ket{K_0}~,\quad \hat{C}_{\mfa}(t)\equiv\frac{2}{\lambda\sqrt{1+\mfa}}\text{sinh}^{-1}\qty[\sqrt{1+\mfa}\sinh\frac{\lambda\,\hn(t)}{2}]~,
\end{eqnarray}
where $\mathfrak{r}$ is a parameterization of the complexity measure, whose value is related to the mean curvature $k$ in the bulk ($\mathfrak{r}=k^2$ for the perturbative branches $\varepsilon=-1$ in both AdS$_2$ and dS$_2$, and $\mathfrak{r}=-k^2$ for the non-perturbative ones, $\varepsilon=+1$, formally, through the continuation $L=\rmi\lambda\expval{\hat{n}}$, see Sec.~\ref{ssec:CMC lenghts main}), $\ket{K_0}$ is the seed state in the Lanczos algorithm and $\hn(t)$ is the Krylov complexity operator (reviewed in Sec.~\ref{ssec:rev Krylov}). The complexity measure \eqref{eq:intro def CMC-Krylov cmplexity} is one member in the family of cost functions in \cite{Balasubramanian:2022tpr} used to defined Krylov complexity, and it reduces to Krylov complexity when $\mathfrak{r}=0$. The specific functional expression in \eqref{eq:intro def CMC-Krylov cmplexity} is selected so that when it is evaluated, with $\ket{K_0}$ as the cyclic separating state in the deformed DSSYK \eqref{eq:intro H PT}, \eqref{eq:intro def CMC-Krylov cmplexity} reproduces the length $L_k$ of a spacelike or timelike CMC slice in an AdS$_2$ black hole or dS$_2$ space respectively. This suggests a connection to the complexity=anything (CAny) framework in \cite{Belin:2021bga,Belin:2022xmt}, which we discuss in Sec.~\ref{ssec:outlook}.

We further extend the construction to include ETW branes \cite{Kourkoulou:2017zaj,Gao:2021tzr}, which is a codimension-one hypersurface where spacetime terminates. This allows us to investigate how the finite-cutoff dictionary changes once boundary conditions in the bulk are modified due to the brane, including the full boundary theory deformation, and its effects on the eternal traversable wormhole solutions \cite{Maldacena:2018lmt}. 

On the boundary side, we first construct a chord Hamiltonian
\begin{equation}\label{eq:def Hamiltonian}
\begin{aligned}
\hH_{\rm ETW}&=\frac{-1}{\sqrt{1-q}}\qty(b_{\rm ETW}(\hat{n})\rme^{-\rmi \hat{p}_n}+\rme^{\rmi \hat{p}_n}b_{\rm ETW}(\hat{n})+a_{\rm ETW}(\hat{n}))~,\\
    a_{\rm ETW}(\hat{n})&=\frac{2\mathcal{Y}\,(\overline{\mathcal{Y}}-\xi\sinh (\lambda \hat{n}/2))}{\cosh^2(\lambda \hat{n}/2)}~,\\
    b_{\rm ETW}(\hat{n})&=\sqrt{\qty(1+\frac{\mathcal{Y}^2}{\cosh^2(\lambda \hat{n}/2)})\qty(1+\frac{\overline{\mathcal{Y}}^2-\xi^2}{\cosh^2(\lambda \hat{n}/2)}-2\xi\overline{\mathcal{Y}}\frac{\tanh (\lambda \hat{n}/2)}{\cosh (\lambda \hat{n}/2)})}~,
\end{aligned}
\end{equation}
where $\mathcal{Y}$ and $\overline{\mathcal{Y}}$ are so-called ``q-Askey'' deformation parameters \cite{Aguilar-Gutierrez:2026nmd} in the boundary theory which are associated to the parameters of the ETW brane in the bulk \cite{Blommaert:2025avl}. \eqref{eq:def Hamiltonian} can be further deformed by the improvement term in \eqref{eq:intro PT improved H}, which allows its low-energy description to have a natural choice of boundary conditions and a well-defined physical Hilbert space.

On the gravitational side, we derive the corresponding classical ADM Hamiltonian in sine dilaton gravity at finite cutoff (appearing in \eqref{eq:H SDG ETW}), which has the same structure as \eqref{eq:def Hamiltonian}. Its low-energy limit \eqref{eq:low energy limit0} results in the JT gravity Hamiltonian, which, after the inclusion of the improvement term, is 
\begin{equation}
    \hH_{\scaleto{\rm JT-ETW}{5pt}}=\frac{\hat{P}^2}{4}+U_{\rm m}(\hat{L})~, \quad U_{\rm m}(\hat{L})\equiv\frac{2m_{\rm JT}\Phi_b\sinh \hat{L}+\Phi_b^2-m_{\rm JT}^2}{\cosh^2\hat{L}}-\frac{\text{csch}^2(2\hat{L})}{4}~,
\end{equation}
where $m_{\rm JT}=\mathcal{Y}+\overline{\mathcal{Y}}$ is the tension of the ETW brane in JT gravity, $\hat{P}$ the canonical momentum, and $\hat{L}$ the geodesic length connecting the cutoff Dirichlet boundary to the ETW brane (see Fig.~\ref{fig:ETW_brane_regions} for the different configurations of interest). $U_{\rm m}(L)$ can be seen as a modification of the type II Scarf \cite{scarf1958new} potential energy term, which can be solved using Heun functions \cite{heun1888theorie} (see Sec.~\ref{ssec:scattering ETW}). By quantizing the system, we provide a Hilbert-space description that incorporates either dynamical branes when the ADM energy is positive, or static ones when its negative. In the static case, it acts as an eternal traversable wormhole \cite{Maldacena:2018lmt} in the bulk geometry, similar to the asymptotic boundary limit \cite{Rajgadia:2026ask,Aguilar-Gutierrez:2026ogo}. A summary of the holographic dictionary relating the two descriptions in this work is shown in Tab.~\ref{tab:holographic_dictionary}.
\begin{table}
    \centering
    \begin{tabular}{C{7.5cm}C{7.5cm}}
    \toprule
    \textbf{Bulk (AdS$_2$ black hole)}&\textbf{Boundary (deformed DSSYK)}\\\midrule
    \footnotesize{Bulk cutoff} $\Phi_b$ \eqref{eq:main text L brane}&\footnotesize{$T^2$-deformation parameter} $y=\frac{4\pi G_N}{1-\cos\Phi_b(y)}$ \eqref{eq:holog dict SDG DSSYK}\\[2ex]
    \footnotesize{Bulk temperature} $\Phi_h/(2\pi)$ \vspace{0.2cm}&\footnotesize{Fake temperature} $\sin\theta/(2\pi)$\\[2ex]
    \footnotesize{Proper time} $t_P$ \eqref{eq:tP}&\footnotesize{Time conjugate to the chord Hamiltonian $\hH_{\rm PT}$ \eqref{eq:chord PT H}} $t\sqrt{\xi^2-\sin^2\theta}$ \\[2ex]
{\footnotesize Length variable $L(t)$ with an ETW brane} (\ref{eq:H SDG ETW}) &{\footnotesize Krylov complexity, $\lambda~C_0(t)/2$ (\ref{eq:complexity DSSYK brane})}\\[2ex]
    {\footnotesize Brane tension} $m_{\rm JT}$ \eqref{eq:main text L brane}&{\footnotesize Combination of q-Askey deformation parameters} $\overline{\mathcal{Y}}+\mathcal Y\cos\theta$ \eqref{eq:complexity simple case}\\[3ex]
    {\footnotesize Brane curvature} $K_{\rm AdS}$ \eqref{eq:main text L brane}&{\footnotesize q-Askey deformation parameter} $-\rmi\mathcal{Y}$ \eqref{eq:complexity simple case}\\
    \bottomrule
    \end{tabular}
    \caption{Holographic dictionary of the finite cutoff AdS black hole with an ETW brane and the deformed DSSYK model using the nomenclature of this section (summarized in App.~\ref{app:notation}). $\mathcal{Y}$ and $\overline{\mathcal{Y}}$ are denoted q-Askey deformation parameters \cite{Aguilar-Gutierrez:2026nmd} that are associated to the ETW brane in the bulk and are chosen such that the Hamiltonian \eqref{eq:def Hamiltonian} is always Hermitian, while $K_{\rm AdS}\in\mathbb{C}$ in sine dilaton gravity.}
    \label{tab:holographic_dictionary}
\end{table}

Towards the end, we speculate about the bulk Lyapunov exponent in finite cutoff JT gravity based on a careful analysis of CMC slices in shockwave geometries (App.~\ref{app:shockwaves}).

\paragraph{Outline}
The remainder of the paper is organized as follows. We first review the DSSYK model in Sec.~\ref{sec:review}, its Krylov-space description, and the finite-cutoff gravitational systems relevant for the subsequent analysis. We then construct in Sec.~\ref{sec:deformation} microscopic SYK deformations, derive their double-scaling limits, and discuss the bulk dual of the deformed DSSYK model. In Sec.~\ref{sec:CMC-Krylov complexity} we study the length of CMC slices in the different regions associated to the perturbative and non-perturbative branches (Fig.~\ref{fig:CMC-Krylov_AdS_BH}). We develop its boundary interpretation as CMC-Krylov complexity. We analyze in Sec.~\ref{sec:bulk with matter} finite-cutoff JT gravity and sine dilaton gravity with an ETW brane, and discuss their boundary interpretation. We study the Hilbert space of the low-energy theory including scattering states and the condition to recover bound states. At last, in Sec.~\ref{sec:disc}, we summarize the work, discuss future directions for finite-cutoff holography and complexity. We provide several appendices covering the technical details in the evaluations.

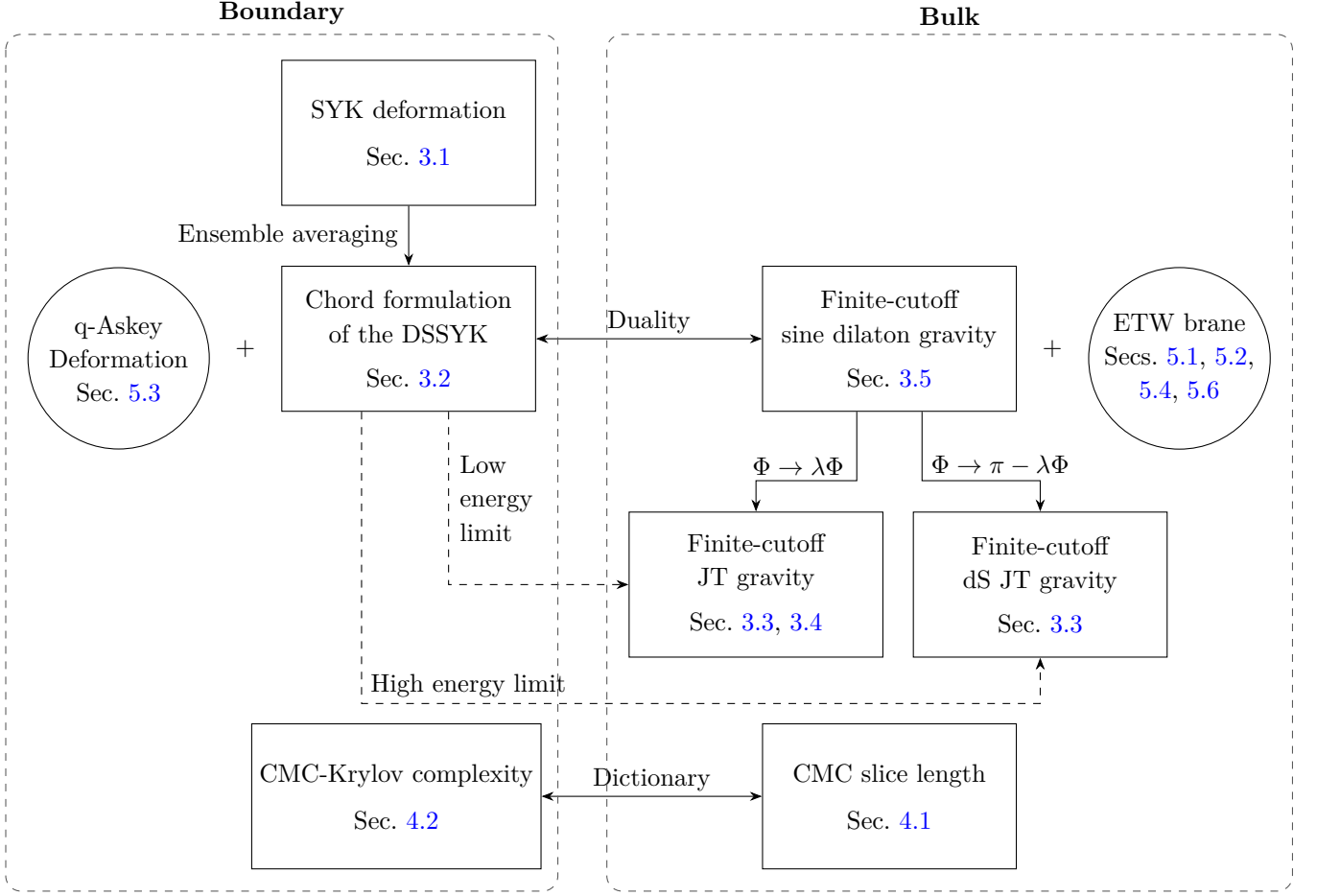
\begin{figure}[t]\centering
\begin{tikzpicture}[
    font=\small,
    >=Stealth,
    box/.style={draw, rectangle, minimum width=3.5cm, minimum height=2.0cm,
                align=center, inner sep=3pt},
    circ/.style={draw, circle, minimum size=2.5cm, align=center, inner sep=1pt},
    lab/.style={font=\small, inner sep=2pt},
    arr/.style={->},
    darr/.style={->, dashed},
    barr/.style={<->}
]

\node[box] (syk)    at (5.27,-1.55)  {SYK deformation\\[1.2ex]Sec.~\ref{ssec:SYK defs}};
\node[box] (chord)  at (5.27,-4.39)  {Chord formulation\\of the DSSYK\\[1.0ex]Sec.~\ref{ssec:chord descripttion 3}};
\node[box] (sine)   at (11.90,-4.39) {Finite-cutoff\\sine dilaton gravity\\[1.0ex]Sec.~\ref{ssec:sine dilaton}};
\node[box] (jt)     at (10.06,-7.78)  {Finite-cutoff\\JT gravity\\[1.0ex]Sec.~\ref{ssec:bulk interpretation in (dS) JT at finite cutoff}, \ref{ssec:improvement JT}};
\node[box] (ds)     at (13.98,-7.78) {Finite-cutoff\\dS JT gravity\\[1.0ex]Sec.~\ref{ssec:bulk interpretation in (dS) JT at finite cutoff}};
\node[box] (cx)     at (5.10,-10.70) {CMC-Krylov complexity\\[1.2ex]Sec.~\ref{ssec:def CMC-K complexity}};
\node[box] (cmc)    at (11.90,-10.70){CMC slice length\\[1.2ex]Sec.~\ref{ssec:CMC lenghts main}};

\node[circ] (qa)  at (1.27,-4.66)  {q-Askey\\Deformation\\Sec.~\ref{ssec:chord interpretation ETW brane}};
\node[circ] (etw) at (15.90,-4.66) {ETW brane\\Secs.~\ref{ssec:ETW branes}, \ref{ssec:SDG ETW},\\
\ref{ssec:1st approach QJT}, \ref{ssec:traversable wormholes}};
\node at ($(qa.east)!0.5!(chord.west)$)  {$+$};
\node at ($(sine.east)!0.5!(etw.west)$) {$+$};

\draw[arr] (syk.south) -- node[left=2pt, lab] {Ensemble averaging} (chord.north);
\draw[barr] (chord.east) -- node[above, lab] {Duality} (sine.west);
\draw[barr] (cx.east) -- node[above, lab] {Dictionary} (cmc.west);

\coordinate (s1) at ($(sine.south)+(-0.45,0)$);
\coordinate (s2) at ($(sine.south)+(0.45,0)$);
\draw[arr] (s1) |- ($(s1)+(0,-0.95)$) -| (jt.north);
\draw[arr] (s2) |- ($(s2)+(0,-0.95)$) -| (ds.north);
\node[lab, anchor=south east] at ($(s1)+(-0.1,-0.95)$) {$\Phi\to\lambda\Phi$};
\node[lab, anchor=south west] at ($(s2)+(0.05,-0.95)$) {$\Phi\to\pi-\lambda\Phi$};

\coordinate (c1) at ($(chord.south)+(0.55,0)$);
\coordinate (c2) at ($(chord.south)+(-0.65,0)$);
\draw[darr] (c1) |- (jt.west);
\draw[darr] (c2) -- (c2|-0,-9.42) -| (ds.south);

\node[lab, anchor=south west] at ($(c2|-0,-9.42)+(0.05,0.02)$) {High energy limit};

% ---- dashed groups: boundary and bulk
\coordinate (bdyNW) at ($(qa.west|-syk.north)+(-0.3,0.35)$);
\coordinate (bdySE) at ($(chord.east|-cx.south)+(0.3,-0.3)$);
\coordinate (blkNW) at ($(jt.west|-syk.north)+(-0.3,0.35)$);
\coordinate (blkSE) at ($(etw.east|-cmc.south)+(0.3,-0.3)$);
\draw[dashed, rounded corners=6pt, gray!70!black] (bdyNW) rectangle (bdySE);
\draw[dashed, rounded corners=6pt, gray!70!black] (blkNW) rectangle (blkSE);
\node[font=\bfseries\small, anchor=south] at ($(bdyNW)!0.5!(bdyNW-|bdySE)$) {Boundary};
\node[font=\bfseries\small, anchor=south] at ($(blkNW)!0.5!(blkNW-|blkSE)$) {Bulk};
\node[lab, anchor=west, align=left] at ($(c1)!0.5!(c1|-jt.west)+(0.08,0)$) {Low\\energy\\limit};
\end{tikzpicture}     \caption{Structure of this work. We highlight the connection between the different topics, ranging from SYK deformations to ETW branes in sine dilaton gravity, and their location in the main text.}\label{fig:structure}
   \end{figure}

We present a summary figure with the structure of the work in Fig.~\ref{fig:structure}.

\section{Review}\label{sec:review}
In this section, we briefly review the DSSYK model (Sec.~\ref{ssec:rev DSSYK}), its Hamiltonian deformations in finite-cutoff holography (Sec.~\ref{ssec:finite cutoff holo}), Krylov complexity (Sec.~\ref{ssec:rev Krylov}), and CAny with CMC slices (Sec.~\ref{ssec:CMCC-Any slices}). Additional review material is referred to in each section.

\subsection{Double-scaled SYK model}\label{ssec:rev DSSYK}
A great introduction on this topic can be consulted in \cite{Berkooz:2024lgq}.
\paragraph{The SYK model} Originally introduced by \cite{Sachdev_1993} and shaped to its more modern form by \cite{kitaevTalks1,kitaevTalks2,kitaevTalks3}, the SYK model is a system with $2N$ Majorana fermions in $(0+1)$-dimensions, and $p$-body all-to-all interactions. The theory is given by
\begin{equation}\label{eq:SYK Hamiltonian}
    \hat{H}_{\rm SYK}=\rmi^{p/2}\sum_{I}\mathcal{J}_{I}\hat{\psi}_{I}~,
\end{equation}
where $I=i_1,\dots,~i_p$ is a collective index, $1\leq i_1<\dots<i_p\leq 2N$; and similarly $\hat{\psi}_I\equiv\hat{\psi}_{i_1}\dots\hat{\psi}_{i_p}$ is a string of Majorana fermions, obeying the Clifford algebra
\begin{equation}\label{eq:Majo fer}
    \qty{\hat{\psi}_{i},~\hat{\psi}_{j}}=2\delta_{ij}\mathbbm{1}~;
\end{equation}
and the coupling constants $\mathcal{J}_I=\mathcal{J}_{i_1\dots i_p}$ follow a Gaussian distribution
\begin{equation}\label{eq:GEA J}
\overline{\mathcal{J}_{I}}=0~,\quad\overline{\mathcal{J}_{I}\mathcal{J}_{J}}=\begin{pmatrix}
        2N\\
        p
    \end{pmatrix}^{-1}\mathcal{J}^2\delta_{IJ}~,
\end{equation}
with $\mathcal{J}\in\mathbb{R}$ being a constant, which we will set $\mathcal{J}=1$, and the expectation values denote averaging over the couplings. The infrared part of the spectrum in this theory has important connections with JT gravity, as emphasized several works, e.g.~\cite{kitaevTalks1,kitaevTalks2,kitaevTalks3,Maldacena:2016hyu}.

Now, we consider the double scaling regime of \eqref{eq:SYK Hamiltonian}, where:
\begin{equation}\label{eq:double scaling}
    N,~ p \rightarrow \infty~,\quad q=\rme^{-\lambda} \equiv \rme^{-\frac{p^2}{N}}~ \text{fixed}~.
\end{equation}
The SYK model becomes analytically solvable even away from the low-energy regime; its ensemble averaged observables can be described with chord diagrams techniques \cite{Berkooz:2018jqr,Berkooz:2018qkz,Cotler:2016fpe,Erdos:2014zgc} which leads to an auxiliary quantum mechanical theory containing thermal information about the physical system.\footnote{Other models in the same universality class as the DSSYK model can be solved using the same type of chord counting techniques \cite{Parisi_1994,Berkooz:2024evs,Berkooz:2024ofm,Almheiri:2024xtw,Gao:2024lem}.} The auxiliary chord Hilbert space without matter insertions is 
\begin{equation}\label{eq:basis}
\mathcal{H}_0={\rm span}\qty{\ket{n}\quad|n\in\mathbb{Z}_{\geq0}}~,
\end{equation}
where $n$ denotes the number of open chords at a given time slice in a chord diagram (see e.g.~\cite{Berkooz:2024lgq} for a review), and from now on, we will denote the empty state ($n=0$) as $\ket{\Omega}$. The corresponding Hamiltonian of the auxiliary quantum system can be expressed in terms of creation and annihilation operators acting on the chord number basis as
\begin{equation}\label{eq:H chord ortho Krylov}
    \hH=-\qty(a+a^\dagger)~,\quad\hat{a}^\dagger\ket{n}=\sqrt{[n+1]_{q}}\ket{n+1}~,\quad \hat{a}\ket{n}=\sqrt{[n]_{q}}\ket{n-1}~,
\end{equation}
with $[n]_q=(1-q^n)/(1-q)$, and they satisfy a chord algebra \cite{Lin:2023trc}, which in the matterless case takes the form
\begin{align}
        \label{eq:chord algebra}
    [\hat{n},~\hat{a}^\dagger]=\hat{a}^\dagger~,\quad [\hat{n},~\hat{a}]=-\hat{a}~,\quad[\hat{a},~\hat{a}^\dagger]_{q}&\equiv \hat{a} \hat{a}^\dagger - q \hat{a}^\dagger \hat{a} =1~,
    \end{align}
    where $[x,\,y]_q\equiv xy-qyx$, and we included the number operator acting on the $\ket{n}$ basis as $\hat{n}\ket{n}=n\ket{n}$. This can be used to find a momentum operator $\hat{p}_n$ conjugate to the chord number \begin{align}\label{eq:p n def}
    \hat{a}^\dagger=\sqrt{[\hat{n}]_q}\rme^{-\rmi\hat{p}_n}~,\quad \hat{a}=\rme^{\rmi\hat{p}_n}\sqrt{[\hat{n}]_q}~,
\end{align}
so that $\hat{p}_n$ acts as a canonical conjugate of $\hat{n}$ by the chord algebra \eqref{eq:chord algebra}, i.e. $[\hat{n},\rme^{\mp\rmi\hat{p}_n}]=\pm\rme^{\mp\rmi\hat{p}_n}$. In this way, one finds that in the limit \eqref{eq:double scaling} $\overline{\Tr[\hH_{\rm SYK}^m]}/\Tr[\mathbbm{1}]=\bra{\Omega}\hH^m\ket{\Omega}$, where the trace and the expectation value \eqref{eq:GEA J} on the left-hand side are defined in the original SYK system.

\paragraph{Kourkoulou-Maldacena states}
The Kourkoulou-Maldacena (KM) \cite{Kourkoulou:2017zaj} states are eigenstates of a spin operator
\begin{equation}\label{eq:spin op}
\qty(\hat{S}_k-s_k)\ket{\rm KM}=0~,\quad \hat{S}_k\equiv \rmi\hat{\psi}_{2k-1}\hat{\psi}_{2k}~.
\end{equation}
We define creation and annihilation operators in the spin basis
\begin{equation}
\hat{c}_j=\frac{1}{2}\qty(\hat{\psi}_{2j-1}+\rmi\hat{\psi}_{2j})~,\quad \hat{c}_j{}^\dagger=\frac{1}{2}\qty(\hat{\psi}_{2j-1}-\rmi\hat{\psi}_{2j})~,
\end{equation}
which obey the commutation relations $[\hat{S}_j,~\hat{c}_i^\dagger]=2\delta_{ij}\hat{c}_j{}^\dagger$, $[\hat{S}_j,~\hat{c}_i]=-2\delta_{ij}\hat{c}_j$, and allow us to express a product of creation or annihilation operators of the form
\begin{equation}\label{eq:def M Mdagger}
\hat{M}\equiv\rmi^{w/2}\sum_{I''}R_{i_1,\dots,i_{w}}\hat{c}_{i_1}\dots \hat{c}_{i_{w}}~,\quad\hat{M}^\dagger\equiv(-\rmi)^{w/2}\sum_{I''}R_{i_1,\dots,i_{w}}^*\hat{c}_{i_w}{}^\dagger\dots \hat{c}_{i_{1}}{}^\dagger~,
\end{equation}
where $I''=\qty{i_1,\dots,i_{w}}$ with $1\leq i_1<\cdots< i_{w}\leq N$.
The random couplings $R_{i_1,\dots,i_{w}}$ and $R_{i_1,\dots,i_{w}}^*$ can be chosen to be Gaussian-distributed
\begin{equation}\label{eq:gaussian dist R}
\overline{R_{i_1,\dots,i_{w}}}=\overline{R_{i_1,\dots,i_{w}}^*}=0~,\quad \overline{R_{i_1,\dots,i_{w}}R_{i_1,\dots,i_{w}}^*}=R^2\begin{pmatrix}
N\\
w
\end{pmatrix}^{-1}
\end{equation}
where $R\in\mathbb{R}$ is a constant, which we will take $R=1$.
The double-scaling limit of KM states and the $\hat{M}$ and $\hat{M}^\dagger$ operators above has been recently studied by Rajgadia and Xu (RX) \cite{Rajgadia:2026ask}, who find a relationship to the chord Hilbert space. The relevant relationship between microscopic and chord operators can be summarized by
\begin{equation}\label{eq:dictionary SYK chord KM}
\begin{aligned}
    &\lim_{\rm DS}\overline{\bra{\omega}P(\hH_{\rm SYK},\hH_{\rm RX})\ket{\omega}}=\bra{\Omega}P(\hH,q^{\Delta_w\hat{n}})\ket{\Omega}~,
\end{aligned}
\end{equation}
where the bar indicates ensemble averaging, $\ket{\omega}$ is a KM state such that $\hat{S}_i\ket{\omega}=\pm\ket{\omega}$ (i.e.~all the spins are either ``up'' or ``down''), $\Delta_w\equiv w/p$, and $P$ again represents a general polynomial function, and we introduced an interaction term
\begin{equation}
    \hH_{\rm RX}\equiv\qty{\hat{M},\hat{M}^\dagger}~,\label{seq:H int}
\end{equation}
where $\qty{,~}$ denotes an anticommutator.

\paragraph{Two-copy SYK}Consider a two-copy SYK system with total Hilbert space
\begin{align}\label{eq:SYK D}
    \mH_{\rm SYK^2}\equiv&\mH_{\rm SYK}\otimes\mH_{\rm SYK}~,\\
    \mathcal{H}_{\rm SYK}={\rm span}\qty{\ket{E_n}}&~,\quad \bra{E_n}\ket{E_m}=\delta_{nm}~,
\end{align}
where we denote its eigenstates of the SYK Hamiltonian as $\ket{E_n}$:
\begin{equation}\label{eq:SYK H}
    \hH_{\rm SYK}\ket{E_n}=E_n\ket{E_n}~.
\end{equation}
In particular, we will consider the infinite-temperature TFD state to evaluate expectation values, with appropriate normalization
\begin{equation}\label{eq:inf temp TFD}
    \ket{\rm TFD_\infty}\equiv\frac{1}{\sqrt{\dim\mH_{\rm SYK}}}\sum_n\ket{E_n}\otimes\ket{E_n}~.
\end{equation}
The pair of SYK models can be coupled in different ways to describe rich holographic systems. A coupling of interest was proposed by Maldacena and Qi (MQ) \cite{Maldacena:2018lmt} and it can be expressed after a rescaling as
\begin{equation}\label{eq:operator size}
\hat{s}\equiv\frac{1}{p}\sum_{j=1}^N\qty(1+\rmi ~\psi^{(L)}_{j}\psi^{(R)}_{j})~,   
\end{equation}
with $\qty{\psi_i^{(L)},\psi_i^{(R)}}=0$. As explained in \cite{Lin:2022rbf}, $\hat{s}$ corresponds to the total chord number $\hat{n}$ in the double-scaling limit. It is also convenient to define a generalization of $\hat{s}$
\begin{equation}\label{eq:H MQ}
    \hH_{\rm MQ}\equiv\begin{pmatrix}
        N\\s
    \end{pmatrix}^{-1}\rmi^{s}\sum_{i_1\dots i_{s}}\prod_{j=1}^s\psi^{(L)}_{i_j}\psi^{(R)}_{i_j} ~,
\end{equation}
which leads to \eqref{eq:operator size} when $s=1$. The double-scaling limit \eqref{eq:double scaling} of the above SYK operators in the doubled formalism has been studied in \cite{Lin:2022rbf}, and they can be summarized by the expression,\footnote{The relationship \eqref{eq:dictionary SYK and chord TFD} can be extended with matter operators \cite{Lin:2022rbf}, which will not be necessary for the present discussion.}
\begin{eqnarray}\label{eq:dictionary SYK and chord TFD}
    \lim_{\rm DS}\overline{\bra{\rm TFD_\infty}P\qty(\hH_{\rm SYK}^{(L/R)},~\hH_{\rm MQ})\ket{\rm TFD_\infty}}=\bra{\Omega}P\qty(\hH,q^{\Delta_s\hat{n}})\ket{\Omega}~.
\end{eqnarray}
where $\Delta_s\equiv s/p$; $\hH_{\rm SYK}^{(L)}\equiv\hH_{\rm SYK}\otimes\mathbb{1}$ and $\hH_{\rm SYK}^{(R)}\equiv\mathbb{1}\otimes\hH_{\rm SYK}$, and $P$ indicates a general polynomial function.

In the following section, we apply the KM and TFD formulations above to describe new microscopic Hamiltonian-deformations of the DSSYK model based on those reviewed below for the auxiliary quantum mechanical system.

\subsection{Finite-cutoff dilaton gravity and \texorpdfstring{$T^2(+\Lambda_1)$}{}-deformations}\label{ssec:finite cutoff holo}
We briefly review T$^2$ and T$^2+\Lambda_1$ deformations in the DSSYK model after ensemble averaging. In general, both types of deformations are described by a flow in the energy spectrum $E_y$ of the type \cite{Aguilar-Gutierrez:2026ogo}
\begin{equation}\label{eq:flow eq}
    \partial_{y} E_y=\frac{(E_y)^2+(\eta-1)/y^2}{2(1-y E_y)}~,
\end{equation}
where $y$ corresponds to a deformation associated with moving a finite radial cutoff with Dirichlet boundary conditions in a dilaton gravity theory, and $\eta=\pm1$ is another deformation parameter associated with a modification in the asymptotic boundary conditions in the bulk.

The $\eta=+1$ case describes a lower-dimensional analog of $\TT$-deformations in 2D CFTs, which we refer to as $T^2$ deformations from now on. For general $y$, this results in the following chord Hamiltonian:
\begin{equation}\label{eq:E(theta) eta+1}
  \eta=+1:~~  \hH_y=\frac{1}{y}\qty(1+\varepsilon\sqrt{1-2y\hH})~,
\end{equation}
here $\varepsilon=\pm1$ represent different solutions to the flow equation \eqref{eq:flow eq}, where $\varepsilon=-1$, denoted perturbative branch, is connected to $E_{y=0}=E$, while $\varepsilon=+1$ is not, and it will be referred to as a non-perturbative solution. 
Meanwhile, $\eta=-1$ in \eqref{eq:flow eq} corresponds to a lower dimensional version of $\TT+\Lambda_2$ deformations \cite{Gorbenko:2018oov,Coleman:2020jte,Silverstein:2022dfj,Batra:2024kjl,AliAhmad:2025kki}, which we denote $T^2+\Lambda_1$ deformations from now on, and the Hamiltonian is given by
\begin{equation}\label{eq:E(theta) eta-1}
    \eta=-1:~~\hH_y=\frac{1}{y}\qty(1+\varepsilon\sqrt{2y \hH-1})~.
\end{equation}
where $\varepsilon=\pm1$ and we assume initial conditions connecting the $T^2$ deformation to the $T^2+\Lambda_1$ when the square-root term in \eqref{eq:E(theta) eta+1} vanishes, corresponding to $y=y_0\equiv1/(2E)$. Due to this gluing, the flow parameter is state-dependent. Moreover, the corresponding energies \eqref{eq:E(theta) eta-1} would be imaginary for $E(\theta)<1/(2y)$, which is in principle allowed (see \eqref{eq:E theta seed}). We have this discussed this case in \cite{Aguilar-Gutierrez:2024oea,Aguilar-Gutierrez:2026ogo}. In the rest of the work we focus on the regime where $\hH_y$ is Hermitian. At last, the branch associated with the ADM energy in the bulk still corresponds to $\varepsilon=-1$ \cite{AliAhmad:2025kki}; we will consider both of them in this work.

\paragraph{Quantum JT gravity at finite cutoff}
The Hilbert space of JT without matter at finite cutoff has been recently discussed by \cite{Griguolo:2026eoi}. Let $\hH_{\rm JT-PT}$ denote the ADM Hamiltonian, $\Phi_b$ the boundary value of the dilaton, and $\ket{L}$ the geodesic length basis, which we previously referred to as wormhole state, in the corresponding Hilbert space. In the zero-matter case, the Hamiltonian is the only operator available generating its algebra of observables, and it is diagonalized by states $\ket{s}$ such that $\hH_{\rm JT-PT}\ket{s}=s^2\ket{s}$. By imposing a specific operator ordering in the canonical quantization of the system motivated by $\mathfrak{sl}(2,\mathbb{R})$ representation theory and that the Schrödinger equation only admits scattering states, one finds that the wavefunctions solutions take the form \cite{Griguolo:2026eoi}
\begin{eqnarray}\label{eq:psi s fried}
    \psi_s(L)
    =\bra{s}\ket{L}=
    \frac{\sqrt{\tanh(L/2)}}{\cosh(L/2)^{2is}} \, _2F_1\!\left( \frac{1}{2} + is+ i\Phi_b, \frac{1}{2} + is - i\Phi_b, 1 ; \tanh^2(L/2)  \right).
\end{eqnarray}
where $_2F_1$ is the hypergeometric function, and \eqref{eq:psi s fried} is real-valued. The corresponding renormalized partition function in this theory takes the form \cite{Griguolo:2026eoi},
\begin{align}\label{eq:orthogonality}
    Z(\beta)&=(\lim_{L\rightarrow0}\psi_s(L)^2)\int_0^\infty\rmd s \,\rho_0(s)\,\rme^{-\beta s^2}~,\quad \rho_0(s) \equiv 
    \frac{2 s\sinh(2\pi s)}
    {\cosh(2\pi s)+\cosh(2\pi\Phi_b)}~,
\end{align}
The two-point function of a bulk free scalar field with mass $m^2\equiv\Delta(\Delta-1)$ evaluated between the finite boundaries of the AdS$_2$ black hole has been conjectured in \cite{Griguolo:2026eoi} to be given by
\begin{align}
    \nonumber\tilde{G}_2(\tau_1,\tau_2)&=\includegraphics[width=0.18\linewidth, valign=c]{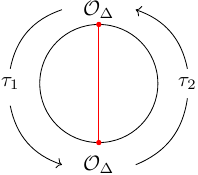}\equiv \lim_{L\rightarrow0}\bra{L}\rme^{-(\beta-\tau_1-\tau_2)\hH_{\rm JT-PT}}\frac{\Phi_b^{2\Delta}}{\cosh^{2\Delta}\frac{\hat{L}}{2}}\rme^{-\tau_1\hH_{\rm JT-PT}}\ket{L}\\
    &=\frac{\Phi_b^{2\Delta}}{Z(\beta)} \int_0^\infty \prod_{i=1}^2\rmd s_i\rho_0(s_i)\, e^{-(\beta-\tau_2+\tau_1) E(s_2)} \, e^{-(\tau_2-\tau_1) E(s_1)} G_2(s_1,s_2)~,\label{eq:2pt}\\
     G_2(s_1,s_2)&\equiv \Gamma(\Delta \pm i s_1 \pm i s_2) \, \mathbb{W}(s_1, \Phi_b; \textstyle{\frac{1}{2}} + i s_2, \textstyle{\frac{1}{2}} - i s_2, \Delta - i \Phi_b, \Delta + i \Phi_b )~,
\end{align}
where $\mathbb{W}(s_1, \Phi_b; \textstyle{\frac{1}{2}} \pm i s_2, \Delta \mp i \Phi_b)\equiv \mathbb{W}(s_1, \Phi_b; \textstyle{\frac{1}{2}} + i s_2, \textstyle{\frac{1}{2}} - i s_2, \Delta - i \Phi_b, \Delta + i \Phi_b )$ is the Wilson function \cite{koekoek2010hypergeometric}. Note that albeit $\lim_{L\rightarrow0}\ket{L}\rightarrow0$, the partition function in the denominator of \eqref{eq:2pt} allows a cancellation between the vanishing norms involved in this expression.

\subsection{Krylov complexity}\label{ssec:rev Krylov}
Modern reviews in this subject can be found in \cite{Nandy:2024evd,Baiguera:2025dkc,Rabinovici:2025otw,Jeong:2026gdc}.
To define Krylov complexity \cite{Balasubramanian:2022tpr}, we consider an ordered basis:
\begin{equation}
    \qty{\ket{B_0},\ket{B_1},\dots,\ket{B_n},\dots}~,
\end{equation}
and cost function with respect to a reference state $\ket{\psi(\tau)}$:
\begin{equation}\label{eq:cost}
    C(t)\equiv\sum_{n=0}^\infty c_n\frac{\abs{\bra{B_n}\ket{\psi(\tau)}}^2}{\bra{\psi(\tau)}\ket{\psi(\tau)}}~,
\end{equation}
where $\qty{c_n\in\mathbb{R}}$ is a monotonically increasing sequence, and we allow for the reference state to evolve in complex-valued time $\tau\equiv\frac{\beta}{2}+\rmi t$ with $\beta$ and $t\in \mathbb{R}$ as \cite{Erdmenger:2023wjg}:
\begin{equation}\label{eq:liouv}
    \ket{\psi(\tau)}\equiv\rme^{-\tau\hat{\mathcal{L}}}\ket{\psi(0)}~,
\end{equation}
with $\hat{\mathcal{L}}$ being the generator of evolution parametrized by $t$ of the system. In particular, when $\hat{\mathcal{L}}$ is the Hamiltonian of the system, one describes standard Schrödinger evolution when $\beta=0$, while $\hat{\mathcal{L}}$ can also be the super-Liouvillian operator describing Heisenberg evolution of operators. In the latter case, $\qty{\ket{B_n}}$ correspond to an operator basis mapped to states through a Choi–Jamiołkowski isomorphism \cite{Choi:1975nug,Jamiolkowski:1972pzh}. Krylov complexity is then defined by minimizing the cost function \eqref{eq:cost} over all possible basis sets, which occurs when $\qty{\ket{B_n}}$ is the Krylov basis $\qty{\ket{K_n}}$ where $\ket{K_0}=\ket{\psi(\tau=0)}$ is the initial state in the basis, which we may choose as the zero-chord state $\ket{\Omega}$. The other elements in the Krylov basis are obtained recursively through the Lanczos algorithm in the form:
\begin{align}
    \ket{K_{n+1}}&=\frac{1}{b_{n+1}}\qty(\hat{\mathcal{L}}\ket{K_n}-b_n\ket{K_{n-1}}-a_n\ket{K_n})~,\label{eq:lanczos new}
\end{align}
where the algorithm is initialised by the condition $b_0=0$, $\ket{K_{-1}}\equiv0$, and $a_{n}=\bra{K_n}\hat{\mathcal{L}}\ket{K_n}$; with $b_n$ and $a_n$ the Lanczos coefficients, which are determined from \eqref{eq:lanczos new} from the condition $\bra{K_n}\ket{K_m}=\delta_{nm}$. Then, Krylov complexity is defined by \eqref{eq:cost} with $\qty{B_n}=\qty{\ket{K_n}}$ and $c_n=n$ as:
\begin{equation}\label{eq:n hat def intro}
    C_0(t)\equiv\frac{\bra{K_0}\hat{n}(t)\ket{K_0}}{\bra{\psi(t)}\ket{\psi(t)}}~,\quad \hat{n}(t)\equiv \rme^{-\tau^*\hat{\mathcal{L}}}\sum_{n=0}^\infty n\ket{K_n}\bra{K_n}\rme^{-\tau\hat{\mathcal{L}}}~,
\end{equation}
where $\hat{\mathcal{L}}$ is the generator of the evolution, and $\tau\equiv\beta/2+\rmi t$ is a complexified time parameter. In Sec.~\ref{ssec:def CMC-K complexity} we discuss another specific choice of cost function \eqref{eq:cost}.

\subsection{Complexity=anything and CMC slices}\label{ssec:CMCC-Any slices}
The CAny framework proposes a large number of bulk observables \cite{Belin:2021bga,Belin:2022xmt} conjectured to measure complexity on the boundary side. Consider the codimension-one  gravitational observables in CAny, which are generally of the form \cite{Belin:2021bga,Belin:2022xmt}
\begin{eqnarray}\label{eq:OF1}
    \mathcal{C}^\epsilon\equiv\int_{\Sigma_{\epsilon}}\rmd^{d}x\sqrt{-h}~F[g_{\mu\nu},X^\mu]~,
\end{eqnarray}
where $F$ is a scalar functional of the bulk metric and embedding functions $X^\mu(\sigma^a)$ of the hypersurfaces $\Sigma_{\epsilon}$ which live at the boundaries of a spacetime region $\mathcal{M}$ determined by the extremization of a functional evaluated at the spacetime region $\mathcal{M}$. 
A convenient choice used in \cite{Belin:2022xmt,Jorstad:2023kmq,Aguilar-Gutierrez:2023pnn,Aguilar-Gutierrez:2023zqm} is a combination of codimension-one and codimension-zero volumes with different weights
\begin{equation}\label{eq:regions CMC}
\begin{aligned}
    \mathcal{C}_{\rm CMC}&=\frac{1}{G_N}\biggl[\alpha_+\int_{\Sigma_+}\rmd^d x\,\sqrt{h}+\alpha_-\int_{\Sigma_-}\rmd^d x\,\sqrt{h}+\alpha_B\int_{\mathcal{M}}\rmd^{d+1}x\sqrt{-g}\biggr] \, ,
\end{aligned}
\end{equation}
where $\mathcal{M}$ is a $(d+1)$--dimensional bulk region with future (past) boundaries $\Sigma_{\pm}$ anchored at the Dirichlet boundaries, such that $\partial\mathcal{M}=\Sigma_+\cup\Sigma_-$, and we take the relevant (A)dS length scale to be order $1$,\footnote{We will label with $\epsilon=\lbrace +,- \rbrace$ the quantities defined on the codimension-one surfaces $\Sigma_{\pm}$.}
and we denote with $h$ the determinant of the induced metric on $\Sigma_{\pm}$. 
The coefficients $\alpha_{\pm}$ and $\alpha_B$ are dimensionless constants. The extremization of the functional \eqref{eq:regions CMC} defines CMC slices, \ie the extrinsic curvature on these surfaces is given by \cite{MARSDEN1980109,Belin:2022xmt}
\begin{equation}
    K_\epsilon\equiv\eval{K}_{\Sigma_\epsilon} = -\epsilon \, \frac{\alpha_B}{\alpha_\epsilon}~,
    \label{eq:K_CMC}
\end{equation}
where $K^{\mu\nu}=h^{\mu\alpha}\nabla_\alpha n^\nu$ is the extrinsic curvature, and we consider the convention where $n^\mu$ is a future-pointing normal vector for either $\Sigma_\epsilon$. 
Then, $\mathcal{C}^{\epsilon}$ is defined on the codimension-one CMC slice $\Sigma_{\varepsilon}$ determined by the extremization of the functional \eqref{eq:regions CMC} with fixed Dirichlet boundary conditions. Different choices for the $F$ function correspond to different observables. 

\section{Microscopic origin of the deformed DSSYK}\label{sec:deformation}
Our work in this section is to elucidate the microscopic interpretation of the deformed DSSYK model in Sec.~\ref{ssec:finite cutoff holo}. In previous work \cite{Aguilar-Gutierrez:2026ogo} (App E), we proposed chord Hamiltonians representing a deformation with respect to those in JT gravity \cite{Griguolo:2025kpi}, namely
\begin{equation}\label{eq:chord PT H}
\begin{aligned}
    \hH_{\rm PT}&=-\frac{1}{\sqrt{1-q}}\qty(\rme^{\rmi \hat{p}_n}\,b_{\rm PT}(\hat{n})+b_{\rm PT}(\hat{n})\rme^{-\rmi\hat{p}_n})~,
\end{aligned}
\end{equation}
where $\hat{p}_n$ is defined in \eqref{eq:p n def}, and the coefficient $b_{\rm PT}(\hat{n})$ depends on the chord number basis representation; e.g.~using the Krylov orthogonal basis of the Hamiltonian with undeformed energy spectrum $E_0(\theta)=-2\cos\theta/\sqrt{1-q}$, we can represent
\begin{equation}\label{eq:case 1 b}
b_{\rm PT}(\hat{n})\equiv\sqrt{1-\xi^2\sech^2\frac{\lambda\hat{n}}{2}}~,
\end{equation}
which is real only when $\xi<\cosh(\lambda n/2)$, with $n$ an eigenvalue of $\hat{n}$, and we will set the domain of the constant parameter to be $\xi\in[0,1]$. The spectrum of \eqref{eq:chord PT H} is that of the seed theory, as we will find in Sec.~\ref{ssec:SYK defs} (see \eqref{eq:E theta seed}), while the $T^2$
deformation then enters separately in accordance to \eqref{eq:E(theta) eta+1} and \eqref{eq:E(theta) eta-1},
\begin{eqnarray}\label{eq:PT y}
    \hH^{(\eta)}_{{\rm PT},y}&\equiv \frac{1}{y}\qty(1+\varepsilon\sqrt{\eta\qty(1-2y\hH_{\rm PT})})~.
\end{eqnarray}
This relationship, in the JT gravity context, has been recently emphasized by \cite{Griguolo:2026eoi}. 
As we will show below (Sec.~\ref{ssec:sine dilaton}), the parameter $\xi$ in \eqref{eq:case 1 b} can be identified with the boundary value of the dilaton $\Phi=\Phi_b$ in finite cutoff sine dilaton gravity; which implies from the holographic dictionary relating the deformation parameter with the radial cutoff in general dilaton gravity theories that \cite{Aguilar-Gutierrez:2024oea,Aguilar-Gutierrez:2026ogo},\footnote{See (2.18) in \cite{Aguilar-Gutierrez:2026ogo}, where $G(\Phi)=-2\cos\Phi$ in sine dilaton gravity, applied to $G(\Phi_b)-G(0)=2(1-\cos\Phi_b)$, i.e.~to the energy measured from the ground state, as explained below.}
\begin{eqnarray}\label{eq:holog dict SDG DSSYK}
    y=\frac{4\pi G_N}{1-\cos\Phi_b(y)}~,
\end{eqnarray}
where $G_N$ is Newton's constant in the sine dilaton gravity action (Sec.~\ref{ssec:sine dilaton}), and the flow \eqref{eq:flow eq} is understood to act on the energy measured from the ground state of the seed theory, $E\rightarrow E+2/\sqrt{1-q}=2(1-\cos\theta)/\sqrt{1-q}$ (the same shift as in \eqref{eq:corrected H JT}). With \eqref{eq:dict 1} this gives $2yE=(1-\cos\theta)/(1-\cos\Phi_b)$, so that the perturbative $T^2$ branch is real for $\theta\leq\Phi_b$, i.e.~for a cutoff outside the horizon, consistently with the assignment of regions in Sec.~\ref{ssec:bulk interpretation in (dS) JT at finite cutoff}; in the limit \eqref{eq:low energy limit}, $y\rightarrow8\pi G_N/(\lambda^2\Phi_b^2)$, $2yE\rightarrow s^2/\Phi_b^2$ with $\theta=\lambda s$, and $E_y\rightarrow2\lambda^{3/2}\Phi_b^2(1-\sqrt{1-s^2/\Phi_b^2})$, the finite-cutoff JT spectrum \cite{Griguolo:2026eoi}. \eqref{eq:holog dict SDG DSSYK} is to be understood as an equation for the flow parameter $y$ of \eqref{eq:flow eq} at a given $\Phi_b$, which we write as $\Phi_b(y)$ when we need its inverse.

In the following, we will be interesting in the limit
\begin{equation}\label{eq:low energy limit}
    \xi\rightarrow\lambda\Phi_b~,\quad \hat{p}_n\rightarrow\lambda\hat{P}~,\quad \lambda\hat{n}\rightarrow\hat{L}~:~~{\rm fixed}~,
\end{equation}
where $\Phi_b\sim\mathcal{O}(\lambda^0)$ is a fixed parameter, and $\hat{L}$ and $\hat{P}$ are canonical variables whose spectrum in the low-energy limit is $\mathcal{O}(\lambda^0)$ as $\lambda\rightarrow0$, and satisfying $[\hat{L},\,\hat{P}]=\rmi$ due to the relation $[\hat{n},\rme^{\mp\rmi \hat{p}_n}]=\pm\rme^{\mp\rmi\hat{p}_n}$ \eqref{eq:chord algebra}.

Under the limit \eqref{eq:low energy limit}, one finds that \eqref{eq:chord PT H} reproduces the ADM Hamiltonian in JT gravity \cite{Griguolo:2025kpi} with a PT potential,
\begin{eqnarray}\label{eq:JT PT}
    \hH_{\rm JT-PT}=\hat{P}^2+\Phi_b^2\sech^2\frac{\hat{L}}{2}~.
\end{eqnarray}
One should note that the finite-cutoff JT Hamiltonian \eqref{eq:JT PT} can receive quantum corrections by performing a different operator ordering discussed by \cite{Griguolo:2026eoi}, which is motivated by $\mathfrak{sl}(2,\mathbb{R})$ representation theory. It plays an important role in selecting a natural choice of self-adjoint boundary conditions which give a well-defined Hilbert space in finite-cutoff JT gravity. We propose how to incorporate it in the chord description in Sec.~\ref{ssec:improvement JT}.\footnote{Additional discussion appears in Sec.~\ref{ssec:1st approach QJT} and \ref{ssec:scattering ETW}.}

The main aim of this section is to seek a microscopic model whose double-scaling limit reproduces the chord Hamiltonian \eqref{eq:chord PT H}, \eqref{eq:case 1 b}, as well as to interpret it in the bulk, seen as finite-cutoff JT and dS JT gravity \cite{Svesko:2022txo,Alonso-Monsalve:2024oii,Dey:2025osp} in the low and high energy limits, and finite-cutoff sine dilaton gravity for the full energy spectrum of the deformed DSSYK. We also stress that the auxiliary system \eqref{eq:chord PT H} is defined only from the moment matching with the microscopic theory (Sec.~\ref{ssec:SYK defs}); we are identifying a dictionary between observables of the physical SYK system in the double-scaling limit, to those in the auxiliary quantum mechanical system, there is no equivalence between the two theories; it is only a matching between observables, in the same way as in the original work in the undeformed DSSYK model by \cite{Berkooz:2018jqr}.

\paragraph{Outline}In Sec.~\ref{ssec:SYK defs} we construct the explicit deformations based on KM and TFD states with MQ interaction terms. We analyze their effective description in the double-scaling limit in Sec.~\ref{ssec:chord descripttion 3}. Supporting details about the double-scaling limit of the microscopic deformations are found in App.~\ref{app:Krylov micro H}. In Sec.~\ref{ssec:bulk interpretation in (dS) JT at finite cutoff} we show that finite-cutoff (dS) JT gravity can be recovered from the proposed boundary theory. We also discuss the semiclassical geometric interpretation for the different branches of the $T^2$ and $T^2+\Lambda_1$ flows described in Sec.~\ref{ssec:finite cutoff holo}. Sec.~\ref{ssec:improvement JT} we incorporate an improvement term in the chord Hamiltonian whose low-energy limit can be understood in terms of the operator ordering choice proposed by \cite{Griguolo:2026eoi}.
In Sec.~\ref{ssec:sine dilaton} we verify that the ADM Hamiltonian of sine dilaton gravity at finite cutoff has the same form as that of the deformed chord Hamiltonian resulting from the microscopic model in Sec.~\ref{ssec:SYK defs}.

\subsection{Explicit SYK deformations}\label{ssec:SYK defs}
\paragraph{KM formulation}
For the KM formulation, we define the following deformation
    \begin{align}\label{eq:def H SYK PT KM}
    \hH_{0}=&\frac{1}{2}\qty(\text{b}_{\rm PT}(\hH_{\rm RX})~{\hH_{\rm def}^\dagger}+{\hH_{\rm def}}~\text{b}_{\rm PT}(\hH_{\rm RX}))~,
\end{align}
where $\hH_{\rm RX}$ is given in \eqref{seq:H int} while
\begin{subequations}\label{eq:part_2}
    \begin{align}\label{eq:H def KM}
    \hH_{\rm def}&\equiv\hH_{\rm SYK}+\qty[\hH_{\rm SYK},~\frac{\rmi}{4p}\sum_{j=1}^N\qty[\hat{\psi}_{2j-1},~\hat{\psi}_{2j}]]~,\\
        \text{b}_{\rm PT}(\HRX)&=\qty(1-\HRX)^{-1/2}\sqrt{1-4\xi^2\HRX(1+\HRX)^{-2}}~,\label{eq:part_2b}
    \end{align}
\end{subequations}
where $\text{b}_{\rm PT}(\HRX)$ is defined by functional calculus (the spectral theorem for the Hermitian operator $\HRX$) on the spectral subspace where $\HRX<1$ and the radicand is non-negative, where it is Hermitian. As a function, $\text{b}_{\rm PT}(x)$ has a simple pole at $x=-1$ and branch points at $x=1$ and at $x=2\xi^2-1\pm2\xi\sqrt{\xi^2-1}$; the branch point at $x=1$ is the image of $[\hat{n}]_q^{-1/2}$ in the double-scaling limit and is harmless on $\ket{\Omega}$ (see the footnote after \eqref{eq:P PT series}).

Here and below we assume that the dictionaries \eqref{eq:dictionary SYK chord KM} and \eqref{eq:dictionary SYK and chord TFD}, derived for polynomials, extend to continuous functions of $\HRX$ (respectively $\hH_{\rm MQ}$) such as $\text{b}_{\rm PT}$; this follows from the convergence of the joint moments of $(\hH_{\rm SYK},\HRX)$ together with the fact that the spectrum of $\HRX$ is bounded uniformly at the level of the double-scaled moments, i.e.~with $\HRX\rightarrow q^{\hat{n}}$ \eqref{eq:dictionary SYK chord KM} applied before evaluating the function $\text{b}_{\rm PT}$, (see footnote \ref{fnt:one can }), by the same argument as for $P^{(\eta)}_{\rm PT}$ below \eqref{eq:P PT series}.
As detailed in App.~\ref{app:detalssww}, we find that in the double-scaling limit, the following holds
\begin{eqnarray}\label{eq:DS H PT}
    \lim_{\rm DS}\overline{\bra{\omega}\hH_{0}^m\ket{\omega}}=\bra{\Omega}\hH_{\rm PT}^m\ket{\Omega}~,
\end{eqnarray}
where $m\in 2\mathbb{Z}_{\geq0}$ because odd moments
vanish on both sides, and $\hH_{\rm PT}$ is the deformed chord Hamiltonian in \eqref{eq:chord PT H}. In deriving \eqref{eq:DS H PT} we applied the dictionary \eqref{eq:dictionary SYK chord KM} with $\Delta_w=1$.

In particular, as seen below, the low-energy limit of the chord Hamiltonian \eqref{eq:chord PT H} reproduces the ADM Hamiltonian of canonically quantized JT gravity at finite cutoff \eqref{eq:JT PT}.

Moreover, following \cite{Lin:2022rbf}, the continuous spectrum of the undeformed chord Hamiltonian at this point can be derived when $\lambda n\rightarrow\infty$ and replacing $\hat{a}$ and $\hat{a}^\dagger$ with eigenvalues $\rme^{\pm\rmi\theta}$ with $\theta\in[0,\pi]$, so that the continuous spectrum is bounded, and the corresponding eigenstates can be represented by $\ket{\theta}$ in
\begin{eqnarray}\label{eq:E theta seed}
    \hH_{\rm PT}\ket{\theta}=E(\theta)\ket{\theta}~,\quad E(\theta)=\frac{-2}{\sqrt{1-q}}\cos\theta~.
\end{eqnarray}

\paragraph{TFD formulation}
We now show that the chord Hamiltonian \eqref{eq:chord PT H} can be recovered from the TFD formulation in \eqref{eq:dictionary SYK and chord TFD}. We begin from
    \begin{align}\label{eq:def H SYK PT TFD}
    \hH_{0'}=\frac{1}{2}\qty(\text{b}_{\rm PT}(\hH_{\rm MQ}){\hH_{\rm def'}^\dagger}+{\hH_{\rm def'}}\text{b}_{\rm PT}(\hH_{\rm MQ}))~.
\end{align}
where we use $\hH_{\rm MQ}$ in \eqref{eq:H MQ} and we defined the deformation operator
\begin{equation}
    \label{eq:H def TFD}
    \hH_{\rm def'}\equiv\hH_{\rm SYK}^{(R)}+\qty[\hH_{\rm SYK}^{(R)},~\frac{\rmi}{p}\sum_{j=1}^N\psi_{j}^{(L)}\psi_{j}^{(R)}]~.
\end{equation}
Using the dictionary between SYK and chord operators \eqref{eq:dictionary SYK and chord TFD} for $\Delta_s=1$, we recover (see App.~\ref{app:detalssww} for details)
\begin{eqnarray}\label{eq: H TFD Omega}
    \lim_{\rm DS}\overline{\bra{\rm TFD_\infty}\hH_{0'}^m\ket{\rm TFD_\infty}}=\bra{\Omega}\hH^m_{\rm PT}\ket{\Omega}~,
\end{eqnarray}
where $m\in2\mathbb{Z}_{\geq0}$ since, again, odd moments
vanish on both sides.

\paragraph{Double-scaling limit of deformed chord Hamiltonians}
Let us collect the results. The relation between the Hamiltonian moments \eqref{eq:DS H PT} \eqref{eq: H TFD Omega} in both cases, we find that the $T^2$ and $T^2+\Lambda_1$ deformations of the deformed SYK model in the KM \eqref{eq:def H SYK PT KM} or TFD \eqref{eq:def H SYK PT TFD} formulations can be described, respectively, by
\begin{subequations}\label{eq:P PT series}
    \begin{align}
            \bra{\Omega}(\hH^{(\eta)}_{{\rm PT},y})^m\ket{\Omega}&=\lim_{\rm DS}\overline{\bra{\omega}[P^{(\eta)}_{\rm PT}(2y~\hH_{0})]^m\ket{\omega}}=\lim_{\rm DS}\overline{\bra{{\rm TFD}_\infty}[P^{(\eta)}_{\rm PT}(2y~\hH_{0'})]^m\ket{{\rm TFD}_\infty}}~,\\
            P^{(\eta)}_{\rm PT}(x)&\equiv\frac{1}{y}\qty(1+\varepsilon\sqrt{\eta(1-x)})~,\label{eq:2nd}
    \end{align}
\end{subequations}
where $\hH^{(\eta)}_{{\rm PT},y}$ appears in \eqref{eq:PT y}, $\varepsilon=\pm1$, $m\in 2\mathbb{Z}_{\geq0}$, and $P^{(\eta)}_{\rm PT}(2y\hH_0)$ is defined by functional calculus, i.e.~by the spectral theorem for the Hermitian operator $\hH_0$ (and likewise for $\hH_{0'}$), on the spectral subspace where $\eta(1-2y\hH_0)\geq0$. This is the same subspace on which $\hH^{(\eta)}_{{\rm PT},y}$ is Hermitian (Sec.~\ref{ssec:finite cutoff holo}), which we focus on throughout. For the $T^2$ deformation, $\eta=+1$, this is the whole Hilbert space whenever $2y\norm{\hH_0}<1$, and $P^{(+1)}_{\rm PT}(x)$ then admits the binomial expansion $\frac{1}{y}\qty(1+\varepsilon\sum_{n=0}^\infty\binom{1/2}{n}(-x)^n)$, so that \eqref{eq:P PT series} follows order by order in $y$ from the polynomial dictionaries \eqref{eq:dictionary SYK chord KM} and \eqref{eq:dictionary SYK and chord TFD}.\footnote{\label{fnt:one can }One can see that the would-be $x=1$ case in \eqref{eq:case 1 b} is harmless since the divergence of the
$(1 - q^{\hat{n}})^{-1/2}$ term acting on $\ket{\Omega}$ cancels against that of the $\sqrt{[\hat{n}]_q}$ term in $\hat{a}$.} For the $T^2+\Lambda_1$ deformation, $\eta=-1$, the square root $\sqrt{2y\hH_0-1}$ has no convergent expansion around $x=0$ on the spectrum of $\hH_0$, but \eqref{eq:P PT series} still holds. The reason is that the equality of all the moments \eqref{eq:DS H PT}, \eqref{eq: H TFD Omega}, together with the boundedness of the spectrum \eqref{eq:E theta seed}, implies that the ensemble-averaged spectral measure of $\hH_0$ (or $\hH_{0'}$) in the KM (or TFD) state converges, in the double-scaling limit, to the spectral measure of $\hH_{\rm PT}$ in $\ket{\Omega}$; a measure with bounded support is determined by its moments \cite{schmudgen2012unbounded}. Hence $\lim_{\rm DS}\overline{\bra{\omega}f(\hH_0)\ket{\omega}}=\bra{\Omega}f(\hH_{\rm PT})\ket{\Omega}$ for any bounded function $f$ that is continuous on the spectrum, up to a set of zero measure with respect to the continuous spectral density of \eqref{eq:E theta seed}; the choice $f=[P^{(\eta)}_{\rm PT}(2y\,\cdot)]^m$ restricted to the subspace above, which is continuous except at the threshold $E=1/(2y)$, gives \eqref{eq:P PT series} for both $\eta=\pm1$. \eqref{eq:P PT series} shows that the proposed microscopic deformations reproduce the annealed chord-theory moments of \cite{Aguilar-Gutierrez:2026ogo} in the chosen KM/TFD state, for both the $T^2$ and $T^2+\Lambda_1$ deformations. We present an alternative derivation in App.~\ref{app:Krylov micro H}. Note that the dictionaries \eqref{eq:dictionary SYK chord KM} and \eqref{eq:dictionary SYK and chord TFD} are derived for polynomial functions; the extension to the non-polynomial functions $\text{b}_{\rm PT}$ \eqref{eq:part_2b} and $P^{(\eta)}_{\rm PT}$ relies on the convergence of the spectral measures just described.

\subsection{Chord Hamiltonian, Hilbert space and recursion}\label{ssec:chord descripttion 3}
We now study the Hilbert space corresponding to \eqref{eq:chord PT H}. Similar to the undeformed DSSYK model, we select a cyclic separating state $\ket{\Omega}$ to build the space. Acting with the raising and lowering operators in \eqref{eq:chord PT H}, one generates the chord number basis $\qty{\ket{n}}$ for the auxiliary Hilbert space $\mathcal{H}_{\rm PT}$, which is orthonormal in the representation where the Hamiltonian \eqref{eq:chord PT H} is symmetric. We can express the eigenvalue problem \eqref{eq:E theta seed} in term of the wavefunctions $\psi_n(\theta)\equiv\bra{\theta}\ket{n}$ as
\begin{eqnarray}\label{eq:recurrence_1}
    2\cos\theta\psi_n(\theta)=\sqrt{1-\xi^2\sech^2\qty(\frac{\lambda}{2} n)}\psi_{n-1}(\theta)+\sqrt{1-\xi^2\sech^2\qty(\frac{\lambda}{2} (n+1))}\psi_{n+1}(\theta)~,
\end{eqnarray}
with initial condition $\psi_{-1}(\theta)=0$ and $\psi_0(\theta)=\langle\theta|\Omega\rangle=1$, since the chord number
basis starts at $|\Omega\rangle=|n=0\rangle$ \eqref{eq:basis}. Note that,
unlike the undeformed case where $[0]_q=0$, here $b_{\rm PT}(0)=\sqrt{1-\xi^2}\neq0$, so this boundary condition
has to be imposed explicitly. In particular, when $\xi=1$, the chord Hamiltonian \eqref{eq:chord PT H} at the classical level has the same functional form as the finite-cutoff JT gravity Hamiltonian in Schwarzschild time discussed by \cite{Griguolo:2025kpi} (7.6)
$H_{\rm PT}\propto \tanh\frac{\lambda n}{2}\cos p_n$.\footnote{It is interesting that one does not need to take a small $p_n$ limit to connect the chord Hamiltonian to that of JT gravity in this formulation, but rather only set $\xi=1$ and take the classical limit where we fix $\lambda n$ as $\lambda\rightarrow0$. This should be further explored in the future.} At the quantum level, the recurrence relation is simpler
\begin{eqnarray}
   \xi=1:~~ \frac{1-q^{n+1}}{1+q^{n+1}}\psi_{n+1}(\theta)+\frac{1-q^{n}}{1+q^{n}}\psi_{n-1}(\theta)=2\cos\theta\psi_n(\theta)~.
\end{eqnarray}
The recurrence relation for $\xi\geq0$ \eqref{eq:recurrence_1} determines the polynomial solution for $\psi_n(x)$ with $x=2\cos\theta$, where the first few terms are $\psi_1(x) = x/ a_1$, $\psi_2 = (x^2-a_1^2)/(a_1 a_2)$, {$\psi_3 = (x^3- x(a_1^2+a_2^2))/(a_1 a_2 a_3)$} and so on where $a_n\equiv b_{\rm PT}(n)$. As we saw in the analysis \eqref{eq:chord PT H}, given that its asymptotic limit reduces to that of the undeformed DSSYK model, the polynomials solving \eqref{eq:recurrence_1} provide a finite-cutoff generalization of the q-Hermite polynomials. A complementary treatment of the recurrence relation \eqref{eq:recurrence_1} can be found in App.~\ref{app:finite cutoff ASC} (see \eqref{eq:sol q Heun} for $\mathcal{Y}=\overline{\mathcal{Y}}=0$).

\subsection{Low- and high-energy limits: finite-cutoff (dS) JT gravity}\label{ssec:bulk interpretation in (dS) JT at finite cutoff}
We now verify that the low-energy description of the PT chord Hamiltonian \eqref{eq:chord PT H} corresponds to the JT gravity ADM Hamiltonian in \eqref{eq:JT PT} in the classical limit, while its canonical quantization may introduce additional quantum corrections \cite{Griguolo:2026eoi}. In the regime of interest $\lambda \hat{n}\equiv \hat{\text{L}}$ is fixed (i.e.~$\mathcal{O}(\lambda^0)$), and the spectrum of the operator $\hat{\text{P}}\equiv\hat{p}_n/\lambda$ is $\mathcal{O}(\lambda^0)$ as we take $\lambda\rightarrow0$ in the low-energy limit \cite{Lin:2022rbf}, which in the finite cutoff setting takes the form of \eqref{eq:low energy limit}. By expanding up to order $\lambda^2$, the chord Hamiltonian in \eqref{eq:chord PT H} becomes
\begin{eqnarray}\label{eq:corrected H JT}
  \sqrt{1-q}\, \hH_{\rm PT}+2=\lambda^2\qty(\hat{\text{P}}^2+\Phi_b^2\sech^2\frac{\hat{\text{L}}}{2})+\mathcal{O}(\lambda^3)~.
\end{eqnarray}
The chord Hamiltonian then takes the form of $\lambda^2\hH_{\rm JT-PT}$ \eqref{eq:JT PT}. The $\mathcal{O}(\lambda^3)$ term, $$-\frac12\lambda^3\Phi_b^2\sech^2\frac{\hat{\text{L}}}{2}\tanh\frac{\hat{\text{L}}}{2}~,$$ comes from the asymmetric placement of $b_{\rm PT}(\hat{n})$ in \eqref{eq:chord PT H} (an $\hat{n}\rightarrow\hat{n}+\frac12$ shift) and is removed by the symmetric ordering $\rme^{\rmi\hat{p}_n/2}b_{\rm PT}(\hat{n}+\frac12)\rme^{\rmi\hat{p}_n/2}+{\rm h.c.}$, after which the corrections are $\mathcal{O}(\lambda^4)$, as in the asymptotic boundary limit \cite{Lin:2022rbf}. Below we use $\sqrt{1-q}\simeq\sqrt{\lambda}$ when only the leading order matters. 

Let us now describe the high-energy limit, corresponding to dS JT gravity at finite cutoff. We know from \cite{Heller:2025ddj} that the canonical quantization of dS JT proceeds in the same way as in the AdS$_2$ case; however, the corresponding length variables are evaluated on timelike geodesics (Fig.~\ref{fig:CMC-Krylov_AdS_BH}), and they enter \eqref{eq:JT PT} in the same way. In our conventions, timelike geodesics are real valued; there is then a conserved quantity of the same form as \eqref{eq:corrected H JT}, which is reproduced by making an expansion around the value $E(\theta=\pi-\theta_{\rm dS})$ for $\theta_{\rm dS}\ll1$, where $\hat{p}_n\equiv\pi-\lambda\hat{P}$ and the spectrum of $\hat{P}$ is $\mathcal{O}(\lambda^0)$ in this limit, as justified in \cite{Blommaert:2024whf,Heller:2025ddj}, so that
\begin{equation}
   2-\sqrt{1-q}\,\hH_{\rm PT}=\lambda^2\qty(\hat{\text{P}}^2+\Phi_b^2\sech^2\frac{\hat{\text{L}}}{2})+\mathcal{O}(\lambda^3)~.
\end{equation}
which leads to a simple identification between a rescaling chord number operator by $\lambda$ and the geodesic length operator, in the same way as in the asymptotic boundary case. The key difference is that the spectrum of $\hat{L}$ is $L\geq0$ while in the asymptotic boundary limit, defined as
\begin{equation}\label{eq:ABL}
    \Phi_b\rightarrow\infty~,\quad \lambda n\rightarrow\infty~,\quad \Phi_b\rme^{-L/2}~:~~{\rm fixed}~,
\end{equation}
so that \eqref{eq:chord PT H} recovers the undeformed chord Hamiltonian \eqref{eq:H chord ortho Krylov} with \eqref{eq:p n def}, one has that $L\in \mathbb{R}$ due to the regularization by the factor $\Phi_b$ above. This means that one has to implement appropriate choice of self-adjoint boundary conditions at $L=0$ to quantize the theory. We investigate this in more detail in Sec.~\ref{ssec:improvement JT}.

\begin{figure}
    \centering
    \subfloat[]{\includegraphics[width=0.3\linewidth]{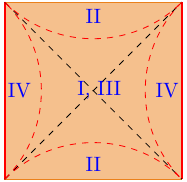}}\hspace{1cm}\subfloat[]{\includegraphics[width=0.3\linewidth]{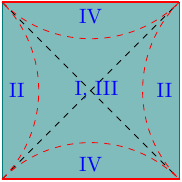}}
    \caption{Asymptotic Dirichlet boundaries (solid red lines) in the AdS$_2$ black hole (left) and in dS$_2$ space (right) can be moved to different locations (red dashed lines), marked I-IV, in the Penrose diagram through a dual $T^2$ ($\eta=+1$) or $T^2+\Lambda_1$ ($\eta=-1$) deformation in the boundary theory. Each region represents a combination of the $\eta$ and $\varepsilon$ parameters in the energy spectrum \eqref{eq:E(theta) eta+1}, \eqref{eq:E(theta) eta-1}, where $\varepsilon=-1$ corresponds to the perturbative solutions and $\varepsilon=+1$ the non-perturbative ones. The regions are denoted: (I) $\eta=+1$, $\varepsilon=-1$; (II) $\eta=-1$, $\varepsilon=-1$, (III) $\eta=-1$, $\varepsilon=+1$; (IV) $\eta=+1$, $\varepsilon=+1$. CMC slices in cases I and II are space/timelike in AdS/dS$_2$, while III and IV are time/spacelike.}
    \label{fig:summary_Penrose_diagrams}
\end{figure}

\paragraph{Classical geometry of the (non-)perturbative branches}
Next, we seek to identify the regions in the bulk that each of the $T^2$ and $T^2+\Lambda_1$ deformations for both $\varepsilon=\pm1$ cases describe in JT gravity and dS$_2$ gravity, which is illustrated in Fig.~\ref{fig:summary_Penrose_diagrams}.

In both cases, we use the notion of ADM energy
\begin{equation}\label{eq:general energy dS}
    E_{\rm ADM}=\int\rmd x^{d-1}\sqrt{\abs{\zeta}}~n^\sigma J_\sigma n^\mu n^\nu T_{\mu\nu}~,
\end{equation}
where $\zeta_{ab}$ is the metric of a codimension-2 surface, $n^\mu$ is a normal vector to the surface, $J^\mu$ is a Killing vector associated with translations along the asymptotic boundary, which is timelike in asymptotically AdS spacetimes, and spacelike for asymptotically dS spacetimes. This leads to the identification of a mass parameter in Schwarzschild-de Sitter geometries.

In the AdS$_2$ black hole case, seen as a dimensional reduction of the BTZ, the ADM energy \eqref{eq:general energy dS} takes the form,\footnote{There is an overall constant prefactor discussed in \cite{AliAhmad:2025kki} that depends on the definition of boundary time.}
\begin{eqnarray}\label{eq:ADM energy}
    E_{\rm ADM}=\eval{\frac{1}{4G_N}(\Phi-\partial_n\Phi)}_{\partial\mathcal{M}}~,
\end{eqnarray}
where $\partial_n=\sqrt{\abs{f(\Phi)}}\,\partial_{\Phi}$ is the outward-pointing normalized derivative along the outward-pointing spacelike or timelike normal direction, and $f$ is the blackening factor in \eqref{eq:Schwarzschild coordinates}. From \eqref{eq:ADM energy} it follows that the regions with $0\leq\Phi\leq\Phi_b$ are described by $-$ sign factors in the boundary theory spectrum; while the $+$ sign corresponds to the complement of the geometry which is removed by the finite cutoff between $\Phi_b<\Phi\leq \infty$, corresponding to the non-perturbative solutions, $\varepsilon=+1$ in the boundary theory spectrum. Since the $T^2$ ($\eta=+1$) deformation is continuously connected to the seed (i.e.~undeformed) spectrum $E_{y=0}=E$, while the $T^2+\Lambda_1$ ($\eta=-1$) is not, it follows that the former is valid for a radial cutoff at $\Phi=\Phi_b$ outside the horizon at $\Phi=\Phi_h$ (where the normal vector to the cutoff surface is spacelike), i.e.~$\Phi_h\leq\Phi\leq\Phi_b$, while the latter is valid for the black hole interior, i.e. $0\leq\Phi< \Phi_b$ where $\Phi_b<\Phi_h$ and the normal is timelike. In both cases, illustrated in Fig.~\ref{fig:summary_Penrose_diagrams}, the spectrum is given by \eqref{eq:ADM energy}.\footnote{Note there is a difference with respect to the identification in \cite{Griguolo:2026eoi} (v1) between the bulk regions and the energy spectrum; for us there is a $-$ relative sign in the spectrum for the perturbative $T^2+\Lambda_1$ case, which is $+$ in \cite{Griguolo:2026eoi}. The difference is that we work in an ensemble where the black hole temperature for $\Phi_b<\Phi<\Phi_h$ is negative, and the partition function is thus finite. For $\eta=-1$, $\varepsilon=-1$ the deformed energy $E_y=\frac{1}{y}\big(1-\sqrt{2yE-1}\big)$ decreases without bound as $E\rightarrow\infty$, as happens in the JT limit, so that $\int_0^\infty\rmd s\,\rho_0(s)\,\rme^{-\beta E_y(s^2)}$ converges only for $\beta<0$. I thank Alex Tarana for discussions about this point.}

One can similarly derive the analog of ADM energy for dS$_2$ space as a dimensional reduction of dS$_3$ space, which was considered \cite{Balasubramanian:2001nb} as well as higher-dimensional cases. One can verify that after the dimensional reduction, \eqref{eq:general energy dS} becomes of the same form as \eqref{eq:ADM energy}, although one should remember that the roles of time and space are inverted and thus the conserved charge associated with translations along the boundaries is not an energy when the Killing vector is spacelike rather than timelike. 

The identification of the deformed boundary theory spectrum with $\varepsilon=\pm1$ and $\eta=\pm1$ in the (A)dS$_2$ JT limits of the DSSYK model are displayed in Fig.~\ref{fig:summary_Penrose_diagrams}. In particular, we identify that in the dS$_2$ case, $\eta=-1$, $\varepsilon=+1$ corresponds to the cosmic horizon patch construction in \cite{Coleman:2021nor}, while $\eta=-1$, $\varepsilon=-1$ to the pole patch proposal in \cite{Gorbenko:2018oov}. Meanwhile, the $\eta=-1$ and $\varepsilon=-1$ cases in AdS have been explored in \cite{AliAhmad:2025kki,Griguolo:2026eoi}. 
In other works, the physical interpretation of the $\varepsilon=+1$ solutions in the flow equation has been debated (e.g.~\cite{Rosso:2020wir}). We discuss classical aspects of the non-perturbative branches in Sec.~\ref{ssec:CMC lenghts main} and their quantization in App.~\ref{app:non perturbative wavefunctions}. 

\subsection{Alternative operator ordering}\label{ssec:improvement JT}
\paragraph{Operator ordering in \cite{Griguolo:2026eoi}}
Following up on the discussion of Sec.~\ref{ssec:bulk interpretation in (dS) JT at finite cutoff}, we are interested in building a different deformation of the chord Hamiltonian whose low-energy description reproduces the finite-cutoff JT gravity Hamiltonian with the operator ordering considered in \cite{Griguolo:2026eoi}. Their proposal in our notation amounts to replacing $\hat{P}\rightarrow \mathcal{G}\qty(\hat{L},\,\hat{P})$ where 
\begin{eqnarray}\label{eq:sl2R op ordering}
    \mathcal{G}\qty(\hat{L},\,\hat{P})&\equiv\frac{1}{2}\qty({\sqrt{\text{csch} \hat{L}}}~\hat{P}~{\sqrt{\sinh \hat{L}}}~\hat{P}+\hat{P}~{\sqrt{\sinh \hat{L}}}~\hat{P}~{\sqrt{\text{csch} \hat{L}}})~.
\end{eqnarray}
This results in an additional correction term $-(1/4)\csch^2 \hat{L}$ in \eqref{eq:corrected H JT} after using the commutation relation $[\hat{L},\hat{P}]=\rmi$. The resulting factor can also be interpreted as a quantum correction term \cite{Griguolo:2026eoi} once one restores dimensionful units with $\hbar\neq 1$ (we set $\hbar=1$ in \eqref{eq:corrected H JT2}, \eqref{eq:PT improved H} and App.~\ref{app:log corrections}), which amounts to a rescaling of the momentum variable (see Sec.~\ref{ssec:sine dilaton} and \cite{Blommaert:2024ymv}), for which one has to consistently represent $\hat{P}^2=-\hbar^2\partial_L^2$, as discussed below.

\paragraph{Motivation for the operator ordering choice}
To motivate considering the choice in \cite{Griguolo:2026eoi}, note that for $L\ll1$, the resulting Schrödinger equation for the ADM Hamiltonian of finite cutoff JT in \cite{Griguolo:2026eoi} $\hat{P}^2+\Phi_b^2\sech^2\frac{\hat{L}}{2}-\frac{\hbar^2}{4}\csch^2\hat{L}$ (the term in parentheses in \eqref{eq:corrected H JT2} below) takes the form
\begin{eqnarray}\label{eq:inverese coupling problem}
    (L\ll1)~:~~\qty(-\partial_L^2+\frac{\Phi_b^2}{\hbar^2}-\frac{1}{4L^2})\psi_s(L)=\frac{s^2}{\hbar^2}\psi_s(L)~,
\end{eqnarray}
which is a critical inverse-square coupling problem in quantum mechanics. The differential operator $-\partial_L^2+\tilde{\kappa}/L^2$, with $\tilde{\kappa}$ some constant, defined on smooth functions with compact support away from the origin at $L=0$ is symmetric but not self-adjoint, so it needs to be extended to be self-adjoint. When $\tilde{\kappa}\geq-1/4$, the operator is bounded from below, and the Friedrichs extension \cite{friedrichs1934spektraltheorie} is a unique canonical choice that preserves the lower bound of the operator and keeps its domain as small as possible \cite{schmudgen2012unbounded}, and the specific case $\tilde{\kappa}=-1/4$ in \eqref{eq:inverese coupling problem} sits exactly on the threshold of stability in the sense of energy boundedness.

We will consider more general self-adjoint extensions in Sec.~\ref{ssec:1st approach QJT}.

\paragraph{DSSYK deformations consistent with \eqref{eq:sl2R op ordering}}
The arguably simplest deformation of the DSSYK Hamiltonian \eqref{eq:chord PT H} which reproduces the correction in the low-energy theory due to the choice of operator ordering \eqref{eq:sl2R op ordering} is to simply include an additional term $\frac14\hbar ^2\lambda^2\csch^2(\lambda\hat{n})$ inside the factor $b_{\rm PT}(\hat{n})$, i.e.
\begin{equation}\label{eq:specifif replacement}
    b_{\rm PT}(\hat{n})\rightarrow\sqrt{1-\xi^2\sech^2\frac{\lambda\hat{n}}{2}+\frac{\hbar^2\lambda^2}{4}\csch^2{\lambda\hat{n}}}~,
\end{equation}
and to keep the same structure for the chord Hamiltonian in \eqref{eq:chord PT H}. One can easily verify from the same steps as \eqref{eq:corrected H JT} that the resulting deformation of \eqref{eq:chord PT H} becomes
\begin{eqnarray}\label{eq:corrected H JT2}
  \sqrt{1-q}\, \hH_{\rm PT}+2\rightarrow\lambda^2\qty(\hat{\text{P}}^2+\Phi_b^2\sech^2\frac{\hat{\text{L}}}{2}-\frac{\hbar^2}{4}\csch^2\hat{\text{L}})+\mathcal{O}(\lambda^4)~,
\end{eqnarray}
which then reproduces the improvement term in \cite{Griguolo:2026eoi} with $\hat{P}^2=-\hbar^2\partial_L^2$. However, we find the specific replacement \eqref{eq:specifif replacement} to be too ad hoc and is not naturally connected to the specific operator ordering choice in \eqref{eq:sl2R op ordering}. 

There are different possibilities to deform the chord Hamiltonian \eqref{eq:chord PT H} in such a way that its low-energy limit incorporates the replacement $\hat{P}\rightarrow \mathcal{G}\qty(\hat{L},\,\hat{P})$. We find that the most promising one in this work is to consider
\begin{equation}\label{eq:PT improved H}
    \begin{aligned} H_{\rm PT'}&=\frac{-1}{\sqrt{1-q}}\left( \rme^{\rmi\hat{p}_n}b_{\rm PT}(\hat{n}) +b_{\rm PT}(\hat{n})\rme^{-\rmi\hat{p}_n}+\mathcal K_\infty(\hat{n},\hat{p}_n)-\mathcal K_0(\hat{n},\hat{p}_n)\right) \\ 
    \mathcal K_0&=\frac12\left( \sqrt{\text{csch}(\lambda\hat{n})}\sin\hat{p}_n\sqrt{\sinh (\lambda\hat{n})} \sin\hat{p}_n+\sin\hat{p}_n \sqrt{\sinh (\lambda\hat{n})} \sin\hat{p}_n \sqrt{\text{csch}(\lambda\hat{n})} \right)~,\\
    \mathcal K_\infty &=\frac12\left( \rme^{-\lambda\hat{n}/2}\sin\hat{p}_n\rme^{\lambda\hat{n}/2} \sin\hat{p}_n+\sin\hat{p}_n\rme^{\lambda\hat{n}/2} \sin\hat{p}_n\rme^{-\lambda\hat{n}/2} \right)~,
    \end{aligned} 
\end{equation}
where $b_{\rm PT}$ appears in \eqref{eq:case 1 b}.

The resulting Hamiltonian has some useful features:
\begin{itemize}
    \item The term $\mathcal{K}_0=\lambda^2\mathcal{G}(\hat{L},\hat{P})+\mathcal{O}(\lambda^4)$ while $\mathcal{K}_\infty=\lambda^2\hat{P}^2+\mathcal{O}(\lambda^4)$, so that \eqref{eq:PT improved H} reproduces the Hamiltonian with the $\csch^2 L$ term in \eqref{eq:corrected H JT2} and manifestly incorporates the operator ordering factor $\mathcal{G}$ in \eqref{eq:sl2R op ordering} in the low-energy regime.
\item It preserves the momentum shift symmetry discussed in \cite{Blommaert:2024whf} $\hat{p}_n\rightarrow\hat{p}_n+2\pi$, and it reproduces the DSSYK chord Hamiltonian \eqref{eq:H chord ortho Krylov} in the asymptotic boundary limit \eqref{eq:ABL}. In addition, \eqref{eq:Zdef} reproduces the density of states in the JT gravity partition function in \eqref{eq:orthogonality} found by \cite{Griguolo:2026eoi} up to $\mathcal{O}((\log\lambda)^{-1})$ as we show in App.~\ref{app:log corrections}.
\item In the classical limit, one can see that $\mathcal K_0=\mathcal K_{\infty}$, so it reproduces the classical ADM Hamiltonian of finite cutoff sine dilaton gravity (Sec.~\ref{ssec:sine dilaton}). This means that \eqref{eq:PT improved H} could be interpreted as an operator ordering choice in finite cutoff sine dilaton gravity, in a similar way as $\mathcal G$ \eqref{eq:sl2R op ordering} in the JT case.
\end{itemize}
We first note what the minimal ordering gives. For \eqref{eq:chord PT H}, $\ket{\Omega}$ is in the domain of $\hH_{\rm PT}$, and $\bra{\Omega}\rme^{-\beta\hH_{\rm PT}}\ket{\Omega}$ is well defined. In the low-energy limit \eqref{eq:low energy limit}, the recursion \eqref{eq:recurrence_1} with $\psi_{-1}(\theta)=0$ becomes the Schrödinger problem $-\psi''+\Phi_b^2\sech^2(L/2)\,\psi=s^2\psi$ on $L\geq0$ with a Dirichlet condition at $L=0$, which is a regular endpoint of \eqref{eq:JT PT}. The spectral density of $\ket{\Omega}$ then grows as $s^2$ for $s\gg\Phi_b$, as for a free particle with a Dirichlet wall, whereas $\rho_0(s)\rightarrow2s$ in \eqref{eq:orthogonality}. Thus, the minimal ordering does not reproduce \eqref{eq:orthogonality}, which motivates \eqref{eq:PT improved H}.

\paragraph{Partition function}To evaluate the partition function of \eqref{eq:PT improved H} one should notice that due to the factor $\csch (\lambda\hat{n})$, $\ket{\Omega}$ is not in the domain of $\mathcal{K}_0$, and neither for $\hH_{\rm PT'}$. This means that we should restrict the chord space to $\qty{\ket{n\geq1}}$, which sets the amplitude $b_{\rm PT}(1)$ of $\rme^{\rmi\hat{p}_n}b_{\rm PT}(\hat{n})$ between $\ket{1}$ and $\ket{0}$ to zero by hand; this is a Dirichlet-type condition at the lattice scale, whose relation to the Friedrichs extension is analysed in App.~\ref{app:log corrections}. We may use $\ket{n=1}$ as the corresponding building block of the chord number basis. This could be interpreted as a lattice realization of the $\ket{L=0}$ length states $L$ for which $\psi_s(0)\equiv \bra{L=0}\ket{s}=0$, using a lattice spacing $L=\lambda$. The corresponding partition function in this proposal is then
\begin{eqnarray}\label{eq:Zdef}
    Z_{\rm PT'}(\beta)=\bra{n=1}\rme^{-\beta\hH_{\rm PT'}}\ket{n=1}~,
\end{eqnarray}
which is evaluated in App~\ref{app:log corrections}, and leads to
\begin{equation}
  Z_{\rm PT'}\big(\beta\big)\propto\frac{\lambda^2{\rm e}^{\frac{2}{\sqrt{1-q}}\beta}}{\log^2(1/\lambda)}
  \int_0^\infty{\rm d}s\,\rho_0(s)\left[1+\frac{2\,{\rm Re}\big(\tilde{\mathcal A}_s/\mathcal A_s\big)}{\log(1/\lambda)}
  +\mathcal O\qty(\frac{1}{\log^2(1/\lambda)})\right]{\rm e}^{-\lambda^{3/2}\beta s^2}\,,
  \label{eq:Zmain}
\end{equation}
where $\tilde{\mathcal A}_s/\mathcal A_s$ is a ratio of asymptotic amplitudes of the two solutions of the low-energy
Schrödinger equation, defined in App.~\ref{app:log corrections}, and the proportionality factor does not depend on $s$ (it contains the lattice-scale factor $\log^2(1/n_0)$ of
\eqref{eq:Zexact} and depends on $\lambda$ only as $1+\mathcal O(1/\log(1/\lambda))$ beyond the explicit prefactor). Therefore, the
partition function \eqref{eq:orthogonality} is recovered as $\lambda\to0$ with $\beta_{\rm JT}=\lambda^{3/2}\beta$, but the approach is
logarithmic in $\lambda$ rather than a power law, and the leading correction is an $s$-dependent distortion of the
density of states $\rho_0(s)$. The partition function of the deformed theory \eqref{eq:PT y} follows in the same way: replacing $\hH_{\rm PT'}\rightarrow\hH^{(\eta)}_{{\rm PT'},y}\equiv\frac{1}{y}\qty(1+\varepsilon\sqrt{\eta(1-2y\hH_{\rm PT'})})$ in \eqref{eq:Zdef}, and using that both operators are diagonal in the same basis, $Z^{(\eta)}_{{\rm PT'},y}(\beta)=\bra{n=1}\rme^{-\beta\hH^{(\eta)}_{{\rm PT'},y}}\ket{n=1}$ is obtained from \eqref{eq:Zmain} by replacing the Boltzmann factor $\rme^{-\beta E}$ of each eigenstate, $E=(\lambda^2s^2-2)/\sqrt{1-q}$, by $\rme^{-\beta E_y(E)}$ with $E_y(E)=\frac{1}{y}\qty(1+\varepsilon\sqrt{\eta(1-2yE)})$; for instance, for the perturbative branch of the $T^2$ deformation, $E_y(\theta)=\frac{1}{y}\qty(1-\sqrt{1-2yE(\theta)})$ in terms of the undeformed energies \eqref{eq:E theta seed}. Thus, \eqref{eq:Zdef} is the partition function \eqref{eq:orthogonality} of \cite{Griguolo:2026eoi} in the time conjugate to the ADM Hamiltonian, with the $\Phi_b$-dependent density $\rho_0(s)$ and unflowed energies $s^2$, while $Z^{(\eta)}_{{\rm PT'},y}$ weighs the same $\rho_0(s)$ with the flowed energies $E_y$, i.e.~with the finite-cutoff JT spectrum quoted below \eqref{eq:holog dict SDG DSSYK}. The cutoff is not counted twice: $\xi=\lambda\Phi_b$ fixes the density of states, while $y$ in \eqref{eq:holog dict SDG DSSYK} converts ADM energies into energies conjugate to the time of the deformed theory.

It would be interesting to investigate whether there is a symmetry argument in favor of the specific operator ordering choice in \eqref{eq:PT improved H} in finite-cutoff sine dilaton gravity, in analogy with the JT case. For this reason, we will not present a SYK deformation that reproduces \eqref{eq:PT improved H}; however, the same techniques we presented in Sec.~\ref{ssec:SYK defs} could be applied to describe the deformation in \eqref{eq:PT improved H}.

\subsection{Finite-cutoff sine dilaton gravity and the chord Hamiltonian}\label{ssec:sine dilaton}
To close the section, we verify that the construction of the chord Hamiltonian \eqref{eq:chord PT H} can be interpreted in terms of sine dilaton gravity \cite{Blommaert:2023opb,Blommaert:2024ymv} at finite cutoff. Instead of limiting the study to the low/high-energy sectors, where $\cos\theta\simeq 1-\theta^2/2$, we use the full energy spectrum as the DSSYK model \eqref{eq:E theta seed}. We consider the following theory \cite{Blommaert:2023opb,Blommaert:2024ymv,Aguilar-Gutierrez:2026ogo}
\begin{eqnarray}\label{eq:action SDG}
    I=-\frac{1}{16\pi G_N}\qty(\int_{\mathcal{M}}\rmd^2 x\sqrt{g}\qty(\Phi\mathcal{R}+2\sin\Phi)+2\int_{\partial\mathcal{M}}\rmd x\sqrt{h}\qty(\Phi_b K-\rmi\rme^{-\rmi\Phi_b/2}))~,
\end{eqnarray}
where $\mathcal{R}$ is the Ricci scalar, $h_{\mu\nu}$ the induced metric, and $\mathcal{M}$ the spacetime manifold. The solutions to the equations of motion in terms of the dilaton $\Phi$ are
\begin{equation}
    \rmd l^2=F(\Phi)\rmd \tau^2+\frac{\rmd \Phi^2}{F(\Phi)}~,\quad F(\Phi)=2(\cos\theta-\cos\Phi)~,
\end{equation}
with $\Phi=\theta$ a parametrization of the black hole horizon. Following \cite{Blommaert:2023opb,Blommaert:2024ymv,Aguilar-Gutierrez:2026ogo}, we implement Dirichlet boundary conditions at $r=\xi$ in the effective AdS$_2$ geometry \eqref{eq:sine dilaton metric} below, where the dilaton takes the value $\Phi_b=\frac{\pi}{2}+\rmi\log(\xi+\rmi\cos\theta)$; in the low-energy limit \eqref{eq:low energy limit}, $\Phi_b=\xi+\mathcal{O}(\xi^2)$, which is the identification used in Secs.~\ref{ssec:SYK defs} and \ref{ssec:bulk interpretation in (dS) JT at finite cutoff}. The two-sided geodesic between the AdS$_2$ finite cutoff boundaries in the effective geometry of sine dilaton gravity \cite{Blommaert:2024ymv}
\begin{equation}\label{eq:sine dilaton metric}
    \rmd l_{\rm eff}^2\equiv\rme^{-\rmi\Phi}\rmd l^2=(r^2-\sin^2\theta)\rmd\tau^2+\frac{\rmd r^2}{r^2-\sin^2\theta}~,\quad \Phi=\frac{\pi}{2}+\rmi\log(r+\rmi \cos\theta)~.
\end{equation}
In this geometry, we know that the geodesic length corresponding to a Hartle-Hawking \cite{hartle1983wave,Lehners:2023yrj} (HH) state preparation $\tau=\beta/2+\rmi t$ is \cite{Chapman:2021eyy} (see also App.~\ref{app: CMC lengths})
\begin{equation}\label{eq:FC AdS length SDG}
\sinh \frac{L}{2}=\sqrt{\frac{\xi^2}{\sin^2\theta}-1}\cosh(\sin\theta~t)~, 
\end{equation}
where $r=\xi$ is the location of the Dirichlet wall in the effective geometry \eqref{eq:sine dilaton metric}. We want to find a momentum $P$ conjugate to $L$ which satisfies the following symplectic form
\begin{equation}\label{eq:simpletic vanilla}
    \omega=\frac{\sin\theta}{8\pi G_N}\rmd t\wedge\rmd \theta=\frac{1}{16\pi G_N}\rmd L\wedge\rmd P~.
\end{equation}
The above relation is solved by
\begin{equation}
\tan P=\frac{\sqrt{\xi^{2}-\sin^{2}\theta}\;\sinh(\sin\theta\,t)}{\cos\theta\,\cosh(L/2)}~,
\end{equation}
which implies that the corresponding ADM Hamiltonian is
\begin{equation}\label{eq:vanilla SDG finite cutoff}
    H(L,P)=-\frac{1}{8\pi G_N}\sqrt{1-\frac{\xi^{2}}{\cosh^{2}(L/2)}}\;\cos P .
\end{equation}
This Hamiltonian indeed agrees with the chord Hamiltonian \eqref{eq:chord PT H} in terms of $\hat{n}$ and $\hat{p}_n$ \eqref{eq:p n def} after taking the classical limit, and taking
\begin{equation}\label{eq:dict 1}
    8\pi G_N=\frac{\sqrt{1-q}}{2}~,
\end{equation}
which agrees with the sine dilaton gravity/DSSYK holographic dictionary \cite{Blommaert:2024ymv,Aguilar-Gutierrez:2025pqp} after implementing our normalizations. Note that \eqref{eq:simpletic vanilla} gives $\{L,P\}=16\pi G_N=\sqrt{1-q}$, whereas the chord variables satisfy $\{\lambda\hat{n},\hat{p}_n\}=\lambda$; the bulk time $t$ is therefore related to the time conjugate to $\hH_{\rm PT}$ by the factor $\sqrt{1-q}/\lambda\simeq\lambda^{-1/2}$, which is the factor $1/\sqrt{\lambda}$ appearing in the saddle-point equations \eqref{eq:EOM H} of Sec.~\ref{ssec:def CMC-K complexity} and in the entry $t\sqrt{\xi^2-\sin^2\theta}$ of Tab.~\ref{tab:holographic_dictionary}. As a consequence, one recovers the JT gravity ADM Hamiltonian \eqref{eq:corrected H JT} by implementing the rescaling 
\begin{equation}\label{eq:low energy 2sided}
\xi\rightarrow\lambda\Phi_b~,\quad P\rightarrow\lambda P~,
\end{equation}
with $L=\lambda n$, which reproduces \eqref{eq:corrected H JT} with $\sqrt{1-q}\simeq\sqrt{\lambda}$. As explained in Sec.~\ref{ssec:improvement JT}, to perform the canonical quantization, one might need to adopt the operator ordering choice \eqref{eq:PT improved H} in sine dilaton gravity; however, whether there exists a group symmetry argument for doing so still needs to be further investigated.

\section{Finite-cutoff CMC slices and CMC-Krylov complexity}\label{sec:CMC-Krylov complexity}
In this section, we seek a systematic analysis of the CMC slice lengths associated to the perturbative and non-perturbative branches in the flow equation \eqref{eq:flow eq} labeled by $\eta$ and $\varepsilon$, motivated by recent findings on finite causal diamonds in JT gravity \cite{Blommaert:2026lvp}. In particular, we study a boundary interpretation of CMC slice lengths in finite-cutoff holography by identifying a complexity measure in Krylov space, which we denote CMC-Krylov complexity. It is part of the cost functions in \cite{Balasubramanian:2022tpr} used to define Krylov complexity. It reproduces some properties of CAny proposals in AdS \cite{Belin:2022xmt,Jorstad:2023kmq,Caceres:2024edr} and dS \cite{Aguilar-Gutierrez:2023pnn,Aguilar-Gutierrez:2023zqm,Aguilar-Gutierrez:2024rka,Aguilar-Gutierrez:2024xfi} spacetimes. 

We present a summary of the  different lengths studied in this section, as well as App.~\ref{app:shockwaves} and \ref{app:distance brane} in Tab.~\ref{tab:lengths}.
\begin{table}[t]
    \centering
    \begin{tabular}{cccccc}
        \toprule
      Length& $\eta$&$\varepsilon$&$k$&Spacetime&Appearance  \\
        \midrule
      \eqref{eq:FC AdS length SDG}, \eqref{eq:geodesic length etaplus epsminus} &$+1$ &$-1$ &$0$ &AdS, dS&\cite{Chapman:2021eyy} for AdS\\
      \eqref{eq:geodesic length etaminus epsminus}&$-1$ &$-1$ &$0$ &AdS, dS&\cite{Griguolo:2026eoi} for AdS\\
      \eqref{eq:geodesic length etaminus epsplus}&$-1$ &$+1$ &$0$ &AdS, dS&\cite{Jorstad:2022mls} for dS\\
      \eqref{eq:geodesic length etaplus epsplus}&$+1$ &$+1$ &$0$ &AdS, dS&New\\
      \eqref{eq:AdS geo vs CMC}, \eqref{eq:dS geo vs CMC}&$\pm1$ &$-1$ &$\mathbb{R}$ &AdS, dS&\cite{Blommaert:2026lvp} for AdS, $\eta=+1$\\
      \eqref{eq:CMC eps plus closed}&$\pm1$ &$+1$ &$\mathbb{R}$ &AdS, dS&\cite{silke} for dS, $\eta=-1$\\
\eqref{eq:full sw length}&$+1$ &$-1$ &$0$ &AdS with a shockwave&New\\
\eqref{eq:length new}&$-1$ &$+1$ &$0$ &AdS with a shockwave&New\\
\eqref{eq:total length shocks}&$+1$ &$-1$ &$\mathbb{R}$ &AdS with a shockwave&New\\
\eqref{eq:L total}, \eqref{eq:shift length}&$+1$ &$-1$ &$0$ &AdS with an ETW brane&New\\
\eqref{eq:L transformed}&$-1$ &$-1$ &$0$ &AdS with an ETW brane&New\\
\eqref{eq:evol Length}&$+1$ &$-1$ &$\mathbb{R}$&AdS with an ETW brane&New\\
        \bottomrule
    \end{tabular}
    \caption{Summary of the different finite cutoff CMC lengths \eqref{eq:length CMC} in this work, including the deformation parameters $\eta$, $\varepsilon$ in \eqref{eq:flow eq} associated to them, the mean curvature $k$, the spacetime (AdS or dS) where they are valid, and we indicate whether they have been previously derived. Entries with two equation numbers (first and antepenultimate rows) indicate sine dilaton gravity and (dS) JT gravity evaluations. The last rows include finite-cutoff lengths with a shockwave (App. \ref{app:shockwaves}), and an ETW brane (App.~\ref{app:distance brane}).}
    \label{tab:lengths}
\end{table}

\paragraph{Outline}In Sec.~\ref{ssec:CMC lenghts main} we collect geodesic and CMC slice lengths in (A)dS spaces representing the different bulk regions accessible to the perturbative and non-perturbative solutions of the energy flow equation \eqref{eq:flow eq} for $T^2$ and $T^2+\Lambda_1$ deformations, including the black hole interior and the static patch of dS space. In Sec.~\ref{ssec:def CMC-K complexity} we identify CMC-Krylov complexity as a boundary dual observable that reproduces the growth of the finite-cutoff CMC slices in the context of the DSSYK/sine dilaton duality, and evaluate it using saddle point methods. 

\subsection{Geodesic and CMC slice lengths in (A)dS\texorpdfstring{$_2$}{2}}\label{ssec:CMC lenghts main}
We now collect old and new results on the CMC lengths (including geodesic ones) between two timelike or spacelike Dirichlet boundaries in AdS$_2$ black holes \cite{Maldacena:2016upp} and dS$_2$ space at finite cutoff, shown in Fig.~\ref{fig:summary_Penrose_diagrams} respectively. The formulas are based on App.~\ref{app: CMC lengths} where we review relevant results from the previous literature, and we derive new ones, such as the CMC length in the case $\varepsilon=+1$. A summary of the different lengths in this section is presented in Tab.~\ref{tab:lengths}.

\paragraph{(A)dS geodesics at finite cutoff}
Let us denote the metric in Schwarzschild coordinates
\begin{eqnarray}\label{eq:Schwarzschild coordinates}
    \rmd l^2=-f(\Phi)\rmd t^2+\rmd \Phi^2/f(\Phi)~,
\end{eqnarray}
where $f(\Phi)=\Phi^2-\Phi_h^2$ for Rindler-AdS$_2$ and $f(\Phi)=\Phi_h^2-\Phi^2$ in static patch coordinates of dS$_2$ space, where $\Phi=\Phi_h$ is the horizon radius of the black hole AdS$_2$, and $\Phi_h$ is a conical defect parameter of Schwarzschild-dS$_3$ space reduced to dS$_2$ (see e.g.~\cite{Svesko:2022txo}). For later calculation, we specify below the definition of the length of a boundary-anchored CMC slice,
\begin{equation}\label{eq:length CMC}
    L_k=\int\sqrt{\abs{g_{\mu\nu}\dv{{x}^\mu}{\chi}\dv{{x}^\nu}{\chi}}}\rmd\chi~,
\end{equation}
with $\chi$ a general parameterization of the coordinates $x^\mu(\chi)$ of the CMC slice.

In both AdS$_2$ and dS$_2$ space, the following relations hold for the geodesic lengths ($k=0$ in \eqref{eq:length CMC}) of the different regions with a radial boundary cutoff, as well as finite causal diamonds with Dirichlet boundaries,\footnote{\eqref{eq:geodesic length etaminus epsplus} and \eqref{eq:geodesic length etaplus epsplus} can be alternatively written as \begin{equation}
    \begin{aligned}
         \eta=-1,\varepsilon=+1:\quad \cot\frac{L}{2}&=\sqrt{\frac{\bbr^2+1-(1-\bbr^2)\cosh(2 u)}{2(1-\bbr^2)}}\sech(u)~,\\
    \eta=+1,\varepsilon=+1:\quad \cot\frac{L}{2}&=\sqrt{\frac{\bbr^2+1-(\bbr^2-1)\cosh(2u)}{2(\bbr^2-1)}}\abs{\csch(u)}~,
    \end{aligned}
\end{equation} where the first relation is consistent with \cite{Jorstad:2022mls} obtained by a different method, see App.~\ref{app: CMC lengths}, and the second one is simply an analytic continuation between the black hole interior and the Rindler-AdS$_2$ patch. Here $L\in(0,\pi)$ is taken on the principal branch. In particular, the critical time at which the hyperfast growth ends, $\tau_\infty$ in (A.6) of \cite{Jorstad:2022mls}, is reproduced when $\sin(L/2)=1$ in \eqref{eq:geodesic length etaminus epsplus}.}
{\allowdisplaybreaks\begin{subequations}\label{eq:geodesic length growth}
    \begin{align}
    \eta=+1,\varepsilon=-1:\quad \sinh\frac{L}{2}&=\sqrt{\bbr^2-1}\cosh(u)~,\label{eq:geodesic length etaplus epsminus}\\
    \eta=-1,\varepsilon=-1:\quad \sinh\frac{L}{2}&={\sqrt{1-\bbr^2}}\sinh(u)~,\label{eq:geodesic length etaminus epsminus}\\
    \eta=-1,\varepsilon=+1:\quad \sin\frac{L}{2}&=\sqrt{1-\bbr^2}\cosh(u)~,\label{eq:geodesic length etaminus epsplus}\\
    \eta=+1,\varepsilon=+1:\quad \sin\frac{L}{2}&=\sqrt{\bbr^2-1}\sinh(u)~,\label{eq:geodesic length etaplus epsplus}
    \end{align}
\end{subequations}}
where $\bbr\equiv\Phi_b/\Phi_h$ with $\bbr\geq1$ for $\eta=+1$, $0\leq\bbr\leq1$ for $\eta=-1$, and $u=\Phi_h t$. There is a conserved quantity in all cases
\begin{eqnarray}\label{eq:conserved quantity}
    \Phi_h^2=\begin{cases}
        P^2+\Phi_b^2\sech^2(L/2)~,&\varepsilon=-1~,\\
        -P^2+\Phi_b^2\sec^2(L/2)~,&\varepsilon=+1~,
    \end{cases}
\end{eqnarray}
where $P=(1/2)\rmd L/\rmd t $. When the boundaries are timelike (spacelike) and slices spacelike (timelike) in AdS$_2$ (dS$_2$) the conserved quantity is an ADM Hamiltonian. Moreover, instead of using Schwarzschild time, one can work with proper time at the boundary location, 
\begin{equation}\label{eq:tP}
t_P\equiv t\sqrt{f(\Phi_b)}~,
\end{equation}
with $f(\cdot)$ the blackening factor.

At last, let us comment on the non-perturbative branches, where $\varepsilon=+1$; we see from \eqref{eq:conserved quantity} that there are similar conserved quantities with respect to the ADM Hamiltonians, albeit with a different interpretation. In terms of the boundary theory, given that the geodesic length in both cases is real valued, the identification between the geodesic length with the expectation value of the chord number operator $\hat{n}$ requires that there is an overall factor $\rmi$ for the non-perturbative solutions, which is absent in the perturbative ones, i.e.
\begin{equation}
\varepsilon=-1~,~~ L=\lambda\expval{\hat{n}}~,\qquad
\varepsilon=+1~,~~ L=\rmi\, \lambda\expval{\hat{n}}~,
\end{equation}
where the expectation value is taken with respect to the HH state of the respective branch (including the factor $u\rightarrow u+\rmi\pi/2$ for the $T^2+\Lambda_1$ deformation). Note that, since $L$ is always real-valued in our conventions, the non-perturbative solutions, where $\varepsilon=+1$, require an analytic continuation where $\expval{\hat{n}}\in \rmi \mathbb{R}$ with respect to the $\varepsilon=-1$ branch, which is impossible for $\hat{n}$ Hermitian. Thus, while the $\varepsilon=+1$ branches might provide a useful semiclassical description of certain bulk regions; at the quantum gravity level, they are ill-behaved. We provide further details about their features in App.~\ref{app:non perturbative wavefunctions}. This can already be seen from \eqref{eq:conserved quantity}, the corresponding canonically quantized Hamiltonian as the form
\begin{equation}\label{eq:ADM non-perturbative}
\hat{H}_{\rm ADM}=-\hat{P}^2+\frac{\Phi_b^2}{\cos^2(\hat{L}/2)}+\frac{\csc^2\hat{L}}{4}~,
\end{equation}
which is unbounded from below due to the kinetic term, even with the improvement term in \cite{Griguolo:2026eoi}, and it is thus ill-behaved at the quantum mechanical level. The nontrivial statement is that $-\hat{H}_{\rm ADM}$ is also unbounded from below, since the endpoint $L=\pi$ carries a supercritical inverse-square potential (App.~\ref{app:non perturbative wavefunctions}). Their Hilbert space construction will not be further discussed in the main text.

\begin{figure}
    \centering
    \subfloat[]{\includegraphics[width=0.21\linewidth]{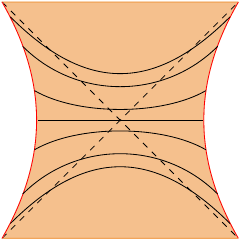}}\hfill\subfloat[$k=0$]{\includegraphics[width=0.36\linewidth]{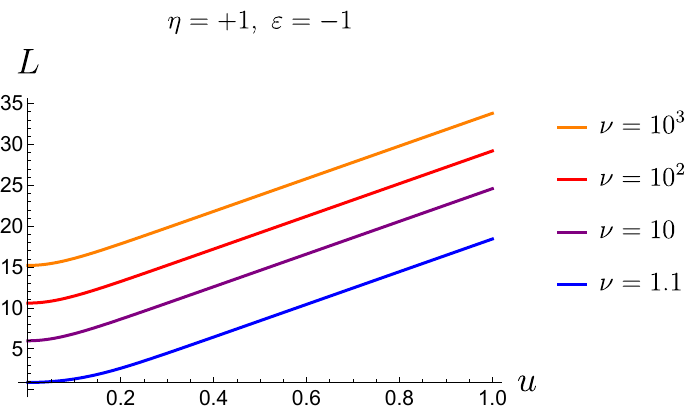}}\hfill\subfloat[$k=100$]{\includegraphics[width=0.36\linewidth]{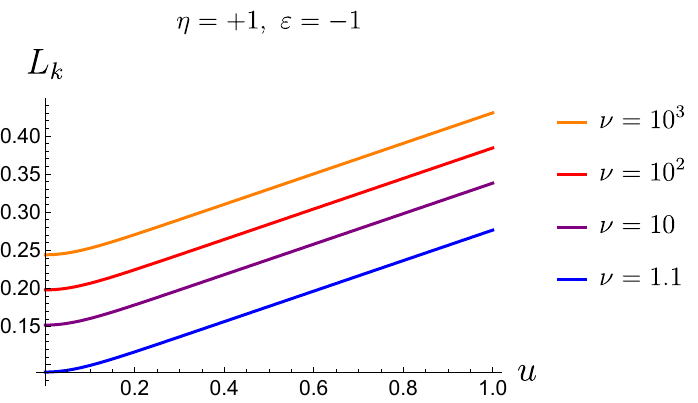}}\\
    \subfloat[]{\includegraphics[width=0.21\linewidth]{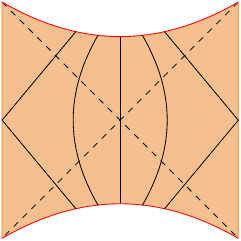}}\hfill\subfloat[$k=0$]{\includegraphics[width=0.36\linewidth]{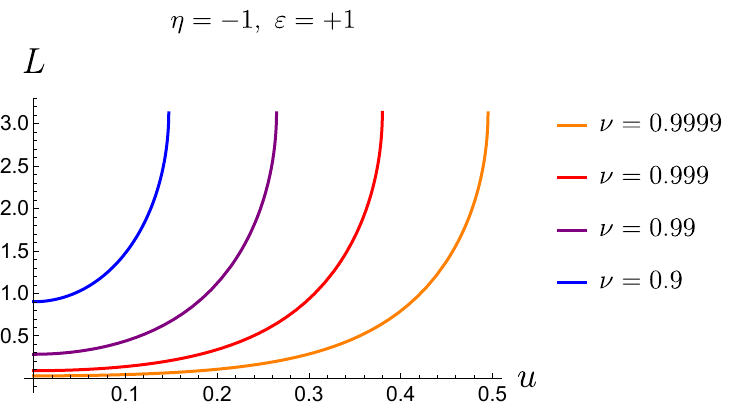}}\hfill\subfloat[$k=100$]{\includegraphics[width=0.36\linewidth]{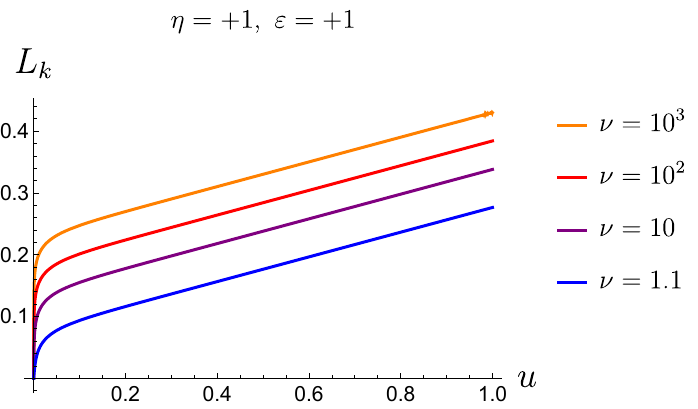}}\\
    \subfloat[]{\includegraphics[width=0.21\linewidth]{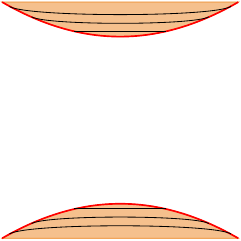}}\hfill\subfloat[$k=0$]{\includegraphics[width=0.36\linewidth]{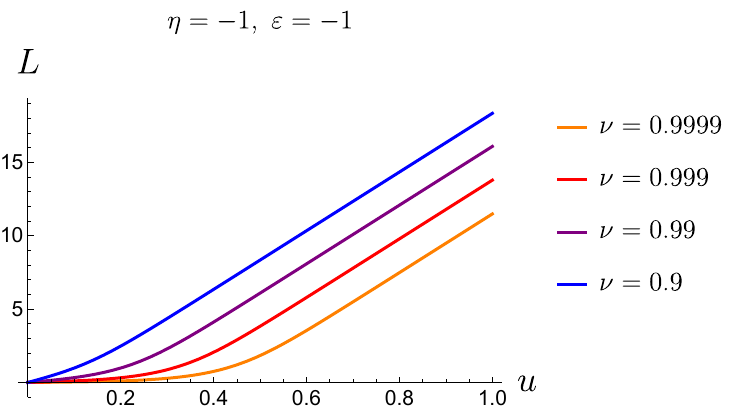}}\hfill\subfloat[$k=100$]{\includegraphics[width=0.36\linewidth]{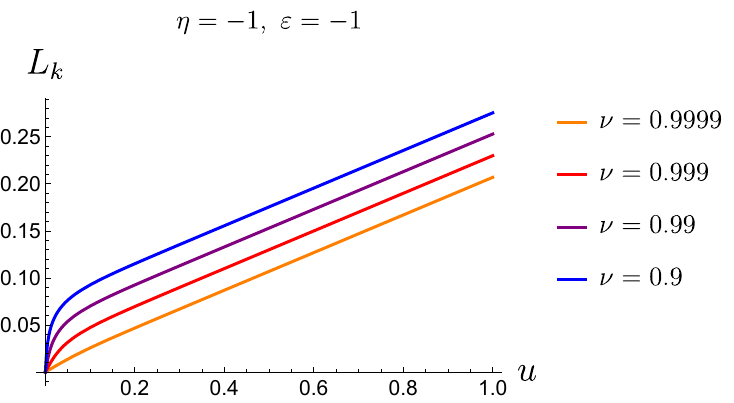}}\\
    \subfloat[]{\includegraphics[width=0.21\linewidth]{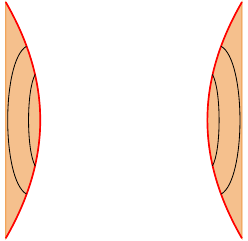}}\hfill\subfloat[$k=0$]{\includegraphics[width=0.36\linewidth]{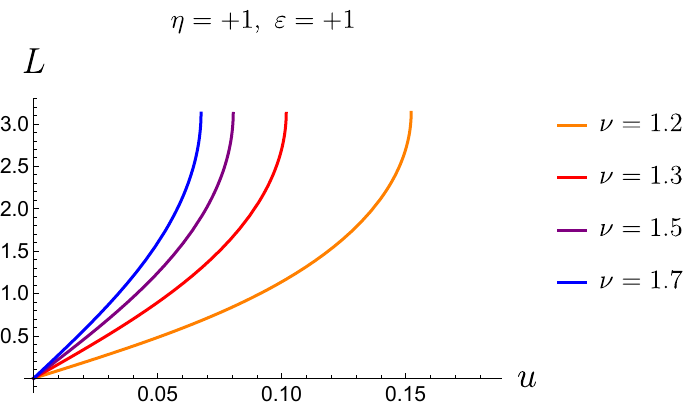}}\hfill\subfloat[$k=100$]{\includegraphics[width=0.36\linewidth]{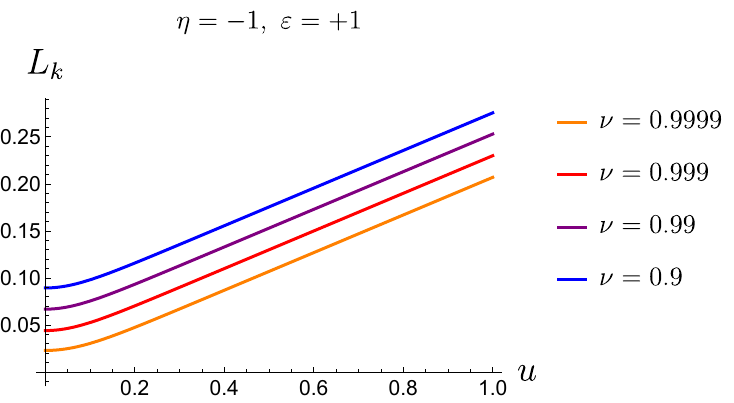}}
    \caption{Left: AdS$_2$ black hole Penrose diagrams describing the low-energy bulk dual of the deformed DSSYK for (a) $\eta=+1$, $\varepsilon=-1$; (d) $\eta=-1$, $\varepsilon=+1$; (g) $\eta=-1$ $\varepsilon=-1$, (j) $\eta=+1$, $\varepsilon=+1$. Black curves represent geodesic lines. Middle and right: Comparison between the geodesic length $L$ (middle column) and the CMC slice length $L_k$ when $k=100$ (right column) as a function of the rescaled time $u$ for different values of $\bbr$ (displayed in the figure). The convexity of $L_k(t)$ increases with increasing $\abs{k}$.}
    \label{fig:L_geo_vs_CMC_AdS}
\end{figure}

\paragraph{AdS CMC slices}
Let us now consider CMC slices between two finite boundaries, $L_k$. For AdS$_2$ black holes, one finds:
\begin{equation}\label{eq:AdS geo vs CMC}
    \eta=\pm1,~\varepsilon=-1:\quad \sinh\frac{L}{2}=\frac{1}{\sqrt{1+k^2}}\sinh(\frac{\sqrt{1+k^2}}{2}L_k)~.
\end{equation}
For the non-perturbative branches, $\varepsilon=+1$, the slices are timelike in AdS$_2$ (and spacelike in dS$_2$, see below), and the corresponding relation is obtained from \eqref{eq:AdS geo vs CMC} by the same continuation that relates \eqref{eq:geodesic length etaplus epsminus}, \eqref{eq:geodesic length etaminus epsminus} to \eqref{eq:geodesic length etaminus epsplus}, \eqref{eq:geodesic length etaplus epsplus}, namely $L\rightarrow\rmi L$, $L_k\rightarrow\rmi L_k$, $k^2\rightarrow-k^2$:
\begin{equation}\label{eq:CMC eps plus closed}
    \eta=\pm1,~\varepsilon=+1:\quad
    \sin\frac{L}{2}=\begin{cases}\dfrac{1}{\sqrt{1-k^2}}\sin\qty(\dfrac{\sqrt{1-k^2}}{2}L_k)~,&\abs{k}<1~,\\[2ex]
    \dfrac{1}{\sqrt{k^2-1}}\sinh\qty(\dfrac{\sqrt{k^2-1}}{2}L_k)~,&\abs{k}>1~,\end{cases}
\end{equation}
with $L$ given by \eqref{eq:geodesic length etaminus epsplus} or \eqref{eq:geodesic length etaplus epsplus}. In dS$_2$ this follows from the fact that every spacelike curve of constant geodesic curvature $\abs{k}<1$ is a boosted global slice, $k=\tanh\tau'_0$, so that for time-symmetric anchors $\sin\frac{L}{2}=\cosh\tau'_0\sin\theta'_0$ and $L_k=2\theta'_0\cosh\tau'_0$; the AdS$_2$ case follows by $g_{\mu\nu}\rightarrow-g_{\mu\nu}$, which exchanges the two spacetimes and preserves geodesic curvature and length (see the discussion of dS$_2$ below). We have verified \eqref{eq:CMC eps plus closed} against the parametric expressions $L_k(\mathfrak{c})$, $t(\mathfrak{c})$ derived in App.~\ref{app: CMC lengths}, where $\mathfrak{c}$ is an integration constant. \eqref{eq:CMC eps plus closed} shows that the growth is hyperfast for $\abs{k}<1$, ending at $\sqrt{1-k^2}\sin\frac L2=1$, where $L_k=\pi/\sqrt{1-k^2}$, and linear at late times for $\abs{k}>1$. A comparison between AdS$_2$ geodesic lengths and CMC lengths is shown in Fig.~\ref{fig:L_geo_vs_CMC_AdS}.\footnote{CMC slices with $k\neq0$ are tilted with respect to the $k=0$ slices in Fig.~\ref{fig:L_geo_vs_CMC_AdS} and do not generically evolve in a symmetric way, as discussed in \cite{Jorstad:2023kmq,Aguilar-Gutierrez:2024nau} among others.} 
Noticeably, the cases with $\varepsilon=+1$ and $\abs{k}<1$ show hyperfast growth, i.e.~they evolve only up to a finite time range,\footnote{However, one can consider an analytic continuation after the CMC slice reaches the region where the time derivative of its length diverges, this approach has been used by \cite{Jorstad:2022mls} in the analogous dS space case with $k=0$. Due to the non-uniqueness of the length growth after the analytic continuation, their prescription will not be used it in our evaluations.} and their time derivative is infinite at the end of the evolution. It can also be seen that $k$ modifies the convexity of the length, which we identify with convexity of CMC-Krylov complexity in the $\varepsilon=-1$ cases. $k$ is a tunable parameter in our case; in contrast, the convexity of Krylov complexity itself is associated with the integrable/chaotic properties of the system \cite{Alishahiha:2024vbf,Nandy:2024evd,Rabinovici:2025otw,Baiguera:2025dkc}. Further, given that the cases with $\varepsilon=-1$ display late time linear growth, one identifies the existence of an accumulation surface in the holographic complexity growth (that is a radial surface which is asymptotically reached at late times) for any of the CMC slices, which we illustrate in Fig.~\ref{fig:accumulation}. 

\begin{figure}
    \centering
    \subfloat[]{\includegraphics[width=0.21\linewidth]{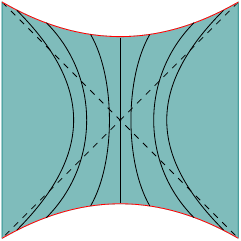}}\hfill\subfloat[$k=10$]{\includegraphics[width=0.36\linewidth]{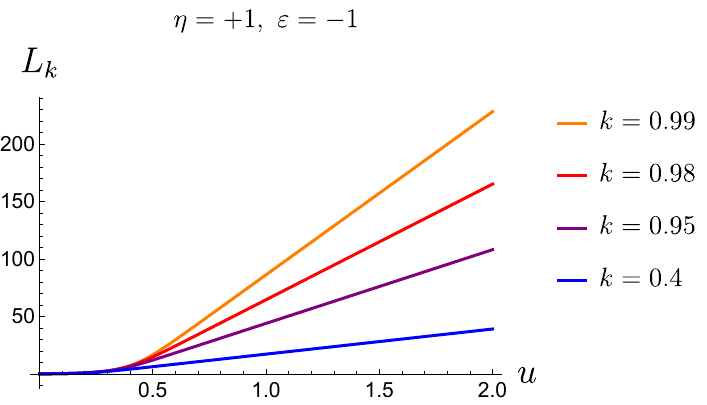}}\\
\subfloat[]{\includegraphics[width=0.21\linewidth]{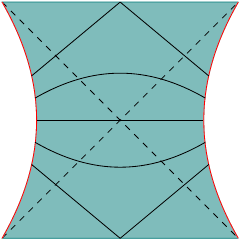}}\hfill\subfloat[$k<1$]{\includegraphics[width=0.36\linewidth]{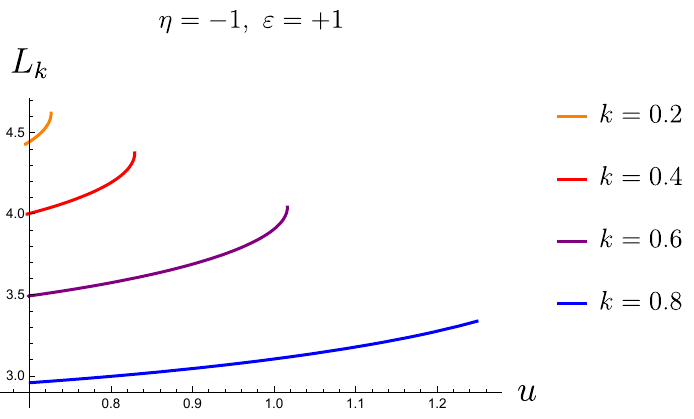}}\hfill\subfloat[$k>1$]{\includegraphics[width=0.36\linewidth]{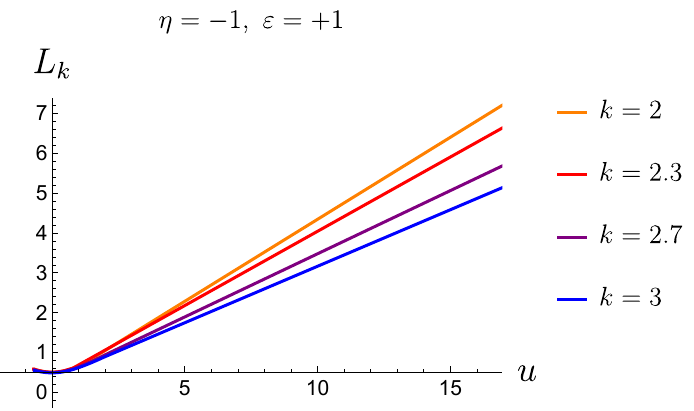}}\\
\subfloat[]{\includegraphics[width=0.21\linewidth]{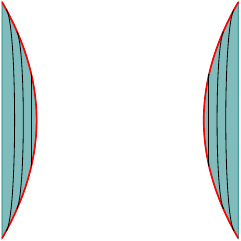}}\hfill\subfloat[$k=10$]{\includegraphics[width=0.36\linewidth]{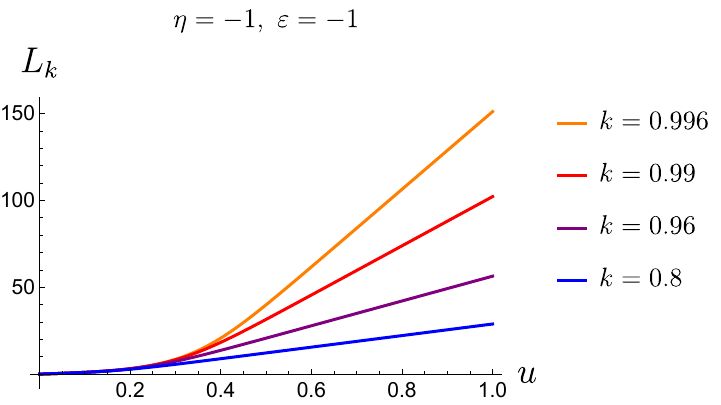}}\\
\subfloat[]{\includegraphics[width=0.21\linewidth]{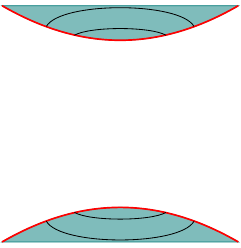}}\hfill\subfloat[$k<1$]{\includegraphics[width=0.36\linewidth]{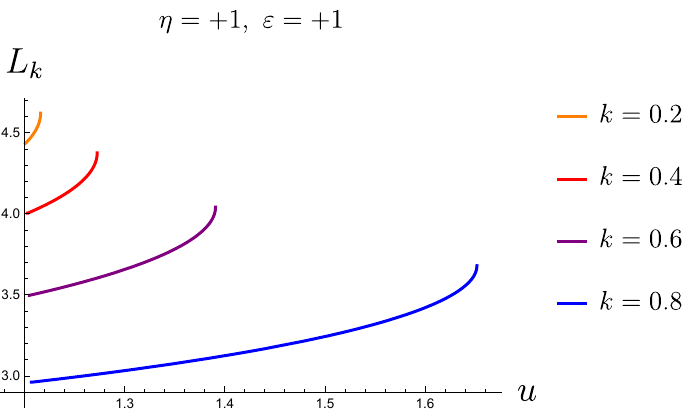}}\hfill\subfloat[$k>1$]{\includegraphics[width=0.36\linewidth]{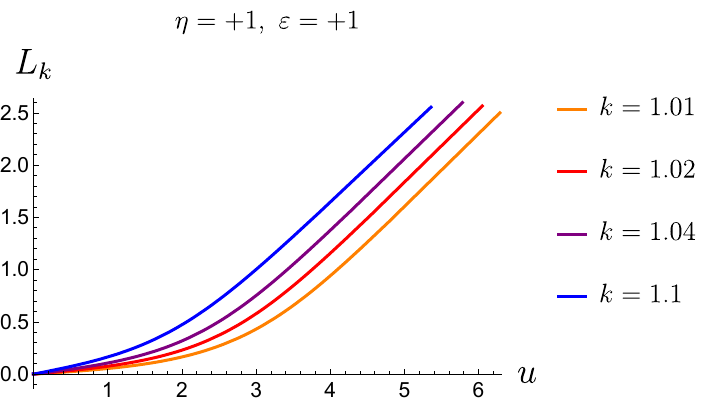}}
    \caption{Left: dS$_2$ Penrose diagrams for (a) $\eta=+1$, $\varepsilon=-1$; (c) $\eta=-1$, $\varepsilon=+1$; (f) $\eta=-1$ $\varepsilon=-1$, (h) $\eta=+1$, $\varepsilon=+1$. Black curves denote geodesic lines. Middle and right: corresponding plots of the CMC slice length $L_k$ for fixed $\bbr=0.6$ (when $\eta=-1$) and $\bbr=1.1$ ($\eta=+1$); for $\varepsilon=-1$ the relation \eqref{eq:dS geo vs CMC} is the same as in AdS$_2$ and the growth is linear at late times for every $k$, while for $\varepsilon=+1$ the growth is hyperfast when $\abs{k}<1$ and linear when $\abs{k}>1$ \eqref{eq:CMC eps plus closed}. The convexity of $L_k(t)$ increases as $\abs{k}$ increases.}
    \label{fig:L_geo_vs_CMC_dS}
\end{figure}

\paragraph{dS CMC slices}
For dS$_2$ space the relation is the same as in AdS$_2$. The dS$_2$ metric in the regions relevant for $\varepsilon=-1$ is minus the AdS$_2$ metric in the corresponding regions, $f\rightarrow-f$ in \eqref{eq:Schwarzschild coordinates}; under $g_{\mu\nu}\rightarrow-g_{\mu\nu}$ the Christoffel symbols, the geodesics, the curves of constant geodesic curvature and the length \eqref{eq:length CMC} are unchanged, and so is $\abs{k}$, since the unit normal is the same vector. The timelike CMC slices between the spacelike Dirichlet boundaries of dS$_2$ are therefore, point by point, the spacelike CMC slices of the AdS$_2$ black hole, and
\begin{equation}\label{eq:dS geo vs CMC}
    \eta=\pm1,~\varepsilon=-1:\quad \sinh\frac{L}{2}=\frac{1}{\sqrt{1+k^2}}\sinh(\frac{\sqrt{1+k^2}}{2}L_k)~,
\end{equation}
with $L$ now given by \eqref{eq:geodesic length etaplus epsminus} or \eqref{eq:geodesic length etaminus epsminus} in the dS$_2$ regions. For the same reason, \eqref{eq:CMC eps plus closed} holds for the spacelike slices with $\varepsilon=+1$ in dS$_2$; the parametric expressions $L_k(\mathfrak{c})$, $t(\mathfrak{c})$ from which we verified it are collected in App.~\ref{app: CMC lengths}.
We display the growth of the CMC slices in dS space for the different $\varepsilon=\pm1$, $\eta=\pm1$ in Fig.~\ref{fig:L_geo_vs_CMC_dS}.

As seen in the figure and in \eqref{eq:CMC eps plus closed}, there are cases that show hyperfast growth \cite{Jorstad:2022mls} (in the same sense as described for AdS black holes above) when $\abs{k}<1$ and $\varepsilon=+1$. In contrast, $L_k$ evolves linearly with time at late times as $\abs{k}>1$. For $\varepsilon=-1$ the growth is linear at late times for every $k$, as in AdS$_2$, with the rate $\sqrt{1+k^2}$ of \eqref{eq:ABL complexity} below.

\begin{figure}
    \centering
    \subfloat[]{\includegraphics[width=0.4\linewidth]{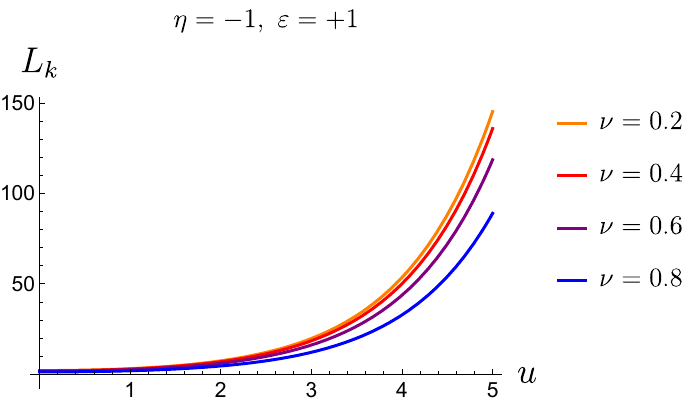}}\hspace{0.5cm}\subfloat[]{\includegraphics[width=0.23\linewidth]{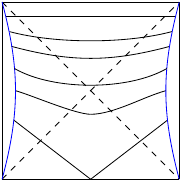}}\\
    \subfloat[]{\includegraphics[width=0.4\linewidth]{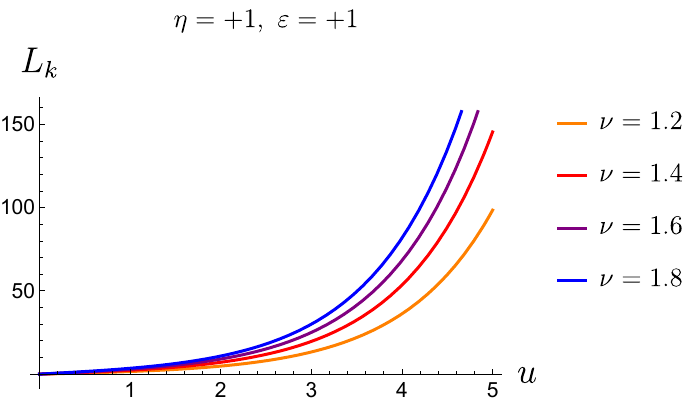}}\hspace{0.5cm}\subfloat[]{\includegraphics[width=0.23\linewidth]{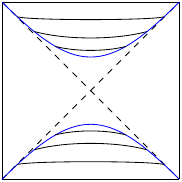}}
    \caption{Left: Exponential late-time growth of $L_{\abs{k}=1}$ in dS$_2$ space for $\varepsilon=+1$ and (a) $\eta=-1$, (c) $\eta=+1$. Right: Corresponding Penrose diagram (b) for a fixed CMC slice with curvature $k=-1$, and (d) $k=-1$ on the top and $k=+1$ bottom Milne patches with a Dirichlet boundary.}
    \label{fig:Lk_dS_k_eq_1}
\end{figure}

As special case, note that when $\abs{k}=1$, for $\eta=\pm1$:
\begin{subequations}\label{eq:L k eps pm1}
\begin{align}
    \varepsilon=-1:\quad L_{\abs{k}=1}=&\sqrt{2}\sinh^{-1}\qty(\sqrt{2}\sinh \frac{L}{2})~,\label{eq:Lk_k1_epsminus}\\
\eta=-1,~\varepsilon=+1:\quad L_{\abs{k}=1}=&2\sin\frac{L}{2}=2\sqrt{1-\bbr^2}\cosh u~,\label{eq:eta minus eps plus Lk}\\
\eta=+1,~\varepsilon=+1:\quad L_{\abs{k}=1}=&2\sin\frac{L}{2}=2\sqrt{\bbr^2-1}\sinh u~,\label{eq:eta plus eps plus Lk}
\end{align}
\end{subequations}
where the first line follows from \eqref{eq:AdS geo vs CMC}, \eqref{eq:dS geo vs CMC} and the last two from \eqref{eq:CMC eps plus closed} (the same expressions before simplification appear in App.~\ref{app: CMC lengths}). For $\varepsilon=-1$, $L_{\abs{k}=1}$ displays late-time linear growth, since $L(t)$ does \eqref{eq:geodesic length growth}. For $\varepsilon=+1$ and $\abs{k}=1$ the CMC length is twice the sine of half the geodesic length, so it grows exponentially, $L_{\abs{k}=1}\simeq\sqrt{\abs{1-\bbr^2}}\,\rme^{u}$, and does not terminate: the CMC slices for the critical value $\abs{k}=1$ reach $\mathcal{I}^\pm$ asymptotically, parallel to it, rather than an accumulation surface, so that the length scales with time in terms of the exponential scale factor of dS space. The case $\eta=-1$ was first encountered in \cite{Aguilar-Gutierrez:2023zqm} albeit not in the closed form above, while \eqref{eq:eta plus eps plus Lk} is new. We show the growth of the lengths \eqref{eq:eta minus eps plus Lk}, \eqref{eq:eta plus eps plus Lk} in Fig.~\ref{fig:Lk_dS_k_eq_1}.

\subsection{CMC-Krylov complexity}\label{ssec:def CMC-K complexity}
We now search for a boundary interpretation of the length of CMC slices \eqref{eq:AdS geo vs CMC} in terms of complexity. The identification of the complexity measure associated to CMC slices in the bulk is also motivated by recent progress in new complexity measures in Krylov space based on the Lanczos algorithm \cite{Camargo:2026szl}.

Consider a complexity measure,
\begin{eqnarray}\label{eq:def CMC-Krylov cmplexity}
    C_{\mfa}(t)\equiv\bra{K_0} \hat{C}_{\mfa}(t)\ket{K_0}~,\quad \hat{C}_{\mfa}(t)\equiv\frac{2}{\lambda\sqrt{1+\mfa}}\text{sinh}^{-1}\qty[\sqrt{1+\mfa}\sinh\frac{\lambda\,\hn(t)}{2}]~,
\end{eqnarray}
where, again, $\lambda$ is taken as a small parameter in the classical limit of the theory of interest; $\ket{K_0}$ $\mathcal{L}$, $\tau$, $\hn(t)$ were introduced in Sec.~\ref{ssec:rev Krylov}; and $\mfa\in\mathbb{R}$ is a free parameter of our complexity carrier.
Several comments are in place.
\begin{itemize}
\item The specific functional expression of $C_{\mfa}$ in
\eqref{eq:def CMC-Krylov cmplexity} is selected so that it mimics the relationship between geodesic and CMC slice lengths when $\varepsilon=-1$ in \eqref{eq:AdS geo vs CMC}. In this way, $\hat{C}_{\mfa}$ \eqref{eq:def CMC-Krylov cmplexity} can be holographically identified with a CMC slice length operator, where $\mfa$ in \eqref{eq:def CMC-Krylov cmplexity} corresponds to the square of the mean curvature of a (spacelike or timelike) CMC slice, at least when $\hat{n}$ admits a holographic interpretation as a geodesic length. We discuss this in Sec.~\ref{ssec:def CMC-K complexity} for the boundary theory in Sec.~\ref{sec:CMC-Krylov complexity}.
\item \eqref{eq:def CMC-Krylov cmplexity} is simply a specific choice of the cost function used in \cite{Balasubramanian:2022tpr} which appears in \eqref{eq:cost}. In this case, one needs to substitute $\ket{B_n}$ by the Krylov basis $\ket{K}_n$, and take 
\begin{eqnarray}
    c_n=\frac{2}{\lambda\sqrt{1+\mathfrak{r}}}\sinh^{-1}\qty(\sqrt{1+\mathfrak{r}}\sinh\frac{\lambda n}{2})~,
\end{eqnarray}
which is a monotonically growing function when $\mathfrak{r}>-1$.
\item {One should also contrast our choice of $\mfa$ in \eqref{eq:def CMC-Krylov cmplexity} with the parameter $K$ in the CAny conjecture \eqref{eq:regions CMC}. In the latter case, $K$ is a parameter determined by a theory-independent gravitational functional determining the precise bulk observable. Similarly, $\mfa$ in \eqref{eq:def CMC-Krylov cmplexity} is again a parameter that defines the specific boundary theory observable.}
\item The definition \eqref{eq:def CMC-Krylov cmplexity} can be evaluated in terms of moments of the chord number operator (i.e.~the Krylov complexity operator with respect to the HH state) from its power series expansion as
\begin{equation}\label{eq:hbar expansion Ca}
    C_{\mfa}=\sum_{m=0}^{\infty}\frac{\lambda^{2m}\expval{\hn(t)^{2m+1}}}{2^{2m}(2m+1)!}
\sum_{k=0}^{m}
\sum_{j=0}^{2k+1}\frac{(-1)^{k+j}\binom{2k}{k}(1+\mfa)^k}
{(2k+1)\,2^{4k+1}}
\binom{2k+1}{j}
(2k+1-2j)^{2m+1}~,
\end{equation}
where the expectation value is taken with respect to $\ket{K_0}$. One should be careful that the above moment expansion also contains powers of $\lambda$. Given that $\expval{n(t)^m}$ generically can be expressed in terms of a $\lambda$ expansion itself, one needs to be careful in the evaluation. It shows, for instance, that at leading order in the semiclassical expansion, CMC-Krylov complexity is completely determined by the semiclassical Krylov complexity; however, its first-order quantum correction is determined by both the first-order correction to $\expval{\hn(t)}$ and the semiclassical value of $\expval{\hn(t)^2}$, corresponding to $\expval{\hn(t)}^2$. A similar reasoning holds for higher $\lambda$ corrections. Thus, Krylov moment methods are required to extract information about the complexity measure \eqref{eq:def CMC-Krylov cmplexity} beyond its classical limit.
\item In the asymptotic boundary limit \eqref{eq:ABL} where the spectrum of $\hat{n}$ and $C_\mfa$ are large, \eqref{eq:def CMC-Krylov cmplexity} simplifies to
\begin{equation}\label{eq:ABL complexity}
    C_\mfa(t)\simeq\frac{1}{\lambda\sqrt{1+\mfa}}\qty(\lambda\hat{n}(t)+\log(1+\mfa))~,
\end{equation}
with the expectation value being with respect to $\ket{K_0}$ in the Lanczos algorithm. This means that $C_{\mathfrak{r}}$ is simply a linear function of Krylov complexity in the asymptotic boundary limit, and one can only find a non-trivial functional difference between them in the finite cutoff case.
\item Next, note that when $\mfa<-1$ there is a cancellation between $\rmi$ factors on the right hand side of \eqref{eq:def CMC-Krylov cmplexity} which takes the form
\begin{eqnarray}\label{eq:Hermitian C a}
    \hat{C}_{\mfa}(t)\equiv\frac{2}{\lambda\sqrt{-(1+\mfa)}}\text{sin}^{-1}\qty[\sqrt{-(1+\mfa)}\sinh\frac{\lambda\,\hn(t)}{2}]~.
\end{eqnarray}
The operator above is non-Hermitian when the argument in $\sin^{-1}$ is larger than 1. This is the branch relevant for the non-perturbative solutions, $\varepsilon=+1$, where $L=\rmi\lambda\expval{\hat{n}}$ turns \eqref{eq:def CMC-Krylov cmplexity} into the relation \eqref{eq:CMC eps plus closed}; the hyperfast growth of \eqref{eq:geodesic length etaminus epsplus}, \eqref{eq:geodesic length etaplus epsplus} discussed in Sec.~\ref{ssec:CMC lenghts main} is reached when the argument of $\sin^{-1}$ exceeds $1$. We adopt a similar perspective to \cite{Aguilar-Gutierrez:2023zqm} and take \eqref{eq:Hermitian C a} to be undefined when the hyperfast behavior is reached. Thus, the late-time behavior of $C_{\mfa}$ changes  substantially with respect to Krylov complexity (i.e.~$\mfa=0$).
\item The definition \eqref{eq:def CMC-Krylov cmplexity} has a wider range of applicability than for holography; it is motivated by bulk considerations with or without a finite cutoff. \eqref{eq:def CMC-Krylov cmplexity} is a valid definition of complexity for any quantum theory, independent of DSSYK or holography, as long as the Krylov basis $\ket{K_n}$ can be built and \eqref{eq:Hermitian C a} remains Hermitian. Moreover, \eqref{eq:AdS geo vs CMC} continues to hold in shockwave geometries, with $L$ the shockwave geodesic length (App.~\ref{sapp:JT shcowkaves}), since the JT metric remains locally AdS$_2$ and the shockwave only shifts the boundaries. Hence $C_{\mfa}$ reproduces the CMC slice length with a shockwave whenever $\lambda\expval{\hat{n}}$ reproduces the corresponding geodesic length, and it inherits the switchback behavior of the latter \cite{Belin:2021bga,Belin:2022xmt}. For the same reason, shockwaves do not distinguish \eqref{eq:def CMC-Krylov cmplexity} from Krylov complexity beyond the functional relation \eqref{eq:AdS geo vs CMC}.
\end{itemize}
\paragraph{Semiclasical evaluation}
In the following, we evaluate CMC-Krylov complexity \eqref{eq:def CMC-Krylov cmplexity} for the HH state 
\begin{equation}\label{eq:HH def }
    \begin{aligned}
\ket{\psi(\tau)}=\rme^{-\tau\hH_{\rm PT}}\ket{\Omega}
    \end{aligned}
\end{equation}
as the reference state in the Lanczos algorithm, with $\ket{K_0}=\ket{\Omega}$ as the seed state, where the evolution is parametrized by complexified time $\tau$, whose generator $\hH_{\rm PT}$ is the chord Hamiltonian of the deformed DSSYK model in the PT formulation \eqref{eq:chord PT H}. We stress that the Lanczos algorithm and the time evolution are generated by $\hH_{\rm PT}$, whose spectrum is the undeformed one \eqref{eq:E theta seed}, and not by $\hH^{(\eta)}_{{\rm PT},y}$ \eqref{eq:PT y}; the latter is not tridiagonal in the chord basis. Since both operators share the eigenstates $\ket{\theta}$, the Krylov basis is the same, and the only difference is the time variable: $t$ below is conjugate to $\hH_{\rm PT}$, which on the bulk side corresponds to Schwarzschild time (see \eqref{eq:tP} and Tab.~\ref{tab:holographic_dictionary}), rather than to the time of the deformed theory conjugate to $E_y(\theta)$. The HH state prepared with $\hH^{(\eta)}_{{\rm PT},y}$ differs from \eqref{eq:HH def } by the replacement $\rme^{-\tau E(\theta)}\rightarrow\rme^{-\tau E_y(\theta)}$ in the $\ket{\theta}$ basis. The Lanczos coefficients $b_n$ take the form $b_n=b(n)$ \eqref{eq:case 1 b}
for both the $T^2$ and $T^2+\Lambda_1$ deformations. For the relevant evaluation, we wish to solve the path integral based on the PT Hamiltonian \eqref{eq:chord PT H},
\begin{eqnarray}
    Z=\int[\rmd \text{L}][\rmd p_n]\exp[\int\rmd\tau\qty(\frac{\rmi}{\sqrt{\lambda}} p_n\,\partial_\tau \text{L}-H_{\rm PT})]~,
\end{eqnarray}
where we take the $\lambda\rightarrow0$ limit to compute the saddle point equations of motion
\begin{align}\label{eq:EOM H}
     \frac{1}{\sqrt{\lambda}}\dv{\text{L}}{t}&=\pdv{H_{\rm PT}}{p_n}~,\quad\frac{1}{\sqrt{\lambda}}\dv{p_n}{t}=-\pdv{H_{\rm PT}}{\text{L}}~,
\end{align}
Note that we are working at the saddle point level of the path integral to recover \eqref{eq:EOM H}, i.e.~the evaluation is valid at $\mathcal{O}(\lambda^0)$ in terms of the parameter $\text{L}$. Thus, the evaluation is insensitive to the quantum correction in the PT potential \eqref{eq:corrected H JT2}.

Using the explicit form of the Hamiltonian in \eqref{eq:chord PT H} we recover
\begin{align}\label{eq:dL dt}
    \frac{1}{\sqrt{\lambda}}\dv{\text{L}}{t}&=\frac{2}{\sqrt{\lambda}}\sin p_n~b(\text{L}),\quad \frac{1}{\sqrt{\lambda}}\dv{p_n}{t}=\frac{2}{\sqrt{\lambda}}\cos p_n~b'(\text{L})~,
\end{align}
Combining \eqref{eq:dL dt} results in
\begin{equation}
\frac{1}{\lambda}\dv[2]{\text{L}}{t}=\frac{2}{\lambda}\dv{\qty(b(\text{L})^2)}{\text{L}}~.\label{eq:ODEs important}
\end{equation}
The differential equation \eqref{eq:ODEs important} is supplemented by boundary conditions
\begin{equation}\label{eq:BDRY cond cutoff}
    \text{L}(0)\equiv \lambda\, C_0(0)~,\quad \text{L}'(0)=0~,
\end{equation}
where the second condition follows from the fact that $\hH_{\rm PT}$ \eqref{eq:chord PT H} is real-symmetric in
the chord basis and thus time-reversal invariant, so that $\expval{\hat{n}(t)}$ is even in $t$. This means that its time derivative at $t=0$ vanishes. See \cite{Aguilar-Gutierrez:2026ogo} for a similar argument. Meanwhile, $ C_0(0)$ in \eqref{eq:BDRY cond cutoff} is a constant determined from initial conditions and conservation of energy. Namely, consider \eqref{eq:dL dt} and the second initial condition in \eqref{eq:BDRY cond cutoff} for $t=0$ implies $p_n(0)=0$. Then, by introducing $p_n(0)=0$ and the first initial condition in \eqref{eq:BDRY cond cutoff} to the conserved energy of the deformed theory, and the parametrization \eqref{eq:E theta seed}, it follows that
\begin{equation}\label{eq:L0}
   \cos\theta=\sqrt{1-\frac{\xi^2}{\cosh^2(\lambda C_0(0)/2)}}~.
\end{equation}
The solution to \eqref{eq:ODEs important} subject to the boundary conditions \eqref{eq:BDRY cond cutoff} allows us to evaluate \eqref{eq:CMC-Krylov length} as
\begin{eqnarray}\label{eq:CMC-Krylov length}
C_{\mfa}(t)=\frac{2}{\lambda\sqrt{1+\mfa}}
    {\rm sinh}^{-1}\qty(\sqrt{\qty(1+\mfa)\qty(\frac{\xi^2}{\sin^2\theta}-1)}~\cosh(u))~,\quad\eta=+1~,
\end{eqnarray}
where $u=\sin\theta~t$. The time dependence of \eqref{eq:CMC-Krylov length} indeed agrees with that of the CMC length \eqref {eq:AdS geo vs CMC} and the corresponding geodesics \eqref{eq:geodesic length etaplus epsminus}.\footnote{\eqref{eq:CMC-Krylov length} can be used to evaluate the semiclassical two-point in \eqref{eq:2pt} as
\begin{equation} 
\eval{G_2(\tau^*,\tau)}_{\tau=\frac{\beta}{2}+\rmi t}=\xi^{2\Delta}/\cosh^{2\Delta}(L(t)/2)~,
\end{equation}
which, similar to the asymptotic boundary case, describes the expectation value of the generator of chord number moments \cite{Xu:2024gfm}.} This might be expected since the chord Hamiltonian \eqref{eq:chord PT H} reproduces the finite-cutoff sine dilaton gravity ADM Hamiltonian, which relies on the canonical length for the $\varepsilon=-1$ cases.

\paragraph{Case $\varepsilon=-1$, $\eta=-1$}
On the other hand, we also wish to holographically describe the length associated to the $T^2+\Lambda_1$ deformations (where $\eta=-1$) in \eqref{eq:geodesic length etaminus epsminus} and \eqref{eq:AdS geo vs CMC}. For this purpose, consider the HH state preparation  where we shrink in half the size of the thermal circle, i.e.~$\beta\rightarrow\beta/2$, and we evolve in real time without any other modification $t\rightarrow t$, which is described by the state 
\begin{eqnarray}\label{eq:eta minus state}
    \ket{\psi}=Z(\beta/2)^{-1/2}\exp(-\qty(\frac{\beta}{4}-\rmi t)\hH_{\rm PT})\ket{\Omega}~.
\end{eqnarray}
Alternatively, this preparation of state can be seen as setting $\beta\rightarrow\beta$ and $t\rightarrow t+\rmi \beta/4$ in the HH state. The state \eqref{eq:eta minus state} is an HH state at twice the temperature, and its chord-number growth is again of the form \eqref{eq:CMC-Krylov length} with $\xi/\sin\theta>1$. The interior length \eqref{eq:geodesic length etaminus epsminus} is reached from \eqref{eq:CMC-Krylov length} by the continuation $u\rightarrow u+\rmi\pi/2$ together with $\bbr\rightarrow\bbr<1$, i.e.~$\xi<\sin\theta$, which lies outside the range of the saddle \eqref{eq:L0}. On the boundary side, this corresponds to the preparation of the state via $t\rightarrow t+\rmi\beta/4$ with $\beta=2\pi/\sin\theta$, which takes the semiclassical trajectory from Rindler-AdS to the black hole interior (and similarly from the static patch to the Milne patch in dS space); we leave a derivation of the $\xi<\sin\theta$ regime from the chord description to future work. Note that this is a general feature when relating physical time in the boundary theory to the boundary time in the bulk in Schwarzschild coordinates.\footnote{Relating the boundary state preparation outside vs inside the AdS black hole does not rely on the particular features of lower dimensions; it is not restricted to the DSSYK model, nor to $T^2$ deformations.}

\paragraph{Summary and comments on holographic complexity}
To summarize, in this section we started from a specific bulk observable, the length of a CMC slice without matter at finite cutoff \eqref{eq:length CMC} for different regions in the bulk (Figs.~\ref{fig:L_geo_vs_CMC_AdS} and \ref{fig:L_geo_vs_CMC_dS}), and we showed that it can be holographically interpreted in the boundary as a specific complexity measure in Krylov space \eqref{eq:def CMC-Krylov cmplexity}, inside the family of cost functions introduced by \cite{Balasubramanian:2024lqk}, at least in the semiclassical limit and without matter. The important point is that the geometric observable need not be restricted to the volume of a maximal slice. A choice of the cost function other than Krylov complexity can select a different geometric carrier from the volume of a maximal slice, corresponding to the length of a CMC slice in our case. 

The CMC slice and the volume are functions of each other in this simple case \eqref{eq:AdS geo vs CMC}, which is reflected in the dual observable at the semiclassical level.  Nevertheless, depending on the parameter and on the region of the gravitational geometry being probed, the resulting complexity can display late-time linear, exponential, or hyperfast growth. In particular, the boundary deformation can also grant access to regimes associated with the dS static patch \cite{Gorbenko:2018oov,Coleman:2020jte,Silverstein:2022dfj,Batra:2024kjl,Chang:2025ays} and with the interior of an AdS black hole \cite{AliAhmad:2025kki,Griguolo:2026eoi,Aguilar-Gutierrez:2026ogo}, thereby extending the usual geodesic/Krylov correspondence to other geometric observables/complexity measures, in agreement with bulk observables conjectured to be dual to complexity in higher dimensions in the asymptotically (A)dS settings \cite{Jorstad:2023kmq,Aguilar-Gutierrez:2023pnn,Aguilar-Gutierrez:2023zqm,Aguilar-Gutierrez:2023tic,Aguilar-Gutierrez:2024xfi}, and regions beyond the standard asymptotic boundary description.

This discussion suggests a route toward identifying analogous observables in higher-dimensional holographic complexity \cite{Belin:2021bga,Belin:2022xmt}, and timelike holographic complexity measures \cite{Alishahiha:2025xml}. We provide further discussion about these points in the outlook Sec.~\ref{ssec:outlook}.

\section{Finite cutoff wormhole states with an end-of-the-world brane}\label{sec:bulk with matter}
In this section, we make first steps towards describing finite-cutoff holography with an ETW brane. ETW branes are an important ingredient to model the microstates of evaporating AdS$_2$ black holes \cite{Penington:2019kki}, and display several interesting properties.  Static ETW branes in AdS$_2$ space appear when its ADM energy becomes negative, which generate eternal traversable wormholes \cite{Maldacena:2018lmt} rather than AdS$_2$ black holes with dynamical ETW brane behind the horizon \cite{Gao:2021uro}. For $m_{\rm JT}>0$ the geodesic brane \eqref{eq:q imp} lies behind the horizon, in the left patch; for $m_{\rm JT}<0$, the case relevant for bound states, it lies in the same exterior as the cutoff boundary, reaching $\Phi_{\max}=\sqrt{\Phi_h^2+m_{\rm JT}^2}$ at $t=0$, and a positive geodesic length requires $\Phi_b^2>\Phi_h^2+m_{\rm JT}^2$, i.e.~$E_{\rm ADM}<U(0)$ in the notation of Sec.~\ref{ssec:1st approach QJT}. From the boundary side, they can be described by deformations of the DSSYK model \cite{Rajgadia:2026ask,Cao:2025pir,Aguilar-Gutierrez:2026ogo}. This makes ETW branes in the finite-cutoff bulk theory a natural setting to investigate the effect of deformations of the boundary theory in the holographic dictionary. There are several choices of boundary conditions one might consider \cite{Goel:2020yxl}. We will focus on CMC ETW branes with Neumann boundary conditions, which are the most studied ones in sine dilaton gravity \cite{Blommaert:2025avl,Cui:2025sgy} and DSSYK \cite{Okuyama:2023byh,Cao:2025pir,Aguilar-Gutierrez:2025hty,Aguilar-Gutierrez:2026nmd,Berkooz:2025ydg}. A summary of the finite-cutoff holographic dictionary identified in this section is shown in Tab.~\ref{tab:holographic_dictionary}.

\paragraph{Outline}
We begin Sec.~\ref{ssec:ETW branes} by studying the ADM Hamiltonian of finite-cutoff JT gravity with an ETW brane for the analog of the $\varepsilon=-1$ branches in the previous sections. We move on to the sine dilaton gravity generalization in Sec.~\ref{ssec:SDG ETW} (complementary technical details are presented in App.~\ref{app:distance brane}). This motivates our follow-up chord interpretation in Sec.~\ref{ssec:chord interpretation ETW brane} in terms of a deformation of the DSSYK model, whose microscopic SYK description without double-scaling limit is explained in App.~\ref{app:chord ETW brane}. We then analyze quantum aspects of its JT gravity limit in Sec.~\ref{ssec:1st approach QJT}, in particular regarding the choice of self-adjoint extension for the respective Hamiltonian. We incorporate an improvement term proposed by \cite{Griguolo:2026eoi} which provides a natural set of boundary conditions in Sec.~\ref{ssec:scattering ETW}, and study the resulting Hilbert space. At last, we analyze bound state solutions and eternal traversable wormholes in Sec.~\ref{ssec:traversable wormholes}; additional details are spelled out in App.~\ref{sapp:bound states}.

\subsection{Classical ETW brane in finite-cutoff JT gravity}\label{ssec:ETW branes}
We work with JT gravity with a CMC ETW brane (see e.g.~\cite{Gao:2021uro,Iliesiu:2020qvm})
\begin{equation}\label{eq:SDG with ETW brane}
\begin{aligned}
    I=I_0-\frac{1}{2\kappa^2}\bigg(&\int _{\mathcal{M}}\rmd^2x\sqrt{g}\qty(\Phi\mathcal{R}+2\Phi)+2\int _{\partial\mathcal{M}}\rmd x\sqrt{h}\qty(\Phi_{b} (K-1))\\
    &-\int_{\rm brane }\rmd x\sqrt{h}(\Phi (K-K_{\rm AdS})-m_{\rm JT})\bigg)~,
\end{aligned}
\end{equation}
where $I_0$ is a topological term, $\mathcal{M}$ is the Euclidean spacetime manifold, $\kappa^2\equiv 8\pi G_N$, $h_{ab}$ the induced metric at either the cutoff boundary $\partial\mathcal{M}$ or the ETW brane with tension $m_{\rm JT}$, and $K_{\rm AdS}$ is the fixed mean curvature value for the trajectory of the ETW brane. The extremization of the action, $\delta I=0$ leads to the usual equations of motion in JT gravity plus boundary conditions in the radial cutoff (where we impose Dirichlet boundary conditions), and for the brane (see e.g.~App.~A in \cite{Gao:2021uro})
\begin{eqnarray}\label{eq:brane conditions}
    n^b\partial_b\Phi=m_{\rm JT}+K_{\rm AdS}\Phi~,\quad K=K_{\rm AdS}~,
\end{eqnarray}
where $n^a$ denotes the unit-norm vector orthogonal to the brane trajectory, which we take to point away from the cutoff boundary, i.e.~outward from the region bounded by the brane and the Dirichlet boundary (in the global coordinates \eqref{eq:coordinate useful} used in Sec.~\ref{ssec:traversable wormholes}, $n^b=(-1,0)$). This orientation fixes the signs of $K_{\rm AdS}$ and $m_{\rm JT}$ throughout the paper. As in previous parts, the AdS black hole has a Dirichlet boundary at $\Phi=\Phi_b$ and a black hole horizon at $\Phi=\Phi_h$.

\begin{figure}
    \centering
    \subfloat[$\varepsilon=-1$, $\eta=+1$]{\includegraphics[width=0.3\linewidth]{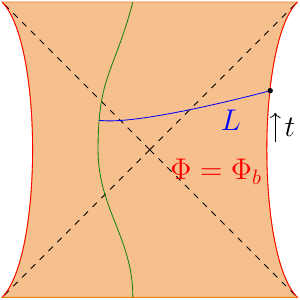}}\hspace{2cm}\subfloat[$\varepsilon=-1$, $\eta=-1$]{\includegraphics[width=0.3\linewidth]{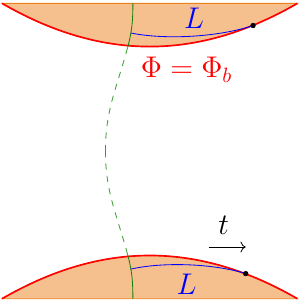}}\\
    \subfloat[$\varepsilon=+1$, $\eta=-1$]{\includegraphics[width=0.3\linewidth]{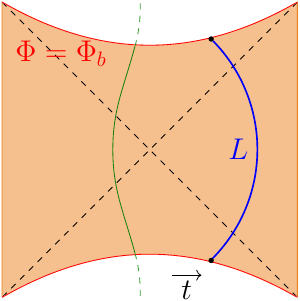}}\hspace{2cm}\subfloat[$\varepsilon=+1$, $\eta=+1$]{\includegraphics[width=0.3\linewidth]{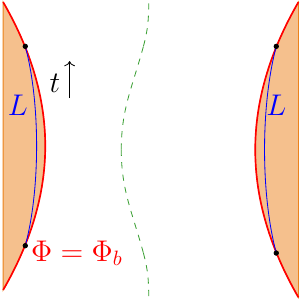}}
\caption{Geodesic lengths in AdS$_2$ for the $\varepsilon=\pm1$, $\eta=\pm1$ configurations with a timelike ETW brane, where the boundary anchoring point is translated symmetrically in the (b, c, d) cases. The dashed green line represents the continuation of the brane trajectory past the cutoff boundary at $\Phi=\Phi_b$ (red solid curve). Note that the cases with $\varepsilon=+1$ do not interact with the brane.}
\label{fig:ETW_brane_regions}
\end{figure}

In the following, we derive the ADM Hamiltonian of finite-cutoff JT with an ETW brane from the geodesic length between a Dirichlet boundary and the ETW brane with tension for the different $\varepsilon=\pm1$ and $\eta=\pm 1$ regions as in the previous sections, which we illustrate in Fig.~\ref{fig:ETW_brane_regions}.\footnote{In dS JT gravity, the ETW branes we consider are spacelike, while for a timelike brane there would be violation of the null energy condition \cite{deBoer:2025tmh}. Other connections between dS$_2$ gravity and ETW branes appear in \cite{Fujiki:2026pzs,Hoback:2026yqj,Aguilar-Gutierrez:2023tic}.}
We also work with more generic CMC slice lengths in App.~\ref{sapp:CMC slice length}; however, it does not result in analytic expressions, and they will not be considered in this section.

First, consider the configuration with $\varepsilon=-1$, $\eta=+1$ (i.e.~the perturbative branch of the $T^2$ deformation). In App.~\ref{app:distance brane}, we find that the geodesic length between the finite cutoff boundary and the brane is given by
\begin{align}\label{eq:main text L brane}
    &\sinh (L+\sigma_0)=\frac{m_{\rm JT}\Phi_b+\sqrt{(\Phi_b^2-\Phi_h^2)(m_{\rm JT}^2+(1-K_{\rm AdS}^2)\Phi_h^2)}~\cosh(\Phi_h t)}{\Phi_h^2\sqrt{1-K_{\rm AdS}^2}}~,
\end{align}
where we use Schwarzschild coordinates \eqref{eq:Schwarzschild coordinates}, and $K_{\rm AdS}=-\tanh\sigma_0$, with $\sigma_0$ the signed geodesic distance from the geodesic with the same charge to the CMC brane, positive when the brane is displaced towards the cutoff boundary (App.~\ref{app:distance brane}); the sign follows from the orientation of $n^a$ in \eqref{eq:brane conditions}. For a geodesic ETW brane, $\sigma_0=0$.

Next, we search for a canonical momentum $P$ such that the conserved energy of the system remains $H=\Phi_h^2$, that is
\begin{equation}
2\Phi_h=\pdv{L}{t}\pdv{P}{\Phi_h}-\pdv{L}{\Phi_h}\pdv{P}{t}~,
\end{equation}
such that $\rmd L\wedge \rmd P=2\Phi_h\,\rmd t\wedge \rmd\Phi_h$. We find that the corresponding momentum takes the form
\begin{equation}
P\equiv2\dv{L}{t}= \frac{2\Phi _h\sqrt{\left(\Phi _b^2-\Phi _h^2\right) \left(\Phi _h^2(1-K_{\rm AdS}^2)+m_{\rm JT} ^2\right)}\,\sinh \left(\Phi_h\,t \right)}{\sqrt{\Phi _h^4\qty(1-K_{\rm AdS}^2)+\left(\sqrt{\left(\Phi
   _b^2-\Phi _h^2\right) \left(\Phi _h^2(1-K_{\rm AdS}^2)+m_{\rm JT} ^2\right)} \cosh \left(\Phi _h\,t\right)+m_{\rm JT}  \Phi _b\right){}^2}} ~.
\end{equation}
The corresponding ADM Hamiltonian contains a Scarf \cite{scarf1958new} type II shape-invariant potential (see e.g.~\cite{Ioffe:2012br,Levai:2017vdu,Ioffe:2011wr})
\begin{equation}\label{eq: Hamilt ETW}
    H_{\rm JT-ETW}=\frac{P^2}{4}+U(L+\sigma_0)~, \quad U(L)\equiv\frac{2\tilde{m}_{\rm JT}\Phi_b\sinh L+\Phi_b^2-\tilde{m}_{\rm JT}^2}{\cosh^2L}~,
\end{equation}
where we denoted $\tilde{m}_{\rm JT}\equiv m_{\rm JT}/\sqrt{1-K_{\rm AdS}^2}$. Thus, a CMC brane with $K_{\rm AdS}\neq0$ only shifts the argument of the potential by the constant $\sigma_0$ and rescales the tension; in the rest of this section we focus on geodesic ETW branes, $K_{\rm AdS}=0$, where $\sigma_0=0$ and $\tilde{m}_{\rm JT}=m_{\rm JT}$. \eqref{eq: Hamilt ETW} has a conserved energy $H_{\rm JT-ETW}=\Phi_h^2$. In particular, we recover the JT Hamiltonian with a Morse potential studied by \cite{Gao:2021uro} in the asymptotic boundary limit when $L\gg1$, for which we need to regularize the length $L\rightarrow L+\log(2\Phi_b)$ and perform a rescaling $m_{\rm JT}\rightarrow m_{\rm JT}/2$,
\begin{equation}\label{eq: H brane GJK}
H_{\rm JT-ETW}\rightarrow\frac{P^2}{4}+m_{\rm JT}\rme^{-L}+\rme^{-2L}~.
\end{equation}
Similar to Sec.~\ref{ssec:CMC lenghts main}, we find the geodesic length between a finite cutoff Dirichlet boundary and the ETW brane for the $\varepsilon=-1$, $\eta=-1$ configuration (i.e.~the perturbative branch of the $T^2+\Lambda_1$ deformation) is given by the $t\rightarrow t+\rmi \pi/(2\Phi_h)$ analytic continuation of \eqref{eq:main text L brane}
\begin{eqnarray}\label{eq:L transformed}
    \sinh (L+\sigma_0)=\frac{m_{\rm JT}\Phi_b+\sqrt{(\Phi_h^2-\Phi_b^2)(m_{\rm JT}^2+(1-K_{\rm AdS}^2)\Phi_h^2)}~\sinh(\Phi_h t)}{\Phi_h^2\sqrt{1-K_{\rm AdS}^2}}~.
\end{eqnarray}
The conjugate momentum is again $P=2\rmd L/\rmd t$, and the ADM Hamiltonian is the same one as for the $\eta=+1$ case in \eqref{eq: Hamilt ETW}.

In contrast, we find that for the configurations with $\varepsilon=+1$, while they still exist, the presence of a timelike ETW brane does not modify the corresponding geodesic lengths (one would need to introduce a spacelike brane instead, see Fig.~\ref{fig:ETW_brane_regions}), and thus, the ADM Hamiltonian is still the one in previous sections (namely \eqref{eq:conserved quantity} for $\varepsilon=+1$).

\subsection{Classical ETW brane in finite-cutoff sine dilaton gravity}\label{ssec:SDG ETW}
Similar to Sec.~\ref{ssec:bulk interpretation in (dS) JT at finite cutoff}, we wish to upgrade the lessons from JT gravity to sine dilaton gravity with an ETW brane \cite{Blommaert:2025avl} at finite cutoff, for which we use the full energy spectrum in \eqref{eq:E theta seed}. This will allow us to identify the boundary dual theory as a deformation of the DSSYK model (Sec.~\ref{ssec:chord interpretation ETW brane}). The theory of interest is the same as \eqref{eq:action SDG} with the addition of ETW brane terms as \cite{Blommaert:2025avl},\footnote{From the boundary perspective, there are many more extensions that one can consider \cite{Aguilar-Gutierrez:2026nmd}; we will take the one with the best understood bulk interpretation \cite{Blommaert:2025avl}.} 
\begin{eqnarray}\label{eq:SDG action ETW}
\begin{aligned}
    I=&-\frac{1}{16\pi G_N}\qty(\int_{\mathcal{M}}\rmd^2 x\sqrt{g}\qty(\Phi\mathcal{R}+2\sin\Phi)+2\int_{\rm bdry}\rmd x\sqrt{h}\qty(\Phi_b K-\rmi~\rme^{-\rmi\Phi_b/2}))\\
    &-\frac{1}{8\pi G_N}\qty(\overline{\mathcal{Y}}\int_{\rm brane }\rmd x\sqrt{h}~\rme^{\rmi\Phi/2}+\mathcal{Y}\int_{\rm brane }\rmd x\sqrt{h}~\rme^{-\rmi\Phi/2})~,
\end{aligned}
\end{eqnarray}
where ``bdry'' denotes the $\Phi=\Phi_b$ Dirichlet boundary, while $\mathcal{Y}$ and $\overline{\mathcal{Y}}$ are related to the brane tensions $m_{\rm JT}$.  The solutions to the equations of motion in terms of the dilaton $\Phi$ are exactly of the same form as \eqref{eq:sine dilaton metric}. As shown by \cite{Blommaert:2025avl} in the ``semi-open channel'', the problem of geodesic lengths between an AdS$_2$ boundary and the ETW brane takes exactly the same form as in JT gravity with $\Phi_h=\sin\theta$ and $\tau=\beta/2+\rmi t$. Then, based on the JT geodesic length \eqref{eq:main text L brane} and using the substitution $m_{\rm JT}\rightarrow\overline{\mathcal{Y}}+\mathcal{Y}\cos\theta$ in \cite{Blommaert:2025avl} (3.20), we have that in sine dilaton gravity the finite-cutoff geodesic length between a Dirichlet boundary and the ETW brane is (see App.~\ref{sapp:details length SDG ETW} for details),
\begin{align}\label{eq:L total}
    &\sinh L=\frac{\xi(\overline{\mathcal{Y}}+\mathcal{Y}\,\cos\theta)+\sqrt{(\xi^2-\sin^2\theta)((\overline{\mathcal{Y}}+\mathcal{Y}\,\cos\theta)^2+(1+\mathcal{Y}^2)\sin^2\theta)}~\cosh(\sin\theta~t)}{\sqrt{1+\mathcal{Y}^2}\sin^2\theta}~.
\end{align}
Here $L$ stands for the combination $L_{\rm geo}+\sigma_0$ of \eqref{eq: Hamilt ETW}, with $L_{\rm geo}$ being the geodesic length, and $\sigma_0\equiv\rmi\tan^{-1}\mathcal{Y}$ for $K_{\rm AdS}=-\rmi\mathcal{Y}$ \eqref{eq: subst SDG brane}, so that $\sinh(L_{\rm geo}+\sigma_0)=(\sinh L_{\rm geo}+\rmi\mathcal{Y}\cosh L_{\rm geo})/\sqrt{1+\mathcal{Y}^2}$. For the canonical analysis, we find it convenient to work instead with the real variable $L$ defined by the canonical transformation $\sinh L\rightarrow\sqrt{1+\mathcal{Y}^2}\sinh L$ of \eqref{eq:L total}, i.e.~$\sinh L\equiv\sinh L_{\rm geo}+\rmi\mathcal{Y}\cosh L_{\rm geo}$, which satisfies \eqref{eq:L total} without the factor $\sqrt{1+\mathcal{Y}^2}$; this is the variable appearing in \eqref{eq:H SDG ETW} and the one reproduced by Krylov complexity in Sec.~\ref{ssec:chord interpretation ETW brane}. We can then find the momentum $P$ conjugate to $L$ (i.e.~satisfying the classical Poisson brackets $\qty{L,P}_{\rm P.B}=1$), which satisfies the following symplectic form \eqref{eq:simpletic vanilla} with $L\rightarrow 2L$ since we evaluate the one-sided geodesic length \eqref{eq:main text L brane} instead of the two-sided one in Sec.~\ref{ssec:bulk interpretation in (dS) JT at finite cutoff}. Similar to footnote 29 of \cite{Blommaert:2025avl}, by writing $\rmd L/\rmd t=\rmd H/\rmd P$ in terms of $L$ and $\cos\theta$, we find that the solution is,\footnote{Similar to Sec.~\ref{ssec:bulk interpretation in (dS) JT at finite cutoff} there is an ambiguity $P\rightarrow P+f(L)$ with $f(L)$ any real function that only depends on $L$; we have chosen the one that leads to agreement with the chord Hamiltonian \eqref{eq:brane chord H}.}
\begin{equation}
\tan P=\frac{\sqrt{(\xi^2-\sin^2\theta)((\overline{\mathcal{Y}}+\mathcal{Y}\,\cos\theta)^2+(1+\mathcal{Y}^2)\sin^2\theta)}}{\sin\theta\qty(\cos\theta\,\cosh^2L-\mathcal{Y}(\overline{\mathcal{Y}}-\xi\sinh L))}\cosh L\,\sinh(\sin\theta~t) ~,
\end{equation}
which then, by inverting $\cos\theta$ in terms of $L$ and $P$ implies that the corresponding ADM Hamiltonian with conserved energy $-\cos\theta/(8\pi G_N)$ is,
\begin{equation}\label{eq:H SDG ETW}
    H(L,P)=\frac{\mathcal{Y}\,(\overline{\mathcal{Y}}-\xi\sinh L)+\sqrt{\big(\cosh^2L+\mathcal{Y}^2\big)\big(\cosh^2L-\xi^2+\overline{\mathcal{Y}}^2-2\xi\overline{\mathcal{Y}}\sinh L\big)}\;\cos P}{-8\pi G_N\cosh^2L}~.
\end{equation}
There are several comments in place.
\begin{itemize}
    \item \eqref{eq:H SDG ETW}  indeed agrees with the two-sided finite-cutoff sine dilaton gravity case \eqref{eq:vanilla SDG finite cutoff} when $\mathcal{Y}=\overline{\mathcal{Y}}=0$, and $L\rightarrow L/2$ since the geometry in this section is one-sided.
    \item Once we consider the asymptotic boundary limit, $L\gg1$, which can be regularized by taking $2\xi\rme^{-L}\rightarrow \rme^{-L}$, so that we obtain
\begin{equation}\label{eq:new H SDG ETW}
H(L,P)\rightarrow\frac{-1}{8\pi G_N}\qty(\sqrt{1-2\overline{\mathcal{Y}}\rme^{-L}-\rme^{-2L}}\;\cos P-\mathcal{Y}\rme^{-L})~,
\end{equation}
which is in exact agreement with the previous sine dilaton gravity literature \cite{Blommaert:2025avl} (3.35).
\item From \cite{Berkooz:2025ydg,Blommaert:2025avl,Aguilar-Gutierrez:2026nmd}, we know that \eqref{eq:new H SDG ETW}  can be put in a form that encodes the recurrence relations of the Al-Salam Chihara polynomials \cite{koekoek2010hypergeometric} by rescaling the length parameter and choosing the $\mathcal{Y}$ and $\overline{\mathcal{Y}}$ constants appropriately. We discuss its generalization in the finite cutoff case in App.~\ref{app:finite cutoff ASC}.
\item We can show that the finite-cutoff JT gravity theory with the ETW brane \eqref{eq: Hamilt ETW} is successfully recovered from the sine dilaton gravity case in a similar way to our discussion in Sec.~\ref{ssec:bulk interpretation in (dS) JT at finite cutoff}. Using $8\pi G_N=\sqrt{1-q}/2\simeq\sqrt{\lambda}/2$ \eqref{eq:dict 1}, we implement the following rescaling of the parameters
\begin{equation}\label{eq:low energy lim ETW}
    \mathcal{Y}\rightarrow\lambda \mathcal{Y}~,\quad \overline{\mathcal{Y}}\rightarrow\lambda \overline{\mathcal{Y}}~,\quad \xi\rightarrow\lambda \Phi_b~,\quad P\rightarrow \lambda \frac{P}{2}~,\quad \lambda\hat{n}\rightarrow2\hat{L}~,
\end{equation}
where the last replacement refers to the chord description of Sec.~\ref{ssec:chord interpretation ETW brane} and is canonical, $\{2L,P/2\}=\{L,P\}$, corresponding to the low-energy dynamics in \eqref{eq: Hamilt ETW} as $\lambda\rightarrow0$. Expanding to $\mathcal{O}(\lambda^2)$ we find
\begin{equation}
    \sqrt{1-q} H+2=\lambda^2\Big[\frac{P^2}{4}+\frac{\Phi_b^2+2\Phi_b(\mathcal Y+\overline{\mathcal Y})\sinh L-(\mathcal Y+\overline{\mathcal Y})^2}{\cosh^2 L}\Big]+\mathcal{O}(\lambda^3)~,
\end{equation}
from which we recognize precisely the term $\lambda^2\hH_{\rm JT-ETW}$ \eqref{eq: Hamilt ETW} under the identification $m_{\rm JT}=\mathcal{Y}+\overline{\mathcal{Y}}$, which agrees with the relation between sine dilaton and JT gravity in App.~\ref{sapp:details length SDG ETW} : $m_{\rm JT}=\mathcal{Y}\cos\theta+\overline{\mathcal{Y}}$ when $\theta\rightarrow0$ in the leading order low-energy regime. Thus, we recover finite-cutoff JT gravity as the low-energy limit of sine dilaton gravity for the specific identification above and at the classical level. \footnote{On the other hand, at the quantum level, one also needs to take into account additional factors $\lambda$ due to operator ordering. This may require promoting $\mathcal{Y}$ and $\overline{\mathcal{Y}}$ to be $\lambda$-dependent, as discussed in the asymptotic boundary case by \cite{Blommaert:2025avl,Aguilar-Gutierrez:2026nmd}.}
\end{itemize}
Below, we will construct a chord Hamiltonian \eqref{eq:brane chord H} in terms of $\hat{n}$ and $\hat{p}_n$ \eqref{eq:p n def} that reproduces the features of \eqref{eq:H SDG ETW}, providing an explicit realization of finite-cutoff holography.

\subsection{Chord Hamiltonian with an ETW brane}\label{ssec:chord interpretation ETW brane}
We now develop a deformation of the DSSYK chord Hamiltonian, which reproduces the ADM Hamiltonian of finite cutoff-sine dilaton gravity with an ETW brane \eqref{eq:H SDG ETW}. The details of the microscopic SYK deformation generating the chord Hamiltonian are presented in App.~\ref{app:chord ETW brane}. 
The proposed Hamiltonian deformation is 
\begin{equation}\label{eq:brane chord H}
\hH_{\rm ETW}=\frac{-1}{\sqrt{1-q}}\qty(b_{\rm ETW}(\hat{n})\rme^{-\rmi \hat{p}_n}+\rme^{\rmi \hat{p}_n}b_{\rm ETW}(\hat{n})+a_{\rm ETW}(\hat{n}))~,
\end{equation}
where
\begin{equation}\label{eq:lanczos ETW}
\begin{aligned}
    a_{\rm ETW}(\hat{n})&=\frac{2\mathcal{Y}\,(\overline{\mathcal{Y}}-\xi\sinh (\lambda \hat{n}/2))}{\cosh^2(\lambda \hat{n}/2)}~,\\
    b_{\rm ETW}(\hat{n})&=\sqrt{\qty(1+\frac{\mathcal{Y}^2}{\cosh^2(\lambda \hat{n}/2)})\qty(1+\frac{\overline{\mathcal{Y}}^2-\xi^2}{\cosh^2(\lambda \hat{n}/2)}-2\xi\overline{\mathcal{Y}}\frac{\tanh (\lambda \hat{n}/2)}{\cosh (\lambda \hat{n}/2)})}~,
\end{aligned}
\end{equation}
with $\mathcal{Y}$ and $\overline{\mathcal{Y}}$ being constants. The argument $\lambda\hat{n}/2$ is chosen such that $\lambda\hat{n}$ is identified with twice the one-sided length $L$ of \eqref{eq:H SDG ETW}, $\lambda\hat{n}\rightarrow2\hat{L}$, $\hat{p}_n\rightarrow\lambda\hat{P}/2$, which is a canonical identification, $\{2L,P/2\}=\{L,P\}$, and makes the classical limit of \eqref{eq:brane chord H} coincide with \eqref{eq:H SDG ETW} with its symplectic form (cf.~\eqref{eq:low energy lim ETW}). We will only consider parameters such that the radicand in \eqref{eq:lanczos ETW} is non-negative $\forall n\in\mathbb{Z}_{\geq0}$ (i.e.~$\xi\leq1$, or $\xi>1$ with $\xi\overline{\mathcal Y}<0$ and $\overline{\mathcal Y}^2\geq\xi^2-1$), and $\hat{n}$, $\hat{p}_n$ are the chord number and momentum operators acting on the chord number basis $\qty{\ket{n}}$, as usual $\hat{n}\ket{n}=n\ket{n}$, $\rme^{\mp\rmi\hat{p}}\ket{n}=\ket{n\pm 1}$. It is important to include improvement terms in the Hamiltonian such that its low-energy limit has a natural choice of boundary conditions. We discuss this in more detail in Secs.~\ref{ssec:1st approach QJT} and \ref{ssec:scattering ETW}. Meanwhile, in App.~\ref{app:finite cutoff ASC} we analyze the wavefunctions $\bra{\theta}\ket{n}$ associated with the eigenvalue problem of \eqref{eq:brane chord H} in terms of orthogonal polynomials.

In the remainder of this subsection, we consider Krylov complexity for the HH state,
\begin{equation}\label{eq:complexity DSSYK brane}
C_0(t)=\frac{1}{\bra{\psi}\ket{\psi}}\sum_{n=0}^\infty n\abs{\bra{\psi}\ket{n}}^2~,\quad \ket{\psi}=\rme^{-\tau\hH_{\rm ETW}}\ket{n=0}~,
\end{equation} 
with $\tau=\beta/2+\rmi t$. We can perform its evaluation using the saddle-point methods in Sec.~\ref{ssec:def CMC-K complexity}. Following the same steps, with $\text{L}\equiv \lambda C_0$, which is dual to twice the one-sided length of \eqref{eq:H SDG ETW}, we find that in the semiclassical limit ($C_0\simeq\mathcal{O}(1/\lambda)$) we have
\begin{equation}\label{eq:ODE ETW}
\dv[2]{\text{L}}{t}=2\dv{\text{L}}\qty(b(\text{L})^2+a(\text{L})\qty(\cos\theta -\frac{1}{4}a(\text{L})))~,
\end{equation}
where, again, we denote $\text{L}\equiv\lambda C_0$, and we represent $b(\text{L})\equiv b_{\rm ETW}(\text{L}/\lambda)$, $a(\text{L})\equiv a_{\rm ETW}(\text{L}/\lambda)$ in the semiclassical limit, where the Lanczos coefficients in this problem were introduced in \eqref{eq:lanczos ETW}. \eqref{eq:ODE ETW} is subject to the initial conditions \eqref{eq:BDRY cond cutoff} for the same arguments, however, the initial value $\text{L}(t=0)$ determined by the inverse temperature $\beta$ is, from energy conservation, given by
\begin{equation}
2\cos\theta=2b(\text{L}(t=0))+a(\text{L}(t=0))~.
\end{equation} 
From \eqref{eq:ODE ETW} with the above initial conditions, the Krylov complexity of the HH state in the semiclassical limit we recover
\begin{equation}\label{eq:complexity simple case}
\begin{aligned}
&\sinh\frac{\lambda C_0(t)}{2}\\
&=\frac{\xi(\overline{\mathcal{Y}}+\mathcal{Y}\,\cos\theta)+\sqrt{(\xi^2-\sin^2\theta)((\overline{\mathcal{Y}}+\mathcal{Y}\,\cos\theta)^2+(1+\mathcal{Y}^2)\sin^2\theta)}~\cosh(\sin\theta~t)}{\sin^2\theta}~,    
\end{aligned}
\end{equation}
which reproduces the trajectory of the length variable $L$ of \eqref{eq:H SDG ETW}, i.e.~\eqref{eq:L total} after the canonical transformation $\sinh L\rightarrow\sqrt{1+\mathcal{Y}^2}\sinh L$ discussed below \eqref{eq:L total}, with $\lambda C_0=2L$. This allows one to identify a finite-cutoff holographic dictionary, shown in Tab.~\ref{tab:holographic_dictionary}. 

At last, similar to Sec.~\ref{ssec:def CMC-K complexity}, we can evaluate the semiclassical Krylov complexity of the HH state with a shift $\beta\rightarrow\beta/2$, corresponding to \eqref{eq:complexity simple case} after $ t\rightarrow t+\rmi\pi/(2\sin\theta)$; which would correspond to the $\varepsilon=-1$, $\eta=-1$ case \eqref{eq:L transformed} with $\Phi_h=\sin\theta$ and $\xi=\Phi_b$, $m_{\rm JT}=\overline{\mathcal{Y}}+\mathcal{Y}\cos\theta$ \eqref{eq: subst SDG brane}.

\subsection{Quantization with the minimal operator ordering}\label{ssec:1st approach QJT}
We are now interested in solving the low-energy description of \eqref{eq:brane chord H}, corresponding to canonically quantizing finite-cutoff JT gravity with the ETW brane, whose ADM Hamiltonian appears in \eqref{eq: Hamilt ETW}. We need to promote the length and momentum variables to operators acting on a Hilbert space, with a canonical choice of basis $\qty{\ket{L}}$. There are different choices of how to carry out the quantization which depend on the operator ordering, which must have compatible self-adjoint boundary conditions. Let us begin considering the simplest choice, where we simply put hats to the operators appearing in \eqref{eq: Hamilt ETW}. The Schrödinger equation for the wavefunctions of \eqref{eq: Hamilt ETW} contains a shape-invariant type II Scarf potential,
\begin{equation}\label{eq:schrodinger ETW}
    \hH_{\rm JT-ETW}\psi_s(L)=-\frac{1}{4}\psi''_s(L)+\qty(\frac{2 m_{\rm JT}  \Phi_b \sinh L+\Phi_b^2-m_{\rm JT} ^2}{\cosh^2L})\psi_s(L)=s^2\psi_s(L)~.
\end{equation}
The potential in \eqref{eq:schrodinger ETW} generates scattering states when $s^2>0$ and bound states when $s^2=-s_n^2<0$ with $n\in\mathbb{Z}_{\geq0}$. The most general solution to the Schrödinger equation with the Scarf II potential \eqref{eq:schrodinger ETW} is displayed in App.~\ref{sapp:bound states} (see \eqref{eq:wavefunct}). 

There are important subtleties about the boundary conditions at $L=0$ which distinguishes between the finite cutoff setting and its asymptotic boundary limit. First, as we have seen, the Schrödinger equation with the Scarf II potential \eqref{eq:schrodinger ETW} recovers the Schrödinger equation with a Morse potential (corresponding to \eqref{eq: H brane GJK} with $P^2=-\partial_L^2$) in the asymptotic boundary limit \eqref{eq:ABL}. However, one does not automatically recover the set of physically allowed wavefunctions in the latter case from the former one. The key difference between these problems is the spectral domain for $L$, in the asymptotic boundary case, $L$ is a regularized length with range over the real line $\mathbb{R}$, while in the finite cutoff setting $L$ is unregularized Riemannian distance, so that $L\geq0$. Square integrability is what selects the allowed wavefunction in asymptotic boundary case, where the wavefunctions are taken to be Whittaker W-functions instead of Whittaker M-functions \cite{Gao:2021uro}. 

In the finite cutoff case with minimal operator ordering, $L=0$ is a regular endpoint of \eqref{eq:schrodinger ETW}, and self-adjointness of
$\hat H_{\rm JT-ETW}$ requires a boundary condition there. The most general local choice is the Robin family
\begin{equation}
  \psi_s'(0)=\gamma\,\psi_s(0)\,,\qquad \gamma\in\mathbb R\,,
  \label{eq:Rrobin}
\end{equation}
where $\gamma=\infty$ and $\gamma=0$ correspond to Dirichlet and Neumann boundary conditions respectively. One might expect
the asymptotic boundary limit to depend on $\gamma$, since the general solution $\psi_s=C_1u_1+C_2u_2$ (see \eqref{eq:wavefunct} for explicit expressions) maps to a
combination of $W$- and $M$-Whittaker functions. However, this is not the case. The potential in \eqref{eq:schrodinger ETW} takes the value
$\Phi_b^2-\mJT^2$ at $L=0$, so the boundary is deep inside a classically forbidden region for $s\ll\Phi_b$. The admixture of
the $M$-Whittaker function generated by (\ref{eq:Rrobin}) is then suppressed by the tunneling factor of this barrier, of
order $e^{-2\pi\Phi_b}$, for any $\gamma$ that does not scale with $\Phi_b$. As a consequence, every such boundary condition
recovers the Whittaker $W$-function solution of \cite{Gao:2021uro} in the asymptotic boundary limit. Moreover, normalizing
$\psi_s(L)$ by its value at $L=0$, we find in App.~\ref{sapp:bound states} that the density of states is
\begin{equation}
  \begin{aligned}
      \rho(s)&=c_0(\Phi_b)\,\rho_\infty(s)\left(1+\frac{c_2(s)}{\Phi_b^2}+\mathcal O\big(\Phi_b^{-4}\big)\right)~,\\
  \rho_\infty(s)&=\frac{4}{\pi^2}\,s\sinh(4\pi s)\left|\Gamma\big(\tfrac12+2\mJT+2{\rm i}s\big)\right|^2~,
  \end{aligned}
  \label{eq:Rmain}
\end{equation}
where $\rho_\infty(s)$ is the density of states of JT gravity with an ETW brane \cite{Gao:2021uro}, $c_0(\Phi_b)$ a constant that does not depend on $s$
and can be absorbed in the normalization of $\psi_s$, and the finite cutoff correction $c_2(s)$ (\ref{eq:Rc2}) is
constant independent of $\gamma$ up to $\mathcal O(e^{-2\pi\Phi_b})$. The correction $c_2(s)$ is determined by the difference
between the Scarf~II potential \eqref{eq:schrodinger ETW} and the Morse potential \eqref{eq: H brane GJK} at fixed $x=8\Phi_b e^{-L}$. Therefore, the minimal
operator ordering also has a well-defined asymptotic boundary limit. What the non-minimal operator ordering which we study in Sec.~\ref{ssec:scattering ETW} (motivated by \cite{Griguolo:2026eoi}) provides is a
natural (Friedrichs) choice among the Robin boundary conditions \eqref{eq:Rrobin}.

\subsection{Improved ordering: Heun solutions and density of states}\label{ssec:scattering ETW}
In what follows, we consider the operator ordering introduced by \cite{Griguolo:2026eoi} in the analogous case without the ETW brane. This amounts to promoting the momentum operator in \eqref{eq: Hamilt ETW} to $\hat{P}\rightarrow\mathcal{G}\qty(2\hat{L},\,\hat{P})$  where $\mathcal{G}$ appears in \eqref{eq:sl2R op ordering}. Similar to Sec.~\ref{ssec:improvement JT}, in terms of the chord description, this can be realized from the low-energy description of the following deformed Hamiltonian \eqref{eq:brane chord H} into
\begin{eqnarray}
   \sqrt{1-q} \hH_{\rm ETW}\rightarrow-\qty(b_{\rm ETW}(\hat{n})\rme^{-\rmi \hat{p}_n}+\rme^{\rmi \hat{p}_n}b_{\rm ETW}(\hat{n})+a_{\rm ETW}(\hat{n}))+\mathcal{K}_0-\mathcal{K}_\infty~,
\end{eqnarray} 
with $\mathcal{K}_0,~\mathcal{K}_\infty$ appear in \eqref{eq:PT improved H} and the other operators in \eqref{eq:lanczos ETW}. Applying \eqref{eq:sl2R op ordering} to the kinetic term of \eqref{eq: Hamilt ETW}, $\hat{P}^2/4\rightarrow\mathcal{G}(\hat{\ell},\hat{P}_\ell)$ with the canonical pair $\ell=2L$, $P_\ell=P/2$, and $[\hat{L},\hat{P}]=\rmi$, we find the Schrödinger equation with the ETW brane \eqref{eq:schrodinger ETW} is modified as
\begin{equation}\label{eq:schr Scarf mod}
\begin{aligned}
   &-\frac{1}{4}\partial_L^2\psi_s(L)+U_{\rm m}\qty(L)\psi_s(L)=s^2\psi_s(L)~,\\
    &U_{\rm m}(L)\equiv\frac{2 m_{\rm JT}  \Phi_b \sinh L+\Phi_b^2-m_{\rm JT} ^2}{\cosh^2L}-\frac{1}{4\sinh^2(2L)}~.
\end{aligned}
\end{equation}
There is now a modification to the Scarf II potential term in \eqref{eq:schrodinger ETW} by the $-\csch^2(2L)/4$ term, which is clearly singular at $L=0$, which we discuss in more detail below. In particular, note that for $L\ll1$ one again obtains the inverse critical coupling differential equation \eqref{eq:inverese coupling problem} (with an overall multiplicative factor $1/4$) after substituting $\Phi_b^2/\hbar^2\rightarrow4(\Phi_b^2-m_{\rm JT}^2)$ and $s^2/\hbar^2\rightarrow4s^2$ in \eqref{eq:inverese coupling problem}, since we are setting $\hbar=1$ in \eqref{eq:schr Scarf mod}.

\paragraph{Classical analysis}Let us begin with its classical analysis, and move on to the quantum one. Doing a substitution $w=\sinh L$, one can see that there are extrema of the potential $U_{\rm m}(L)$ satisfy 
\begin{eqnarray}\label{eq:polynomial}
    16m_{\rm JT}\Phi_b w^5+16(\Phi_b^2-m_{\rm JT}^2)w^4-16m_{\rm JT}\Phi_b w^3-2w^2-1=0~.
\end{eqnarray}
There are 5 roots. Assuming that $m_{\rm JT}<0$ in the case of interest, one can see from Fig.~\ref{fig:ETW_potential} that the potential has a local maximum after $L=0$, then decreases to a local minimum and reaches $U_{\rm m}=0$ asymptotically at $L\rightarrow\infty$. In particular, in the limit where $\abs{m_{\rm JT}}/\Phi_b\ll1$, we find from \eqref{eq:polynomial} that the location of the minimum and the value of the potential are 
\begin{equation}
L_{\rm min}=\log\frac{2\Phi_b}{\abs{m_{\rm JT}}}+\frac{m_{\rm JT}^2}{4\Phi_b^2}+\mathcal{O}\qty(\frac{m_{\rm JT}^4}{\Phi_b^4})~,\quad  U_{\rm m}(L_{\rm min})=-m_{\rm JT}^2\qty(1+\frac{m^2_{\rm JT}}{16\Phi_b^4}+\mathcal{O}\qty(\frac{m^4_{\rm JT}}{\Phi_b^6}))~.
\end{equation}
Thus, this suggests there might exist bound states with discrete negative energies for suitable values of the brane tension, as we discuss next.

\paragraph{Quantum analysis}
We want to find solutions to \eqref{eq:schr Scarf mod}. By performing the substitution $z=-\rmi\sinh L$ and $f(z)\equiv \psi_s(L)$, \eqref{eq:schr Scarf mod} can be put in the form
\begin{equation}
    0=\partial_z^2f(z)+\frac{z\partial_zf}{z^2-1}+\qty(\frac{4s^2}{z^2-1}+4\frac{\Phi_b^2-m_{\rm JT}^2+2\rmi\Phi_b m_{\rm JT}z}{(z^2-1)^2}+\frac{1}{4z^2(z^2-1)^2})f~,
\end{equation}
where we used $\partial_L^2=z\partial_z-(1-z^2)\partial_z^2$; the last term is the image of the improvement term $-\frac14\csch^2(2L)$ and produces a regular singular point at $z=0$ with double exponent $\frac12$. The singular points are $z=0,\,\pm1,\,\infty$, with exponents $\{\frac12,\frac12\}$, $\{\frac14\pm(\rmi\Phi_b-m_{\rm JT})\}$, $\{\frac14\pm(\rmi\Phi_b+m_{\rm JT})\}$ and, up to the prefactor below, $\{\pm2\rmi s\}$; this is the Heun equation \cite{heun1888theorie,ronveaux1995heun,Hortacsu:2011rr}, which has two independent solutions. The one we will be mainly focused on is
\begin{equation}\label{eq:f1 sol}
\begin{aligned}
    f_1(z)=&\sqrt{z}(z-1)^{\frac{1}{4}-m_{\rm JT}+\rmi \Phi_b}(z+1)^{\frac{1}{4}+m_{\rm JT}+\rmi \Phi_b}\cdot\\
    &\cdot{\rm H\ell}(-1,2m_{\rm JT},1+2\rmi (\Phi_b-s),1+2\rmi(\Phi_b+s),1,1-2m_{\rm JT}+2\rmi \Phi_b;z)~,
\end{aligned}
\end{equation}
while from the method of reduction of order \cite{boyce2009elementary} the second linearly independent solution is
\begin{eqnarray}\label{eq:f2 sol}
    f_2(z)=f_1(z)\int^z\frac{\rmd w}{\sqrt{w^2-1}f_1(w)^2}~.
\end{eqnarray}
Considering $z\rightarrow0$ (corresponding to $L\rightarrow0$), one can see that $f_1(z)\sim \sqrt{z}$ in \eqref{eq:f1 sol}, so that $f_2(z)\propto \sqrt{z}\log z$ in \eqref{eq:f2 sol}. Following the Friedrichs self-adjoint extension \cite{friedrichs1934spektraltheorie,schmudgen2012unbounded} we should pick the solution that behaves as $\sqrt{z}$ for the boundary problem at $z=0$ ($L=0$). Thus, by adding the term $-\frac14\csch^2(2L)$ in the modified Scarf II potential \eqref{eq:schr Scarf mod}, we find a natural selection for the self-adjoint boundary conditions, resulting in the wavefunction $\psi_s(L)=\mathcal{N}f_1(z(L))$ with $f_1$ in \eqref{eq:f1 sol}, $z(L)=-\rmi\sinh L$ and $\mathcal{N}$ being a normalization constant.

We can now determine the corresponding density of states associated with $\psi_s$. First note that,
\begin{equation}\label{eq:energy difference}
    (s_1^2-s_2^2)\int_0^\infty\rmd L\psi_{s_1}(L)^*\psi_{s_2}(L)=\frac{1}{4}\eval{\qty(\psi_{s_1}^*\partial_L\psi_{s_2}-\partial_L\psi_{s_1}^*\psi_{s_2})}_{L=0}^{L\rightarrow\infty}
\end{equation}
where we applied \eqref{eq:schr Scarf mod}. Using the Friedrichs self-adjoint extension, where $\psi_s(L)\sim \sqrt{L}$ as $L\rightarrow0$, implies that the lower bound in \eqref{eq:energy difference} must vanish. Meanwhile, given that the eigenfunctions with distinct eigenvalues of a self-adjoint operator are orthogonal,
\begin{equation}
    \int_0^\infty \rmd L\psi_{s_1}^*(L)\psi_{s_2}(L)=\frac{\delta(s_1-s_2)}{\rho(s_1)}~,
\end{equation}
with $\rho(s)$ being the density of states. We can explicitly determine $\rho(s)$ using the asymptotics of the Heun function \eqref{eq:f1 sol} for $L\gg1$ where $z^{2\rmi s}\rightarrow\rme^{2\rmi s(L-\rmi\pi/2)}/2^{2\rmi s}$. We have that
\begin{eqnarray}\label{eq:asymptotics of psi s}
    \eval{\psi_s(L)}_{L\gg1}\simeq \mathcal{A}_+\rme^{2\rmi sL}+\mathcal{A}_-\rme^{-2\rmi sL}~,
\end{eqnarray}
where $\mathcal{A}_+=\mathcal{N}\rme^{s\pi}\mathcal{C}_{1+2\rmi(\Phi_b-s)}/2^{2\rmi s}$, $\mathcal{A}_-=\mathcal{N}\rme^{-s\pi}\mathcal{C}_{1+2\rmi(\Phi_b+s)}/2^{2\rmi s}$, where $\mathcal{C}_u$ denotes the Heun connection coefficients \cite{schafke1980connection,fiziev2016new}. We then have 
\begin{equation}\label{eq:density coeffs}
    \rho(s)=\frac{2}{\pi}(\abs{\mathcal{A}_+}^2+\abs{\mathcal{A}_-}^2)^{-1}~.
\end{equation}
Due to the self-adjointness of the Hamiltonian leading to a Schrödinger equation with a real potential in \eqref{eq:schr Scarf mod}, as well as the Friedrichs extension at $L=0$, the wavefunction is real up to multiplication by an arbitrary constant phase factor. This means that $\mathcal{A}_+=\mathcal{A}_-^*$, which in terms of the Heun connection coefficients below \eqref{eq:asymptotics of psi s} leads to the following identity:
\begin{equation}
    \abs{\rme^{\pi s}\mathcal{C}_{1+2\rmi(\Phi_b-s)}}=\abs{\rme^{-\pi s}\mathcal{C}_{1+2\rmi(\Phi_b+s)}}~.
\end{equation}
The resulting density of states, in terms of the Heun connection coefficients below \eqref{eq:asymptotics of psi s}, is then
\begin{eqnarray}\label{eq:rho Heun}
    \rho(s)=\frac{1}{\pi\abs{\mathcal{N}}^2}\frac{\rme^{-2\pi s}}{\abs{\mathcal{C}_{1+2\rmi (\Phi_b-s)}}^2}~.
\end{eqnarray}
Here the constant $\mathcal{N}$ depends on the choice of the normalization of the $\psi_s(L)$, such that it can be interpreted as a probability density
\begin{eqnarray}
    \int_0^\infty\rmd s~\rho(s)\psi_s^*(L)\psi_s(L')=\delta(L-L')~.
\end{eqnarray}
This holds in the absence of discrete states; when they exist (Sec.~\ref{ssec:traversable wormholes}), $\sum_n\psi_{s_n}^*(L)\psi_{s_n}(L')$ must be added to the left-hand side, as in \eqref{eq:resolution identity ETW}.
In particular, we can see that $\rho(s)$ must be proportional to the density of states associated with the Whittaker W-function in the asymptotic boundary limit. This can be seen from the expansion around the origin in $\psi_s(L)\sim \sqrt{\sinh L}$ \eqref{eq:f1 sol}. 

First, we use that when $1\ll L\ll\log 8\Phi_b$ the potential in \eqref{eq:schr Scarf mod} reduces to that of
\eqref{eq: H brane GJK}, and in terms of $x\equiv 8\Phi_b e^{-L}\gg1$ the two independent solutions behave as \cite{whittaker1945course}
\begin{equation}\label{eq:W asymp}
  W_{\kappa,m}(x)\sim x^{\kappa}e^{-x/2},\qquad M_{\kappa,m}(x)\sim x^{-\kappa}e^{x/2}.
\end{equation}
For $s\ll\Phi_b$ this window is classically forbidden. In it, $W$ grows towards large $L$ and $M$
grows towards the wall at $L=0$. The solution \eqref{eq:f1 sol}, and more generally the solution obeying any
self-adjoint condition at $L=0$ that does not scale with $\Phi_b$, enters the barrier with
comparable $W$ and $M$ components. Crossing the barrier suppresses the $M$ component relative to
$W$ by the tunneling factor
\begin{equation}
  \mathfrak t(s)=\exp\Big[-2\int_{L_1}^{L_2}2\sqrt{U_{\rm m}(L)-s^2}\,dL\Big]\simeq e^{-2\pi\Phi_b},
\end{equation}
where $U_{\rm m}(L)$ is the potential in \eqref{eq:schr Scarf mod} and $L_1<L_2$ are the two turning points enclosing the barrier (the region near $L=0$, where $U_{\rm m}\rightarrow-\infty$, is classically allowed). Therefore,
\begin{equation}\label{eq:kimit crucial}
  \lim_{\rm ABL}\psi_s(L)\propto\frac{1}{\sqrt x}\,W_{-2m_{\rm JT},2is}(x)\,\big(1+O(e^{-2\pi\Phi_b})\big),
\end{equation}
for $s\ll\Phi_b$. The only exception is the boundary condition $\psi'(0)=\gamma\psi(0)$ with
$\gamma<-2\sqrt{U(0)}\simeq-2\Phi_b$, which is discussed in App.~\ref{sapp:bound states}. We stress that the operator ordering choice \eqref{eq:schr Scarf mod} is distinguished by
providing a natural boundary condition compatible with the Friedrichs extension at $L=0$. Similarly, we show in App.~\ref{sapp:bound states} that the minimal operator ordering \eqref{eq:schrodinger ETW} with any boundary condition
\eqref{eq:Rrobin} with $\gamma$ independent of $\Phi_b$ also leads to \eqref{eq:kimit crucial} up to a $\mathcal O(e^{-2\pi\Phi_b})$ correction; meaning that there is no specific reason to pick a specific one in that case.

\eqref{eq:kimit crucial} thus implies that the probability density associated with the Heun function \eqref{eq:f1 sol} in \eqref{eq:rho Heun} corresponds to that of the W-function in the validity limit of \eqref{eq:kimit crucial}.

The solution \eqref{eq:f1 sol} can then be used to evaluate two-point correlation functions according to the proposal by \cite{Griguolo:2026eoi} (summarized in Sec.~\ref{ssec:finite cutoff holo}) as,
\begin{equation}\label{eq:L int}
\begin{aligned}
    &\frac{1}{Z(\beta)}\bra{L_{\rm reg}=0}\rme^{-\tau_2\hH_{\rm JT-ETW}}\frac{\Phi_b^{2\Delta}}{\cosh^{2\Delta}\hat{L}}\rme^{-\tau_1\hH_{\rm JT-ETW}}\ket{L_{\rm reg}=0}\\
    &=\frac{1}{Z(\beta)}\int_0^\infty\rmd s_1\rmd s_2~\rho(s_1)\rho(s_2)\,\psi_{s_2}(L_{\rm reg}=0)\psi^*_{s_1}(L_{\rm reg}=0)\, \rme^{-\tau_1 s_1^2-\tau_2 s_2^2}\int_0^\infty \rmd L~\frac{\Phi_b^{2\Delta}}{\cosh^{2\Delta}L}\psi_{s_2}^*(L)\psi_{s_1}(L)~,
\end{aligned}
\end{equation}
where $\ket{L_{\rm reg}}\equiv\ket{L}/\sqrt{L}$ so that it is normalizable as $L\rightarrow0$,\footnote{\label{fnt:constrat griguolo}In contrast, \cite{Griguolo:2026eoi} does not work with regularized states; in principle, one can carry the vanishing norm as an overall constant of the partition function \eqref{eq:orthogonality}.} and
\begin{eqnarray}
    Z(\beta)=\int_0^\infty\rmd s~\rho(s)\rme^{-\beta s^2}+\sum_{n=0}^{n_{\rm max}}\rme^{\beta s_n^2}~,
\end{eqnarray}
where $\rho(s)$ is the density of states \eqref{eq:rho Heun}, the sum runs over the discrete states $E_{\rm ADM}=-s_n^2$ of Sec.~\ref{ssec:traversable wormholes} when they exist, and $\psi_s(L)$ is the Heun function in \eqref{eq:f1 sol}, with $\psi_s(L_{\rm reg}=0)\equiv\lim_{L\rightarrow0}\psi_s(L)/\sqrt{L}$; when bound states exist, the discrete contributions of \eqref{eq:resolution identity ETW} are added to the integrals in the same way. There are no analytic simplifications to the $L$ integral in \eqref{eq:L int} to our knowledge.

Geometrically, $L=0$ is the configuration in which the cutoff boundary touches the ETW brane, so that $\ket{L_{\rm reg}=0}$ corresponds to the corner where the Dirichlet boundary meets the brane in the Euclidean disk. The factor $\Phi_b^{2\Delta}/\cosh^{2\Delta}\hat{L}$ is the semiclassical two-point function between a boundary point and its image across the brane, at geodesic distance $2L$, so that \eqref{eq:L int} computes the one-point function of the boundary operator in the presence of the brane.

\subsection{Bound states and eternal traversable wormholes}\label{ssec:traversable wormholes}
Next, we explore the existence and interpretation of discrete energy states, where the ADM energy takes the form $E_{\rm ADM}=s^2=-s_n^2<0$, in the Schrödinger equation with the modified Scarf II potential \eqref{eq:schr Scarf mod} using the Friedrichs extension, while those for the minimal operator ordering choice \eqref{eq:schrodinger ETW} appear in App.~\ref{sapp:bound states}.

First, notice from the asymptotic expansion of the Heun wavefunction in \eqref{eq:asymptotics of psi s} that for such states
\begin{eqnarray}
    \eval{\psi_s(L)}_{L\gg1}=\mathcal{A}_+\rme^{-2s_n L}+\mathcal{A}_-\rme^{2s_n L}~.
\end{eqnarray}
However, square integrability for $L\geq0$ requires $\mathcal{A}_-(s)=0$
\begin{equation}\label{eq:connection poles}
 \mathcal{A}_-(s)=\mathcal{N}4^{-\rmi s}\rme^{-\pi s}\mathcal{C}_{1+2(\rmi \Phi_b-s_n)}(s_n)\Big|_{s=\rmi s_n}=0~.
\end{equation}
Thus, the existence of discrete states is determined by the roots $s_n$ of the Heun connection coefficient $\mathcal{C}_{1+2(\rmi \Phi_b-s_n)}(s_n)=0$. 

\paragraph{Asymptotic boundary limit}
Let us start from the asymptotic boundary limit and then discuss the finite $\Phi_b$ corrections. The Heun function becomes in the Whittaker W-function by \eqref{eq:kimit crucial}, and there indeed exist bound energy states \cite{Rajgadia:2026ask}, namely
\begin{equation}
    \lim_{\rm ABL}\psi_{\rmi s_n}(L)\propto\frac{W_{-2m_{\rm JT},-2 s_n}(x(L))}{\sqrt{x(L)}}\propto \frac{\Gamma(-4s_n)~x(L)^{\frac{1}{2}+2s_n}}{\Gamma(1/2-2s_n+2m_{\rm JT})}+\frac{\Gamma(4s_n)~x(L)^{\frac{1}{2}-2s_n}}{\Gamma(1/2+2s_n+2m_{\rm JT})}~,
\end{equation}
where again we denote $x(L)\equiv 8\Phi_b\rme^{-L}$. This means that for $L\gg1$, the growing branch is the term $\Gamma(1/2+2s_n+2m_{\rm JT})$ whose poles then determines the allowed energy value for $s_n$,
\begin{equation}\label{eq:Nsn}
    s_n=-\frac{n}{2}-{m_{\rm JT}}-\frac{1}{4}~,\quad n\in\mathbb{Z}_{\geq0}~,
\end{equation}
which thus implies there are normalizable bound states when $m_{\rm JT}<-1/4$. The same result follows even semiclassically (without the correction term in \eqref{eq:schr Scarf mod}). The minimum of $U$ in
\eqref{eq:schrodinger ETW} sits at $\sinh L_{\rm min} = -\Phi_b/m_{\rm JT}$ (as we also examine in App.~\ref{sapp:bound states}) with $U(L_{\rm min}) = -m^2_{\rm JT}$ and $U''(L_{\rm min}) = 2m^2_{\rm JT}$.
For $H = P^2/4 + U$ the oscillation frequency is therefore $\abs{m_{\rm JT}}$, independent of $\Phi_b$. By quantizing the Hamiltonian, its energy levels follow as
$E_n = -m^2_{\rm JT} + \abs{m_{\rm JT}}(n + 1/2)$, which gives $s_n \simeq -m_{\rm JT} - 1/4 - n/2$. Then, the leading term
$s_n=-\mJT$ is fixed by the classical energy, while $-\frac14$ is the zero-point energy of the oscillations.

Once we include the discrete energy states, the resolution of the identity in energy basis becomes
\begin{equation}\label{eq:resolution identity ETW}
    \int_0^\infty\rmd s~\rho(s)\ket{s}\bra{s}+\sum_{n=0}^{n_{\rm max}}\ket{s_n}\bra{s_n}~,
\end{equation}
where $n_{\rm max}$ denotes the largest allowed index, so that there are $n_{\rm max}+1$ discrete states; in the case of \eqref{eq:Nsn}, $n_{\rm max}=\lfloor -2m_{\rm JT}-1/2\rfloor$, or $\lceil-2\mJT-\frac12\rceil-1$ when $-2\mJT-\frac12\in\mathbb{Z}$, since a state at threshold, $s=0$, is not normalizable.

\paragraph{Finite-cutoff corrections}
Let us consider finite $\Phi_b$ corrections using the condition \eqref{eq:connection poles}. There are no analytic formulas for this, however, we can apply numerical methods \cite{lay1998asymptotic,lay1994central} or perturbative connection formulas \cite{lisovyy2022perturbative} to inquire about this. We show in App.~\ref{app:numerics} that there are discrete energy states at finite cutoff using numerical methods. We also show in App.~\ref{sapp:bound states} there are bound state solutions to the Scarf II potential term \eqref{eq:schrodinger ETW} with precise analytic conditions.

\paragraph{Eternal traversable wormholes}Let us interpret the results in the semiclassical picture. Below, we will assume that there exist discrete energy states even at finite $\Phi_b$ based on the numerical results at finite cutoff in App.~\ref{app:numerics}, and the analytic ones in the asymptotic boundary limit.

\begin{figure}
    \centering
    \subfloat[]{\includegraphics[width=0.6\linewidth]{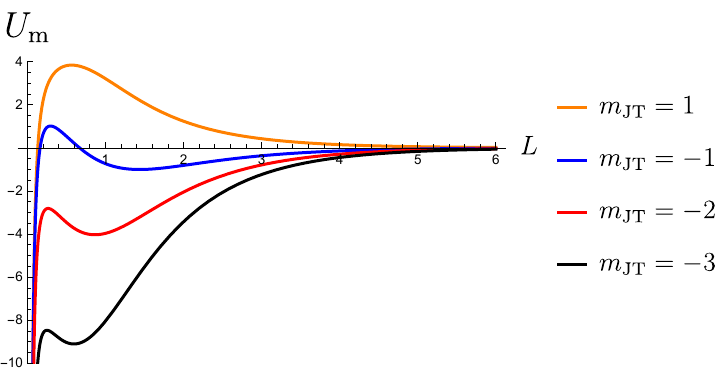}}\hspace{0.5cm}\subfloat[]{\includegraphics[width=0.25\linewidth]{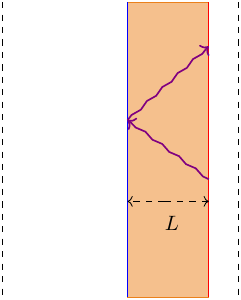}}
    \caption{(a) Modified Scarf II potential $U_{\rm m}$ \eqref{eq:schr Scarf mod} for different values of the brane tension (displayed in the figure) at a finite cutoff parameter $\Phi_b=2$. When $m_{\rm JT}<0$, local minima of $U_{\rm m}$ emerge before the singularity at $L=0$. (b) Semiclassical interpretation of the negative energy bound states $E_{\rm ADM}=-s_n^2$ in \eqref{eq:schr Scarf mod}, where the Penrose diagram uses $\sigma$, $T$ global coordinates \eqref{eq:coordinate useful}. The ETW brane (blue line) is now static in global AdS$_2$ space, and light rays (purple oscillating curves) emitted at the finite boundary (red line) bounce back. Dashed black lines represent the portion of spacetime that is removed by the cutoff and the ETW brane.}
    \label{fig:ETW_potential}
\end{figure}

We start from Lorentzian embedding coordinates for AdS$_2$ (App.~\ref{sapp:geodesic shockwaves}),
\begin{eqnarray}\label{eq: X coords}
    \rmd l^2=\eta_{IJ}\rmd X^I\rmd X^J\equiv-\rmd X_{0}^2-\rmd X_{1}^2+\rmd X_{2}^2~,\quad X\cdot X=-1~,
\end{eqnarray}
where we denote $X\cdot Y\equiv \eta_{IJ}X^IY^J$. The dilaton and the ADM energy $E_{\rm ADM}$ in JT gravity can be expressed as \cite{Maldacena:2016upp}
\begin{equation}\label{eq:dilaton coord}
    \Phi=V\cdot X~,\quad E_{\rm ADM}=-V\cdot V~.
\end{equation}
From our analysis, when there exist roots $s_n$ in \eqref{eq:connection poles} it follows that $V\cdot V>0$. In terms of the embedding coordinates \eqref{eq: X coords}, we can express $V^I=(0,\,0,\,s_n)$. This means that there is a generator $\mathbb{J}$ of a Killing flow which makes the dilaton invariant $\mathbb{J}\cdot V=0$, which in the embedding coordinates \eqref{eq: X coords} lives in the ($X^0$, $X^1$) plane. Using global AdS$_2$ coordinates 
\begin{eqnarray}\label{eq:coordinate useful}
    \rmd l^2=\rmd\sigma^2-\cosh^2\sigma~\rmd T^2~.
\end{eqnarray}
We then have that $\mathbb{J}^I=\mathcal{N}\cosh\sigma(-\sin T,\cos T,0)=\mathcal{N}\partial_T X^I$, where $\mathcal{N}$ is a constant, generates rotations in the ($X_0$, $X_1$) plane parameterized by global time displacement $T\rightarrow T+$const, and satisfies $\mathbb{J}\cdot V=0$. The dilaton \eqref{eq:dilaton coord} is indeed static in terms of global time \eqref{eq:coordinate useful}, and it is given by $\Phi=s_n\sinh\sigma$. In contrast, the dilaton is time dependent for the AdS$_2$ black hole, and given by $\Phi_{\rm BH}=\Phi_h\cos T\cosh\sigma$ \cite{Penington:2023dql}. Thus, when there exist roots in \eqref{eq:connection poles}, which simplifies to $m_{\rm JT}<-1/4$ in the asymptotic boundary limit \eqref{eq:Nsn}, the brane backreacts on the dilaton, in contrast to an ETW brane behind the horizon \cite{Gao:2021uro}. Given that the dilaton is static, the brane stays at the same value $\sigma=\sigma_0$ for any constant $T$ slice. Since the ETW brane stays at $\sigma=\sigma_0$ constant, its mean curvature is $K=-\tanh \sigma_0$, where the sign is fixed by the unit normal $n^b=(-1,0)$ in global coordinates \eqref{eq:coordinate useful}, pointing away from the cutoff boundary as specified in \eqref{eq:brane conditions}; the brane condition \eqref{eq:brane conditions} then gives $s_n=-m_{\rm JT}\cosh\sigma_0$ for the static dilaton $\Phi=s_n\sinh\sigma$, so that $m_{\rm JT}<0$, consistently with \eqref{eq:Nsn}. For instance, for geodesic ETW branes (i.e.~$K=0$) we have $\sigma_0=0$. This means that at the $T=0$ slice the would-be horizon at $\sigma=0$ is shielded by the brane; and thus, no AdS$_2$ black hole forms. Light rays from the finite cutoff boundary will travel towards the ETW brane and will be reflected back, as in the MQ case \cite{Maldacena:2018lmt}, which is by definition an eternal traversable wormhole. We illustrate the Modified Scarf II potential and the semiclassical interpretation of negative energy states in Fig.~\ref{fig:ETW_potential}.
The same conclusion holds for more general CMC ETW branes, where $\sigma_0>0$. This also means that the length between the boundary and the ETW brane (which is a $T=$constant slice) is $L=\sigma_b-\sigma_0$ due to the brane being static for negative energy states, which is consistent with \cite{Chapman:2021eyy}. Thus, \eqref{eq:L total} is only valid for scattering states. This means that the semiclassical limit of Krylov complexity for the discrete energy states is expected to be trivial even beyond the asymptotic boundary limit.

\section{Discussion}\label{sec:disc}
A summary of ADM Hamiltonians and the bulk geodesic lengths studied in the previous sections is displayed in Tab.~\ref{tab:main resultss}.
\begin{table}[t!]
\begin{center}
\begin{tabular}{C{7.5cm}C{7.5cm}}
\toprule
\textbf{Bulk length}&\textbf{Chord Hamiltonian}\\\midrule 
\multicolumn{2}{c}{Perturbative branches ($\varepsilon=-1$) in the flow equation \eqref{eq:flow eq}}\\[2ex]
$\begin{aligned}\sinh\frac{L}{2}&=
\sqrt{\frac{\Phi_b^2}{\Phi_h^2}-1}\abs{\cosh(\Phi_ht+\delta_\eta)}\\
\delta_\eta&\equiv \rmi(1-\eta)\frac{\pi}{4}
\end{aligned}$ \eqref{eq:geodesic length growth}&$\hH_{\rm PT}=b_1(\hat{n})\rme^{-\rmi \hat{p}}+\rme^{\rmi \hat{p}}b_1(\hat{n})$ \eqref{eq:chord PT H}\\[2ex]
&$b_1(\hat{n})\equiv \frac{-1}{\sqrt{1-q}}\sqrt{1-\xi^2\sech^2(\lambda\hat{n}/2)}$\\[2ex]\hline
\multicolumn{2}{c}{Non-perturbative branches ($\varepsilon=+1$)}\\[2ex]
$\sin\frac{L}{2}=
\sqrt{1-\frac{\Phi_b^2}{\Phi_h^2}}\abs{\cosh(\Phi_ht+\delta_\eta)}$ \eqref{eq:geodesic length growth}&None\\[2ex] \hline
\multicolumn{2}{c}{Perturbative branch with an ETW brane}\\[2ex]
$\sinh L=
\frac{m_{\scaleto{\rm JT}{0.8ex}}\Phi_b+\sqrt{(\Phi_b^2-\Phi_h^2)(m_{\scaleto{\rm JT}{0.8ex}}^2+\Phi_h^2)}~\cosh(\Phi_h t)}{\Phi_h^2}$ \eqref{eq:main text L brane}&$\begin{aligned}
\hH_{\scaleto{\rm ETW}{0.8ex}}=&b_{2}(\hat{n})\rme^{-\rmi\hat{p}_n}+\rme^{\rmi \hat{p}_n}b_{2}(\hat{n})+a_2(\hat{n})~\text{\eqref{eq:brane chord H}}\\
a_2(\hat{n})=&\frac{-2}{\sqrt{1-q}}\sech^2(\lambda \hat{n})\mathcal{Y}\,(\overline{\mathcal{Y}}-\xi\sinh (\lambda \hat{n}))\\
 b_2(\hat{n})=&\frac{-1}{\sqrt{1-q}}\sqrt{1+\frac{\mathcal{Y}^2}{\cosh^2(\lambda \hat{n})}}\cdot\\
&\cdot\sqrt{1+\frac{\overline{\mathcal{Y}}^2-\xi^2}{\cosh^2(\lambda \hat{n})}-\frac{2\xi\overline{\mathcal{Y}}\tanh (\lambda \hat{n})}{\cosh (\lambda \hat{n})}}
    \end{aligned}$\\\bottomrule
\end{tabular}
\end{center}
\caption{Left: Bulk geodesic length in the finite cutoff between (top and middle) two Dirichlet boundaries, and (bottom) one Dirichlet boundary and one ETW brane dual of the DSSYK model. Right: Deformed chord Hamiltonian encoding the dual length. We defined $\delta_\eta$ as the phase factor decoding the $T^2$ ($\eta=+1$) and $T^2+\Lambda_1$ ($\eta=-1$) deformations. The cell ``none'' indicates that there is no boundary interpretation for the $\varepsilon=+1$ case due to its bulk Hilbert space being ill-defined. There is a deformation of the chord Hamiltonian \eqref{eq:PT improved H} encoding a different operator ordering in the low-energy theory.}
\label{tab:main resultss}
\end{table}

In this work, we have developed a boundary and bulk framework for studying finite-cutoff holography in DSSYK. On the boundary side, we constructed microscopic deformations of KM and thermofield-double states with MQ-type interactions. After the double-scaling limit, these deformations reproduce the $T^2$ and $T^2+\Lambda_1$ deformations of the DSSYK chord Hamiltonian. The resulting energy-flow equation has both perturbative and non-perturbative branches, whose bulk interpretation was associated to finite-cutoff sine dilaton gravity, which in some limits reduces to finite-cutoff foliations of (A)dS$_2$ space. In particular, the non-perturbative branch provides access to regions that are not described by the usual asymptotic-boundary description, but its quantum mechanical formulation is ill-defined, in contrast to the perturbative branch. Self-adjoint extensions have a crucial role to solve the system, which includes a boundary condition not present in the asymptotic boundary limit. This required an appropriate deformation of the chord Hamiltonian.

{We extended the analysis of perturbative and non-perturbative branches by considering the CMC slices anchored to the Dirichlet boundaries in the corresponding bulk geometries. We found a variety of evolutions depending on the branch and the mean curvature of the slice. We investigated the boundary interpretation of the CMC slice length using a class of cost functions in \cite{Balasubramanian:2022tpr}, which we denoted CMC-Krylov complexity. Depending on the branch and the bulk region, the resulting CMC
lengths exhibit linear, exponential, or hyperfast growth. This might allow progress on identifying concrete holographic complexity measures beyond the CV.}

We then generalized the finite-cutoff construction in sine dilaton gravity and JT gravity by including a timelike, dynamical ETW brane in the finite-cutoff bulk and investigated the combined deformation in the DSSYK model. This provides a starting point for testing the finite cutoff dictionary for increasingly more general gravitational systems. We encountered important subtleties in the quantization of the low-energy system, which were resolved by incorporating the operator ordering prescription in \cite{Griguolo:2026eoi} resulting in an improvement term in the potential energy. The scattering states describe spacetime wormhole states in an AdS black hole with a dynamical ETW brane behind its horizon, while bound states with negative energies have a semiclassical interpretation in terms of eternal traversable wormholes with a static ETW brane in global AdS. The finite cutoff leads to a modification on the conditions for normalizability of the bound states as well as its spectrum.

Taken together, the results in this work suggest that the deformed DSSYK is a useful laboratory in which the boundary deformation, bulk cutoff, quantum Hilbert space, and geometric notion of complexity can be studied within a single solvable framework.

\subsection{Outlook}\label{ssec:outlook}
There are several opportunities for future developments. 

In Sec.~\ref{ssec:improvement JT} we presented a chord Hamiltonian deformation in \eqref{eq:PT improved H} which reproduces the operator ordering choices of \cite{Griguolo:2026eoi} in the low-energy effective description. However, a first-principle reason for choosing it is not addressed in this work. We expect that one needs to understand more about representation theory in the analog of $\mathfrak{sl}(2,\mathbb{R})$ for sine dilaton gravity, corresponding to its quantum group \cite{Berkooz:2018jqr,Berkooz:2022mfk,Blommaert:2023opb, Almheiri:2024ayc,Belaey:2025ijg,vanderHeijden:2025zkr, Schouten:2025tvn}. It would be very interesting to investigate about the origin of the proposed deformation, or whether any other one should be preferred over it, in detail in future work, which is outside the scope of the present manuscript.

Next, it would be interesting to formulate the associated von Neumann algebra \cite{Xu:2024hoc} for the deformed DSSYK models in this work and study in what ways they might differ qualitatively from that of the asymptotic limit. Determining its precise type and relation to the DSSYK algebra could provide an operator-algebraic characterization of the finite-cutoff theory. It would also be interesting to connect our study of $T^2+\Lambda_1$ deformations with other flows through the interior of black holes and their relation to trans-IR constructions \cite{Caceres:2022smh,Caceres:2026mug,Caceres:2025xzl,Caceres:2024edr,Caceres:2023zhl,Caceres:2022hei}.

Similarly, for finite-cutoff JT and sine dilaton gravity with ETW branes in Sec.~\ref{sec:bulk with matter}, it would be worth to (i) evaluate higher-point correlation functions, and (ii) extend our Hilbert space analysis with matter on the spirit of \cite{Cao:2025pir}. 

In general, it would be worth to connect our boundary analysis of the finite-cutoff JT gravity with the Riccati-type equation discussed by \cite{Griguolo:2025kpi}. The authors studied a boundary curve of the hyperbolic disk in finite cutoff JT gravity, which is not part of this work.

Also, the double-scaling limit of the SYK deformation in our work suggests extensions beyond the AdS and dS limits considered here. The intermediate-energy regime of DSSYK is related to flat-space JT gravity \cite{Blommaert:2024whf}, providing a possible route toward finite-cutoff flat-space holography, building on the asymptotic boundary case \cite{Blommaert:2023opb,Blommaert:2024whf,Mariani:2026jyp}.

Another direction is to replace the Dirichlet boundary condition by more general boundary conditions. In particular, conformal or fixed-mean-curvature boundary conditions \cite{An:2021fcq,Anninos:2023epi,Banihashemi:2024yye, Anninos:2024wpy,Anninos:2024xhc, Galante:2025tnt,Banihashemi:2025qqi,Galante:2025emz,Anninos:2025zgr,Hamdan:2026olq,Caceres:2026yvs, Liu:2024ymn,Liu:2025xij} may provide a more natural setting for treating the CMC parameter itself as a dynamical variable. Rather than regarding the mean curvature $k$ only as a label of the bulk slices, one could ask whether translations in $k$ generate a useful notion of quantum evolution and whether a corresponding Krylov construction defines a complexity associated with CMC flow as physical time \cite{York:1973ia,York:1972sj,Parrikar:2025xmz,Blommaert:2026lvp}. 

Next, in the evaluation of CMC-Krylov complexity in Sec.~\ref{ssec:def CMC-K complexity} was computed first at the semiclassical level and with finite cutoff effects. However, the complexity measure reduces to a simple function of the Krylov complexity \eqref{eq:hbar expansion Ca} rather than capturing more interesting information about its higher moments. It might be interesting to confirm if this structure persists exactly in a different quantum regime, or whether the finite-cutoff realization produces other differences. One could then check if the holographic dictionary relating complexity and lengths hold quantum mechanically incorporating the mean curvature and the finite cutoff. It would also be useful to understand whether the relation survives beyond the particular cost function and initial states considered here. Studying the connection between CMC-Krylov complexity with proper radial momentum of probe particle \cite{Caputa:2024sux,Caputa:2025dep,Fan:2024iop,He:2024pox,Fatemiabhari:2025usn,Fatemiabhari:2025poq,Fatemiabhari:2026goj,Fatemiabhari:2026rob,Nastase:2026lhz,Li:2025fqz,Li:2026pdh,Nunez:2026kwr,Nunez:2026vhw,Chatzis:2026oou,Nongmaithem:2026rop} would also be a direction worth taking.

Extending the above ideas to higher-dimensional gravity, and determining which features of CMC-Krylov complexity survive in that setting, would provide a useful test of whether the need for self-adjoint extensions and special choices of operator order are generic in finite cutoff holography at the quantum level or special to JT gravity, sine dilaton gravity and the DSSYK model.

In the following, we elaborate on two specific research directions in the hope that it sparks new developments in this area.

\subsection{Towards finite-cutoff holography with matter}\label{ssec:H JT with matter}
An important next step in finite-cutoff holography is to incorporate matter in the Hilbert space formulation. This could allow us to evaluate the crossed four-point function in the finite cutoff bulk, and to make contact with recent proposals for correlators in finite-cutoff JT gravity \cite{Griguolo:2026eoi}. Moreover, this type of correlator may also provide a direct boundary representation for the generator of Krylov complexity in the presence of matter \cite{Aguilar-Gutierrez:2025hty,Aguilar-Gutierrez:2025mxf}.

One way to approach this problem is the Penington-Witten construction \cite{Penington:2023dql} of the ADM Hamiltonian of JT gravity with matter. We will consider similar arguments as \cite{Penington:2023dql}, however, we do not take an infinite cutoff. Consider AdS$_2$ in global coordinates as \eqref{eq:coordinate useful}.

The Lagrangian describing a geodesic between left and right finite cutoff boundaries, denoted by coordinates $(\sigma_{L/R},\,T_{L/R})$, as $\mathcal{L}_{L/R}=\sqrt{\abs{(\rmd l_{L/R}/\rmd \chi)^2}}$, with $\chi$ as a general parametrization along each curve. In the coordinate choice \eqref{eq:coordinate useful}, geodesics with matter are described by $\sigma_R-\sigma_L+f(T_L,T_R)$, where $f(T_L,T_R)$ is a function that vanishes in the matterless case, where one can set $\sigma_R=-\sigma_L=L/2$, with $L$ being the geodesic length.

The corresponding classical Hamiltonian of the system is \cite{Penington:2023dql},\footnote{Note we are adding a finite constant term in each Hamiltonian with respect to \cite{Penington:2023dql} to agree with the conventions in \cite{Griguolo:2026eoi} in the matterless space case.} 
\begin{eqnarray}\label{eq:PW JT}
    H_L=\frac{P_{\sigma_L}^2}{4}+\xi^2-\qty(\frac{P_{T_L}}{\cosh \sigma_L}-\xi\tanh \sigma_L)^2~,\\
    H_R=\frac{P_{\sigma_R}^2}{4}+\xi^2-\qty(\frac{P_{T_R}}{\cosh \sigma_R}-\xi\tanh \sigma_R)^2~.
\end{eqnarray}
In particular, in the matterless case, the classical JT-PT Hamiltonian \eqref{eq:JT PT} is recovered from \eqref{eq:PW JT}, where $\sigma_R=-\sigma_L=L/2$, $P_{\sigma_R}=-P_{\sigma_L}=2P$, $\xi\rightarrow\Phi_b$ and $P_{T_{L/R}}=0$. 

One can canonically quantize the system by promoting the canonical variables in \eqref{eq:PW JT} to operators, with appropriate ordering discussed below, acting on a corresponding Hilbert space of the form
\begin{equation}\label{eq:tot H}
    \mathcal{H}\equiv\mathcal{H}_{\rm grav}\otimes \mathcal{H}_{\rm matter}~,\quad \mathcal{H}_{\rm grav}=L^2(\mathbb{R}_{\geq0})~,
\end{equation}
with $\mathcal{H}_{\rm grav}$ and $\mathcal{H}_{\rm matter}$ the gravitational and matter subspaces respectively. The matter subspace corresponds to multiple particle states of a free real scalar with mass $m^2=\Delta(\Delta-1)$. Using this construction, one could then express $G_4$ \eqref{eq:G4 semiclassical} as a semiclassical approximation to the two-point function in the one-particle space as
\begin{equation}
   G_4(t_L,t_R) \equiv\bra{\Psi_\Delta}\rme^{-\tau_R^*\hH_R}\sech^2\frac{\hat{L}}{2}\rme^{-\tau_L\hH_L}\ket{\Psi_\Delta}~,
\end{equation}
where $\ket{\Psi_\Delta}$ is the HH state with an operator insertion, defined in the corresponding bulk Hilbert space with matter, $\hH_{L/R}$ are the corresponding two-sided Hamiltonians with matter, and $\tau_{L/R}=\beta/2+\rmi t_{L/R}$ is the time parameter in the HH state preparation.

It would be interesting to continue this discussion. However, we should keep in mind that for a fixed radial
surface, the full AdS2 isometry group does not obviously preserve the cutoff surface. As a consequence, due to the finite cutoff, the construction of one-particle states in \cite{Iliesiu:2024cnh} does not apply. In particular, the normal components of the Killing
vectors move the cutoff surface, while finite Dirichlet boundary conditions generally break part
of the asymptotic symmetry structure. 

As a special case of matter, in App.~\ref{app:shockwaves}, we examine the two-sided CMC slice length for shockwave geometries in finite cutoff JT gravity. In the case of geodesic lengths, the result heuristically suggest that, assuming of the validity of the proposal by \cite{Griguolo:2026eoi} on two-point functions, and its application to the one-particle space of finite cutoff JT gravity, the expected Lyapunov exponent in the OTOC of the auxiliary theory is unchanged by the finite cutoff according to the same boundary time variable as in the asymptotic limit, while according to the proper time along the boundary, the expected redshift occurs. It is worth systematically confirming this evaluation. It might suggest that the early-time Lyapunov exponent remains unchanged, while late-time quantities such as spectral statistics may be modified due to the flow in the energy spectrum. Establishing this distinction directly in the microscopic SYK deformations would be an interesting test of the proposed finite-cutoff dictionary.

For the Hilbert space construction, we need to better understand the ADM Hamiltonian of JT gravity in the CMC slice foliation with shockwaves (App.~\ref{app:shockwaves}). One should also show that the switchback effect is present from the boundary side. There has been interesting work on the switchback effect in Krylov complexity in DSSYK \cite{Ambrosini:2025hvo,Aguilar-Gutierrez:2025mxf} which might be relevant for this problem. We hope this might provide a better understanding of the problem in some detail.

\subsection{Towards complexity=anything from Krylov space}
We now want to relate the discussion in Sec.~\ref{ssec:def CMC-K complexity} to the CAny proposals \cite{Belin:2021bga, Belin:2022xmt} for more general measures.

One may define complexity measures based on the Krylov complexity operator \eqref{eq:n hat def intro} representing geometric invariants in the bulk that are functions of the geodesic length as
\begin{eqnarray}\label{eq:generalized measures}
    C_{\rm KAny}(t)\equiv \bra{K_0}f(\hat{n}(t))\ket{K_0}~,
\end{eqnarray}
which by the arguments in Sec.~\ref{ssec:def CMC-K complexity} is again a special type of cost functional \eqref{eq:cost} where $c_n=f(n)$ a monotonic increasing coefficient set, and $\ket{B_n}=\ket{K_n}$. In particular, one might require that \eqref{eq:generalized measures} scales as $1/\lambda^\alpha$ as $\lambda\rightarrow0$ with $\alpha>0$ so that they admit a semiclassical description and a saddle point evaluation akin to the one presented in Sec.~\ref{sec:CMC-Krylov complexity}. Moreover, the holographic interpretation of \eqref{eq:generalized measures} might depend on whether one considers the perturbative ($\varepsilon=-1$) or non-perturbative ($\varepsilon=+1$) branches in the boundary theory, as we saw in Sec.~\ref{ssec:def CMC-K complexity}, due to modifications in the geometric relation between geodesics and more general boundary-anchored bulk observables of the type in \eqref{eq:OF1}. Restricting the present discussion to the $\varepsilon=-1$ branches, it is natural to then ask what modifications we need to include in \eqref{eq:generalized measures} such that we recover \eqref{eq:OF1}; i.e.~what are the parameters in the complexity measure which should be identified with $F$, $\Sigma_\epsilon$ in the codimension-one proposals \eqref{eq:OF1}. One could speculate based on our construction that just modifying the coefficients $c_n$ in the cost function \eqref{eq:cost} leads to a different slice $\Sigma_\epsilon$ since that determines the difference between Krylov complexity and the measure \eqref{eq:def CMC-Krylov cmplexity}. Moreover, given that CMC-Krylov complexity corresponds to a length in the bulk, i.e.~$F=1$ in \eqref{eq:OF1}, it might indicate that one needs to use a basis other than the Krylov basis $\ket{K_n}$ in the cost function \eqref{eq:cost} to reproduce a different factor $F$ in \eqref{eq:OF1}. It would be interesting to connect this to recent work \cite{Bhattacharya:2026ybq} in a higher-dimensional settings, where it appears that $F$ in \eqref{eq:OF1} has an infinite number of curvature terms resulting in quadratic late-time growth rather than a linear one in \cite{Belin:2021bga,Belin:2022xmt}.

Nevertheless, there are several more possibilities besides changing the system itself to recover a non-trivial factor $F$ in our setting; e.g.~picking a different reference state from $\ket{\Omega}$ and evolving in complex-valued time $\tau$, changing the type of Liouvillian operator $\hat{\mathcal{L}}$, considering more general types of cost functions than the one proposed \cite{Balasubramanian:2022tpr}. It might be worthwhile to carry out a more detailed analysis of the freedom in the different boundary parameters that result used to identify a given complexity measure with a gravitational observable within the CAny class in \eqref{eq:OF1}.

\paragraph{Note added}While this work was nearly completed, we noticed potential partial overlap between Sec.~\ref{ssec:1st approach QJT} and forthcoming work by \cite{BMRT}, who also study ETW branes in finite-cutoff JT gravity with other techniques. There might also be potential partial overlap between our results in Sec.~\ref{ssec:sine dilaton} and those of another work in progress \cite{GPRST}. It would be worth exploring the similarities and differences between our works. For instance, see footnote \ref{fnt:constrat griguolo}.

\section*{Acknowledgments}
I thank Gonçalo Araujo-Regado, Matteo Beccaria, Micha Berkooz, Rathindra N.~Das, Michal P.~Heller, Masamichi Miyaji, Vyshnav Mohan, Shoy Ouseph, Philipp A.~Höhn, Jacopo Papalini, Francesco Sartini, Tim Schuhmann, Watse Sybesma and Ming Zhang for useful discussions; especially Andreas Blommaert, Hugo Camargo, Lorenzo Russo, Silke van der Schueren, Alex Tarana, Jiayue Yang and Jiuci Xu. The author is grateful to the Yukawa Institute for Theoretical Physics at Kyoto University for support during development of this work in the ``Holographic Universe'' 2026 conference, and Deutsches Elektronen-Synchrotron (DESY), where this work was presented, and near completed during the workshop ``Quantum gravity: Strings, Mathematical Models, Black Holes''. 
SEAG is supported by the Okinawa Institute of Science and Technology Graduate University. This work was made possible through the support of the WOST, WithOut SpaceTime project (\hyperlink{https://withoutspacetime.org}{https://withoutspacetime.org}), supported by Grant ID\# 63683 from the John Templeton Foundation (JTF), and ID\#62312 grant from the JTF, as part of the ‘The Quantum Information Structure of Spacetime’ Project (QISS), as well as Grant ID\# 62423 from the JTF. The opinions expressed in this work are those of the author(s) and do not necessarily reflect the views of the John Templeton Foundation. The author acknowledges ChatGPT and Claude for mock referee reports on a previous version which allowed me to improve the final version, proofreading it, and for assistance in formulating some of the arguments. All mathematical derivations and the written presentation were produced and verified by the author, who remains accountable for the accuracy and content of the work.

\appendix

\section{Notation and glossary}\label{app:notation}
\paragraph{Glossary}
\begin{itemize}[noitemsep]
\item $\hH_0$ \eqref{eq:def H SYK PT KM}, $\hH_{0'}$ \eqref{eq:def H SYK PT TFD}: Microscopic SYK deformations whose ensemble-averaged moments in the double-scaling limit reproduce those of $\hH_{\rm PT}$.
\item $\text{b}_{\rm PT}(\hH_{\rm RX})$ \eqref{eq:def H SYK PT KM}, $\text{b}_{\rm PT}(\hH_{\rm MQ})$ \eqref{eq:def H SYK PT TFD}: Hermitian operators, defined by functional calculus \eqref{eq:part_2b} on the spectral subspace $\HRX<1$, used in the KM and TFD deformations.
\item $\hH_{\rm MQ}$ \eqref{eq:H MQ}: Generalized Maldacena--Qi interaction term.
\item $\hH_{\rm def}$ \eqref{eq:H def KM}, $\hH_{\rm def'}$ \eqref{eq:H def TFD}: Deformation operators of the SYK Hamiltonian, which reproduce an annihilation operator in the double-scaling limit.
\item $\ket{K_n}$ \eqref{eq:lanczos new}, $\ket{n}$ \eqref{eq:H chord ortho Krylov}: Krylov basis of a generic system, and chord number basis of DSSYK, respectively.
\item $\hH_{\rm PT}$ \eqref{eq:chord PT H}: Chord Hamiltonian reproducing $\hH_{\rm JT-PT}$ in the low-energy limit.
\item $q=\rme^{-\lambda}\in[0,1)$ \eqref{eq:double scaling}: DSSYK parameter.   
\item $\xi$ \eqref{eq:case 1 b}: Boundary deformation parameter which reproduces the value of the dilaton at the finite cutoff $\Phi_b$.
\item $\hat{n}$, $\text{L}=\lambda\expval{\hat{n}}$ and $\hat{p}_n$ \eqref{eq:p n def}: Chord number operator, a rescaled variable, and its canonical conjugate momentum. In the low-energy limit \eqref{eq:low energy limit}, $\hat{L}\equiv\lambda\hat{n}$ and $\hat{P}\equiv\hat{p}_n/\lambda$ denote the corresponding operators. Similarly, $L$, $ P$ bulk length between finite cutoff boundaries and its conjugate momentum in (dS) JT gravity.
\item $\lambda$ \eqref{eq:double scaling}: Double-scaling parameter, $\lambda=p^2/N$, whose classical limit is $\lambda\rightarrow0$.
\item $\hH_{\rm PT'}$, $\mathcal{K}_0$, $\mathcal{K}_\infty$ \eqref{eq:PT improved H}: Chord Hamiltonian with the improvement term, and the operators implementing the operator ordering \eqref{eq:sl2R op ordering}.
\item $\Phi_b$: Boundary value of the dilaton.
\item $\Phi$: Dilaton field.
\item $\eta$ \eqref{eq:flow eq}: Deformation parameter whose value labels $T^2$ ($\eta=+1$) or $T^2+\Lambda_1$ ($\eta=-1$) deformations.
\item $\varepsilon$ \eqref{eq:E(theta) eta+1}: Spectral parameter labeling perturbative ($\varepsilon=-1$) and non-perturbative ($\varepsilon=+1$) solutions to the energy flow equation \eqref{eq:flow eq}.
\item $L_k$, $\tilde{L}_k$ \eqref{eq:length CMC}: Lengths of CMC slices with mean curvature $k$ for $\varepsilon=-1$ and $\varepsilon=+1$, respectively, with $L\equiv L_{k=0}$ and $\tilde{L}\equiv\tilde{L}_{k=0}$.
\item $K$ \eqref{eq:K_CMC}: Extrinsic curvature, which is also denoted $k$ when it is fixed.
\item $\hat{H}_{\rm ADM}$ \eqref{eq:ADM H non}: ADM Hamiltonian of the non-perturbative branches.
\item $C_{\mfa}(t)$ \eqref{eq:def CMC-Krylov cmplexity}: CMC-Krylov complexity.
\item $\mfa$ \eqref{eq:def CMC-Krylov cmplexity}: Parameter of CMC-Krylov complexity, with $\mfa=k^2$ in AdS$_2$ and $\mfa=-k^2$ in dS$_2$.
\item $\hH_{\rm ETW}$, $a_{\rm ETW}(\hat{n})$, $b_{\rm ETW}(\hat{n})$ \eqref{eq:def Hamiltonian}: Deformed chord Hamiltonian with an ETW brane and its coefficients.
\item $\mathcal{Y}$, $\overline{\mathcal{Y}}$ \eqref{eq:def Hamiltonian}: q-Askey deformation parameters associated with the ETW brane.
\item $\hH_{\rm JT-ETW}$, $U_{\rm m}(\hat{L})$ \eqref{eq:schr Scarf mod}: Finite-cutoff JT gravity Hamiltonian with an ETW brane and the improvement term, and its modified Scarf II potential.
\item $m_{\rm JT}$ \eqref{eq:main text L brane}: Tension of an ETW brane.
\item $y$ \eqref{eq:flow eq}: Deformation parameter of the $T^2$ and $T^2+\Lambda_1$ flows, associated with a finite radial cutoff; $y=4\pi G_N/(1-\cos\Phi_b(y))$ \eqref{eq:holog dict SDG DSSYK}.
\item $G_N$ \eqref{eq:holog dict SDG DSSYK}: Newton's constant in the sine dilaton gravity action.
\item $\Phi_h$: Dilaton at the horizon.
\item $\theta\in[0,\pi]$ \eqref{eq:E theta seed}: Energy parameter of the chord Hamiltonian, where $\sin\theta/(2\pi)$ is the fake temperature.
\item $t_P$ \eqref{eq:tP}: Proper time at the cutoff boundary.
\item $C_0(t)$ \eqref{eq:n hat def intro}: Krylov complexity.
\item $K_{\rm AdS}$ \eqref{eq:brane conditions}: Extrinsic curvature of the ETW brane, with respect to the unit normal pointing away from the cutoff boundary ($K_{\rm AdS}=-\tanh\sigma_0$).
\item $2N$, $p$ \eqref{eq:SYK Hamiltonian}: Number of Majorana fermions $\hat{\psi}_i$ and order of the $p$-body all-to-all interaction, respectively.
\item $\hH_{\rm SYK}$, $\hat{\psi}_i$, $\mathcal{J}_I$ \eqref{eq:SYK Hamiltonian}, \eqref{eq:GEA J}: SYK Hamiltonian, Majorana fermions, and Gaussian random couplings.
\item $\ket{\Omega}$ \eqref{eq:basis}: Empty chord state.
\item $\hat{a}$, $\hat{a}^\dagger$ \eqref{eq:H chord ortho Krylov}: Annihilation and creation operators acting on the chord number states $\ket{n}$.
\item $\hat{S}_k$ \eqref{eq:spin op}: Spin operators whose eigenstates are the KM states.
\item $\hat{c}_j$, $\hat{c}_j^\dagger$: Annihilation and creation operators in the spin basis.
\item $\hat{M}$, $\hat{M}^\dagger$ \eqref{eq:def M Mdagger}: Product of creation and annihilation operators in the spin basis.
\item $w$, $R_{i_1,\dots,i_w}$ \eqref{eq:def M Mdagger}, \eqref{eq:gaussian dist R}: Number of fermion operators in $\hat{M}$, and their Gaussian random couplings.
\item $\ket{\omega}$ \eqref{eq:dictionary SYK chord KM}: KM state.
\item $P$ \eqref{eq:dictionary SYK chord KM}: Generic polynomial; not to be confused with the momentum $\hat{P}$ conjugate to $\hat{L}$). $P^{(\eta)}_{\rm PT}(x)$ \eqref{eq:2nd}: The function $\frac{1}{y}(1+\varepsilon\sqrt{\eta(1-x)})$ defining the deformed microscopic Hamiltonians by functional calculus.
\item $\hH_{\rm RX}$ \eqref{seq:H int}: Rajgadia--Xu interaction term acting on KM states.
\item $\Delta_w\equiv w/p$ \eqref{eq:dictionary SYK chord KM}: Parameter of the KM dictionary, set to $\Delta_w=1$.
\item $\mathcal{H}_{\rm SYK}$ \eqref{eq:SYK D}: Seed theory SYK Hilbert space.
\item $\ket{E_n}$ \eqref{eq:SYK D}: Microscopic energy basis.
\item $\ket{\rm TFD_\infty}$ \eqref{eq:inf temp TFD}: Infinite-temperature TFD state.
\item $\hat{s}$ \eqref{eq:operator size}: Maldacena--Qi size operator, which corresponds to the total chord number $\hat{n}$ in the double-scaling limit.
\item $s$, $\Delta_s\equiv s/p$ \eqref{eq:H MQ}, \eqref{eq:dictionary SYK and chord TFD}: Number of fermion pairs in $\hH_{\rm MQ}$ and parameter of the TFD dictionary, set to $\Delta_s=1$ (not to be confused with the energy variable $s$).
\item $E_y$ \eqref{eq:flow eq}: Deformed theory energy spectrum.
\item $\hH_y$ \eqref{eq:E(theta) eta+1}, \eqref{eq:E(theta) eta-1}: $T^2$ and $T^2+\Lambda_1$ deformed chord Hamiltonian. 
\item $y_0\equiv1/(2E)$ \eqref{eq:E(theta) eta-1}: Value of $y$ where the $T^2$ and $T^2+\Lambda_1$ branches are glued.
\item $E(\theta)=-2\cos\theta/\sqrt{1-q}$, $\ket{\theta}$ \eqref{eq:E theta seed}: Energy spectrum of $\hH$ and $\hH_{\rm PT}$, and corresponding eigenstate.
\item $\hH_{\rm JT-PT}$ \eqref{eq:JT PT}: PT JT gravity Hamiltonian.
\item $\ket{L}$, $\ket{s}$, $\psi_s(L)=\bra{s}\ket{L}$ \eqref{eq:psi s fried}: Length basis, energy basis of $\hH_{\rm JT-PT}$ with eigenvalue $s^2$, and wavefunctions selected by the Friedrichs extension.
\item $Z(\beta)$ \eqref{eq:orthogonality}: Partition function.
\item $\rho_0(s)$ \eqref{eq:orthogonality}: Density of states of finite-cutoff JT gravity without matter.
\item $m$ \eqref{eq:2pt}: Mass of the bulk free scalar field in AdS, related to the conformal dimension $\Delta $ of a boundary operator $m^2=\Delta(\Delta-1)$.
\item $\tilde{G}_2(\tau_1,\tau_2)$, $G_2(s_1,s_2)$ \eqref{eq:2pt}: Two-point function and its kernel in the energy basis.
\item $\tau=\beta/2+\rmi t$ \eqref{eq:n hat def intro}: Complexified time parameter in the HH preparation of state. 
\item $\hat{\mathcal{L}}$ \eqref{eq:liouv}: Generalized Liouvillian operator.
\item $a_n$, $b_n$ \eqref{eq:lanczos new}: Lanczos coefficients.
\item $\mathcal{C}^\epsilon$ \eqref{eq:OF1}: CAny observable on the CMC slice $\Sigma_\epsilon$, where $\epsilon=\pm$ (not to be confused with the branch label $\varepsilon$).
\item $\Sigma_{\epsilon}$, $\epsilon=\pm$ \eqref{eq:regions CMC}: Future/past boundaries of $\mathcal{M}$.
\item $F[g_{\mu\nu},X^\mu]$ \eqref{eq:OF1}: Scalar functional of the bulk metric and of the embedding functions of $\Sigma_\epsilon$.
\item $\mathcal{M}$ \eqref{eq:regions CMC}: Spacetime manifold.
\item $h_{\mu\nu}$ \eqref{eq:regions CMC}: Induced metric at a surface.
\item $b_{\rm PT}(\hat{n})$ \eqref{eq:case 1 b}: Coefficient of the deformed chord Hamiltonian \eqref{eq:chord PT H}.
\item $\hH^{(\eta)}_{{\rm PT},y}$ \eqref{eq:PT y}: $T^2$ ($\eta=+1$) and $T^2+\Lambda_1$ ($\eta=-1$) deformations of $\hH_{\rm PT}$.
\item $E_{\rm ADM}$ \eqref{eq:general energy dS}: ADM energy.
\item $\mathcal{G}(\hat{L},\hat{P})$ \eqref{eq:sl2R op ordering}: Operator ordering of the momentum in finite-cutoff JT gravity.
\item $\hbar$ \eqref{eq:specifif replacement}: Constant parametrizing the classical limit in the bulk as $\hbar\rightarrow0$. 
\item $\rmd l^2=-f(\Phi)\rmd t^2+\rmd \Phi^2/f(\Phi)$ \eqref{eq:Schwarzschild coordinates}: Metric in Schwarzschild coordinates $t$, $\Phi$.
\item $u\equiv\Phi_h t,~ \bbr\equiv\Phi_b/\Phi_h$ \eqref{eq:geodesic length etaplus epsminus}: Useful rescalings.
\item $\mathfrak{c}$ \eqref{eq:int eps plus}: Integration constant in the CMC-slice analysis of App.~\ref{app: CMC lengths} parametrizing the boundary time $t({\mathfrak{c}})$.
\item $n^b$ \eqref{eq:brane conditions}: Unit normal vector to the ETW brane.
\item $s_n$ \eqref{eq:connection poles}, \eqref{eq:Nsn}: Bound-state energies, $E_{\rm ADM}=-s_n^2$, with an ETW brane.
\item $\rho(s)$ \eqref{eq:density coeffs}: Density of states of finite-cutoff JT gravity with an ETW brane.
\item $f_1$, $f_2$ \eqref{eq:f1 sol}, \eqref{eq:f2 sol}: Independent solutions of the Heun equation.
\item $\mathcal{N}$: Normalization constant of $\psi_s(L)$.
\item $\mathcal{A}_\pm$ \eqref{eq:asymptotics of psi s}: Asymptotic amplitudes of $\psi_s(L)$.
\item $\mathcal{A}_s$, $\tilde{\mathcal{A}}_s$, $n_0$ (App.~\ref{app:log corrections}): Asymptotic amplitudes of the Friedrichs solution $\psi_s$ and of the second solution $\tilde{\psi}_s$ of the low-energy Schrödinger equation, and the $\mathcal{O}(1)$ constant locating the effective Dirichlet wall $L=\lambda n_0$.
\item $\mathcal{C}_u$ \eqref{eq:rho Heun}: Heun connection coefficient.
\item $x\equiv 8\Phi_b\rme^{-L}$ \eqref{eq:kimit crucial}: Fixed parameter in the asymptotic boundary limit where $\Phi_b$, $L\rightarrow\infty$.
\item $\ket{L_{\rm reg}}=\ket{L}/\sqrt{L}$ \eqref{eq:L int}: Normalized length state for states defined by the Friedrichs extension, which remains finite as $L\rightarrow0$.
\item $\eta_{IJ}$ \eqref{eq: X coords}: Minkowski metric.
\item $X^I$ \eqref{eq: X coords}: Embedding coordinates in 3d Minkowski space.
\item $V^I$, $\mathbb{J}^I$ \eqref{eq:dilaton coord}, \eqref{eq:coordinate useful}: Vector defining the dilaton, $\Phi=V\cdot X$, and $E_{\rm ADM}=-V\cdot V$; and the Killing vector that leaves $\Phi$ invariant.
\item $\sigma$, $T$ \eqref{eq:coordinate useful}: Global AdS$_2$ coordinates; $\sigma_{L/R}$, $T_{L/R}$ refer to the left/right boundaries.
\item $\sigma_0$, $\sigma_b$ \eqref{eq:coordinate useful}: Positions of the static ETW brane and of the cutoff boundary in global coordinates, with $K=-\tanh\sigma_0$.
\item $G_4$ \eqref{eq:G4 semiclassical}: Suggested crossed four-point function.
\item $L_{L/R}$ \eqref{eq:one sided geodesics}, $L_{k,L/R}$ \eqref{eq:LKL}: One-sided geodesic lengths in finite-cutoff JT gravity with shockwaves, and the corresponding CMC slice lengths, respectively.
\end{itemize}

\paragraph{Acronyms}
\begin{itemize}[noitemsep,
    leftmargin=1.7cm,
    labelwidth=1.6cm,
    labelsep=0.5cm,
    align=left]
    \item[ABL] Asymptotic boundary limit.
\item[ADM]	Arnowitt–Deser–Misner.
\item[(A)dS]	(Anti-)de Sitter / de Sitter.
\item[CAny]	Complexity=Anything.
\item[CMC]	Constant Mean Curvature.
\item[CFT]	Conformal Field Theory.
\item[CV]	Complexity=Volume.
\item[(DS)SYK]	(Double-scaled) Sachdev–Ye–Kitaev.
\item[ETW]	End-of-the-World.
\item[HH]	Hartle–Hawking.
\item[JT]	Jackiw–Teitelboim.
\item[KM]	Kourkoulou–Maldacena.
\item[MQ]	Maldacena–Qi.
\item[OTOC]	Out-of-Time-Order Correlator.
\item[PT]	Pöschl–Teller.
\item[RX]	Rajgadia–Xu.
\item[TFD]	Thermofield double.
\item[WKB]  Wentzel–Kramers–Brillouin.
\end{itemize}
\paragraph{Conventions}
In Tab.~\ref{tab:conventions} we compare normalization conventions used in the main text.
\begin{table}[t]
    \centering
    \begin{tabular}{C{7.5cm}C{7.5cm}}
    \toprule
    \text{Two-sided geometry (Sec.~\ref{sec:deformation}, \ref{sec:CMC-Krylov complexity})}     &Single-sided geometry (Sec.~\ref{sec:bulk with matter})  \\
    \midrule
    \multicolumn{2}{c}{Kinetic term}\\
    $P^2$ \eqref{eq:corrected H JT2}&$P^2/4$ \eqref{eq:schr Scarf mod}\\\hline
    \multicolumn{2}{c}{Low-energy limit}\\
    $P\rightarrow\lambda P$, \eqref{eq:low energy 2sided}&$P\rightarrow \lambda P/2$ \eqref{eq:low energy lim ETW}\\
$\xi\rightarrow\lambda\Phi_b$&$\xi\rightarrow\lambda\Phi_b$, $\mathcal{Y}\rightarrow\lambda \mathcal{Y}$, $\overline{\mathcal{Y}}\rightarrow\lambda \overline{\mathcal{Y}}$\\\hline
    \multicolumn{2}{c}{Length variable}\\
    $\lambda\hat{n}\rightarrow\hat{L}$ \eqref{eq:low energy limit}&$\lambda\hat{n}\rightarrow2\hat{L}$ \eqref{eq:low energy lim ETW}\\\hline
    \multicolumn{2}{c}{Prefactor of the chord Hamiltonian}\\
    $-1/\sqrt{1-q}$ \eqref{eq:chord PT H}, \eqref{eq:PT improved H}&$-1/\sqrt{1-q}$ \eqref{eq:brane chord H}\\
    \bottomrule
    \end{tabular}
    \caption{Comparison between different normalizations of the kinetic term and the scaling of the parameters used to take the low-energy limit in the two-sided and one-sided spacetime geometries of the main text.}
    \label{tab:conventions}
\end{table}

\section{Double-scaling limit of the microscopic deformations}\label{app:Krylov micro H}
In this appendix, we provide technical details and extensions of the SYK deformations associated with finite-cutoff holography in Sec.~\ref{ssec:SYK defs}.\footnote{Throughout the appendix, we use standard operator ordering in quantum mechanics.} In App.~\ref{app:detalssww} we give details about the double-scaling limit of the deformation operators $\hH_{\rm def}$ in \eqref{eq:H def KM} and $\hH_{\rm def'}$ in \eqref{eq:H def TFD} based on the dictionaries \eqref{eq:dictionary SYK chord KM} and \eqref{eq:dictionary SYK and chord TFD} between microscopic SYK operators and chord algebra operators \eqref{eq:chord algebra}. In App.~\ref{sapp:alternative H}, we introduce a different SYK deformation whose low-energy limit reproduces JT gravity at finite cutoff in Schwarzschild time \cite{Griguolo:2025kpi}.

\subsection{Deformation operators}\label{app:detalssww}
We now apply the dictionary between SYK and chord operators in the KM and TFD \eqref{eq:dictionary SYK and chord TFD} formulations to derive the double-scaling limit of $\hH_{\rm def}$ \eqref{eq:H def KM} and $\hH_{\rm def'}$ \eqref{eq:H def TFD}.

\paragraph{KM formalism}
Let us start with $\hH_{\rm def}$ \eqref{eq:H def KM}. Using the dictionary between SYK and chord operators in the KM formulation \eqref{eq:dictionary SYK chord KM} when $w=1$, which means that $\lambda\Delta_w=p/N$, and keeping $1/N$ terms in the expansion, one finds that \cite{Aguilar-Gutierrez:2026nmd} (see alternative derivation in \cite{Rajgadia:2026ask})
\begin{eqnarray}\label{eq:towards chord number}
    \lim_{\rm DS}\overline{\bra{\omega}P\qty(\hH_{\rm SYK},\sum_{j=1}^{N}\frac{1}{2p}\qty(1-\rmi\psi_{2j-1}\psi_{2j}))\ket{\omega}}=\bra{\Omega}P\qty(-(\ha+\ha^\dagger),~\hat{n})\ket{\Omega}~.
\end{eqnarray}
Now, using \eqref{eq:dictionary SYK chord KM}, we find,\footnote{Note the constant additive factors inside the sum on the left-hand side of \eqref{eq:towards chord number} do not contribute in computing the commutator used in the definition of $\hH_{\rm def}$ \eqref{eq:H def KM}.}
\begin{eqnarray}\label{eq:Double scaling H def}
\begin{aligned}
    \lim_{\rm DS}\overline{\bra{\omega}P\qty(\hH_{\rm SYK},\hH_{\rm def})\ket{\omega}}&=\bra{\Omega}P\qty(-\ha-\ha^\dagger,~[\hn,\ha+\ha^\dagger]-\ha-\ha^\dagger)\ket{\Omega}\\
    &=\bra{\Omega}P\qty(-\ha-\ha^\dagger,-2\ha)\ket{\Omega}~,
\end{aligned}
\end{eqnarray}
which implies $\hH_{\rm def}$ \eqref{eq:H def KM} can be identified with $-2\ha$ in the double-scaling limit. 
Explicitly, with $\HRX\rightarrow q^{\hat{n}}$ and $\hH_{\rm def}\rightarrow-2\ha$, \eqref{eq:def H SYK PT KM} becomes $-\big(\text{b}_{\rm PT}(q^{\hat{n}})\ha^\dagger+\ha\,\text{b}_{\rm PT}(q^{\hat{n}})\big)$, whose only nonzero matrix elements are
\[
\bra{n+1}\hH_0\ket{n}=\bra{n}\hH_0\ket{n+1}=-\text{b}_{\rm PT}(q^{n+1})\sqrt{[n+1]_q}=-\frac{b_{\rm PT}(n+1)}{\sqrt{1-q}}~,
\]
where we used \eqref{eq:part_2b} and $\sech^2(\lambda n/2)=4q^n/(1+q^n)^2$. They coincide with those of \eqref{eq:chord PT H}, with $\rme^{-\rmi\hat{p}_n}\ket{n}=\ket{n+1}$. Since $\hH_0$ has no diagonal matrix elements in the chord basis, the odd moments vanish on both sides of \eqref{eq:DS H PT}. The same holds for \eqref{eq:def H SYK PT TFD} by \eqref{eq:Double scaling H def prime}.

\paragraph{TFD formalism}
Let us now proceed with $\hH_{\rm def'}$ \eqref{eq:H def TFD}. As mentioned in \eqref{eq:operator size}, $\hat{s}$ can be identified with $\hat{n}$ in the double-scaling limit, as it follows immediately from the same procedure explained for $\hH_{\rm def}$ just before, so that applying the chord algebra \eqref{eq:chord algebra}, we recover
\begin{eqnarray}\label{eq:Double scaling H def prime}
\begin{aligned}
    \lim_{\rm DS}\overline{\bra{\rm TFD_{\infty}}P\qty(\hH_{\rm SYK},\hH_{\rm def'})\ket{\rm TFD_\infty}}&=\bra{\Omega}P\qty(-\ha-\ha^\dagger,-2\ha)\ket{\Omega}~.
\end{aligned}
\end{eqnarray}
Therefore, applying the dictionary between SYK and chord operators in the KM formalism \eqref{eq:dictionary SYK chord KM} for $\Delta_w=1$ and in TFD case \eqref{eq:dictionary SYK and chord TFD} for $\Delta_s=1$, we find that \eqref{eq:DS H PT} and \eqref{eq: H TFD Omega} are satisfied,
\begin{eqnarray}\label{eq:excat match}
     \bra{\Omega}\hH^m_{\rm PT}\ket{\Omega}=\lim_{\rm DS}\overline{\bra{\omega}\hH^m_0\ket{\omega}}=\lim_{\rm DS}\overline{\bra{\rm TFD_\infty}\hH^m_{0'}\ket{\rm TFD_\infty}}~,
\end{eqnarray}
where $\hH_{\rm PT}$ is given in \eqref{eq:chord PT H}. Note that \eqref{eq:excat match} implies the ensemble-averaged moments of the microscopic SYK deformation reproduce the chord Hamiltonian moments of $\hH_{\rm PT}$ exactly in terms of $\lambda$, not just semiclassically, with $\text{b}_{\rm PT}(\HRX)$ understood through the double-scaled moments, i.e.~with $\HRX\rightarrow q^{\hat{n}}$ applied before the function $\text{b}_{\rm PT}$ (see the remark below \eqref{eq:part_2b}).

\subsection{Alternative SYK deformation}\label{sapp:alternative H}
We are interested in a chord Hamiltonian \cite{Aguilar-Gutierrez:2026ogo} that recovers JT gravity at finite cutoff expressed in terms of Schwarzschild time \cite{Griguolo:2025kpi}
\begin{subequations}
    \begin{align}
        \hH_y&=
        \frac{1}{y}\qty(1-\frac{1}{2}\qty(\tanh\qty(\frac{\lambda \hat{n}_y}{2})\rme^{-\rmi\hat{p}_y}+\rme^{\rmi\hat{p}_y}\tanh\qty(\frac{\lambda \hat{n}_y}{2})))~,
        \label{eq:def H Krylov basis2}\\
        \hat{n}_y&\equiv\sum_nn\ket{K^{(y)}_n}\bra{K^{(y)}_n}~.\label{eq:n_y2}
    \end{align}
\end{subequations}
For this, we introduce new SYK deformations whose double-scaling limit recovers \eqref{eq:def H Krylov basis2}
where the operators appearing above act as in \eqref{eq:chord algebra},\footnote{In particular, the low-energy limit of \eqref{eq:def H Krylov basis2} reproduces the JT gravity Hamiltonian at finite cutoff appearing in \cite{Griguolo:2025kpi}
\begin{eqnarray}\label{eq:JT Krylov form}
    \hH_{\rm JT'}=\frac{1}{y}\qty(1-\tanh(\hat{L}/2)\cos(\hat{P}))~,
\end{eqnarray}
where $\hat{L}$ and $\hat{P}$ are canonically conjugate operators.} and \eqref{eq:p n def}. We proceed in a similar way as Sec.~\ref{ssec:SYK defs} by proposing SYK deformations that reproduce \eqref{eq:def H Krylov basis2} in the double-scaling limit.

\paragraph{Microscopic KM formulation}
Let us start from the single-copy SYK, and introduce the following Hamiltonians describing the perturbative branch, where $\varepsilon=-1$:
\begin{equation}\label{eq:micro H_K KM}
 \hH_{\rm SYK}''=
        \frac{1}{y}+\frac{\sqrt{1-q}}{4y}\qty(b_1(\hH_{\rm RX}){\hH_{\rm def}^\dagger}+{\hH_{\rm def}}b_1(\hH_{\rm RX}))~,
\end{equation}
where $b_1(x)=\sqrt{1-x}(1+x)^{-1}$ and $b_{1}(\HRX)$ is again defined by functional calculus as \eqref{eq:part_2}. We then find that
\begin{equation}
    \lim_{\rm DS}\overline{\bra{\omega}(\hH_{\rm SYK}'')^m\ket{\omega}}=\bra{\Omega}\hH_y^m\ket{\Omega}~,
\end{equation}
where we applied the dictionary \eqref{eq:dictionary SYK chord KM} with $\Delta_w=1$, and the double-scaling limit of the Hamiltonian deformation \eqref{eq:H def KM} (explained in App.~\ref{app:detalssww}).

\paragraph{Microscopic TFD formulation}
Alternatively, we consider a bipartite system in \eqref{eq:SYK D} to define the following SYK deformation
\begin{equation}\label{eq:micro H_K TFD}
    \hH'_{\rm SYK'}=\frac{1}{y}+\frac{\sqrt{1-q}}{4y}\qty(b_1(\hH_{\rm MQ}){\hH_{\rm def'}^\dagger}+{\hH_{\rm def'}}b_1(\hH_{\rm MQ}))~,
\end{equation}
where $b_1(x)\sqrt{1-x}(1+x)^{-1}$, $\hH_{\rm MQ}$ in \eqref{eq:H MQ}, and $\hH_{\rm def'}$ in \eqref{eq:H def TFD}.

Using the map between SYK and chord operators in \eqref{eq:dictionary SYK and chord TFD} with $\Delta_s=+1$ as well as the double scaling limit of the Hamiltonian deformation in \eqref{eq:H def TFD} (see App.~\ref{app:detalssww}), we find that \eqref{eq:micro H_K TFD} satisfies the following relation 
\begin{eqnarray}
    \lim_{\rm DS}\overline{\bra{{\rm TFD}_\infty}(\hH'_{\rm SYK'})^k\ket{{\rm TFD}_\infty}}=\bra{\Omega}\hH_{y}^k\ket{\Omega}~,
\end{eqnarray}
which thus allows us to identify $\hH'_{\rm SYK'}$ as a microscopic Hamiltonian whose ensemble average reproduces the chord Hamiltonian \eqref{eq:def H Krylov basis2}.

\section{Logarithmic approach of \texorpdfstring{$Z_{\rm PT'}$}{Z\_PT'} to the JT partition function}\label{app:log corrections}
In this appendix, we evaluate the low-energy limit of the chord partition function \eqref{eq:Zdef} in Sec.~\ref{ssec:improvement JT} and estimate its leading-order corrections with respect to the JT gravity case \eqref{eq:orthogonality}. 

Recall that the low-energy limit of the chord Hamiltonian with the improvement term \eqref{eq:PT improved H} is
\begin{equation}
  \sqrt{1-q}\,\hat H_{\rm PT'}+2=\lambda^2\left(\hat P^2+\Phi_b^2\,{\rm sech}^2\frac{\hat L}{2}-\frac{1}{4}\,{\rm csch}^2\hat L\right)
  +\mathcal O(\lambda^4)\,,
  \label{eq:Zlow}
\end{equation}
and we denote its eigenvalues by $\sqrt{1-q}\,\hat H_{\rm PT'}+2=\lambda^2s^2$. The wavefunctions \eqref{eq:psi s fried} solve the corresponding
Schrödinger equation, and near $L=0$ they behave as $\psi_s(L)\simeq\sqrt{L/2}$.

\paragraph{Self-adjoint extensions}
Near $L=0$, $-\frac14{\rm csch}^2L=-\frac{1}{4L^2}+\frac{1}{12}+\mathcal O(L^2)$, which is the critical value of an
inverse-square potential. The indicial equation has the double root $\frac12$, so the second solution is logarithmic,
\begin{equation}
  \psi_s(L)=\sqrt{\frac L2}\,\big(1+\mathcal O(L^2)\big)\,,\qquad
  \tilde\psi_s(L)=\sqrt{\frac L2}\,\left(\log\frac2L+\mathcal O(L^2\log L)\right).
  \label{eq:Zlocal}
\end{equation}
Since both solutions are square integrable at $L=0$, the self-adjoint extensions of (\ref{eq:Zlow}) form a one-parameter
family, characterized by a fixed ratio between the $\sqrt L\log L$ and $\sqrt L$ terms of the wavefunction as $L\to0$. The
Friedrichs extension, used in \cite{Griguolo:2026eoi} and in Sec.~\ref{ssec:improvement JT}, is the one without $\sqrt L\log L$ term, which selects $\psi_s(L)$ in
\eqref{eq:psi s fried}. For $L\to\infty$, let us write, as in \eqref{eq:asymptotics of psi s},
\begin{equation}
  \psi_s(L)\simeq\mathcal A_s\,{\rm e}^{{\rm i}sL}+\mathcal A_s^*\,{\rm e}^{-{\rm i}sL}\,,\qquad
  \tilde\psi_s(L)\simeq\tilde{\mathcal A}_s\,{\rm e}^{{\rm i}sL}+\tilde{\mathcal A}_s^*\,{\rm e}^{-{\rm i}sL}\,,
  \label{eq:Zasym}
\end{equation}
so that the density of states \eqref{eq:orthogonality} of the Friedrichs extension is $\rho_0(s)\propto|\mathcal A_s|^{-2}$. Evaluating the
Wronskian $\psi_s\tilde\psi_s'-\psi_s'\tilde\psi_s=-\frac12$ with (\ref{eq:Zlocal}) and with (\ref{eq:Zasym}), one finds
\begin{equation}
  {\rm Im}\,\frac{\tilde{\mathcal A}_s}{\mathcal A_s}=\frac{1}{8s\,|\mathcal A_s|^2}\propto\frac{\rho_0(s)}{s}\,,
  \label{eq:ZIm}
\end{equation}
while ${\rm Re}(\tilde{\mathcal A}_s/\mathcal A_s)$ is not fixed by general arguments. At large $s$, where the potential in
(\ref{eq:Zlow}) can be neglected except for the inverse-square term, it grows as $\log s$.

\paragraph{Boundary condition}
The chord Hamiltonian is defined on $L=\lambda n$ with $n\geq1$. For $s\ll1/\lambda$, the discreteness of the chord
number is only relevant within a distance $\mathcal O(\lambda)$ from $L=0$, and at larger distances it acts as a boundary
condition. We can model it as a Dirichlet boundary condition at $L=\lambda n_0$, where $n_0=\mathcal O(1)$ is a
non-universal constant (any other local structure of size $\lambda$ leads to the same form with a different $n_0$).
The solution vanishing at $L=\lambda n_0$ is
\begin{equation}
  \tilde\psi_s(\lambda n_0)\,\psi_s(L)-\psi_s(\lambda n_0)\,\tilde\psi_s(L)
  \;\propto\;\log\frac{2}{n_0\lambda}\,\psi_s(L)-\tilde\psi_s(L)\,,
  \label{eq:Zwall}
\end{equation}
where we used (\ref{eq:Zlocal}), up to corrections of order $\lambda^2s^2$. The ratio between the two solutions does not
depend on $s$, so (\ref{eq:Zwall}) is a self-adjoint extension of (\ref{eq:Zlow}). Its parameter
$\log\frac{2}{n_0\lambda}$ diverges as $\lambda\to0$, so it approaches the Friedrichs extension, but only logarithmically.
This is the analog of the running of a marginal boundary coupling with the cutoff scale \cite{Ohya:2010zm}.

\paragraph{Spectral density of $\ket{n=1}$}
The state $\ket{n=1}$ probes the wavefunction at $L=\lambda$, namely $\langle n=1|s\rangle\propto\sqrt\lambda\,\psi(\lambda)$
for the wavefunction (\ref{eq:Zwall}). From (\ref{eq:Zlocal}),
\begin{equation}
  \log\frac{2}{n_0\lambda}\,\psi_s(\lambda)-\tilde\psi_s(\lambda)=\sqrt{\frac\lambda2}\,\log\frac{1}{n_0}\,,
\end{equation}
which is independent of $s$. On the other hand, by (\ref{eq:Zasym}), the asymptotic amplitude of (\ref{eq:Zwall}) is
$\log\frac{2}{n_0\lambda}\,\mathcal A_s-\tilde{\mathcal A}_s$, so its density of states is
$\rho_0(s)\,|\log\frac{2}{n_0\lambda}-\tilde{\mathcal A}_s/\mathcal A_s|^{-2}$, up to an $s$-independent factor.
Combining both, we find
\begin{equation}
  Z_{\rm PT'}(\beta)\propto\lambda^2\log^2\frac{1}{n_0}\;{\rm e}^{\frac{2}{\sqrt{1-q}}\beta}\int_0^\infty{\rm d}s\,
  \frac{\rho_0(s)}{\big|\log\frac{2}{n_0\lambda}-\tilde{\mathcal A}_s/\mathcal A_s\big|^2}\,{\rm e}^{-\lambda^{3/2}\beta s^2}
  \Big(1+\mathcal O(\lambda^2s^2)\Big)\,,
  \label{eq:Zexact}
\end{equation}
where the proportionality constant does not depend on $s$ or $\lambda$. This means that the state $\ket{n=1}$ measures
$\rho_0(s)$ through the factor $|\log\frac{2}{n_0\lambda}-\tilde{\mathcal A}_s/\mathcal A_s|^{-2}$, which becomes
independent of $s$ only as $\lambda\to0$.

\paragraph{Logarithmic correction}
For $\log(1/\lambda)\gg|\tilde{\mathcal A}_s/\mathcal A_s|$ we can expand
\begin{equation}
  \frac{1}{\big|\log\frac{2}{n_0\lambda}-\tilde{\mathcal A}_s/\mathcal A_s\big|^2}
  =\frac{1}{\log^2(1/\lambda)}\left[1-\frac{2\log(2/n_0)}{\log(1/\lambda)}
  +\frac{2\,{\rm Re}\big(\tilde{\mathcal A}_s/\mathcal A_s\big)}{\log(1/\lambda)}
  +\mathcal O\Big(\frac{1}{\log^2(1/\lambda)}\Big)\right]~,
\end{equation}
which, inserted in (\ref{eq:Zexact}), gives (\ref{eq:Zmain}). We note the following features:
\begin{itemize}
\item The leading term reproduces \eqref{eq:orthogonality} with $\beta_{\rm JT}=\lambda^{3/2}\beta$. The overall factor is
$\lambda^2/\log^2(1/\lambda)$ instead of $\lambda^2$.
\item The non-universal constant $n_0$ only appears in $s$-independent factors at this order, so it does not change the
shape of the density of states.
\item The shape of the density of states is modified at order $1/\log(1/\lambda)$ by
$2\,{\rm Re}(\tilde{\mathcal A}_s/\mathcal A_s)$, which is fixed by the low-energy Schrödinger equation and is therefore
universal. Since it depends on $s$, it cannot be absorbed in the normalization of $Z_{\rm PT'}(\beta)$, and it is
parametrically larger than any power of $\lambda$. The expansion requires $\log s\ll\log(1/\lambda)$.
\end{itemize}
Therefore, the low-energy limit of $\hat H_{\rm PT'}$ fixes the operator (\ref{eq:Zlow}), but not the boundary condition
at $L=0$, which is instead fixed by the chord basis. Recovering \eqref{eq:orthogonality} with power-law accuracy in $\lambda$ might require
terms at the edge $n=1$ of the chord Hamiltonian that implement the Friedrichs extension directly.

\section{Quantization of the \texorpdfstring{$\varepsilon=+1$}{epsilon=+1} branches}\label{app:non perturbative wavefunctions}
In Sec.~\ref{ssec:CMC lenghts main} we found the classical ADM Hamiltonian describing the non-perturbative branches (i.e.~$\varepsilon=+1$) of the $T^2$ \eqref{eq:E(theta) eta+1} and $T^2+\Lambda_1$ \eqref{eq:E(theta) eta-1} deformations. In this appendix, we canonically quantize the Hamiltonian, and we study its solutions.

First, we need to implement an operator ordering choice. In the following, we work with the prescription in \cite{Griguolo:2026eoi} corresponding to the replacement $\hat{P}^2\rightarrow \mathcal{G}(\hat{L},\hat{P})$ in \eqref{eq:sl2R op ordering}, and at the end of the appendix, we comment on implementing the minimal choice where $\hat{P}^2$ in \eqref{eq:sl2R op ordering} is unchanged. The Hamiltonian of interest is then
\begin{equation}\label{eq:H ADM D}
\hH_{\rm ADM}=-\hat{P}^2+\frac{\Phi_b^2}{\cos^2(\hat{L}/2)}+\frac{\csc^2\hat{L}}{4}~.
\end{equation}
The Hamiltonian is trivially unbounded from below due to the kinetic term with overall minus sign and thus unphysical. For these reasons, we do not discuss the Hilbert space of the non-perturbative branch in the main text.

Nevertheless, a meaningful question is how to solve the quantum mechanical system corresponding to $-\hH_{\rm ADM}$, which is the subject of the remainder of this appendix.

We now solve the corresponding Schrödinger equation as
\begin{equation}\label{eq:non perturbative Sch}
(\partial_L^2+U(L))\psi_s(L)=-s^2\psi_s(L)~,\quad U(L)=\Phi_b^2\sec^2(L/2)+\frac{\csc^2L}{4}~,
\end{equation}
i.e.~an attractive potential, with $-E_{\rm ADM}=s^2$. This equation corresponds to (B.1) in \cite{Griguolo:2026eoi} after the replacements $L\rightarrow\rmi L$ and $s\rightarrow\rmi s$, so that its solutions take the form
\begin{equation}\label{eq:sol to herm}
\psi_s(L)=\sqrt{\sin\frac{L}{2}}\qty(\cos\frac{L}{2})^{\frac{1}{2}+2\rmi\Phi_b}\qty(\mathfrak{w}\,{}_2F_{1}(\tfrac12+\rmi\Phi_b+s,\tfrac12+\rmi\Phi_b-s,1;\sin^2(L/2))+\mathfrak{y}\,F_{\log}(\sin^2(L/2)))~,
\end{equation}
where $\mathfrak{w}$ and $\mathfrak{y}$ are constant coefficients, $F_{\log}(\sin^2(L/2))\sim\sqrt{L}\log L$ as $L\rightarrow0$ while the other solution (i.e.$\mathfrak{y}=0$) goes $\psi_s(L)\sim\sqrt{L}$ as $L\rightarrow0$. The crucial difference with respect to \cite{Griguolo:2026eoi} is that there is a singularity in the potential of the Schrödinger equation not only at $L=0$ but also $L=\pi$ (as well as $L=n\pi$ with $n\in \mathbb{Z}$). 

To solve the problem, we need to specify the boundary conditions in \eqref{eq:non perturbative Sch} at the endpoints of the fundamental domain, which is $L\in[0,\pi]$. 

First, at $L=0$ we can impose either $\mathfrak{w}=0$ or $\mathfrak{y}=0$ or some other combination; in any of the cases, we come to the same conclusion. In particular, \cite{Griguolo:2026eoi} shows that in the problem $L\rightarrow\rmi L$ eliminating the logarithmic branch at the origin, i.e.~$\mathfrak{y}=0$, removes the unbounded from below discrete energy. In our case, from \eqref{eq:sol to herm} one can see that near $L\to\pi$ the wavefunction behaves as
\begin{equation}
\eval{\psi_s(L)}_{L\to\pi}=
\qty(\mathfrak{w}+\mathfrak{y}\log\pi)\qty(C_+(s)\,(\pi-L)^{\frac12+2i\Phi_b}+C_-(s)\,(\pi-L)^{\frac12-2i\Phi_b})+\dots~,
\end{equation}
where the ellipsis $\dots$ denotes subleading terms, and
\begin{equation}
\begin{aligned}
    C_+(s)
&\equiv2^{-\frac12-2i\Phi_b}
\frac{\Gamma(-2i\Phi_b)}
{\Gamma(\frac12-i\Phi_b+s)
 \Gamma(\frac12-i\Phi_b-s)}~,\\
 C_-(s)
&\equiv2^{-\frac12+2i\Phi_b}
\frac{\Gamma(2i\Phi_b)}
{\Gamma(\frac12+i\Phi_b-s)
 \Gamma(\frac12+i\Phi_b+s)}~.
\end{aligned}
\end{equation}
Next, we specify the boundary condition at $L=\pi$ to deduce the allowed energy spectrum, which we take as a generic self-adjoint boundary condition, where
\begin{equation}\label{eq:condition spectrum non pert}
C_+(s)=\rme^{\rmi \tilde{\theta}}C_-(s)~,
\end{equation}
with $\tilde{\theta}\in\mathbb{R}$ a U(1) phase associated with an inverse-square endpoint \cite{Derezi_ski_2016}. The energy spectrum is thus determined from \eqref{eq:condition spectrum non pert} and is no longer continuous, unlike in the perturbative branch case, but now discrete. For $s=\rmi\kappa$ with $\kappa>0$, $C_+/C_-$ is a pure phase that grows as $4\Phi_b\log\kappa$ (Stirling's formula), so that \eqref{eq:condition spectrum non pert} gives 
\begin{equation}\label{eq:sn exp}
s_n=\rmi\kappa_n~,\quad \kappa_n\propto \rme^{+\frac{\pi n}{2\Phi_b}}~,
\end{equation}
and $-E^{(n)}_{\rm ADM}=-\kappa_n^2$; hence $-\hH_{\rm ADM}$ is unbounded from below. For real $s$, the phase grows instead as $4\Phi_b\log s-2\pi s$, which gives ordinary levels with unit spacing in $s$, responsible for the trivial unboundedness of $\hH_{\rm ADM}$. Thus, the energy levels $-E_{\rm ADM}^{(n)}=s_n^2$ form a geometric tower, as in the Efimov effect \cite{efimov1973energy}, with $-E^{(n)}_{\rm ADM}\rightarrow-\infty$: the endpoint $L=\pi$ carries the supercritical inverse-square coupling $\frac14+4\Phi_b^2$.

\begin{figure}
    \centering
    \includegraphics[width=0.5\linewidth]{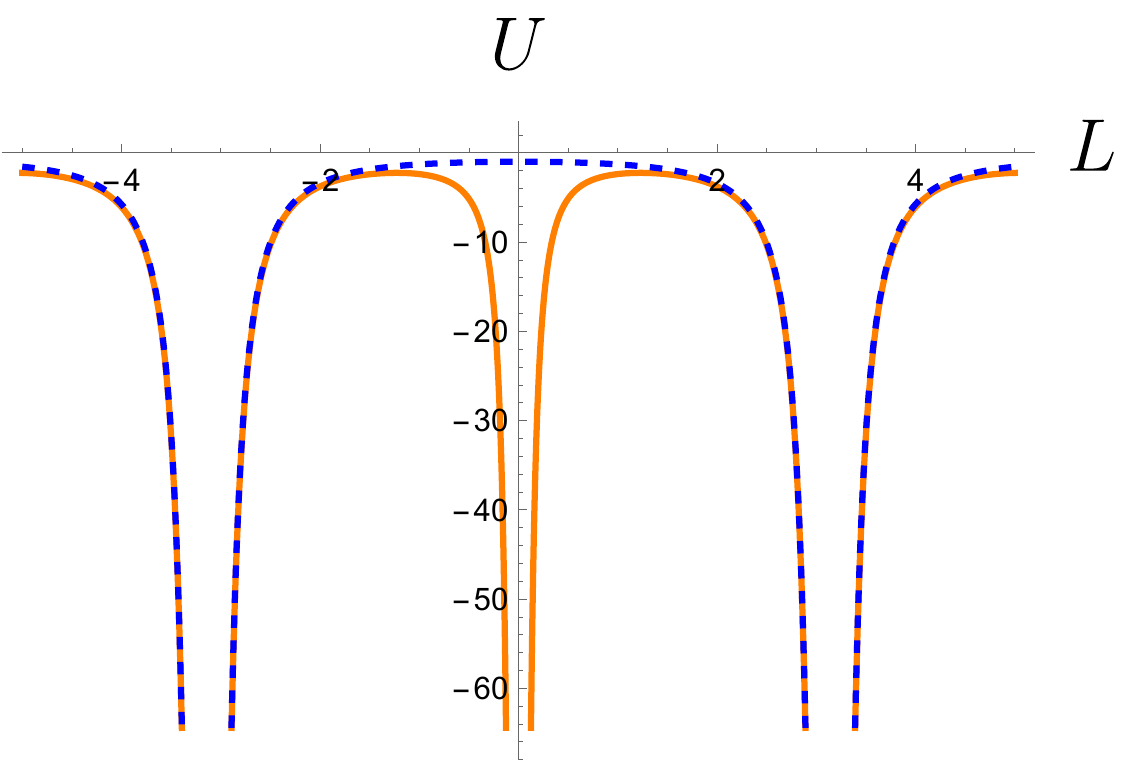}
    \caption{Potential $U(L)$ of the Schrödinger equation associated with the non-perturbative branches in the $T^2$ and $T^2+\Lambda_1$ deformed energy spectrum with the operator ordering prescription by \cite{Griguolo:2026eoi} in \eqref{eq:non perturbative Sch} (orange solid curve), and the minimal case in \eqref{eq:non perturbative Sch2}.}
    \label{fig:potential_non_perturbative_sol}
\end{figure}

Now we want to compare the result with the minimal operator ordering choice where $\hat{P}^2$ in \eqref{eq:sl2R op ordering} is unchanged, the corresponding Schrödinger equation is \begin{equation}\label{eq:non perturbative Sch2}
(\partial_L^2+U(L))\psi_s(L)=-s^2\psi_s(L)~,\quad U(L)=\Phi_b^2\sec^2(L/2)~.
\end{equation}
The potential energy term with and without the improvement term are displayed in Fig.~\ref{fig:potential_non_perturbative_sol}.

The fundamental domain where the potential is bounded is instead $L\in[-\pi,\pi]$. Two linearly independent wavefunctions solving this system are
\begin{equation}
    \begin{aligned}
    \psi_s(L)&=A_1~f_1(L)+A_2~f_2(L)~,\\
f_1(L)&=
\left(\cos\frac L2\right)^{\mathfrak{u}}
{}_2F_1\left(
\frac{\mathfrak{u}}{2}+s,\frac{\mathfrak{u}}{2}-s;
\frac12;
\sin^2\frac L2
\right)~,\\
f_2(L)&=
\sin\frac L2
\left(\cos\frac L2\right)^{\mathfrak{u}}
{}_2F_1\left(
\frac{\mathfrak{u}+1}{2}+s,
\frac{\mathfrak{u}+1}{2}-s;
\frac32;
\sin^2\frac L2
\right)~,
    \end{aligned}
\end{equation}
where $A_{1/2}$ are constant coefficients, $\mathfrak{u}\equiv\frac12+\sqrt{\frac14-4\Phi_b^2}$. The same argument as in the previous part applies. The behavior of the wavefunctions near the edges is
\begin{subequations}
    \begin{align}
    f_1&\sim
\frac{\sqrt\pi\,\Gamma(\frac12-\mathfrak{u})}
{\Gamma(\frac{1-\mathfrak{u}}{2}-s)
 \Gamma(\frac{1-\mathfrak{u}}{2}+s)}
\left(\frac{\epsilon_+}{2}\right)^{\mathfrak{u}}
+\frac{\sqrt\pi\,\Gamma(\mathfrak{u}-\frac12)}
{\Gamma(\frac{\mathfrak{u}}{2}-s)
 \Gamma(\frac{\mathfrak{u}}{2}+s)}
\left(\frac{\epsilon_+}{2}\right)^{1-\mathfrak{u}}~,\\
f_2&\sim
-\frac{\sqrt\pi\,\Gamma(\frac12-\mathfrak{u})}
{2\Gamma(1-\frac{\mathfrak{u}}{2}-s)
 \Gamma(1-\frac{\mathfrak{u}}{2}+s)}
\left(\frac{\epsilon_-}{2}\right)^{\mathfrak{u}}-
\frac{\sqrt\pi\,\Gamma(\mathfrak{u}-\frac12)}
{2\Gamma(\frac{\mathfrak{u}+1}{2}-s)
 \Gamma(\frac{\mathfrak{u}+1}{2}+s)}
\left(\frac{\epsilon_-}{2}\right)^{1-\mathfrak{u}}~,
\end{align}
\end{subequations}
where $\epsilon_{\pm}\equiv\pm(\pi-L)$. Requiring self-adjoint endpoint boundary conditions at $L=\pm \pi$ for either $f_1$ or $f_2$ leads to the requirement that the would-be energy spectrum of $-\hH_{\rm ADM}$ takes the form $-E_{\rm ADM}^{(n)}=(\mathfrak{u}+n)^2/4$, where again $\mathfrak{u}\equiv\frac12+\sqrt{\frac14-4\Phi_b^2}$. Note that when $\Phi_b>1/4$, $\mathfrak{u}\in\mathbb{C}$: the endpoints are then limit-circle oscillatory, so that $-\hH_{\rm ADM}$ is symmetric with a one-parameter family of self-adjoint extensions, all of them unbounded from below, as in the first part of this appendix. This confirms that the non-perturbative branch does not admit a well-defined Hilbert space under the above conditions.

\section{CMC slice lengths in (A)dS\texorpdfstring{$_2$}{2} without matter}\label{app: CMC lengths}
In this appendix, we review and provide new results regarding the different CMC slice lengths used in the main body for $\varepsilon=\pm1$ and $\eta=\pm 1$ displayed in Fig.~\ref{fig:summary_Penrose_diagrams} for AdS$_2$ black holes and   dS$_2$ space. Plots of the boundary time evolution of the lengths studied in this appendix can be found in Figs.~\ref{fig:L_geo_vs_CMC_AdS} and \ref{fig:Lk_dS_k_eq_1}.

\paragraph{$\varepsilon=-1$}
Let us begin the discussion with the $\varepsilon=-1$ case.
For $\eta=+1$, the relevant computation has been done in \cite{Chapman:2021eyy} (more recently discussed in \cite{Griguolo:2025kpi,Griguolo:2026eoi,Aguilar-Gutierrez:2026ogo}). We review it for completeness. Consider the (A)dS$_2$ metric in Schwarzschild coordinates ($t$, $\Phi$) \eqref{eq:Schwarzschild coordinates} and global coordinates ($\sigma$, $T$) in \eqref{eq:coordinate useful}:
\begin{eqnarray}\label{eq:main metric app}
    \rmd l^2=\pm\qty(-f(\Phi)\rmd t^2+\frac{\rmd\Phi^2}{f(\Phi)})=\pm\qty(\rmd\sigma^2-\cosh^2\sigma\rmd T^2)~,
\end{eqnarray}
where $+$ is for AdS$_2$ and $-$ for dS$_2$, $\abs{\sigma}\leq \pi/2$, and $f(\Phi)=\Phi^2-\Phi_h^2$ with $\Phi_h$ a constant. We can evaluate the geodesic length
\begin{eqnarray}\label{eq:geodesic def}
    L\equiv\int\rmd\chi\sqrt{\abs{g_{\mu\nu}\dv{x^\mu}{\chi}\dv{x^\nu}{\chi}}}~.
\end{eqnarray}
with $\chi$ a general parametrization. Notice the absolute value, which makes time- and spacelike geodesics real-valued, while other parts in the literature do not include the absolute value, resulting in complex-valued geodesics in general.

In case $\varepsilon=-1$, the geodesic can be straightforwardly derived using Lorentzian embedding coordinates, as we do in Sec.~\ref{app:shockwaves}. Since the length of a geodesic curve between endpoints at $\sigma=\pm\sigma_{\rm b}$ ($\sigma_{\rm b}$) in (A)dS space corresponds to $T=$ constant lines, \eqref{eq:geodesic def} gives \eqref{eq:geodesic length etaplus epsminus},\footnote{Alternatively, setting $\alpha=0$ and $u_L=-u_R$ in \eqref{eq:full sw length} allows us to readily recover the corresponding length \eqref{eq:geodesic length etaplus epsminus}.}
\begin{eqnarray}\label{eq:relation L sigma bdry}
    \eta=+1,~\varepsilon=-1,\quad\sinh\frac{L}{2}=\sinh\sigma_{\rm b}=\sqrt{\bbr^2-1}\cosh u~,
\end{eqnarray}
where we used the relation $\sinh\sigma=\sqrt{\Phi^2/\Phi_h^2-1}\cosh (\Phi_ht)$ between Schwarzschild coordinates and global coordinates in \eqref{eq:coordinate useful}, and we denote $\bbr=\Phi_b/\Phi_h$, with $\Phi=\Phi_b$ the boundary location $\sigma_{\rm b}$ in Schwarzschild coordinates, and $u\equiv \Phi_h t$. Meanwhile, for the $\eta=-1$ case, still with $\varepsilon=-1$, one can perform the analytic continuation between the Rindler-AdS and interior black hole patch (equivalently between Milne and static patch in dS$_2$) by $u\rightarrow u+\rmi \pi/2$ together with $\bbr\rightarrow\bbr<1$ in Schwarzschild coordinates, which gives \eqref{eq:geodesic length etaminus epsminus},
\begin{eqnarray}
    \eta=-1,~\varepsilon=-1,\quad\sinh\frac{L}{2}=\sqrt{1-\bbr^2}\sinh u~,
\end{eqnarray}
Note again that the results are valid in both AdS and dS space; however, in AdS the geodesic is timelike, while in dS is spacelike.

We can use this to evaluate the length of CMC slices directly. Consider a coordinate change $\sec\tilde{\sigma}=\cosh\sigma$ in \eqref{eq:main metric app}, so that the metric is conformally flat, where the relevant conformal factor is $\pm\cos^2\tilde{\sigma}$. As recently explained in \cite{Blommaert:2026lvp} (based on \cite{Jacobson:2018ahi,Banks:2020tox}), the time dependence of CMC slices in the metric \eqref{eq:main metric app} with boundaries at $\tilde{\sigma}=\tilde{\sigma}_{\rm b}$ can be evaluated with the following coordinates in \cite{Jacobson:2018ahi} (which discusses the higher dimensional case),
\begin{equation}\label{eq:useful metric CMC}
    \rmd l^2=\mp C^2\qty({-\rmd s^2+\rmd x^2})~,\quad C=\frac{\cos\tilde{\sigma}_{\rm b}}{(\cosh s+\sin\tilde{\sigma}_{\rm b}\cosh x)}~,\quad x\in\mathbb{R}~,
\end{equation}
where $+$ is for AdS$_2$ (the Ricci scalar of $C^2(-\rmd s^2+\rmd x^2)$ is $\mathcal{R}=-2$) and $-$ for dS$_2$ \cite{Banks:2020tox}. It was also found in \cite{Jacobson:2018ahi} that CMC slices in \eqref{eq:useful metric CMC} for a timelike normal vector with respect to $s=$constant surfaces correspond to $k=\partial_s C$ where the sign of $k$ depends on the direction of the normal vector; in this work, we choose the outward normal direction. In this way, the mean curvature corresponds to $k=\sec\tilde{\sigma}_{\rm b}{\sinh s}$, so that CMC slices correspond to $s$ fixed, whose length corresponds to \eqref{eq:geodesic def} with \eqref{eq:useful metric CMC} and both $s$ and $\tilde{\sigma}_{\rm b}$ fixed, which gives
\begin{equation}\label{eq:Lk epsm}
    \frac{1}{\sqrt{1+k^2}}\sinh(\frac{\sqrt{1+k^2}}{2}L_k)=\cot\tilde{\sigma}_{\rm b}=\sinh\frac{L}{2}~,
\end{equation}
for both signs of the conformal factor in \eqref{eq:useful metric CMC}, since the overall sign of the metric drops out of the CMC condition, of $\abs{k}$ and of the length; the last relation follows \eqref{eq:relation L sigma bdry}. Since this holds in both $\eta=\pm1$; we thus recover \eqref{eq:AdS geo vs CMC} and \eqref{eq:dS geo vs CMC}. 
Note that \eqref{eq:Lk epsm} contains $1+k^2$ in both AdS$_2$ and dS$_2$: the overall sign of the metric in \eqref{eq:useful metric CMC} does not affect the Christoffel symbols, the curves of constant geodesic curvature, $\abs{k}$ or the length \eqref{eq:length CMC}, so the timelike CMC slices of dS$_2$ are the same curves as the spacelike ones of AdS$_2$.

\paragraph{$\varepsilon=+1$ case}
The geodesics and more general CMC slices when $\varepsilon=+1$, which are timelike in AdS and spacelike in dS, need to be treated differently from the previous case, since we no longer fix the boundaries at $\sigma=\pm\sigma_{\rm b}$. Instead of starting from the geodesic length computation, we perform the CMC slice evaluation, where the relevant surfaces for $\eta=-1$, $\varepsilon=+1$ were considered in \cite{Aguilar-Gutierrez:2023zqm,silke,Aguilar-Gutierrez:2023pnn} in dS space, and take $k=0$ as special case (which in the $\eta=-1$, $\varepsilon=+1$ first appeared in \cite{Jorstad:2022mls} for dS$_2$ space). Our work extends the previous computations to $\varepsilon=+1$ $\eta=-1$ in dS space, and $\varepsilon=+1$, $\eta=\pm1$ in AdS$_2$ space after appropriate analytic continuation. 

Let us at the moment focus on $\eta=-1$, which we will then analytically continue to $\eta=+1$ as in the previous cases by setting $u\rightarrow u+\rmi \pi/2$ in Schwarzschild coordinates. In the following we use the definition of the boundary anchored CMC slice length in \eqref{eq:length CMC}.

\begin{figure}
    \centering
    \subfloat{\includegraphics[width=0.3\linewidth]{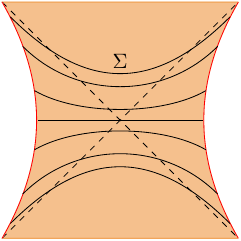}}\hspace{1cm}\subfloat{\includegraphics[width=0.45\linewidth]{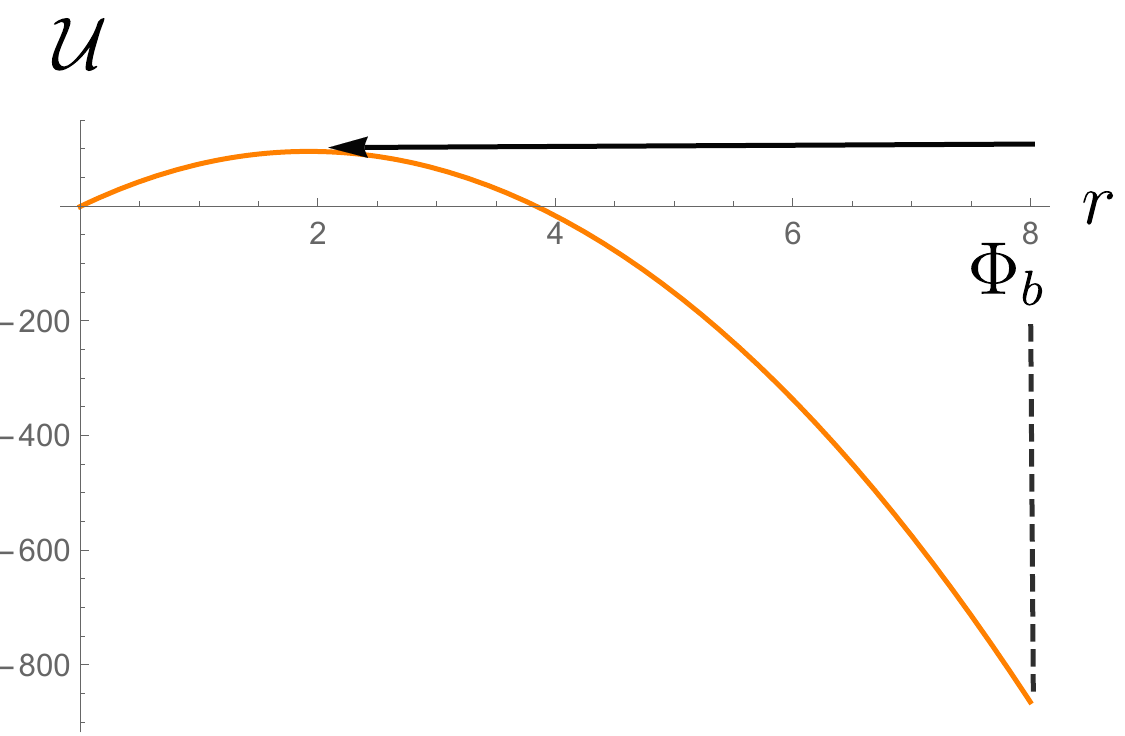}}
    \caption{Accumulation surfaces in finite-cutoff holography. (a) Penrose diagram of the AdS$_2$ black holes, with a CMC slice evolving in boundary time reaching the accumulation surface $\Sigma$ in the late time regime. (b) Illustration in terms of the effective potential \eqref{eq:U pot} for a fixed conserved constant $c$, where $\Phi_h=3$, $\Phi_b=8$, $c=10$. The accumulation surface corresponds to the radial location where the potential reaches its maximum, such that a particle traveling from the cutoff at $r=\Phi_b$ can at most lose all its kinetic energy (i.e.~it is a turning point \eqref{eq:def turning}) when it reaches the $\Sigma$; otherwise, it will reach the black hole interior and eventually the timelike singularity at $\Phi\rightarrow\infty$.}
    \label{fig:accumulation}
\end{figure}

To evaluate \eqref{eq:length CMC} we consider Schwarzschild coordinates ($t$, $\Phi$) in \eqref{eq:main metric app} with $t\rightarrow t_{L/R}$ denoting the boundary times on the left or right boundaries, and we use the following conventions for Eddington-Finkelstein (EF) coordinates
\begin{eqnarray} \rmd l^2=-f(r)\rmd v_{R}^2+2\rmd v_R\rmd r=-f(r)\rmd w_{R}^2-2\rmd w_R\rmd r~,\end{eqnarray}
(similarly for $L\rightarrow R$) as,
\begin{align}
w_{\rm R} &= t_{\rm R} - r^* (r) \, , \qquad
v_{\rm R} = t_{\rm R} + r^* (r) \, ,
\label{eq:general_null_coordinates}\\
w_{\rm L}&=-t_L+r^*(r)~,\quad v_{\rm L}=-t_L-r^*(r)~,    
\end{align}
where we denote $r=\Phi$ and $r^*(r)\equiv\int\rmd r/f(r)$ is the tortoise coordinate. From now on, we impose the gauge-fixing condition
\begin{equation}
    t_L=-t_R:\quad v_R=w_L~,\quad w_R=v_L~.
\end{equation}
The length functional (\ref{eq:length CMC}) and the CMC slice functional (\ref{eq:regions CMC}) in the EF coordinates are
\begin{align}
    L_k=\int_{\Sigma_\epsilon}\rmd\chi\,\sqrt{-f(r)\dot{v}^2+2\dot{v}\dot{r}}\,,\quad \mathcal{C}_{\rm CMC}=\sum_{\epsilon=\pm}\alpha_\epsilon\int_{\Sigma_\epsilon}
    \rmd\chi\,\mathcal{L}_\epsilon~,\label{eq:C epsilon inter}
\end{align}
where, as mentioned in \eqref{eq:K_CMC} the extremization of $\mathcal{C}_{\rm CMC}$ requires $\Sigma_{\epsilon}$ are CMC slices with curvature $k\equiv\eval{K}_{\Sigma_{\epsilon}}=- \epsilon\alpha_{B}/(\alpha_\epsilon)$ where the signs are fixed by requiring the normal vector to be outward oriented future oriented with respect to $\Sigma_\epsilon$, and
\begin{equation}
    \mathcal{L}_\epsilon\equiv \sqrt{-f(r)\dot{v}^2+2\dot{v}\dot{r}}-\epsilon~{k}~\dot{v}~r~.
\end{equation}
For the choice
\begin{equation}
\sqrt{-f(r)\dot{v}^2+2\dot{v}\dot{r}}=1~,
\end{equation}
the Euler-Lagrange equations corresponding of $\mathcal{L}_\epsilon$,
\begin{equation}
    \frac{\partial\mathcal{L}_\epsilon}{\partial r}=\dv{\sigma} \frac{\partial\mathcal{L}_\epsilon}{\partial \dot{r}}~,\quad \frac{\partial\mathcal{L}_\epsilon}{\partial v}=\dv{\sigma} \frac{\partial\mathcal{L}_\epsilon}{\partial \dot{v}}~,
\end{equation}
can be expressed as
\begin{equation}\label{eq:def turning}
    \dot{r}^2+\mathcal{U}(\mathfrak{c},\, k,\,r)=0~,
\end{equation}
where
\begin{align}
\mathfrak{c}&\equiv\pdv{\mathcal{L}_\epsilon}{\dot{v}}=\dot{r}-\dot{v}\,f(r)-\epsilon~{k}\,r~;\\
    \mathcal{U}(\mathfrak{c},\, k,\,r)&\equiv -f(r) - \left(\mathfrak{c}-\epsilon\, k ~r\right)^2~.\label{eq:U pot}
\end{align}
One can notice that the evaluation of holographic complexity reduces to the problem of a classical particle subject to a potential $\mathcal{U}(\mathfrak{c},\, k,\,r)$, as we illustrate in Fig.~\ref{fig:accumulation}.

The length functional for the CMC slices then becomes
\begin{equation}\label{eq:Lk functional}
    L_k=\int_I\frac{1}{\sqrt{-\mathcal{U}(\mathfrak{c},\, k,\,r)}}\rmd r~,
\end{equation}
where the interval $I$ will be later specified. Meanwhile, the Schwarzschild-coordinate time $u$ as 
\begin{equation}\label{eq:time def}
\begin{aligned}
t_L&=\int \rmd r\frac{\dot{t}_L}{\dot{r}}=\int\rmd r\frac{\dot{v}_L-\dot{r}/{f(r)}}{\sqrt{-\mathcal{U}(\mathfrak{c},\, k,\,r)}}= -\int_I \frac{\left(\mathfrak{c} -\epsilon k~r \right)\rmd r}{f(r)\,\sqrt{-\mathcal{U}(\mathfrak{c},\, k,\, r)}}  ~.
\end{aligned}
\end{equation}
Using $f(\bbr)=\bbr^2-1$ for \eqref{eq:Lk functional} and \eqref{eq:time def} with $\epsilon=+1$ results in
\begin{subequations}\label{eq:int eps plus}
\begin{align}
    {\rm AdS_2},~~~L_k&=\int_{\bbr}^{\rho_t}\frac{2\rmd r}{\sqrt{\mathfrak{c}^2-2\mathfrak{c}kr+(k^2+1)r^2-1}}~,\\
    {\rm dS_2},~~~L_k&=\int_{\bbr}^{\rho_t}\frac{2\rmd r}{\sqrt{\mathfrak{c}^2+1+2\mathfrak{c}kr+(k^2-1)r^2}}~,
\end{align}
\end{subequations}
where $\bbr=\Phi_{b}/\Phi_h$, and the turning point satisfies
\begin{eqnarray}
    f(\rho_t)+(\mathfrak{c}-k\rho_t)^2=0~.
\end{eqnarray}
Meaning that
\begin{subequations}\label{eq:truning points}
    \begin{align}
        {\rm AdS}:~~\rho_t&=\frac{\mathfrak{c}k\pm\sqrt{1-\mathfrak{c}^2+k^2}}{1+k^2}~,\\
    {\rm dS}:~~\rho_t&=\frac{\mathfrak{c}k\pm\sqrt{\mathfrak{c}^2-k^2+1}}{1-k^2}~,
    \end{align}
\end{subequations}
where we look for solutions where $\rho_t>0$ so that we keep only the $+$ sign in both cases; in the latter, when $\abs{k}>1$, $c<0$ and bounded by $-\sqrt{k^2-1}$. Recall that in the AdS$_2$ side, $\varepsilon=+1$, $\eta=-1$ corresponds to the black hole interior, while in dS$_2$ to the static patch, which is the reason for the integration limits in \eqref{eq:int eps plus}. One finds that the integrals lead to
\begin{subequations}
    \begin{align}
        {\rm AdS}:\quad L_k=&\frac{-4}{\sqrt{1+k^2}}\Biggl(\text{tanh}^{-1}\left(\frac{\sqrt{\left(\mathfrak{c}^2-1\right)
   \left(k^2+1\right)}}{\sqrt{-\mathfrak{c}^2+k^2+1}+\mathfrak{c} k}\right)\\
   &\qquad\qquad+\text{tanh}^{-1}\left(\frac{\sqrt{\mathfrak{c}^2-2 \mathfrak{c} k \bbr +\left(k^2+1\right)
   \bbr ^2-1}-\sqrt{\mathfrak{c}^2-1}}{\sqrt{k^2+1} \bbr }\right)\Biggr)~,\nonumber\\
{\rm dS}:\quad L_k=&\frac{-4}{\sqrt{1-k^2}}\Biggl(\tan^{-1} \frac{\sqrt{\mathfrak{c}^2-k^2+1}+\mathfrak{c} k}{\sqrt{\mathfrak{c}^2+1}
   \sqrt{1-k^2}}\\
   &\quad\qquad\quad-\tan^{-1}\frac{\sqrt{1-k^2}~\bbr}{
   \sqrt{\mathfrak{c}^2+1}-\sqrt{2 \mathfrak{c} k\bbr+\left(k^2-1\right) \bbr^2+\mathfrak{c}^2+1}}\Biggr)~,\nonumber
    \end{align}
\end{subequations}
The dS$_2$ expression, together with $t(\mathfrak{c})$ below, is the parametric form of \eqref{eq:CMC eps plus closed} in the main text, which we verified numerically resums it (the branch of $\tan^{-1}$ is fixed by continuity in $\mathfrak{c}$ from $\mathfrak{c}=0$; for $\mathfrak{c}$ beyond the value where $\sqrt{\mathfrak{c}^2+1}-\sqrt{\cdots}$ changes sign the second $\arcsin$ branch of \eqref{eq:CMC eps plus closed} is selected). The AdS$_2$ expression, obtained with the spacelike normalization of \eqref{eq:C epsilon inter}, reproduces instead the $\varepsilon=-1$ relation \eqref{eq:AdS geo vs CMC} with the interior geodesic \eqref{eq:geodesic length etaminus epsminus}, and is kept here for reference; the timelike slices of AdS$_2$ with $\varepsilon=+1$ are obtained from the dS$_2$ ones by $g_{\mu\nu}\rightarrow-g_{\mu\nu}$, see Sec.~\ref{ssec:CMC lenghts main}.

Similarly, \eqref{eq:time def} evaluates to
\begin{subequations}
\begin{align}
{\rm AdS}:\quad 2u=&\text{tanh}^{-1} \left(\frac{\bbr\left(\mathfrak{c} k-k^2-1\right)-\mathfrak{c}^2+\mathfrak{c} k+1}{(\mathfrak{c}-k) \sqrt{\left(k^2+1\right)\bbr^2+\mathfrak{c}^2-2 \mathfrak{c} k\bbr-1}}\right) \\
   &-\text{tanh}^{-1} \left(\frac{\bbr\left(\mathfrak{c} k+k^2+1\right)-\mathfrak{c}^2-\mathfrak{\mathfrak{c}} k+1}{(k+\mathfrak{c})
   \sqrt{\left(k^2+1\right)\bbr^2+\mathfrak{c}^2-2 \mathfrak{c} k\bbr-1}}\right)~,\\
     {\rm dS}:\quad 2u=&\text{tanh}^{-1}\frac{(1+k \mathfrak{c}-k^2)\bbr+1-k\mathfrak{c}+\mathfrak{c}^2}{(k-\mathfrak{c})\sqrt{(k^2-1)\bbr^2+2k\mathfrak{c}\bbr+\mathfrak{c}^2+1}}\\&+\text{tanh}^{-1}\frac{(-1+k \mathfrak{c}+k^2)\bbr+1+k\mathfrak{c}+\mathfrak{c}^2}{(k+\mathfrak{c})\sqrt{(k^2-1)\bbr^2+2k\mathfrak{c}\bbr+\mathfrak{c}^2+1}}~,
     \end{align}
\end{subequations}
for $\eta=-1$. Note that there is a cancellation of divergences in the evaluation at the turning point $\rho_t$ \eqref{eq:truning points}, which is regulated through the formula $\text{tanh}^{-1}(a)-\text{tanh}^{-1}(b)=\text{tanh}^{-1}(\frac{a-b}{1-ab})$ where both $a$, $b$ are $\mathcal{O}(\delta^{-1})$ as $\delta\rightarrow0$ and of the same sign. 

It can be seen that when $k=0$ we can invert the above relations for $\mathfrak{c}(t)$, which we substitute in $L_{k=0}(\mathfrak{c})$. This leads to
\begin{equation}
    \eta=-1,\varepsilon=+1:\quad \cot\frac{L}{2}=-\sqrt{\frac{\bbr^2+1-(1-\bbr^2)\cosh(2 u)}{2(1-\bbr^2)}}\sech(u)~,
\end{equation}
in both cases, corresponding to \eqref{eq:geodesic length etaminus epsplus}. This is in agreement with \cite{Jorstad:2022mls} for dS$_2$ space.

Let us also discuss the evaluation of the $L_k(\mathfrak{c})$ and $t(\mathfrak{c})$ more generally. For the parametrization analysis for both cases $\eta=\pm1$ with $\varepsilon=+1$ in (A)dS$_2$ space, one should note that the local maximum of $t(\mathfrak{c})>0$ is reached when
\begin{eqnarray}
    \mathfrak{c}_{\rm max}=(1-k^2)\frac{\bbr}{k}~,\quad \abs{k}<1~,
\end{eqnarray}
So that the evolution of the CMC slice restricts the range $\mathfrak{c}\in[\mathfrak{c}_{\rm max},\infty)$. In contrast, for dS$_2$ with $\abs{k}>1$, there are poles in $t(\mathfrak{c})$ when $\mathfrak{c}=\pm\sqrt{k^2-1}$ so that the slice reaching the Milne patch requires $\mathfrak{c}\in(-\infty,\, -\sqrt{k^2-1}]$. Further details about the restrictions in the dS case can be found in \cite{silke}.

Next, to discuss the $\eta=+1$ case, we need to perform the analytic continuation from the black hole interior to Rindler-AdS$_2$ and from Milne to the static patch in dS$_2$ space in Schwarzschild coordinates, resulting in
\begin{subequations}
    \begin{align}
    {\rm AdS}:\quad 2u= &\text{coth}^{-1} \left(\frac{\bbr\left(\mathfrak{c} k-k^2-1\right)-\mathfrak{c}^2+\mathfrak{c} k+1}{(\mathfrak{c}-k) \sqrt{\left(k^2+1\right)\bbr^2+\mathfrak{c}^2-2 \mathfrak{c} k\bbr-1}}\right)\\
   &-\text{coth}^{-1} \left(\frac{\bbr\left(\mathfrak{c} k+k^2+1\right)-\mathfrak{c}^2-\mathfrak{c} k+1}{(k+\mathfrak{c})
   \sqrt{\left(k^2+1\right)\bbr^2+\mathfrak{c}^2-2 \mathfrak{c} k\bbr-1}}\right)~,\nonumber \\
{\rm dS}:\quad 2u= &\text{coth}^{-1} \left(\frac{\bbr\left(\mathfrak{c} k-k^2+1\right)+\mathfrak{c}^2-\mathfrak{c} k+1}{(k-\mathfrak{c})
   \sqrt{2 \mathfrak{c} k\bbr+\left(k^2-1\right)\bbr^2+\mathfrak{c}^2+1}}\right)\\
   &+\text{coth}^{-1} \left(\frac{\bbr\left(\mathfrak{c} k+k^2-1\right)+\mathfrak{c}^2+\mathfrak{c} k+1}{(\mathfrak{c}+k) \sqrt{2 \mathfrak{c} k\bbr+\left(k^2-1\right)\bbr^2+\mathfrak{c}^2+1}}\right)~,
    \end{align}
\end{subequations}
for $\eta=+1$.

As before, the $k=0$ case simplifies substantially, so that we can invert the above relations for $t(\mathfrak{c})$, insert it in $L_{k=0}(\mathfrak{c})$ and recover 
\begin{equation}
\eta=+1,\varepsilon=+1:\quad \cot\frac{L}{2}=-\sqrt{\frac{\bbr^2+1-(\bbr^2-1)\cosh(2u)}{2(\bbr^2-1)}}\abs{\csch(u)}~,    
\end{equation}
reproducing \eqref{eq:geodesic length etaplus epsplus}.

\paragraph{$k=1$ case}
For completeness, let us analyze the CMC slices with $\abs{k}=1$ in dS$_2$ space, where the previous expressions simplify to
\begin{subequations}
    \begin{align}
         L_k&=-\frac{2}{\mathfrak{c}}\sqrt{\mathfrak{c}^2+2\bbr \mathfrak{c}+1}~,\\
    \tanh(2u)&=\frac{2(\bbr \mathfrak{c}+1)\sqrt{1+\mathfrak{c}^2+2\bbr \mathfrak{c}}}{(\bbr^2+1)\mathfrak{c}^2+4\bbr \mathfrak{c}+2}~,
    \end{align}
\end{subequations}
One finds that the only solution for $\mathfrak{c}(t)<0$ with $\bbr<1$ is
\begin{eqnarray}
    \mathfrak{c}=\frac{-2\bbr+2\sqrt{1-\bbr^2}\sinh (u)}{1+\bbr^2-(1-\bbr^2)\cosh(2u)}
\end{eqnarray}
which gives
    \begin{align}
    \eta=-1,~\varepsilon=+1,\quad &L_k= \frac{ \cosh \left(u \right) \left(\bbr ^2+1-\left(1-\bbr ^2\right)
   \cosh \left(2 u \right)\right) }{\left(\bbr -\sqrt{1-\bbr ^2} \sinh \left(u \right)\right) }\\
   &\quad\frac{\sqrt{2\left(1-\bbr
   ^2\right) \left(\left(1-\bbr ^2\right) \cosh \left(2 u \right)+3 \bbr
   ^2-4 \sqrt{1-\bbr ^2} \bbr  \sinh \left(u \right)-1\right)}}{\left| \bbr ^2-\left(1-\bbr ^2\right) \cosh \left(2 u\right)+1\right|}~,\nonumber
    \end{align}
which simplifies to \eqref{eq:eta minus eps plus Lk}, i.e.~$L_{k=1}=2\sin(L/2)$ with \eqref{eq:geodesic length etaminus epsplus}.
    
Similarly, for $ \eta=+1,~\varepsilon=+1$, since the boundaries are located in the Milne patch instead of the static patch of dS$_2$, we perform $u\rightarrow u-\rmi\pi/2$ in static patch coordinates (with $\sqrt{1-\bbr^2}=\rmi\sqrt{\bbr^2-1}$; the sign of the shift is fixed by requiring $L_k>0$), which gives
\begin{align}
    \eta=+1,~\varepsilon=+1,\quad &L_k= \frac{\sinh \left(u \right) \left(\bbr ^2+1-\left(\bbr ^2-1\right)
   \cosh \left(2 u \right)\right) }{\left(\bbr -\sqrt{\bbr ^2-1} \cosh \left(u \right)\right) }\\
   &\quad\frac{\sqrt{2\left(\bbr
   ^2-1\right) \left(\left(\bbr ^2-1\right) \cosh \left(2 u \right)+3 \bbr
   ^2-4 \sqrt{\bbr ^2-1} \bbr  \cosh \left(u \right)-1\right)}}{\left| \bbr ^2-\left(\bbr ^2-1\right) \cosh \left(2 u
   \right)+1\right|}~,\nonumber
\end{align}
which simplifies to \eqref{eq:eta plus eps plus Lk}.

\section{Shockwaves in finite cutoff JT gravity}\label{app:shockwaves}
In this appendix, we study the effect of shockwave backreaction in the finite cutoff JT gravity. There are different motivations. First, we want to understand JT gravity with matter, shockwaves are arguably some of the simplest solvable backreacted geometries. However, the corresponding ADM Hamiltonian of JT gravity with shockwaves has not been spelled out in the previous literature. Next, we want to systematically characterize CMC slices in (A)dS$_2$ geometries, shockwave being a special case. After that, one of the defining properties of the CAny framework, besides late-time linear growth, is that they display the switchback effect \cite{Belin:2021bga,Belin:2022xmt}, which is a decrease in the complexity growth when one introduces operator insertions at alternating positive and negative times in the preparation of the state. In the bulk, this is expected to be described by inserting shockwaves, whose backreaction leads to a decrease in the magnitude of the complexity growth. Its boundary interpretation in the DSSYK model is sufficiently well-understood \cite{Ambrosini:2024sre,Ambrosini:2025hvo,Aguilar-Gutierrez:2025mxf}. At last, there are new explicit formulas for OTOCs in finite-cutoff JT gravity \cite{Griguolo:2026eoi}, which so far still require an explicit derivation; it is even unclear what the semiclassical expectation is for the corresponding Lyapunov exponent. We discuss a semiclassical estimation for the Lyapunov exponent associated with finite-cutoff JT gravity with a shockwave, in the hope that it accelerates progress towards its quantum derivation.

\paragraph{Geometric set up}
To set the stage, we are interested in evaluating the following observable
\begin{equation}\label{eq:G4 semiclassical}
    G_4(t_L,t_R)\simeq \xi^{2\Delta}\sech^{2\Delta}\frac{L(t_L,t_R)}{2}~,
\end{equation}
where $L(t_L, t_R)$ is the total geodesic length between the finite cutoff boundaries with matter, and $t_L$, $t_R$ are the boundary times in Schwarzschild coordinates. In the asymptotic boundary case, it is known that the shockwave encodes the Lyapunov exponent of the corresponding OTOC when the boundary times flow in opposite directions \cite{Shenker:2013pqa}. 

\begin{figure}
    \centering
    \begin{minipage}{0.5\linewidth}
    \includegraphics[width=\linewidth]{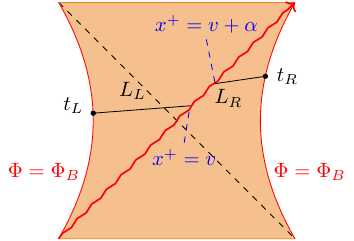}    
    \end{minipage}\hspace{1cm} \vspace{0.5cm}\begin{minipage}{0.345\linewidth}
        \includegraphics[width=\linewidth]{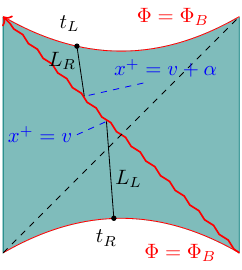}
    \end{minipage}
    \caption{Representation of the shockwave geometries in the AdS$_2$ black hole and dS$_2$ space, including the one-sided lengths \eqref{eq:one sided geodesics} and Schwarzschild coordinates \eqref{eq:Schwarzschild coordinates}.}
    \label{fig:shockdiagrams}
\end{figure}

Motivated by the above, we consider shockwave geometries in an AdS$_2$ black hole (and similarly for timelike CMC slices in dS, see Fig.~\ref{fig:shockdiagrams}), which are described by (see e.g.~\cite{Stanford:2014jda,Shenker:2013pqa,Shenker:2013yza}),
\begin{equation}
\label{eq:discont coordinates}
\rmd l^2=\frac{-4\rmd x^+~\rmd x^-}{(1+x^-(x^++\alpha H(x^-)))^2}~.
\end{equation}
where $H(x)$ is the Heaviside step function, reflecting that the shockwave is sent from $x^-=0$, and $\alpha$ is called the shockwave parameter.

\subsection{Geodesic lengths}\label{sapp:geodesic shockwaves}
For concreteness, let us consider embedding coordinates in AdS to describe the length of spacelike geodesics between its boundaries; an analogous treatment can be done in dS space. The geodesic length between spacelike points in AdS (and similarly dS) can be computed in terms of the Lorentzian embedding coordinates $X_I$ as,
\begin{eqnarray}\label{eq:embedding for length}
    \cosh{L}=-X\cdot X'=X_0X_0'+X_1X_1'-X_2X_2'~,
\end{eqnarray}
which are related to Schwarzschild coordinates \eqref{eq:Schwarzschild coordinates}
by
\begin{eqnarray}\label{eq:embeeding Minkowski and AdS}
    X_0=\sqrt{\frac{\Phi^2}{\Phi^2_h}-1}~\sinh\Phi_h t~,\quad X_1=\frac{\Phi}{\Phi_h}~,\quad X_2=\sqrt{\frac{\Phi^2}{\Phi^2_h}-1}~\cosh\Phi_h t~,
\end{eqnarray}
where one must perform $\Phi_h t\rightarrow \Phi_h t+\rmi \pi/2$ to describe points on the opposite Rindler-AdS patch. 

As first deduced by \cite{Shenker:2013pqa} the geodesic length between each boundary at finite $\Phi_b=\bbr\,\Phi_h$ for arbitrary one-sided times $t_{L/R}$ is
\begin{equation}\label{eq:one sided geodesics}
\begin{aligned}
    L_{L}&=\cosh ^{-1}\left(\sqrt{\bbr ^2-1} e^{-u_L} (\alpha +v)+\bbr \right)~,\\
    L_{R}&=\cosh ^{-1}\left(\bbr -\sqrt{\bbr ^2-1}~v~e^{-u_R}\right)~,
\end{aligned}
\end{equation}
where $v$ is an arbitrary point, and again, we denote $u_{L/R}\equiv \Phi_ht_{L/R}$. Now, we perform the extremization of the total length with respect to the arbitrary point (i.e.~$\rmd L/\rmd v=0$) without restricting to the asymptotic boundary limit of \cite{Shenker:2013pqa}. We obtain:\footnote{Different orders of limit result in different lengths; for instance, taking first order in alpha and then expanding for large $\bbr$ results in a different length than only taking $\bbr\gg1$.}
\begin{equation}
   v_*=\frac{\sqrt{\bbr^2-1}\,\big(\rme^{2u_R}-\rme^{2u_L}-\alpha^2\big)-2\bbr\alpha\,\rme^{u_L}}{2\big(\bbr(\rme^{u_L}+\rme^{u_R})+\alpha\sqrt{\bbr^2-1}\big)}~.
\end{equation}
Then, we find the dimensionless geodesic length as
\begin{equation}\label{eq:full sw length}
    \cosh L=\frac{\big(\bbr(\rme^{u_L}+\rme^{u_R})+\alpha\sqrt{\bbr^2-1}\big)^2-\rme^{2u_L}-\rme^{2u_R}}{2\,\rme^{u_L+u_R}}~,
\end{equation}
Note that as a special case when $t_L=t_R$, we recover \eqref{eq:length symmetric shock}.

Similarly, one can use embedding coordinates to evaluate the shockwave length for the non-perturbative branches, where we will change the notation $u_{L/R}$ by $u_{t/d}$. We find that it is given by
\begin{eqnarray}\label{eq:length new}
   L= \cos ^{-1}\left(\bbr +\sqrt{1-\bbr ^2}
   e^{-u_d} v\right)+\cos ^{-1}\left(\bbr
   -\sqrt{1-\bbr ^2} e^{-u_t} (\alpha
   +v)\right)
   \end{eqnarray}
The extremization of the length yields
\begin{align}
    v_*=\frac{\sqrt{1-\bbr^2}\,\big(\rme^{2u_d}-\rme^{2u_t}+\alpha^2\big)-2\bbr\alpha\,\rme^{u_t}}{2\big(\bbr(\rme^{u_d}+\rme^{u_t})-\alpha\sqrt{1-\bbr^2}\big)}~,
\end{align}
which can be substituted back in \eqref{eq:length new}, leading, for $u_t=u_d$, to
\begin{equation}
\begin{aligned}
    &\cos L=\frac{\big(\bbr(\rme^{u_d}+\rme^{u_t})-\alpha\sqrt{1-\bbr^2}\big)^2-\rme^{2u_d}-\rme^{2u_t}}{2\,\rme^{u_d+u_t}}~.
\end{aligned}
\end{equation}
In particular, for $u_t=u_d\equiv u$ this simplifies to $L(u,u)=2\cos^{-1}\qty(\bbr-\frac{\alpha}{2}\sqrt{1-\bbr^2}\,\rme^{-u})$, which approaches $2\cos^{-1}\bbr$ at late times; in contrast to the spacelike geodesic case in AdS$_2$, the wormhole length therefore saturates rather than grows linearly in this configuration.

\subsection{Hamiltonian formulation}
We would like to find the canonical momentum conjugate to the two-sided length $L(t_L,t_R)$, to study the corresponding ADM Hamiltonian and quantize it. In principle, we can work it out for the most general finite cutoff length $L(t_L,t_R)$ \eqref{eq:full sw length}, which is quite intricate, making it difficult to find a convenient expression to analyze in detail. For this reason, we focus on the case $t_L=t_R$, where the length simplifies to \eqref{eq:length symmetric shock}, and given that this is the relevant case to study Lyapunov behavior, as we find in App.~\ref{sapp:suggestive Lyap}. However, one should keep in mind that if one simply evaluates the conjugate momentum to $L(t,t)$ assuming a two-dimensional phase space with ADM energy $\Phi_h^2$, as
\begin{eqnarray}
2\Phi_h=\pdv{L}{t}\pdv{P}{\Phi_h}-\pdv{P}{t}\pdv{L}{\Phi_h}~,
\end{eqnarray}
one obtains a canonical momentum;
\begin{eqnarray}\label{eq:new mmometna}
    P=-\frac{\sinh\frac L2}{\cosh^2\frac L2}\Big[\Phi_b\ln\!\big(\Phi_h \cosh\frac{L}{2}-\Phi_b\big)+\Phi_h \cosh\frac{L}{2}\Big]+f(L)
\end{eqnarray}
(up to a constant $\rme^{-1}/\Phi_b$ inside the Lambert function below, which can be absorbed in $f(L)$) with $f(L)$ an arbitrary function of $L$ producing a free canonical shift, which is then very different from starting from the one-sided construction, as seen from the asymptotic boundary limit \eqref{eq:momenta} (with $k=0$ and $t_L=t_R$ so that $P_L=-P_R$). For instance the Hamiltonian corresponding to \eqref{eq:new mmometna} is
\begin{eqnarray}
    H=\Phi_h^2=\frac{\Phi_b^2}{\cosh^2\frac L2}\left[1+W\!\left(\exp\!\left[-\frac{P\cosh^2\frac L2}{\Phi_b\sinh\frac L2}\right]\right)\right]^2~,
\end{eqnarray}
with $W$ being the Lambert W function, which remains in the Hamiltonian even in the asymptotic boundary limit ($L\gg1$) in contrast to \eqref{eq:H ADM propopsal} with $k=0$. The difference lies in evaluating derivatives after setting $t_L=t_R$. For this reason, we will instead consider that the canonical momentum to the two-sided length $L$ is its time derivative, $P=\rmd L/\rmd t$, and eliminate $\Phi_h$ directly from \eqref{eq:length symmetric shock}, which gives $\rmd L/\rmd t=-2\Phi_h\coth\frac{L}{2}+2\Phi_b/\sinh\frac{L}{2}$, and then
\begin{eqnarray}
    \Phi_h=\frac{\Phi_b}{\cosh\frac{L}{2}}-\frac{P}{2} \tanh\frac{L}{2}~,
\end{eqnarray}
which implies that the classical ADM Hamiltonian takes the form
\begin{eqnarray}
    H_{\rm shock}=\frac{P^2}{4}\tanh^2\frac{L}{2}-\frac{\Phi_b\tanh\frac{L}{2}}{\cosh\frac{L}{2}}P+\frac{\Phi_b^2}{\cosh^2\frac{L}{2}}~.
\end{eqnarray}
The shockwave parameter $\alpha$ does not appear in the expression. This means that the wavefunction, $\psi_s(L)$ resulting from promoting the would-be canonical variables to operators as in the previous sections, as well as the spectral density, would be independent of matter operator insertion. This would imply that the observables (such as four-point functions), would not truly capture $\alpha$. In contrast, if one works with the ADM Hamiltonian of a general particle insertions and acts with so-called precursor operators \cite{Susskind:2014rva} to time-evolve their insertion time, one recovers the standard shockwave results \cite{Aguilar-Gutierrez:2025mxf}.
A possible lesson is that one should construct the full Hamiltonian with the operator insertion, and take the shockwave limit to the end, given that naive application of canonical quantization would fail to capture the correct shockwave backreaction.

\subsection{Suggestive Lyapunov exponent}\label{sapp:suggestive Lyap}
The geodesic distance between the finite cutoff boundaries \eqref{eq:full sw length}, results in \eqref{eq:full sw length}; the expression further simplifies for opposite-oriented boundary times (corresponding to $t_L=t_R$ in our conventions) as
\begin{equation}\label{eq:length symmetric shock}
\begin{aligned}
    \cosh \frac{L(t,t)}{2}=&\bbr+\frac{\alpha}{2}\sqrt{\bbr^2-1}~\rme^{-\Phi_ht}~.
\end{aligned}
\end{equation}
One indeed recovers the result of \cite{Shenker:2013pqa} for $\bbr\gg1$, as it should. 

Let us now denote $G_4(t,t)=\xi^{2\Delta}\sech^{2\Delta} (L(t,t)/2)$. Based on the proposal by \cite{Griguolo:2021wgy}, where $\cosh\hat{L}$ is the generator of two-point functions in the auxiliary system, our definition $G_4$ would correspond to a crossed four-point function evaluating OTOCs. Notice that the putative Lyapunov exponent and scrambling time can be deduced from $G_4$ by inspecting the regime of exponential growth. Using the conventions in \cite{Shenker:2013pqa} we express $\alpha=(\tilde{E}/8M)\rme^{\Phi_h t_w}$ with $\tilde{E}$ the energy of a probe particle generating the shockwave, $M$ the black hole ADM mass (proportional to $N$ the number of degrees of freedom that are scrambled by the shockwave), where the shockwave approximation occurs when $\tilde{E}\rightarrow0$ and $t_w\rightarrow\infty$ in a way that $\alpha$ is held fixed. In the following, we consider the early-time regime where one can approximate
\begin{equation}\label{eq:OTOCs}
   \begin{aligned}
       G_4(t,t)&=\xi^{2\Delta}\qty(\bbr+\frac{\alpha}{2}\sqrt{\bbr^2-1}\rme^{-\Phi_ht})^{-2\Delta}\\
       &=\frac{\xi^{2\Delta}}{\bbr^{2\Delta}}\qty(1-\frac{\Delta\tilde{E}}{8M}\sqrt{1-\bbr^{-2}}\rme^{(t_w-t)\Phi_h })+\mathcal{O}(\rme^{-2\Phi_ht})~.
   \end{aligned}
\end{equation}
From the last relation, one can notice that if $G_4(t,t)$ can indeed be interpreted as a semiclassical OTOC in finite-cutoff holography, then since $G_4$ grows exponentially in terms of $t_w-t$, the corresponding (kinematic) Lyapunov exponent is
\begin{eqnarray}
    \lambda_{L}\equiv2\pi/\beta~,
\end{eqnarray}
where we used $\beta=2\pi/\Phi_h$ in JT gravity. Note that the exponent $\Phi_h$ is inherited from the asymptotic Schwarzschild time $t_w$ through $\alpha$; the finite cutoff changes the prefactor $\sqrt{1-\bbr^{-2}}$ and, through the blueshift $\sqrt{f(\Phi_b)}$ of the energy measured at the cutoff, the scrambling time below, but not the exponent. We also emphasize this depends on the time normalization. The suggested exponent (and thus the corresponding black hole temperature) remains the same as in the asymptotic boundary limit only when it is measured with respect to the same boundary time variable as in the asymptotic boundary case. Meanwhile, at finite cutoff, the proper time \eqref{eq:tP} at the cutoff differs by a redshift factor $t_P\equiv\sqrt{\Phi_b^2-\Phi_h^2}\,t$ in the AdS black hole case, which means that the expected exponent with respect to $t_P$, which we denote $\lambda_{L'}$, is related to the Tolman temperature \cite{tolman1933thermodynamics,Tolman:1930ona} of the system as
\begin{eqnarray}\label{eq:Tolman 2}
    \lambda_{L'}=2\pi/\beta_T=\Phi_h/\sqrt{\Phi_b^2-\Phi_h^2}~.
\end{eqnarray}
Some remarks
\begin{itemize}
\item Expressed in the proper time \eqref{eq:tP}, the rate \eqref{eq:Tolman 2} is $2\pi/\beta_T$ with $\beta_T$ the Tolman temperature, which is consistent with, but does not establish, saturation of the chaos bound \cite{Maldacena:2015waa} at the local temperature.
\item \eqref{eq:Tolman 2} can also be expressed in terms of the temperature of the $T^2$ deformed boundary theory. We know that the Tolman temperature is related to the one of the $T^2$ deformed theory by $\beta_T=\beta\rmd E/\rmd E_y=\beta_y/\sqrt{y}$ \cite{Aguilar-Gutierrez:2024oea}, so that $\lambda_{L'}=2\pi\sqrt{y}/\beta_y$.
\end{itemize}
\eqref{eq:Tolman 2} suggests that the corresponding Lyapunov exponent to the OTOC in \cite{Griguolo:2026eoi} is not affected by the finite cutoff in
the bulk, at least with respect to a boundary time clock running in Schwarzschild, while the
expected Lyapunov exponent gets time-diluted when measured with respect to a more-natural
boundary observer clock. However, we stress that the shockwave result is only a heuristic argument for the expected Lyapunov behavior instead of a formal derivation.

\begin{figure}
    \centering
    \subfloat[]{\includegraphics[width=0.45\linewidth]{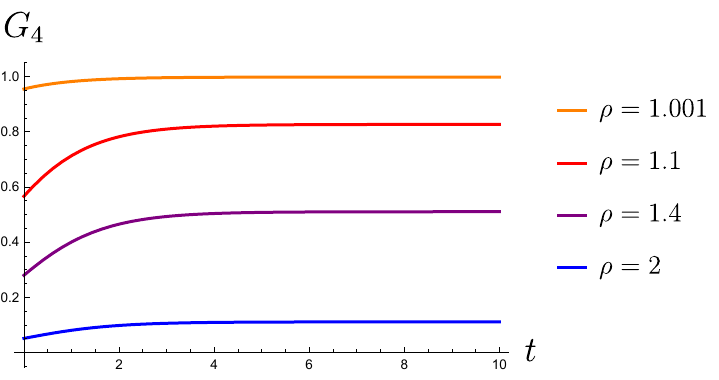}}\hfill\subfloat[]{\includegraphics[width=0.45\linewidth]{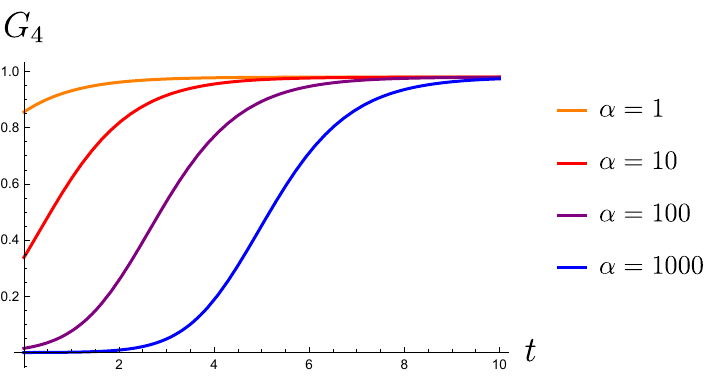}}
    \caption{Evolution of the semiclassical $G_4$ \eqref{eq:OTOCs} when $\Delta$ is fixed, and (a) $\alpha$ is fixed, and (b) $\bbr$ is fixed. Parameters in the plots: $\Delta=1$, $\alpha=1$ (left) $\bbr=1.01$ (right). The magnitude of the generating function of the length increases as we approach the horizon limit (left), and the time to reach its asymptotic value increases with $\alpha$ (right).}
    \label{fig:OTOC}
\end{figure}

\paragraph{Scrambling time}
Next, we note that the approximation where $G_4$ grows exponentially \eqref{eq:OTOCs} is valid until the correction term in \eqref{eq:OTOCs} becomes of order one, corresponding to
\begin{eqnarray}
    t_w-t\approx\frac{1}{\Phi_h}\log\frac{8M\bbr}{\Delta \tilde{E}\sqrt{\bbr^2-1}}~,
    \end{eqnarray}
 which would correspond to the scrambling time at finite cutoff, under the above-stated assumptions. The evolution of the semiclassical generating function of the length with a shockwave insertion \eqref{eq:OTOCs} is displayed in Fig.~\ref{fig:OTOC}.

\subsection{Constant mean curvature slice length with a shockwave}\label{sapp:JT shcowkaves}
We now elaborate on the finite cutoff length of CMC slices. Due to the shockwave backreaction, the geodesics and CMC slices are no longer just given by constant time slices in global coordinates \eqref{eq:coordinate useful}, so that we extremize the length directly. 

The relation \eqref{eq:AdS geo vs CMC} continues to hold in the presence of the shockwave, with $L$ the shockwave geodesic length \eqref{eq:full sw length}. The reason is that the metric \eqref{eq:discont coordinates} is locally AdS$_2$ everywhere (its Ricci scalar has no support at $x^-=0$), as it must be in JT gravity, where matter only sources the dilaton; the gluing $x^+\rightarrow x^++\alpha$ along $x^-=0$ is the restriction of an AdS$_2$ isometry, a null rotation that maps the horizon to itself. Mapping the region on one side of the shockwave by this isometry, the CMC slice becomes an ordinary CMC curve of AdS$_2$ between the right boundary point and the image of the left one, whose geodesic distance is \eqref{eq:full sw length}. Hence, at finite cutoff,
\begin{gather*}
\sinh\frac{L}{2}=\frac{1}{\sqrt{1+k^2}}\sinh\Big(\frac{\sqrt{1+k^2}}{2}L_k\Big)~,\\
\cosh L=\frac{\big(\bbr(\rme^{u_L}+\rme^{u_R})+\alpha\sqrt{\bbr^2-1}\big)^2-\rme^{2u_L}-\rme^{2u_R}}{2\,\rme^{u_L+u_R}}~,
\end{gather*}
whose asymptotic limit $\bbr\gg1$ is \eqref{eq:CMC shockwave length} below.

First, we study the effect of shockwaves (Fig.~\ref{fig:CMC_shockwave}) in codimension-one observables $\mathcal{C}_\epsilon$ in Sec.~\ref{ssec:CMCC-Any slices} evaluated through the extremization of $\mathcal{C}_{\rm CMC}$ in \eqref{eq:regions CMC} in an AdS$_2$ black hole. We set $d=1$ and $F=1$ in \eqref{eq:OF1}, which represents the length of the CMC slices in \eqref{eq:length CMC}. For the Penrose diagram (Fig.~\ref{fig:CMC_shockwave}), 
\begin{figure}
    \centering
    \subfloat[]{\includegraphics[width=0.43\linewidth]{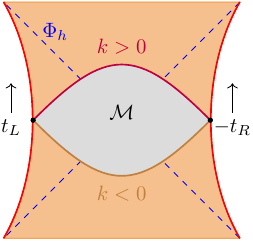}}\hspace{1cm}\subfloat[]{\includegraphics[width=0.43\linewidth]{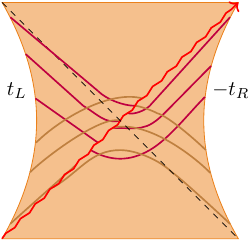}}
    \caption{Different CMC slices $\Sigma_\pm$ anchored to the boundaries of an AdS$_2$ black hole with black hole radius $\Phi_h$ (dashed black line). (a) Spacetime region $\mathcal{M}$ bounded by the CMC slices $\Sigma_{\pm}$ in \eqref{eq:regions CMC}, where the sign of $k$ of the CMC slice is relative to the future-oriented normal vector to the slice. (b) We display the CMC slices for different boundary times when we include a shockwave along one of the horizons, which is represented in the discontinuous coordinates \eqref{eq:special coordinates}. Relative to the asymptotic boundary, a CMC slice with a given $k$ moves further away as we increase the backreaction parameter $\alpha$.}
    \label{fig:CMC_shockwave}
\end{figure}
it is also useful to define the coordinates
\begin{equation}\label{eq:special coordinates}
    U=\text{tan}^{-1}(x^+)~,\quad V=\text{tan}^{-1}(x^-)~.
\end{equation}
We now evaluate the length functional \eqref{eq:length CMC} piecewise, starting from the left boundary to the shockwave surface, and between the shockwave to the right boundary. We can then express (\ref{eq:C epsilon inter}) as \eqref{eq:Lk functional}, and the left-boundary time can be expressed as \eqref{eq:time def}.

Now, consider adding the shockwave in the spacetime. While adding an energy excitation can modify the blackening factor, we work in the strict massless excitation limit of \cite{Shenker:2013pqa} where one does not need to distinguish between the region affected by the shockwave. 

Let $v$ ($\tilde{v}$) denote the EF coordinate of a point in the horizon where the shockwave crosses, which is related to the corresponding Kruskal coordinate as
\begin{equation}
    -\Phi_hv=\log V~,\quad -\Phi_h\tilde v=\log (V+\alpha)~.
\end{equation}
We can then evaluate \eqref{eq:Lk functional} and \eqref{eq:time def} to compute the length between the boundaries at $\Phi=\Phi_b$ to the left or right region with respect to the shockwave (Fig.~\ref{fig:CMC_shockwave}) as,\footnote{In the limit of massless shock waves, the blackening factor used in \eqref{eq:tL tR shocwkave} is unaffected; otherwise, one has to include the modification in the black hole mass \cite{Shenker:2013pqa,Shenker:2013yza}.}
\begin{subequations}\label{eq:tL tR shocwkave}
\begin{align}
    t_L&=v+\int_{\Phi_h}^{\Phi_b}\qty(1-\frac{\mathfrak{c}-\epsilon k r}{\sqrt{-\mathcal{U}(\mathfrak{c},\, k,\,r)}})\frac{\rmd r}{f(r)}~,\quad L_{k,\,L}=\int_{\Phi_h}^{\Phi_b}\frac{\rmd r}{\sqrt{-\mathcal{U}(\mathfrak{c},\, k,\,r)}}~,\label{eq:L k L}\\
    t_R&=\tilde{v}+\int_{\Phi_h}^{\Phi_b}\qty(1-\frac{\tilde{\mathfrak{c}}-\epsilon k r}{\sqrt{-{\mathcal{U}}(\tilde{\mathfrak{c}},\,k,\,r)}})\frac{\rmd r}{f(r)}~,\quad L_{k,\,R}=\int_{\Phi_h}^{\Phi_b}\frac{\rmd r}{\sqrt{-{\mathcal{U}}(\tilde{\mathfrak{c}},\,k,\,r)}}~,\label{eq:L k R}
\end{align}
\end{subequations}
where $\mathfrak{c}$ and $\tilde{\mathfrak{c}}$ denote the conserved charges in the corresponding regions to the left or right of the shockwave.

We will now focus on the case where the blackening factor is $f(r)=r^2-\Phi_h^2$,
\begin{subequations}\label{eq:total length shocks}
\begin{align}
   \label{eq:LKL}L_{k,\,L}= \frac{2}{\sqrt{k^2+1}}&\Biggl(\tanh ^{-1}\left(\frac{\sqrt{-2 \mathfrak{c} k \Phi
   _b+\left(k^2+1\right) \Phi _b^2+\mathfrak{c}^2-1}-\sqrt{\mathfrak{c}^2-1}}{\sqrt{k^2+1}
   \Phi _b}\right)\nonumber\\
   &-\tanh ^{-1}\left(\frac{\sqrt{\mathfrak{c}^2-2 \mathfrak{c}
   k+k^2}-\sqrt{\mathfrak{c}^2-1}}{\sqrt{k^2+1}}\right)\Biggr)~,\\
   \label{eq:LKR}L_{k,\,R}= \frac{2}{\sqrt{k^2+1}}&\Biggl(\tanh ^{-1}\left(\frac{\sqrt{-2 \tilde{\mathfrak{c}} k \Phi
   _b+\left(k^2+1\right) \Phi _b^2+\tilde{\mathfrak{c}}^2-1}-\sqrt{\tilde{\mathfrak{c}}^2-1}}{\sqrt{k^2+1}
   \Phi _b}\right)\nonumber\\
   &-\tanh ^{-1}\left(\frac{\sqrt{\tilde{\mathfrak{c}}^2-2 \tilde{\mathfrak{c}}
   k+k^2}-\sqrt{\tilde{\mathfrak{c}}^2-1}}{\sqrt{k^2+1}}\right)\Biggr)
   \end{align}
\end{subequations}
\begin{equation}
\begin{aligned}
    \Phi_h(t_L-v)=&\tanh ^{-1}\left(\frac{\sqrt{-2 \mathfrak{c} k \Phi _b+\left(k^2+1\right) \Phi
   _b^2+\mathfrak{c}^2-1}-\sqrt{k^2+1} \Phi _b+\mathfrak{c}-k}{\sqrt{k^2+1}}\right)\\
   &+\tanh
   ^{-1}\left(\frac{\sqrt{-2 \mathfrak{c} k \Phi _b+\left(k^2+1\right) \Phi
   _b^2+\mathfrak{c}^2-1}-\sqrt{k^2+1} \Phi _b+\mathfrak{c}+k}{\sqrt{k^2+1}}\right)\\
   &-\tanh^{-1}\left(\frac{k \left(\sqrt{-2 \mathfrak{c} k \Phi _b+\left(k^2+1\right) \Phi
   _b^2+\mathfrak{c}^2-1}-\sqrt{k^2+1} \Phi _b+\mathfrak{c}\right)}{\mathfrak{c} \left(\sqrt{-2 \mathfrak{c} k \Phi
   _b+\left(k^2+1\right) \Phi _b^2+\mathfrak{c}^2-1}-\left(\sqrt{k^2+1}+k\right)
   \Phi _b\right)+A}\right)\\
   &-\tanh ^{-1}\left(\frac{k \left(-2
   \mathfrak{c}+\sqrt{k^2+1}+k\right)}{2 \mathfrak{c}^2-2 \mathfrak{c} \left(\sqrt{k^2+1}+k\right)+k
   \left(\sqrt{k^2+1}+k\right)}\right)\\
   &+\tanh ^{-1}\left(1-\frac{2
   \mathfrak{c}}{\sqrt{k^2+1}}\right)+\tanh ^{-1}\left(\frac{-2 \mathfrak{c}+\sqrt{k^2+1}+2
   k}{\sqrt{k^2+1}}\right)~,
   \end{aligned}
\end{equation}
and similarly for $t_R-\tilde{v}$ in terms of $\tilde{\mathfrak{c}}$, and
\begin{equation}
    A=\Phi _b \left(-\sqrt{\left(k^2+1\right) \left(-2 \mathfrak{c} k
   \Phi _b+\left(k^2+1\right) \Phi _b^2+\mathfrak{c}^2-1\right)}+k^2 \Phi _b+\Phi
   _b\right)+\mathfrak{c}^2-1~.
\end{equation}
For general $\Phi_b>\Phi_h$, we do not find an analytic expression when inverting $\mathfrak{c}(t_L)$ and $\tilde{\mathfrak{c}}(t_R)$. However, we can simplify the expressions considerably by taking the asymptotic boundary limit. For $\Phi_b\gg\Phi_h$, one finds 
\begin{align}\label{eq:tR}
    2\Phi_h (t_L-v)&=\log \left(\frac{2 \sqrt{k^2+1} \mathfrak{c}\Phi_h+\left(k^2+1\right) \Phi_h^2+\mathfrak{c} ^2}{(k
   \Phi_h+\mathfrak{c})^2}\right)~,\\
    2\Phi_h (t_R-\tilde{v})&={\log \left(\frac{2 \sqrt{k^2+1} \tilde{\mathfrak{c}}  \Phi_h+\left(k^2+1\right)
   \Phi_h^2+\tilde{\mathfrak{c}}^2}{(k \Phi_h+\tilde{\mathfrak{c}})^2}\right)}~,\label{eq:tL}
\end{align}
By inverting \eqref{eq:tL} and \eqref{eq:tR} we recover
\begin{align}
    \mathfrak{c}=\Phi_h\big[\sqrt{k^2+1}-kX\pm(\sqrt{k^2+1}-k)\sqrt{X}\big]/(X-1)~,\quad X\equiv\rme^{2\Phi_h(t_L-v)}~,
\end{align}
and similarly for $\tilde{\mathfrak{c}}$ with $t_L-v\rightarrow t_R-\tilde{v}$. Plugging these parameters in the single-sided lengths \eqref{eq:LKL} and \eqref{eq:LKR} with $\Phi_b\gg1$, we find
\begin{subequations}\label{eq:Vtot int}
\begin{align}
   \sqrt{1+k^2}\,L_{k,L}=\log\frac{2\sqrt{1+k^2}\,\Phi_b\big(V+\alpha+\rme^{\Phi_ht_L}\big)\rme^{-\Phi_ht_L}}{\Phi_h}~,\\
\sqrt{1+k^2}\,L_{k,R}=\log\frac{2\sqrt{1+k^2}\,\Phi_b\big(\rme^{\Phi_ht_R}-V\big)\rme^{-\Phi_ht_R}}{\Phi_h}~.
\end{align}
\end{subequations}
Following the argument of this subsection, \eqref{eq:AdS geo vs CMC} with the shockwave geodesic length, we recover the total CMC slice length
\begin{equation}\label{eq:CMC shockwave length}
\begin{aligned}
L_k=\frac{2}{\sqrt{1+k^2}}\log \left(\frac{2\Phi_b}{\Phi_h}\sqrt{k^2+1}\Biggl(\cosh \left(\frac{\Phi_h}{2}(t_L-t_R)\right)+\frac{\alpha}{2} e^{-\Phi_h
   (t_L+t_R)/2}\Biggr)\right)~.
   \end{aligned}
\end{equation}
In the case $\alpha=0$ we recover a previous result by \cite{Parrikar:2025xmz}
\begin{equation}
    \begin{aligned}
    \eval{L_k}_{\alpha=0}&=\frac{2}{\sqrt{1+k^2}}\log\qty(\frac{2\sqrt{1+k^2}}{\Phi_h}\cosh(\frac{\Phi_h}{2}(t_L-t_R)))+L_{\rm reg}~,\\
    L_{\rm reg}&\equiv\frac{2\log(\Phi_b)}{\sqrt{1+k^2}}~.
    \end{aligned}
\end{equation}
Meanwhile, when $k=0$ we recover the result by \cite{Shenker:2013pqa}
\begin{equation}
\begin{aligned}
   L_{k=0}&=2\log\qty(\cosh(\frac{\Phi_h}{2}(t_L-t_R))+\frac{\alpha}{2}\rme^{-\Phi_h(t_L+t_R)/2})+L'_{\rm reg}~,\\ 
    L'_{\rm reg}&\equiv2\log \frac{2\Phi_b}{\Phi_h}~.
\end{aligned}
\end{equation}
We use the results above to define the following variables
\begin{subequations}\label{eq:momenta}
\begin{align}
    &P_L\equiv(1+k^2)\dv{L_k}{t_L}=-\sqrt{k^2+1} \Phi_h \frac{\rme^{\Phi_ht_R}-\rme^{\Phi_ht_L}+\alpha}{\rme^{\Phi_ht_L}+\rme^{\Phi_ht_R}+\alpha}~,\\
    &P_R\equiv-(1+k^2)\dv{L_k}{t_R}=\sqrt{k^2+1} \Phi_h\frac{\rme^{\Phi_ht_L}-\rme^{\Phi_ht_R}+\alpha}{\rme^{\Phi_ht_L}+\rme^{\Phi_ht_R}+\alpha}~.
\end{align}
\end{subequations}
Note that the normalizations were chosen such that when $\alpha=0$ we recover the results of \cite{Parrikar:2025xmz}
\begin{align}
\eval{P_L}_{\alpha=0}=\eval{P_R}_{\alpha=0}=\sqrt{1+k^2}\Phi_h\tanh\qty(\frac{\Phi_h}{2}(t_L-t_R))~.
\end{align}
Meanwhile, we also recover the canonical momenta conjugate to lengths of JT gravity in (4.30) of \cite{Aguilar-Gutierrez:2025pqp} when $k=0$ 
\begin{equation}
\begin{aligned}
    \eval{P_L}_{k=0}=\Phi_h\frac{\rme^{\Phi_ht_L}-\rme^{\Phi_ht_R}-\alpha}{\rme^{\Phi_ht_L}+\rme^{\Phi_ht_R}+\alpha}~,\quad \eval{P_R}_{k=0}=\Phi_h\frac{\rme^{\Phi_ht_L}-\rme^{\Phi_ht_R}+\alpha}{\rme^{\Phi_ht_L}+\rme^{\Phi_ht_R}+\alpha}~.
\end{aligned}
\end{equation}
Similarly, the single sided lengths from \eqref{eq:Vtot int} are then
\begin{subequations}
\begin{align}
    L_{k,\,L}=\frac{L_k}{2}+\frac{\Phi_h(t_R-t_L)}{2\sqrt{k^2+1}}~,\quad L_{k,\,R}=\frac{L_k}{2}+\frac{\Phi_h(t_L-t_R)}{2\sqrt{k^2+1}}~.
   \end{align}
\end{subequations}
From this, one may study the ADM Hamiltonian of the system, based on the conserved spectrum $E_L=-E_R={\Phi_{h}^2}/2$ where $\Phi_h$ is the enlarged horizon radius due to the shockwave.

\subsection{Phase space analysis}
The symplectic form of the phase space is defined by\cite{Harlow:2018tqv}
\begin{equation}\label{eq:xsympletic form}
    \omega=\omega_L+\omega_R~,\quad \omega_{L/R}\equiv\rmd t_{L/R}\wedge \rmd E_{L/R}~,
\end{equation}
which should coincide with $\sum_{i}\rmd L_{k,\,i}\wedge\rmd P_i$ if $P_i$ satisfies Hamilton's equations. 

From the shockwave solutions, we find that
\begin{subequations}
    \begin{align}
    \pdv{L_{k,\,L}}{t_L}\pdv{P_L}{\Phi_h}-\pdv{L_{k,\,L}}{\Phi_h}\pdv{P_L}{t_L}+\pdv{L_{k,\,R}}{t_L}\pdv{P_R}{\Phi_h}-\pdv{L_{k,\,R}}{\Phi_h}\pdv{P_R}{t_L}&=\Phi_h~,\\
    \pdv{L_{k,\,L}}{t_R}\pdv{P_L}{\Phi_h}-\pdv{L_{k,\,L}}{\Phi_h}\pdv{P_L}{t_R}+\pdv{L_{k,\,R}}{t_R}\pdv{P_R}{\Phi_h}-\pdv{L_{k,\,R}}{\Phi_h}\pdv{P_R}{t_R}&=-\Phi_h~,
        \end{align}
\end{subequations}
which confirms that $\omega=\sum_{i}\rmd L_{k,\,i}\wedge\rmd P_i$ so that $P_{L/R}$ is the canonical momenta for $L_{k,\,L/R}$ and the Hamiltonians $H_{L/R}$ have a conserved energy $E_{L/R}=\Phi_h^2/2$.

\paragraph{ADM Hamiltonians}
After performing a constant shift in the length variables,\footnote{Note that the coefficients in the terms involving $L_k$ in the bulk Hamiltonian depend on the choice of regularization shift; the exponents always carry the combination $\sqrt{1+k^2}L_k=L_{k=0}+\log(1+k^2)$ (cf.~\eqref{eq:ABL complexity}). Similarly, the energy spectrum differs by an overall factor $1/2$ with respect to the choices in the main text, which we kept as $\Phi_h^2$; this overall factor can be easily absorbed by an overall rescaling of the Hamiltonian.}
\begin{equation}
    L_{k}\rightarrow L_{k}-\frac{2\log(\sqrt{1+k^2}\Phi_b)}{\sqrt{1+k^2}}~,\quad L_k^{(L/R)}\rightarrow L_k^{(L/R)}-\frac{\log(\sqrt{1+k^2}\Phi_b)}{\sqrt{1+k^2}}~,
\end{equation}
we can write the ADM Hamiltonian of the system as,
\begin{equation}\label{eq:H ADM propopsal}
\begin{aligned}
    \hH^{\rm JT}_{L}&\equiv\frac{1}{2}\qty(\frac{P_{L}^2}{1+k^2}+4\rme^{-\sqrt{1+k^2}L_{k}}+\frac{2(P_R-P_L)}{\sqrt{1+k^2}}\rme^{-\sqrt{1+k^2}L_{k,\,L}})~,\\
    \hH^{\rm JT}_{R}&\equiv-\frac{1}{2}\qty(\frac{P_{R}^2}{1+k^2}+4\rme^{-\sqrt{1+k^2}L_{k}}+\frac{2(P_R-P_L)}{\sqrt{1+k^2}}\rme^{-\sqrt{1+k^2}L_{k,\,R}})~.
    \end{aligned}
\end{equation}
Note that the $-$ sign in  $H_R$ is just for the convention about the direction of $t_R$; if we invert  $t_R \rightarrow -t_R$  in the expressions, we can use a $+$ sign in front of $H_R$. For boundedness of the Hamiltonian, one might prefer to flip the overall sign. Using the explicit solutions for $L_k$ and $P_k$, we confirmed that the Hamiltonians in \eqref{eq:H ADM propopsal} have a conserved spectrum, namely 
\begin{equation}\label{eq:solution JT}
\hH_{L}^{\rm JT}= \Phi_h^2/2~,\quad \hH_{R}^{\rm JT} =-\Phi_h^2/2~.  
\end{equation}
Thus, this shows that \eqref{eq:H ADM propopsal} are indeed ADM Hamiltonian of the system. Moreover, note that the CMC slice length $L_k$ in the asymptotic boundary limit \eqref{eq:CMC shockwave length} is just a rescaling of $\ell_{L/R}$ and the corresponding momentum $P_{L/R}$ which does not affect the matter contribution included in the factor $P_L-P_R$. The result, even when $k=0$, is new with respect to the previous literature.

First, note that \eqref{eq:H ADM propopsal} contains four phase space variables $L_{k,L/R}$, $P_{L/R}$ which in general are functions of to $E_{L/R}$ (the ADM energy to the left/right with respect to the bulk shockwave) and $t_{L/R}$. However, in the shockwave solutions we need to consider $E_L=E_R$ in the relevant approximation where $\tilde{E}\rightarrow0$. The specific form of the two-sided Hamiltonian in \eqref{eq:H ADM propopsal} is fixed by requiring that the momentum appears at most quadratically, which is a reasonable assumption based on JT gravity with matter \cite{Iliesiu:2024cnh}. Second, one can see that the $\alpha=0$ limit of \eqref{eq:H ADM propopsal} is consistent with \cite{Parrikar:2025xmz},
\begin{equation}\label{eq:H ADM CMC-Krylov time}
    \eval{\hH_L}_{\alpha=0}=-\eval{\hH_R}_{\alpha=0}=\frac{1}{2}\qty(\frac{P_{L/R}^2}{1+k^2}+4\rme^{-\sqrt{1+k^2}L_{k}})~.
\end{equation}
Next, one should notice that \eqref{eq:H ADM propopsal} takes a different form from a simple conformal matter operator insertion sourcing matter fields in JT gravity \cite{Penington:2023dql,Kolchmeyer:2023gwa,Iliesiu:2024cnh} where one instead finds a factor $\rmi(P_L-P_R)$ in \eqref{eq:H ADM propopsal}. They are manifestly different Hamiltonians, ours being real valued. The difference might arise because in the shockwave approximation, where the energy of the probe $\tilde{E}\rightarrow0$, we set the left and right ADM energies to be automatically the same $E_L=E_R$, and moreover, we consider an effective description of the conformal matter operator which is strongly blue shifted (by taking the $t_w\rightarrow\infty$ and $\tilde{E}\rightarrow0$ limit stated above) so that it generates shockwave backreaction while the conformal operator insertion for $t_w$ finite does not lead to backreaction. However, the parameter $\alpha$ does not appear explicitly in the two-sided Hamiltonian \eqref{eq:H ADM propopsal}. While at the classical level, one can use energy conservation to reintroduce $\alpha$ in the solution to Hamilton's equations; if one were to canonical quantize \eqref{eq:H ADM propopsal} its scattering wavefunctions would be insensitive to the parameter $\alpha$ and thus to the shockwave backreaction would not appear in the solutions. This would suggest that one should work with the Hamiltonian with a general matter bulk operator insertion and take the shockwave limit in the evaluations until the end.

\subsection{Towards a chord interpretation}Let us now compare \eqref{eq:H ADM propopsal} with the ADM Hamiltonian of JT gravity with matter \cite{Iliesiu:2024cnh}, or equivalently, the triple-scaled Hamiltonian with matter of the DSSYK \cite{Lin:2022rbf}, which takes the form
\begin{subequations}\label{eq:HLR JT final}
    \begin{align}
    \Delta H_L&=\frac{1}{2}\Biggl(\frac{P_L^2}{1+k^2}+\frac{2}{\sqrt{1+k^2}}\rme^{-\sqrt{1+k^2}L_L}\qty(\Delta-1+\rmi(P_L-P_R))\\
    &\qquad\qquad+4\rme^{-\sqrt{1+k^2}L}\Biggr)~,\nonumber\\
    \Delta H_R&=\frac{1}{2}\Biggl(\frac{P_R^2}{1+k^2}+\frac{2}{\sqrt{1+k^2}}\rme^{-\sqrt{1+k^2}L_R}\qty(\Delta-1-\rmi(P_L-P_R))\\
         &\qquad\qquad+4\rme^{-\sqrt{1+k^2}L}\Biggr)~.\nonumber
    \end{align}
\end{subequations}
The main difference with respect to the shockwave case in \eqref{eq:H ADM propopsal} is that here we describe a strongly backreacting particle at late times, instead of a conformal operator with dimension $\Delta$ inserted in the HH preparation of state at $t=0$. While the shockwave can be described as a limiting case of the operator insertion, which is time shifted in the Lorentzian continuation, our findings indicate that its ADM / chord Hamiltonian takes a different form with respect to the particle insertion.

Thus, the difference is that the ADM Hamiltonian above corresponds to the effective description of the AdS$_2$ black hole with the shockwave, while \eqref{eq:HLR JT final} is the fundamental description (in the sense of the ensemble-averaged theory, not the microscopic one which is more fundamental) of a general matter insertion.

The results suggest a chord description of the system of the form (using the notation in Sec.~\ref{sec:CMC-Krylov complexity}),\footnote{We omit the $n$ subscript of $\hat{p}_n$ in \eqref{eq:double chord H shockwave} to avoid clutter.}
\begin{equation}\label{eq:double chord H shockwave}
    H_{L/R}=\rme^{-\rmi \hat{p}_{L/R}}+\rme^{\rmi \hat{p}_{L/R}}+\rmi(\rme^{\rmi \hat{p}_{L}}-\rme^{\rmi \hat{p}_{R}})\rme^{-\hat{\text{L}}_{L/R}}+q^\Delta\rme^{\rmi \hat{p}_{R/L}}\rme^{-\qty(\hat{\text{L}}_L+\hat{\text{L}}_R)}~.
\end{equation}

\section{Geodesic length to an ETW brane}\label{app:distance brane}
\begin{figure}
    \centering
    \subfloat[]{\includegraphics[width=0.4\linewidth]{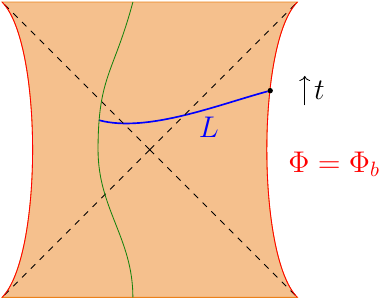}}\hfill\subfloat[]{\includegraphics[width=0.57\linewidth]{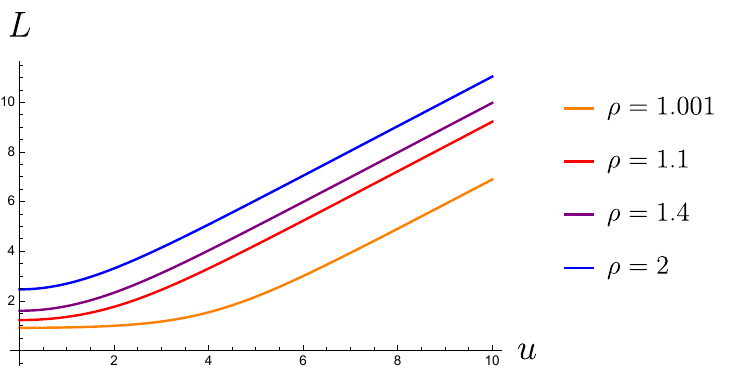}}
    \caption{(a) Penrose diagram of the ETW brane (green solid curve) in the AdS black hole with a finite Dirichlet boundary (red curve). (b) Growth of the geodesic length \eqref{eq:main text L brane} as a function of time for different values of $\bbr$ (shown in the plot), where we set $m_{\rm JT}=1$.}
    \label{fig:ETW}
\end{figure}

In this appendix, we (i) derive the geodesic length \eqref{eq:main text L brane} between a finite cutoff boundary with Dirichlet boundary conditions in AdS$_2$ and an ETW brane with Neumann boundary conditions, illustrated in Fig.~\ref{fig:ETW}, in App.~\ref{sapp:geodesic ETW}; and (ii) generalize the result for CMC slices in App~\ref{sapp:CMC slice length}.

\subsection{Geodesic length}\label{sapp:geodesic ETW}
We begin by deriving the geodesic length between a finite cutoff boundary of an AdS$_2$ black hole and an ETW brane \eqref{eq:main text L brane}. We employ the relation between Minkowski embedding coordinates and the geodesic length in \eqref{eq:embedding for length}, providing an alternative derivation with respect to \cite{Gao:2021uro} and without restricting the calculation to the asymptotic boundary limit.

To evaluate the geodesic length $L$ between the cutoff boundary to the ETW brane, let us consider embedding coordinates of AdS$_2$ in 3-dimensional Minkowski space ($X^I$ in \eqref{eq: X coords}), which are used to evaluate the geodesic length in \eqref{eq:embedding for length}. Similar to \cite{Blommaert:2025avl}, we can describe the dilaton in terms of the ($X_{0},\,X_1,\,X_2$) coordinates by
\begin{eqnarray}
    \Phi=V\cdot X~,\quad V\cdot V=-\Phi_h^2~,
\end{eqnarray}
where $X\cdot Y\equiv \eta_{IJ}X^IY^J$ in the Minkowski embedding. The ETW branes following a CMC worldline can be described with an AdS charge \cite{Maldacena:2016upp} describing the normal vector to the trajectory, which we denote $Q^I$, and the fixed mean curvature \eqref{eq:brane conditions} implies
\begin{eqnarray}\label{eq:q imp}
    Q\cdot X=K_{\rm AdS}~,\quad Q\cdot Q=1-K_{\rm AdS}^2~.
\end{eqnarray}
The normal derivative to the dilaton in \eqref{eq:brane conditions} implies that $Q^I$ can be fixed by a boost to the form \cite{Blommaert:2025avl},
\begin{eqnarray}
    Q=\sqrt{1-K_{\rm AdS}^2}\qty(0,~\frac{m_{\rm JT}}{\Phi_h\sqrt{1-K_{\rm AdS}^2}},~-\frac{\sqrt{\Phi^2_h(1-K_{\rm AdS}^2)+m^2_{\rm JT}}}{\Phi_h\sqrt{1-K_{\rm AdS}^2}})~.
\end{eqnarray}
The unit normal to the brane corresponding to $n^a$ in \eqref{eq:brane conditions} is $n^I=Q^I+K_{\rm AdS}X^I$, which is tangent to AdS$_2$ ($n\cdot X=0$) and has unit norm by \eqref{eq:q imp}; the brane condition \eqref{eq:brane conditions} then reads $Q\cdot V=m_{\rm JT}$, which fixes the second component of $Q^I$ (we use $V^I=(0,-\Phi_h,0)$, so that $\Phi=\Phi_hX_1$ as in \eqref{eq:embeeding Minkowski and AdS}), while the sign of the last component is fixed by requiring $n^I$ to point away from the cutoff boundary, as in \eqref{eq:brane conditions}. It is convenient to introduce the constant unit vector $\hat{W}^I\equiv -Q^I/\sqrt{Q\cdot Q}$, which points from the brane towards the cutoff boundary. For a tangential vector to the brane, one has $\eta_{IJ}\tau^I\hat{W}^J=0$. 

Let us then consider a cutoff boundary point $X'$ with components $(X'_{0},\,X'_1,\,X'_2)$. We want to compute the geodesic length between the cutoff boundary to the brane, which must be perpendicular to the latter. Let us decompose the corresponding vector to $X'$ in terms of perpendicular and orthogonal components with respect to the orthonormal vector $\hat{W}$ to the brane worldline as
\begin{eqnarray}\label{eq:X decomp}
    X'=(\hat{W}\cdot X')\hat{W}+X_{\perp}'~,
\end{eqnarray}
where $X_{\perp}'\cdot \hat{W}=0$. Then, considering there is a geodesic joining the cutoff boundary to the brane and ending
perpendicular to it, we first construct the point closest to $X'$ on the auxiliary geodesic
$\Gamma_0\equiv\{X:\ \hat W\cdot X=0\}$, i.e.\ the geodesic ETW brane with the same charge $Q^I$,
\begin{equation}
  Y_0\equiv\frac{X'_\perp}{\sqrt{-\eta_{IJ}X'^I_\perp X'^J_\perp}}
        =\frac{X'_\perp}{\sqrt{1+(\hat W\cdot X')^2}}\,,
  \label{eq:F5}
\end{equation}
which satisfies $Y_0\cdot Y_0=-1$ and $\hat W\cdot Y_0=0$. The curve
\begin{equation}
  X(\delta)=\cosh\delta\,Y_0+\sinh\delta\,\hat W
  \label{eq:Y0}
\end{equation}
which lies on AdS$_2$, is the intersection of the hyperboloid with the plane spanned by $Y_0$ and
$\hat W$, and has unit-norm tangent. It is therefore the geodesic through $Y_0$ orthogonal to
$\Gamma_0$, parametrized by proper length $\delta$, and $\hat W\cdot X(\delta)=\sinh\delta$.
Hence $\delta=\operatorname{arcsinh}(\hat W\cdot X)$ is the signed geodesic distance to $\Gamma_0$.
Dividing \eqref{eq:q imp} by $-\sqrt{1-K_{\rm AdS}^2}$, the brane satisfies
\begin{equation}
  \hat W\cdot X=-\frac{K_{\rm AdS}}{\sqrt{1-K_{\rm AdS}^2}}=\sinh\sigma_0\,,\qquad
  K_{\rm AdS}=-\tanh\sigma_0\,,
  \label{eq:F7}
\end{equation}
that is, the curve at constant signed distance $\sigma_0$ from $\Gamma_0$, displaced towards the cutoff boundary when $\sigma_0>0$, in agreement with
Sec.~\ref{ssec:traversable wormholes}, where the static brane sits at $\sigma=\sigma_0$ with $K=-\tanh\sigma_0$. Geodesics orthogonal to $\Gamma_0$ are also orthogonal to every curve at constant
distance from it, so the geodesic from $X'$ that ends perpendicular to the brane is
(\ref{eq:Y0}) itself. By \eqref{eq:X decomp} and (\ref{eq:F5}), $X'=X(\delta')$ with
$\sinh\delta'=\hat W\cdot X'$, while the endpoint on the brane is
\begin{equation}
  Y=X(\sigma_0)=\cosh\sigma_0\,Y_0+\sinh\sigma_0\,\hat W\,,
  \qquad Q\cdot Y=K_{\rm AdS}\,,\quad Y\cdot Y=-1\,.
  \label{eq:F8}
\end{equation}
The geodesic length then follows from \eqref{eq:embedding for length}:
\begin{equation}
  \cosh L=-\eta_{IJ}Y^IX'^J=\cosh(\delta'-\sigma_0)\,,\qquad
  L=\operatorname{arcsinh}\big(\hat W\cdot X'\big)-\sigma_0\,,
  \label{eq:F9}
\end{equation}
with
\begin{equation}
  \hat W\cdot X'=\frac{\sqrt{\Phi_h^2(1-K_{\rm AdS}^2)+m_{\rm JT}^2}}{\Phi_h\sqrt{1-K_{\rm AdS}^2}}\,X'_2
  +\frac{m_{\rm JT}}{\Phi_h\sqrt{1-K_{\rm AdS}^2}}\,X'_1\,.
  \label{eq:F10}
\end{equation}
The construction in (\ref{eq:X decomp}, \ref{eq:F5}) alone yields the distance to $\Gamma_0$; the
shift by $\sigma_0$ in (\ref{eq:F8}) is what places the endpoint on the CMC brane. Using
the Schwarzschild coordinates ($\bbr=\Phi_b/\Phi_h$, $u$) \eqref{eq:Schwarzschild coordinates} and the Minkowski embedding \eqref{eq:embeeding Minkowski and AdS},
\begin{equation}\label{eq:shift length}
  \sinh(L+\sigma_0)=\frac{m_{\rm JT}\,\bbr+\sqrt{m_{\rm JT}^2+\Phi_h^2(1-K_{\rm AdS}^2)}\,
  \sqrt{\bbr^2-1}\,\cosh u}{\Phi_h\sqrt{1-K_{\rm AdS}^2}}\,,
\end{equation}
which corresponds to \eqref{eq:main text L brane} in the main text. Note that when $K_{\rm AdS}\neq0$, the right-hand side is still equivalent the geodesic-brane (i.e.~$K_{\rm AdS}=0$)
case after a perfoming a rescaling $m_{\rm JT}\to m_{\rm JT}/\sqrt{1-K_{\rm AdS}^2}$, so the CMC brane changes the
length by the constant $-\sigma_0$ in \eqref{eq:shift length} and rescales the tension. In particular, at $u=0$ we find
$L=\sigma_b-\sigma_0$, consistent with Sec.~\ref{ssec:traversable wormholes}, and in the asymptotic boundary limit the
renormalized length differs from \cite{Gao:2021uro} (2.28) by a constant shift $-\sigma_0$.

\subsection{Constant mean curvature generalization}\label{sapp:CMC slice length}
The generalization of the length for CMC slices connecting an ETW brane to the boundary follows from the same techniques in App.~\ref{app: CMC lengths}.

The length of CMC slices $L_k$ is computed from \eqref{eq:Lk functional}
and boundary time $t$ as \eqref{eq:time def}. The difference is that instead of integrating from the boundary location to the turning point, we evaluate until the intersection with the ETW brane \cite{Aguilar-Gutierrez:2023ccv}. Using Schwarzschild coordinates, we denote the location of the brane $\Phi=\Phi_{\rm JT-ETW}=\Phi_h\rho_{\rm JT-ETW}$, $t=t_{\rm JT-ETW}=u_{\rm JT-ETW}/\Phi_h$, so that \eqref{eq:Lk functional} and \eqref{eq:time def} respectively become 
\begin{subequations}\label{eq:evol Length}
    \begin{align}
    &L_k=\eval{\frac{2}{\sqrt{1-k^2}}\Biggl(\text{tanh}^{-1}\left(\frac{\sqrt{\mathfrak{c}^2-2 \mathfrak{c} k r +\left(k^2+1\right)
   r^2-1}-\sqrt{\mathfrak{c}^2-1}}{\sqrt{k^2+1} r }\right)\Biggr)}_{r=\rho_{\rm JT-ETW}}^{r=\bbr}~,\\
   &2(u-u_{\rm JT-ETW})=\eval{\text{tanh}^{-1} \left(\frac{r\left(\mathfrak{c} k+k^2+1\right)-\mathfrak{c}^2-\mathfrak{c} k+1}{(k+\mathfrak{c})
   \sqrt{\left(k^2+1\right)r^2+\mathfrak{c}^2-2 \mathfrak{c} k r-1}}\right)}_{r=\rho_{\rm JT-ETW}}^{r=\bbr}\\
   &\hspace{3cm}-\eval{\text{tanh}^{-1} \left(\frac{r\left(\mathfrak{c} k-k^2-1\right)-\mathfrak{c}^2+\mathfrak{c} k+1}{(\mathfrak{c}-k) \sqrt{\left(k^2+1\right)r^2+\mathfrak{c}^2-2 \mathfrak{c} \,k\,r-1}}\right)}_{r=\rho_{\rm JT-ETW}}^{r=\bbr}~.\nonumber
\end{align}
\end{subequations}
Next, let us evaluate $\Phi_{\rm JT-ETW}$ and $t_{\rm JT-ETW}$. Consider $(t(\chi),\Phi(\chi))$ the brane worldline parametrized by $\chi$, whose tangent vector is $u^\mu_{\rm br}=(\dot t,\dot\Phi)$. The unit spacelike normal vector is $n^\mu=\Big(\frac{\dot\Phi}{f(\Phi)},\,f(\Phi)\dot t\Big),$ with $f(\Phi)$ the blackening factor. Then $n^b\partial_b\Phi=m_{\rm JT}$ gives $\dot\Phi^{\,2}=m_{\rm JT}^2-f(\Phi)$, which means that
\begin{equation}\label{eq:derivative 1}
    \eval{\frac{dt}{d\Phi}}_{\!\rm brane}=\eval{\frac{\dot t}{\dot\Phi}}_{\!\rm brane}=\frac{m_{\rm JT}}{f(\Phi_{\!\rm brane})\sqrt{m_{\rm JT}^2-f(\Phi_{\!\rm brane})}} .
\end{equation}
The tangent vector at the point where the CMC slice meets the brane is then
\begin{equation}
    u_{\rm CMC}^\mu\equiv\Big(\frac{c-k\Phi_{\rm JT-ETW}}{f(\Phi_{\rm JT-ETW})},\,\sqrt{f(\Phi_{\rm JT-ETW})+(c-k\Phi_{\rm JT-ETW})^2}\Big).
\end{equation}
Since we require that the CMC slice is orthogonal to the brane, $g_{\mu\nu}u_{\rm CMC}^\mu u_{\rm br}^\nu=0$, we then recover
\begin{eqnarray}
    \eval{\frac{dt}{d\Phi}}_{\!\rm brane}=\frac{\sqrt{f(\Phi_{\rm JT-ETW})+(c-k\Phi_{\rm JT-ETW})^2}}{(c-k\Phi_{\rm JT-ETW})\,f(\Phi_{\rm JT-ETW})}.
\end{eqnarray}
Consistency with \eqref{eq:derivative 1} requires
\begin{eqnarray}
    (c-k\Phi_{\rm JT-ETW})^2=\Phi_h^2+m_{\rm JT}^2-\Phi_{\rm JT-ETW}^2~.
\end{eqnarray}
Similarly, we can get $t_{\rm JT-ETW}$ by solving \eqref{eq:derivative 1} with reference $t=0$ when $\Phi_{\rm JT-ETW}=\sqrt{\Phi_h^2+m_{\rm JT}^2}$ (corresponding to choosing the origin of Schwarzschild time to coincide with the brane's turning point), which gives
\begin{eqnarray}
    t_{\rm JT-ETW}=\frac{1}{2\Phi_h}\log\frac{m_{\rm JT}\Phi_{\rm JT-ETW}+\Phi_h\sqrt{\Phi_h^2+m_{\rm JT}^2-\Phi_{\rm JT-ETW}^2}}{m_{\rm JT}\Phi_{\rm JT-ETW}-\Phi_h\sqrt{\Phi_h^2+m_{\rm JT}^2-\Phi_{\rm JT-ETW}^2}}~.
\end{eqnarray}
The evolution of the length as a function of boundary time is then given by \eqref{eq:evol Length}. However, we have not found further simplifications to invert the expressions $L_k(\mathfrak{c})$ and $t(\mathfrak{c})$ as $L_k(t)$ in this case, except when $K=0$, in which case we recover \eqref{eq:main text L brane}.

\subsection{Details on the sine dilaton gravity case}\label{sapp:details length SDG ETW}
In this subappendix, we show that the geodesic length between one finite cutoff boundary and the ETW brane in sine dilaton gravity is \eqref{eq:L total} given in Sec.~\ref{ssec:chord interpretation ETW brane}.

Variation of the sine dilaton gravity action \eqref{eq:SDG action ETW} on the brane gives \cite{Blommaert:2025avl}
\begin{equation}\label{eq:relation SDG ETW}
    n^b\partial_b\Phi=\overline{\mathcal Y}\rme^{\rmi \Phi/2}+{\mathcal Y}\rme^{-\rmi \Phi/2}~,
\end{equation}
where $n^b$ has the same orientation as in \eqref{eq:brane conditions}. We use the following substitutions from \cite{Blommaert:2025avl}
\begin{subequations}\label{eq: subst SDG brane}
    \begin{align}
    \rme^{-\rmi\Phi}&=\cos\theta-\rmi \Phi_{\rm AdS}~,\quad K_{\rm AdS}=-\rmi \mathcal{Y}~,\quad n_{\rm AdS}^b=\rme^{\rmi \Phi/2}n^b~,\\
    m_{\rm JT}&=\overline{\mathcal Y}+{\mathcal Y}\cos\theta~.
\end{align}
\end{subequations}
In this way, \eqref{eq:relation SDG ETW} can be recast in the same way as in JT gravity \cite{Gao:2021uro,Iliesiu:2020qvm}
\begin{equation}\label{eq:brane bdry condt}
    n_{\rm AdS}^b\partial_b\Phi_{\rm AdS}=m_{\rm JT}+\Phi_{\rm AdS}K_{\rm AdS}~.
\end{equation}
Next, we use the effective AdS geometry \eqref{eq:sine dilaton metric} (i.e.~$\rmd l^2_{\rm eff}=\rme^{-\rmi \Phi}\rmd l^2$) to evaluate the geodesic length between the ETW brane to the AdS finite cutoff boundary. Given that the boundary condition at the brane takes exactly the same form as in JT gravity \eqref{eq:brane bdry condt}, it follows that our result from App.~\ref{sapp:geodesic ETW} applies, namely the effective geodesic length is \eqref{eq:main text L brane} with $\Phi_h=\sin\theta$ and the other substitutions in \eqref{eq: subst SDG brane}, i.e.
\begin{equation}
\begin{aligned}
    &\sinh L=\\
    &\frac{\xi(\overline{\mathcal{Y}}+{\mathcal{Y}}\,\cos\theta)+\sqrt{(\xi^2-\sin^2\theta)((\overline{\mathcal{Y}}+\mathcal{Y}\,\cos\theta)^2+(1+\mathcal{Y}^2)\sin^2\theta)}~\cosh(\sin\theta~t)}{\sqrt{1+\mathcal{Y}^2}\sin^2\theta}~,
    \end{aligned}
\end{equation}
which corresponds to \eqref{eq:L total} in the main text, where $L=L_{\rm geo}+\sigma_0$ with $\sigma_0\equiv\rmi\tan^{-1}\mathcal Y$ corresponds to a shift of the geodesic length $L_{\rm geo}$.

\section{Microscopic deformations with an ETW brane}\label{app:chord ETW brane}
In this appendix, we present a chord Hamiltonian which reproduces sine dilaton gravity with the ETW brane \eqref{eq:H SDG ETW}, and we derive the microscopic SYK deformations whose double-scaling limit recovers the corresponding chord theory, thus providing more supporting details for the boundary interpretation of Sec.~\ref{ssec:chord interpretation ETW brane}.

\paragraph{Chord Hamiltonian}
Consider the chord Hamiltonian in \eqref{eq:brane chord H}, 
\begin{equation}
\begin{aligned}\label{eq;importn}
\hH_{\rm ETW}&=\frac{-1}{\sqrt{1-q}}\qty(b_{\rm ETW}(\hat{n})\rme^{-\rmi \hat{p}_n}+\rme^{\rmi \hat{p}_n}b_{\rm ETW}(\hat{n})+a_{\rm ETW}(\hat{n}))~,\\
    a_{\rm ETW}(\hat{n})&=\frac{2\mathcal{Y}\,(\overline{\mathcal{Y}}-\xi\sinh (\lambda \hat{n}/2))}{\cosh^2(\lambda \hat{n}/2)}~,\\
    b_{\rm ETW}(\hat{n})&=\sqrt{\qty(1+\frac{\mathcal{Y}^2}{\cosh^2(\lambda \hat{n}/2)})\qty(1+\frac{\overline{\mathcal{Y}}^2-\xi^2}{\cosh^2(\lambda \hat{n}/2)}-2\xi\overline{\mathcal{Y}}\frac{\tanh (\lambda \hat{n}/2)}{\cosh (\lambda \hat{n}/2)})}~,
\end{aligned}
\end{equation}
with $\mathcal{Y}$ and $\overline{\mathcal{Y}}$ constants. As noted in Sec.~\ref{ssec:SDG ETW}, \eqref{eq;importn} reproduces the finite-cutoff sine dilaton gravity ADM Hamiltonian with the ETW brane, including JT gravity in the low-energy limit, as well as the Al-Salam Chihara recurrence relation in the asymptotic boundary limit. 

We are now interested in the microscopic interpretation of \eqref{eq:brane chord H}. We proceed using the KM and TFD formalisms as in Sec.~\ref{sec:deformation}, with the same dictionary as in App.~\ref{app:detalssww}: $\hH_{\rm def}\rightarrow-2\hat{a}$ \eqref{eq:Double scaling H def}, and $\hH_{\rm RX}\rightarrow q^{\Delta_w\hat{n}}$ \eqref{eq:dictionary SYK chord KM} (respectively $\hH_{\rm MQ}\rightarrow q^{\Delta_s\hat{n}}$ \eqref{eq:dictionary SYK and chord TFD}). Since the coefficients \eqref{eq:lanczos ETW} are functions of $\lambda\hat{n}/2$, it is convenient to take $\Delta_w=\Delta_s=1/2$ (i.e.~$w=s=p/2$), so that the microscopic interaction terms map to $x\equiv q^{\hat{n}/2}$. In terms of $x$, the coefficients in \eqref{eq:lanczos ETW} read
\begin{equation}\label{eq:ab micro functions}
\begin{aligned}
    \text{a}_{\rm ETW}(x)&=\frac{4\mathcal{Y}x\qty(2\overline{\mathcal{Y}}x-\xi(1-x^2))}{(1+x^2)^2}~,\\
    \text{b}_{\rm ETW}(x)&=\frac{\sqrt{\qty(x^4+(2+4\mathcal{Y}^2)x^2+1)\qty(x^4+4\xi\overline{\mathcal{Y}}x^3+(2+4\overline{\mathcal{Y}}^2-4\xi^2)x^2-4\xi\overline{\mathcal{Y}}x+1)}}{(1+x^2)^2}~,
\end{aligned}
\end{equation}
i.e.~$a_{\rm ETW}(\hat{n})=\text{a}_{\rm ETW}(q^{\hat{n}/2})$ and $b_{\rm ETW}(\hat{n})=\text{b}_{\rm ETW}(q^{\hat{n}/2})$. These functions are analytic around $x=0$ and can be expanded as power series using $(1+x^2)^{-2}=\sum_{m\geq0}(-1)^m(m+1)x^{2m}$ and the binomial expansion of the square root, which exhibits the polynomial structure that enters the dictionaries; however, as in Sec.~\ref{ssec:SYK defs}, we define the operator functions below by functional calculus rather than by their series.

Since the hopping term of \eqref{eq:brane chord H} must be generated by $\hH_{\rm def}\rightarrow-2\hat{a}$, with $\hat{a}\ket{n}=\sqrt{(1-q^n)/(1-q)}\ket{n-1}$, we introduce, as in \eqref{eq:part_2b},
\begin{equation}
    \tilde{\text{b}}_{\rm ETW}(x)\equiv\frac{\text{b}_{\rm ETW}(x)}{\sqrt{1-x^2}}~,
\end{equation}
where the factor $(1-x^2)^{-1/2}=(1-q^{\hat{n}})^{-1/2}$ compensates the $\sqrt{1-q^{\hat n}}$ in $\hat{a}$. We then propose the following microscopic realizations of the deformed DSSYK chord Hamiltonian with the ETW brane \eqref{eq:brane chord H},
\begin{equation}\label{eq:micro H ETW}
    \begin{aligned}
\hat{\text{H}}_{\rm ETW}&\equiv \frac{1}{2}\qty(\tilde{\text{b}}_{\rm ETW}(\hH_{\rm RX})\hH_{\rm def}^\dagger+\hH_{\rm def}\tilde{\text{b}}_{\rm ETW}(\hH_{\rm RX}))-\frac{1}{\sqrt{1-q}}\text{a}_{\rm ETW}(\hH_{\rm RX})~,\\
\hat{\text{H}}_{\rm ETW'}&\equiv\frac{1}{2}\qty(\tilde{\text{b}}_{\rm ETW}(\hH_{\rm MQ})\hH_{\rm def'}^\dagger+\hH_{\rm def'}\tilde{\text{b}}_{\rm ETW}(\hH_{\rm MQ}))-\frac{1}{\sqrt{1-q}}\text{a}_{\rm ETW}(\hH_{\rm MQ})~,
    \end{aligned}
\end{equation}
where $\hH_{\rm def}$ and $\hH_{\rm def'}$ appear in \eqref{eq:H def KM} and \eqref{eq:H def TFD} respectively, and the operator functions are defined by functional calculus on the spectral subspace where $\hH_{\rm RX}<1$ (respectively $\hH_{\rm MQ}<1$) and the radicand is non-negative, under the same assumption on the extension of the dictionaries to continuous functions as in Sec.~\ref{ssec:SYK defs}. In the double-scaling limit, the first term of \eqref{eq:micro H ETW} becomes $-(\tilde{\text{b}}_{\rm ETW}(q^{\hat{n}/2})\hat{a}^\dagger+\hat{a}\tilde{\text{b}}_{\rm ETW}(q^{\hat{n}/2}))=-\frac{1}{\sqrt{1-q}}\qty(\rme^{\rmi\hat{p}_n}b_{\rm ETW}(\hat{n})+b_{\rm ETW}(\hat{n})\rme^{-\rmi\hat{p}_n})$, while the second becomes $-\frac{1}{\sqrt{1-q}}a_{\rm ETW}(\hat{n})$, so that
\begin{equation}
\begin{aligned}
    \lim_{\rm DS}\overline{\bra{\omega}{\hat{\text{H}}_{\rm ETW}}^m\ket{\omega}}&=\bra{\Omega}{\hH_{\rm ETW}}^m\ket{\Omega}~,\\
    \lim_{\rm DS}\overline{\bra{\rm TFD_\infty}{\hat{\text{H}}_{\rm ETW'}}^m\ket{\rm TFD_\infty}}&=\bra{\Omega}{\hH_{\rm ETW}}^m\ket{\Omega}~.
\end{aligned}
\end{equation}
In this way, the double-scaling limit of the microscopic SYK deformations generate the moments of the auxiliary Hamiltonian \eqref{eq:brane chord H}.

\subsection{Deformation of the Al-Salam Chihara polynomials}\label{app:finite cutoff ASC}
In this subappendix, we complement the discussion of Sec.~\ref{ssec:chord interpretation ETW brane}. 

\paragraph{Asymptotic boundary case}
First, note that the sine dilaton gravity Hamiltonian with the ETW brane \eqref{eq:new H SDG ETW} can be related to the Al-Salam Chihara (ASC) polynomials by rescaling the length parameter $\rme^{-2L}\rightarrow -AB\rme^{-2L}$ with $A$, $B$ being constants, and choosing $\mathcal{Y}(A,B)$ and $\overline{\mathcal{Y}}(A,B)$ takes the form
\begin{equation}\label{eq:new H SDG ETW2}
H(L,P)\rightarrow\frac{-2}{\sqrt{\lambda}}\qty(\sqrt{(1-\rme^{-L})(1-AB\rme^{-L})}\;\cos P-\frac{A+B}{2}\rme^{-L})~.
\end{equation}
One can promote the above classical variables to quantum operators acting on an auxiliary chord Hilbert space with basis $\ket{n}$ as
\begin{eqnarray}
   \hH_{\rm ASC}=-\qty(b_{\rm ASC}(\hat{n})\rme^{-\rmi \hat{p}_n}+\rme^{\rmi \hat{p}_n}b_{\rm ASC}(\hat{n})+a_{\rm ASC}(\hat{n})) ~,
\end{eqnarray}
where
\begin{eqnarray}
   b_{\rm ASC}(\hat{n})=\sqrt{[\hat{n}]_q}\sqrt{1-AB~ q^{\hat{n}-1}}~,\quad a_{\rm ASC}(\hat{n})=\frac{(A+B)}{\sqrt{1-q}}q^{\hat{n}}~,
\end{eqnarray}
whose wavefunctions $\bra{\theta}\ket{n}$ (where $\hH_{\rm ASC}\ket{\theta}=E(\theta)\ket{\theta}$) are precisely those of the ASC polynomials, i.e.~$\bra{\theta}\ket{n}=Q_n(\cos\theta,A,B|q)/\sqrt{(AB,q;q)_n}$ \cite{Berkooz:2025ydg,Watanabe:2025rwp}.

\paragraph{Finite cutoff case}
Now, we are interested in obtaining the polynomials for the finite-cutoff generalization \eqref{eq:H SDG ETW}. To do this, we start from \eqref{eq:brane chord H} as
\begin{equation}\label{eq:lanczos ETW2}
\begin{aligned}
\hH_{\rm ETW}\ket{n}&=\frac{-1}{\sqrt{1-q}}\qty(b_{\rm ETW}(\hat{n})\rme^{-\rmi \hat{p}_n}+\rme^{\rmi \hat{p}_n}b_{\rm ETW}(\hat{n})+a_{\rm ETW}(\hat{n}))~,
\end{aligned}
\end{equation}
where the coefficients $a_{\rm ETW}$ and $b_{\rm ETW}$ are given in \eqref{eq:lanczos ETW} while $\hat{n}\ket{n}=n\ket{n}$ and $\rme^{\mp\rmi\hat{p}_n}\ket{n}=\ket{n\pm1}$. Denoting $\psi_{n}(\theta)\equiv \bra{\theta}\ket{n}$ where $\theta$ parametrizes the spectrum of $\hH_{\rm ETW}$ as $E(\theta)=-2\cos\theta/\sqrt{1-q}$,\footnote{In the asymptotic boundary case, there are both discrete and continuous energy states, depending on the values of the deformation parameters $\mathcal{Y}$, $\overline{\mathcal{Y}}$; it is not yet understood how they are modified in the finite cutoff case beyond the low-energy limit discussed in App.~\ref{sapp:bound states}.} we have that the eigenvalue problem for \eqref{eq:lanczos ETW2} can be expressed as
\begin{equation}\label{eq:most impr recurrence}
\begin{aligned}
    2\cos\theta\psi_n(\theta)&=b_{\rm ETW}(n+1)\psi_{n+1}(\theta)+b_{\rm ETW}(n)\psi_{n-1}(\theta)+a_{\rm ETW}(n)\psi_{n}(\theta)~,\\
    a_{\rm ETW}(n)&=\frac{4\mathcal{Y}q^{n/2}(2\overline{\mathcal{Y}}q^{n/2}-\xi(1-q^{n}))}{(1+q^{n})^2}~,\\
b_{\rm ETW}(n)&=
\frac{\sqrt{\qty(q^{2n}+(2+4\mathcal Y^2)q^{n}+1)\qty(q^{2n}+4\xi\overline{\mathcal Y}q^{3n/2}
+(2+4\overline{\mathcal Y}^2-4\xi^2)q^{n}
-4\xi\overline{\mathcal Y}q^{n/2}+1)}}
{(1+q^{n})^2}~.
\end{aligned}
\end{equation}
The problem can be simplified by series expanding the coefficients in $x\equiv q^{n/2}$,
\begin{eqnarray}
    a_{\rm ETW}(x)=\sum_{j=0}^{\infty}A_jx^j~,\quad b_{\rm ETW}(x)=\sum_{j=0}^{\infty}B_jx^j~,
\end{eqnarray}
where $B_0=1$ (see \eqref{eq:ab micro functions}). A formal solution of the recurrence relation \eqref{eq:most impr recurrence} at large $n$ is then
\begin{equation}
\begin{aligned}\label{eq:sol q Heun}
    \psi_n(\theta)=\rme^{\rmi n \theta}\sum_{m=0}^\infty c_mx^{m}~,\quad c_m=\frac{\sum_{j=1}^{m}\left[
\rme^{\rmi\theta}q^{m/2}B_j+\rme^{-\rmi\theta}q^{-(m-j)/2}B_j +A_j\right]
}{(1-q^{-m/2})(\rme^{-\rmi\theta}-q^{m/2}\rme^{\rmi\theta})}c_{m-j}~,
\end{aligned}
\end{equation}
together with its $\theta\rightarrow-\theta$ image; the physical wavefunction $\bra{\theta}\ket{n}$ is the combination of the two that satisfies the chord-basis initial conditions $\psi_{-1}(\theta)=0$, $\psi_0(\theta)=1$ of Sec.~\ref{ssec:chord descripttion 3}, as in \eqref{eq:tensionless sol} below. Given that the coefficients $c_m$ above are not a rational function of $q^m$, it follows that the solution $\psi_n(\theta)$ is not a ${}_r\psi_{s}$ hypergeometric series, in contrast to the asymptotic boundary case \cite{Berkooz:2025ydg}, but rather a q-Heun-type series \cite{koekoek2010hypergeometric}. Nevertheless, given that the Hamiltonian \eqref{eq:lanczos ETW2} reduces to the one encoding the Al-Salam Chihara recurrence relation \eqref{eq:new H SDG ETW2} in the asymptotic limit, the q-Heun series solution \eqref{eq:sol q Heun} may be seen as a generalization in the asymptotic boundary limit under the above identifications. 
\paragraph{Tensionless brane}
In the special case $\mathcal{Y}=\overline{\mathcal{Y}}=0$, where $b_{\rm ETW}(n)=b_{\rm PT}(n)$ and the recurrence relation coincides with \eqref{eq:recurrence_1} of Sec.~\ref{ssec:chord descripttion 3} for finite-cutoff sine dilaton gravity without the brane, the above relationship simplifies. Writing $\psi_n(\theta)=\varphi_n(\theta)/\prod_{k=1}^nb_{\rm PT}(k)$, so that $\varphi_{n+1}+b_{\rm PT}(n)^2\varphi_{n-1}=2\cos\theta\,\varphi_n$, one finds
\begin{eqnarray}\label{eq:tensionless sol}
    \varphi_n(\theta)
=\frac{e^{in\theta}\overline{H}(q^{n};\theta)\overline{H}(1;-\theta)-e^{-in\theta}\overline{H}(q^{n};-\theta)\overline{H}(1;\theta)}{e^{i\theta}\overline{H}(q;\theta)\overline{H}(1;-\theta)-e^{-i\theta}\overline{H}(q;-\theta)\overline{H}(1;\theta)},
\end{eqnarray}
where
\begin{subequations}
    \begin{align}
    \overline{H}(x;\theta)&\equiv\sum_{m=0}^{\infty}d_m(\theta)x^m,\\
(1-q^{-m})&(e^{-i\theta}-q^{m}e^{i\theta})d_m=
\qty(2e^{i\theta}q^{m-1}-4\cos\theta-(4\xi^2-2)e^{-i\theta}q^{-m+1})d_{m-1}\nonumber\\
&\hspace{3.6cm}+\qty((q^{-m+2}-1)(e^{-i\theta}-q^{m-2}e^{i\theta}))d_{m-2}~,
\end{align}
\end{subequations}
normalized such that $\varphi_0=0$ and $\varphi_1=1$, i.e.~$\psi_0=0$; the solution with the chord-basis conditions $\psi_{-1}=0$, $\psi_0=1$ is obtained from the two independent solutions $\varphi_n^{(\pm)}=e^{\pm in\theta}\overline{H}(q^n;\pm\theta)$ by the corresponding linear combination.

\section{Robin boundary conditions for the minimal ordering with an ETW brane}\label{sapp:bound states}
In this appendix, we provide complementary details for Sec.~\ref{ssec:scattering ETW}. We use Robin boundary conditions to solve the minimally ordered finite cutoff JT Hamiltonian with the ETW brane \eqref{eq:schrodinger ETW} in Sec.~\ref{ssec:1st approach QJT}.

Recall that \eqref{eq:schrodinger ETW} is
\begin{equation}
  -\frac14\psi_s''(L)+U(L)\,\psi_s(L)=s^2\psi_s(L)\,,\qquad
  U(L)\equiv\frac{2\mJT\Phi_b\sinh L+\Phi_b^2-\mJT^2}{\cosh^2L}\,,
  \label{eq:RU}
\end{equation}
with $L\in[0,\infty)$ and $\psi_s(L)$ is subject to the Robin boundary condition (\ref{eq:Rrobin}). The most general solution to the Schrödinger equation with the Scarf II potential in \eqref{eq:schrodinger ETW} are given by
\begin{subequations}\label{eq:wavefunct}
\begin{align}
\psi_s(L)=&
C_1~u_1(L)+C_2~u_2(L)~,\\
    u_1(L)=&
z(L)^{\frac14-\msw}~
\overline{z}(L)^{\frac14-\bmsw}~{}_2F_1\left(
\frac12-\msw-\bmsw+2\rmi s,\,
\frac12-\msw-\bmsw-2\rmi s;\,
1-2\msw;\,
z(L)\right)~,\\
u_2(L)=&
z(L)^{\frac14+\msw}~
\overline{z}(L)^{\frac14-\bmsw}~{}_2F_1\left(
\frac12+\msw-\bmsw+2\rmi s,\,
\frac12+\msw-\bmsw-2\rmi s;\,
1+2\msw;\,
z(L)
\right)~,\\
z(L)\equiv &\frac{1+\rmi \sinh L}{2}~,\quad \overline{z}(L)\equiv \frac{1-\rmi \sinh L}{2}~,\\
\msw\equiv&\sqrt{\tfrac1{16}-(\Phi_b+\rmi m_{\rm JT})^2},
\quad
\bmsw\equiv\sqrt{\tfrac1{16}-(\Phi_b-\rmi m_{\rm JT})^2}~.
\end{align}
\end{subequations}
In the following, we show that the boundary condition is irrelevant in the
asymptotic boundary limit, and we derive the leading finite cutoff correction (\ref{eq:Rmain}) to the density of states.

For $s^2>0$ we write the asymptotic form of the solutions as in \eqref{eq:asymptotics of psi s},
$\psi_s(L)\simeq\mathcal A_+e^{2{\rm i}sL}+\mathcal A_-e^{-2{\rm i}sL}$ with $\mathcal A_+=\mathcal A_-^*\equiv
|\mathcal A_+|e^{{\rm i}\vartheta(s)}$, so that \eqref{eq:density coeffs} becomes $\rho(s)=1/(\pi|\mathcal A_+|^2)$.

\begin{figure}
    \centering
    \subfloat[]{\includegraphics[width=0.45\linewidth]{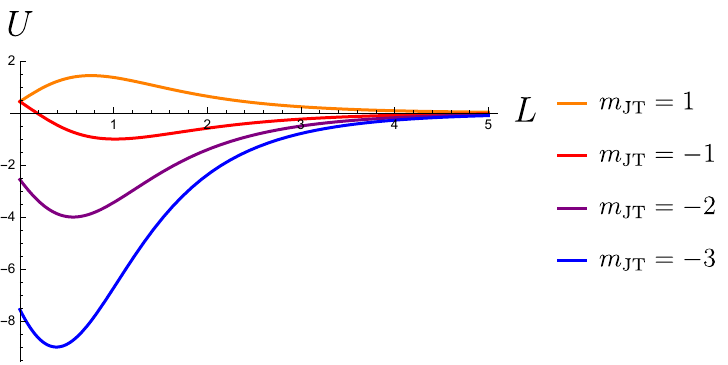}}\hspace{1cm}\subfloat[]{\includegraphics[width=0.45\linewidth]{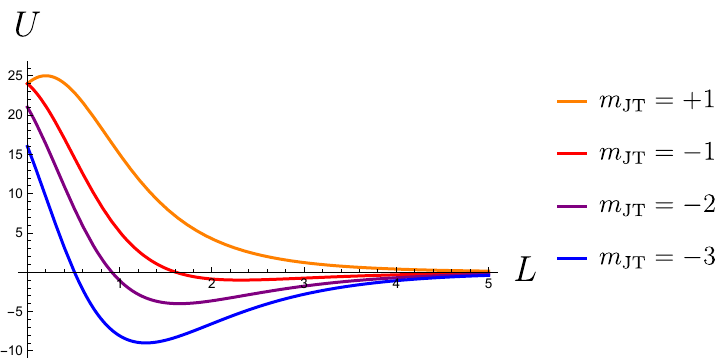}}
    \caption{Potential energy term $U(L)$ \eqref{eq: Hamilt ETW} for different choices of parameters. In (a) we set $\Phi_b=1.2$, while in (b) $\Phi_b=5$ (the maximum of $U$ is $\Phi_b^2$ at $m_{\rm JT}=0$), and different values for the brane tension are displayed in the figure. As expected from \eqref{eq:min max U}, the maximum and minimum are more pronounced as we increase $\Phi_b$ or $\abs{m_{\rm JT}}$, respectively.}
    \label{fig:new ETW figure}
\end{figure}

\paragraph{Potential energy term}We first want to identify the properties of the potential energy term $U(L)$ in the finite cutoff Hamiltonian \eqref{eq: Hamilt ETW}. Its extrema are located at
\begin{align}\label{eq:conditions in l min max}
\sinh L_{\rm min}=-\frac{\Phi_b}{m_{\rm JT}}~,\quad\sinh L_{\rm max}=\frac{m_{\rm JT}}{\Phi_b}
\end{align}
The minimum and maximum values of the potential energy term correspond to
\begin{eqnarray}\label{eq:min max U}
    U(L_{\rm min})=-m_{\rm JT}^2~,\quad U(L_{\rm max})=\Phi_b^2~.
\end{eqnarray}
We illustrate the potential for different parameters in Fig.~\ref{fig:new ETW figure}.

We then need to allow $m_{\rm JT}<0$ for $L_{\rm min}>0$, since in the finite cutoff case, we restrict to $L\geq0$ in finite-cutoff holography.

\paragraph{Asymptotic boundary limit}
At fixed $x=8\Phi_be^{-L}$, we can expand (\ref{eq:RU}) in $e^{-2L}=x^2/(64\Phi_b^2)$, which gives
\begin{equation}
  \begin{gathered}
  U=U_\infty(x)+\frac{U_2(x)}{\Phi_b^2}+\mathcal O\big(\Phi_b^{-4}\big)\,,\qquad
  U_\infty(x)=\frac{\mJT}{2}\,x+\frac{x^2}{16}\,,\\
  U_2(x)=-\frac{1}{64}\left(\frac{x^4}{8}+\frac{3\mJT}{2}\,x^3+4\mJT^2x^2\right).
  \end{gathered}
  \label{eq:RVexp}
\end{equation}
$U_\infty(x)$ is the Morse potential in \eqref{eq: H brane GJK}, and its scattering solution that decays at large $x$ is the Whittaker
function \eqref{eq:kimit crucial}, with density of states $\rho_\infty(s)$ in (\ref{eq:Rmain}) when normalized as
$W_{-2\mJT,2{\rm i}s}(x)\sim x^{-2\mJT}e^{-x/2}$.

\paragraph{Boundary conditions}
Since $U(0)=\Phi_b^2-\mJT^2\gg s^2$, the boundary $L=0$ lies inside the classically forbidden region $0\le L<L_*$, with
$U(L_*)=s^2$. Let us denote by $P(L)\equiv2\sqrt{U(L)-s^2}$ the Euclidean momentum in this region. There, the two
independent solutions behave as
\begin{equation}
  u_\pm(L)\simeq P(L)^{-1/2}\exp\left[\mp\int_L^{L_*}P(L')\,{\rm d}L'\right],
\end{equation}
where $u_+$ decays towards $L=0$ and continues to the $W$-Whittaker function, while $u_-$ grows towards $L=0$ and
continues to the $M$-Whittaker function in \eqref{eq:W asymp}. Writing $\psi_s=u_++r(s)\,u_-$ and imposing (\ref{eq:Rrobin}), we find
\begin{equation}
  r(s)=\frac{P(0)-\gamma}{P(0)+\gamma}\,\mathcal T(s)\,,\qquad
  \mathcal T(s)\equiv\exp\left[-2\int_0^{L_*}P(L)\,{\rm d}L\right]\simeq e^{-2\pi\Phi_b+\mathcal O(|\mJT|)}\,,
  \label{eq:Rr}
\end{equation}
where we used $U(L)\simeq\Phi_b^2\,{\rm sech}^2L$ and $\int_0^\infty{\rm sech}\,L\,{\rm d}L=\pi/2$ for $\Phi_b\gg1$. The
coefficient $r(s)$ measures the admixture of the $M$-Whittaker function, and it is suppressed by the tunneling factor
$\mathcal T(s)$ of the barrier. Through the standard connection formulae, it shifts the phase of $\mathcal A_+$ by
\begin{equation}
  \vartheta(s)=\vartheta_0(s)+\frac12\,\frac{P(0)-\gamma}{P(0)+\gamma}\,\mathcal T(s)+\mathcal O\big(\mathcal T^2\big)\,,
  \label{eq:Rphase}
\end{equation}
where $\vartheta_0(s)$ does not depend on $\gamma$. The scattering data and the bound states (see Sec.~\ref{sapp:bounds}) thus depend on
the boundary condition only at $\mathcal O(e^{-2\pi\Phi_b})$. The only exception is $\gamma<-2\sqrt{U(0)}\simeq-2\Phi_b$,
where $r(s)$ passes through the pole at $\gamma=-P(0)$, and the boundary binds an additional state localized at $L=0$ with energy
$s^2\simeq\gamma^2/4-U(0)$, which has no counterpart in the Morse problem. This requires $\gamma$ to grow with $\Phi_b$.
Therefore, any boundary condition (\ref{eq:Rrobin}) with $\gamma$ independent of $\Phi_b$ leads to the Whittaker
$W$-function solution \eqref{eq:kimit crucial} in the asymptotic boundary limit.

\paragraph{Density of states and tunneling}
Let us now normalize the wavefunctions by their value at the boundary, $\psi_s(0)=1$ and $\psi_s'(0)=\gamma$, and denote by
$u_+(L)\simeq\cos(2sL+\vartheta)$ the solution decaying towards $L=0$ with unit asymptotic amplitude. From
(\ref{eq:Rr}), the $u_-$ branch contributes $\frac{P(0)-\gamma}{P(0)+\gamma}u_+(0)$ at $L=0$, but only
$\mathcal O(\mathcal T)$ at large $L$. As a consequence,
\begin{equation}
  \rho(s)=\frac{4}{\pi}\,\frac{4\,u_+(0;s)^2}{\big(1+\gamma/P(0)\big)^2}\,\big(1+\mathcal O(\mathcal T)\big)\,.
  \label{eq:Rtunnel}
\end{equation}
We then recover the density of states of JT gravity with an ETW brane (corresponding to the $u_+(0;s)^2$ factor) plus finite cutoff corrections. For fixed $\gamma$, the $s$-dependence of the denominator is $\mathcal O(\gamma s^2/\Phi_b^3)$, which is beyond the order
we are interested in. This means that $\rho(s)$ is determined, up to an $s$-independent factor, by the amplitude with which
a unit-amplitude wave tunnels through the barrier to $L=0$. The density of states $\rho_\infty(s)$ has the same structure,
where $L=0$ is replaced by $x\to\infty$: it is the squared amplitude of the unit-amplitude Morse solution relative to the
$s$-independent function $x^{-2\mJT-1/2}e^{-x/2}$ fixed by \eqref{eq:ABL}.

\paragraph{The potential difference}
The logarithmic derivative $v_s\equiv\partial_L\log\psi_s$ satisfies the Riccati equation $v_s'+v_s^2=4(U-s^2)$, which only
involves the potential and the energy. Let us introduce a matching point $x_c$, i.e.\ $L_c=\log(8\Phi_b/x_c)$, with
$1\ll x_c\ll\Phi_b$, which divides the problem into the region $x<x_c$, where $U$ and $U_\infty$ differ only by
$U_2/\Phi_b^2$, and the barrier $x>x_c$. Integrating $v_s$ from $L_c$ to $L=0$ and the corresponding Morse quantity from
$L_c$ to $L\to-\infty$, we find that $\frac12\log(\bbr/\rho_\infty)$ is the sum of an $s$-independent constant, the
difference $\log u_+(L_c)-\log u_+^\infty(L_c)$ in the region $x<x_c$, and the difference of the integrals of $v_s$ and
$v_s^\infty$ across the two barriers. If $U$ and $U_\infty$ were equal, each of these terms would vanish identically, so
every deviation of $\rho(s)$ from $\rho_\infty(s)$ comes from the difference between the potentials.

The question is then which part of this difference depends on $s$. In the barrier, $U\ge U(L_c)\simeq x_c^2/16\gg s^2$,
and $v_s$ is a local function of the potential that can be expanded in $s^2/U$,
\begin{equation}
  v_s=2\sqrt U-\frac{U'}{4U}-\frac{s^2}{\sqrt U}-\frac{s^2U'}{4U^2}+\mathcal O\big(s^4U^{-3/2}\big)\,.
\end{equation}
The first two terms do not depend on $s$. Even though $U$ and $U_\infty$ differ by $\mathcal O(1)$ in the barrier (the
barrier height at $L=0$ is $\Phi_b^2$ instead of $4\Phi_b^2$, and the potential (\ref{eq:RU}) ends at $L=0$), this
difference only contributes to $c_0(\Phi_b)$. The $s$-dependent part of the barrier contribution to
$\frac12\log(\bbr/\rho_\infty)$ is $s^2$ times
\begin{equation}
  \int_0^{L_c}\frac{{\rm d}L}{\sqrt{U}}-\int_{-\infty}^{L_c}\frac{{\rm d}L}{\sqrt{U_\infty}}
  +\frac{1}{4U(0)}-\frac14\left(\frac{1}{U}-\frac{1}{U_\infty}\right)\bigg|_{L=L_c}+\mathcal O\big(x_c^{-3}\big)\,,
  \label{eq:Rtof}
\end{equation}
namely, the difference of the tunneling times through the two barriers plus boundary terms at $L=0$ and $L=L_c$. On the
other hand, the turning point $U=s^2$, where the energy enters at $\mathcal O(1)$, lies in the region $x<x_c$, where the
two potentials differ only by $U_2/\Phi_b^2$ (\ref{eq:RVexp}). There, the difference is a first-order response to $U_2$.
This implies that the $s$-dependence cancels at $\mathcal O(\Phi_b^0)$, i.e.\ $\rho(s)=c_0(\Phi_b)\,\rho_\infty(s)$ at
leading order, and that the first $s$-dependent correction appears at $\mathcal O(\Phi_b^{-2})$.

\paragraph{Leading correction}
The tunneling times in (\ref{eq:Rtof}) can be evaluated in closed form,
\begin{equation}
  \begin{aligned}
  \int_0^{L_c}\frac{{\rm d}L}{\sqrt U}&=\frac{\sqrt{\Phi_b^2-\mJT^2+2\mJT\Phi_b\sinh L_c}-\sqrt{\Phi_b^2-\mJT^2}}{\mJT\Phi_b}\,,\\
  \int_{-\infty}^{L_c}\frac{{\rm d}L}{\sqrt{U_\infty}}&=\frac{\sqrt{1+8\mJT/x_c}-1}{\mJT}\,,
  \end{aligned}
\end{equation}
and their difference is $\big(\frac{\mJT}{4}-\frac{x_c}{16}\big)\Phi_b^{-2}+\mathcal O(x_c^{-1})$. The barrier in
(\ref{eq:RU}) is shorter than the Morse one, since it ends at $L=0$. The boundary terms give $+1/{4\Phi_b^2}$ at
$L=0$, from $U(0)=\Phi_b^2-\mJT^2$, and $-1/{8\Phi_b^2}$ at $L_c$, since
$U(L_c)^{-1}-U_\infty(L_c)^{-1}\rightarrow1/{2\Phi_b^2}$. In the region $x<x_c$, let us define
\begin{equation}
  F_{\mJT}(s)\equiv\lim_{x_c\to\infty}\left[\partial_\epsilon\log u_+^{(\epsilon)}(x_c)\Big|_{\epsilon=0}
  -\frac{s^2x_c}{16}\right],
  \label{eq:RF}
\end{equation}
where $u_+^{(\epsilon)}$ is the unit-amplitude solution for the potential $U_\infty+\epsilon\,U_2$ that matches the Wentzel–Kramers–Brillouin (WKB)
solution at $x_c$. The term $s^2x_c/16$ subtracted in (\ref{eq:RF}) cancels the $x_c$-dependence of the tunneling times,
which reflects the fact that moving $x_c$ only transfers the same part of $U-U_\infty$ between the two regions.
$F_{\mJT}(s)$ is a first-order integral of $U_2$ against Whittaker functions. Collecting these contributions, we find
(\ref{eq:Rmain}) with
\begin{equation}
  c_2(s)=\left(\frac14+\frac{\mJT}{2}\right)s^2+2F_{\mJT}(s)\,,
  \label{eq:Rc2}
\end{equation}
up to a $s$-independent constant, which can be absorbed in $c_0(\Phi_b)$. By (\ref{eq:Rphase}), $c_2(s)$ is the same for
all boundary conditions (\ref{eq:Rrobin}) with $\gamma$ independent of $\Phi_b$, up to $\mathcal O(e^{-2\pi\Phi_b})$.

Note that $\rho(s)$ depends on how $\psi_s$ is normalized at $L=0$. If one normalizes by the derivative instead, as is
natural for Dirichlet boundary conditions ($\psi_s(0)=0$, $\psi_s'(0)=1$), $\rho(s)$ is multiplied by
$P(0)^2=4(U(0)-s^2)$, which replaces $c_2(s)\to c_2(s)-s^2$. Therefore, the normalization at $L=0$ must be specified
when comparing $\rho(s)$ with $\rho_\infty(s)$ beyond leading order.

\subsection{Bound state solutions}\label{sapp:bounds}
We now seek to evaluate bound state solutions to the Schrödinger equation \eqref{eq:schrodinger ETW}. Importantly, since, as we showed, the wall barrier is opaque, the half-line (i.e.~$L>0$) bound states coincide, up to $\mathcal{O}(\tf)$, with those of the full-line ($L\in\mathbb{R}$) problem. This means that we can apply the methods for shape-invariant potentials in
supersymmetric quantum mechanics, \cite{Alvarez-Castillo:2006uzq,morse1946methods,Cooper:1994eh,compean2006trigonometric,Witten:1981nf}, for the Scarf II shape-invariant potential in \eqref{eq: Hamilt ETW} whose superpotential is specified below. In general, one can solve a special class of the Schrödinger equation of the form,\footnote{Note there is a multiplicative factor of $4$ relating $\hat{A}_+\hat{A}_-$ in \eqref{eq:super pot method} with $-\frac{1}{4}\rmd/\rmd L+U(L)$ in \eqref{eq: Hamilt ETW}.}
\begin{eqnarray}\label{eq:super pot method}
    \hH \psi(L)=E_n\psi(L)~,\quad \hH=\hat{A}_+\hat{A}_-+E_0~,\quad \hat{A}_\pm\equiv\pm\dv{L}-\mathscr{U}(L)~,
\end{eqnarray}
where $\hat{A}_+(L)$, $\hat{A}_-(L)$ are creation and annihilation operators, $\mathscr{U}(L)$ is a function called the superpotential, $E_n$ are quantized energy states, $E_0$ the energy of the ground state $\psi_{\rm GS}$, given by 
\begin{equation}
    \psi_{\rm GS}(x)\propto \exp\qty(-\int\rmd x \,\mathscr{U}(x))~,
\end{equation}
and the excited states take the form 
\begin{eqnarray}\label{eq: excited states brane}
    \psi_m\propto(\hat{A}_+(L))^m\psi_{\rm GS}~.
\end{eqnarray}
In our case \eqref{eq:schrodinger ETW} has the form of \eqref{eq:super pot method}, where the superpotential is
\begin{eqnarray}\label{eq:super pot}
    \mathscr{U}(L)=(v-1/2)\tanh L+\frac{4m_{\rm JT}\Phi_b}{\upsilon}\sech L~,
\end{eqnarray}
where $\upsilon$ is a factor that satisfies
\begin{equation}\label{eq:v eq}
    \upsilon^4-\qty(1/4-4\Phi_b^2+4m_{\rm JT}^2)\upsilon^2-16m_{\rm JT}^2\Phi_b^2=0~.
\end{equation}
with $4m_{\rm JT}\Phi_b/\upsilon$ held fixed under $\upsilon\rightarrow \upsilon-1$. This means that the spectrum of discrete energy states is 
\begin{eqnarray}\label{eq:energy condition}
    E_n=-s_n^2,\quad s_n=\tfrac12\big(\upsilon-\tfrac12-n\big)~.
\end{eqnarray}
Note that once we include the discrete energy states $s_n$, the resolution of the identity in the energy basis takes a new form
\begin{eqnarray}
    \delta(L-L')=\int_{0}^\infty \rmd s~\rho(s)~\psi_s^*(L)\psi_s^*(L')+\sum_{n}\psi_{s_n}^*(L)\psi_{s_n}^*(L')~.
\end{eqnarray}
Next, expanding \eqref{eq:v eq} to first order, we have
 \begin{equation}
  \upsilon=-2\mJ-\frac{\mJ}{16\Pb^2}+\mathcal{O}(\Pb^{-4}),
\end{equation}
which implies in \eqref{eq:energy condition} that
\begin{eqnarray} \label{eq:sn}
    s_n=-\mJ-\frac14-\frac n2-\frac{\mJ}{32\Pb^2}+\mathcal{O}(\Pb^{-4}).
\end{eqnarray}
We see that the leading term is \eqref{eq:Nsn}, and the leading-order finite-cutoff correction
$-\mJ/(32\Pb^2)$ does not depend on $n$.

We also can identify that the ground state of \eqref{eq: Hamilt ETW} is
\begin{equation}\label{eq:GS ETW}
    \psi_{\rm GS}(s)=(\cosh L)^{-\upsilon+1/2}\exp[-4m_{\rm JT}\frac{\Phi_b}{\upsilon}\tan^{-1}\sinh L]~,
\end{equation}
Similarly, the excited energy states are just \eqref{eq: excited states brane} with \eqref{eq:super pot}. Next, we notice that for the energies to be  bounded from below, we need that
\begin{eqnarray}
    n=0,\,1,\,\dots,\,\lfloor \upsilon-1/2\rfloor~,
\end{eqnarray}
and for the ground state \eqref{eq:GS ETW} as well as the excited states \eqref{eq: excited states brane} to be normalizable we need to impose $\upsilon\geq1/2$, i.e.~the threshold for the existence of the $n=0$ state. This means that for \eqref{eq:v eq} to be satisfied,
\begin{eqnarray}\label{eq:boun on mu}
    \abs{m_{\rm JT}}\geq \frac{\Phi_b}{\sqrt{1+16\Phi_b^2}}~.
\end{eqnarray}
In particular, for $m_{\rm JT}<0$ one has $L_{\rm min}>0$ in \eqref{eq:conditions in l min max}, where we will only consider $L\geq0$ due to the physical requirement of non-negative geodesic lengths in finite-cutoff holography \cite{Griguolo:2026eoi}. In that case, and considering the asymptotic boundary limit, we find $m_{\rm JT}\leq-1/4$, which agrees with previous work on deformations of the DSSYK model (see \cite{Rajgadia:2026ask} Sec.~7, where $\kappa=2m_{\rm JT}$ with respect to our notation).

Similarly to Sec.~\ref{ssec:traversable wormholes}, the existence of negative energy states in AdS$_2$ \eqref{eq:energy condition} for the brane with negative tension obeying \eqref{eq:boun on mu} implies that there exists an eternal traversable wormhole in this set-up. We emphasize that one must consider $m_{\rm JT}<0$ so that $0<\sigma_0<\sigma<\sigma_b$, with $\sigma_b$ corresponding to the location where $\Phi=\Phi_b$ in the static dilaton case, since we found after \eqref{eq:conditions in l min max} that $L_{\rm min}>0$ only when the brane tension is negative.

\section{Numerical results: Heun spectrum and bound states}\label{app:numerics}
In this appendix, we numerically study the existence of bound states for
the Schrödinger equation \eqref{eq:schr Scarf mod} with the Friedrichs self-adjoint extension at finite $\Phi_b$.

\begin{figure}[t]\centering
\subfloat[]{\includegraphics[height=0.33\textwidth]{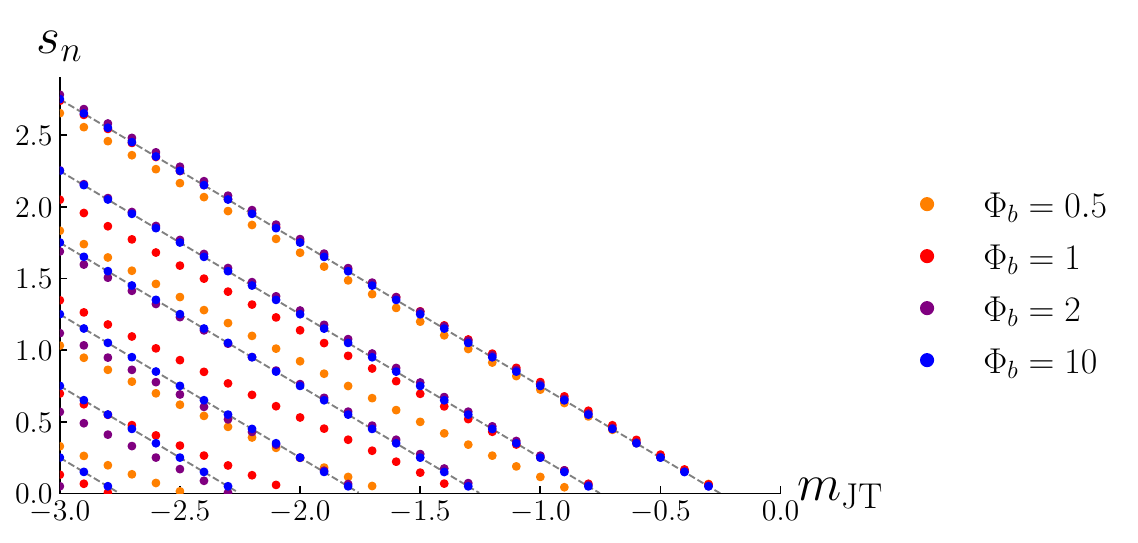}}\\
\subfloat[]{\includegraphics[height=0.3\textwidth]{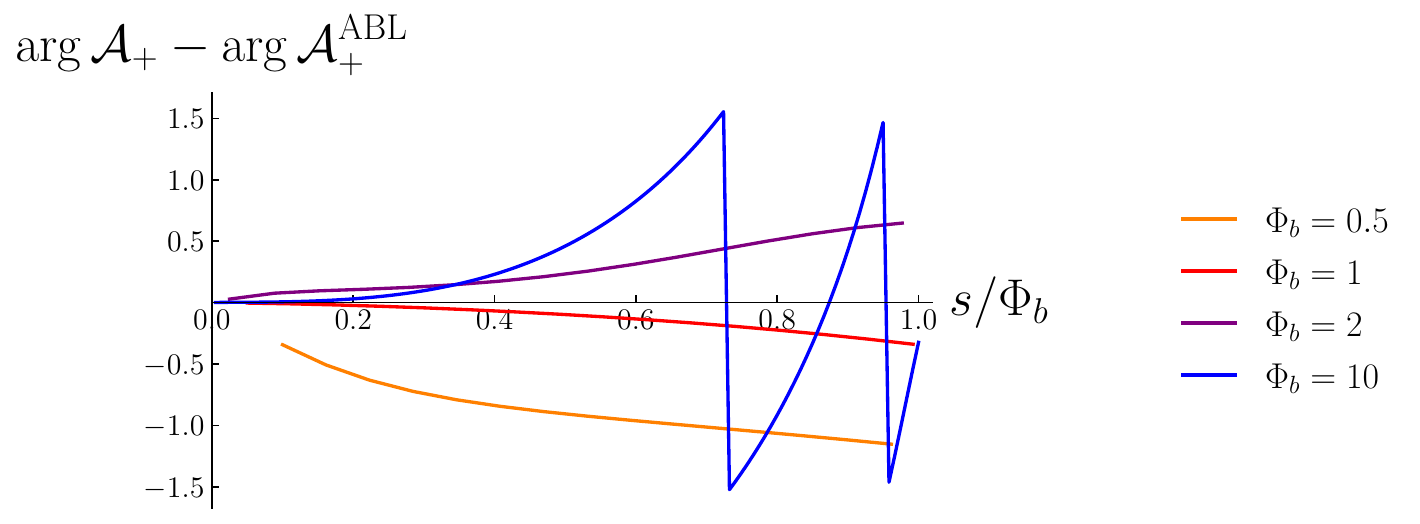}}
\caption{(a) Discrete states $s_n$ of \eqref{eq:schr Scarf mod}, determined numerically from \eqref{eq:connection poles}, as a function of $\mJT$ for different $\Phi_b$; the dashed lines are
(\ref{eq:Nsn}). (b) Difference between the phase of $\mathcal A_+$ in \eqref{eq:asymptotics of psi s} and its asymptotic boundary limit,
$\arg\mathcal A_+^{\rm ABL}=\arg\Gamma(4{\rm i}s)-\arg\Gamma(\frac12+2{\rm i}s+2\mJT)-2s\log(8\Phi_b)$, given by the
Whittaker function \eqref{eq:kimit crucial} (mod $\pi$), at $\mJT=-1$. The difference is small for $s\ll\Phi_b$ and becomes $\mathcal O(1)$ for $s\sim\Phi_b$, where the
region near $L=0$ is no longer classically forbidden.}
\label{fig:Nheun}
\end{figure}

We solve \eqref{eq:schr Scarf mod} on $L\ge0$ numerically and obtain (i) the spectrum $s_n$ at fixed $\Phi_b=2$ with (\ref{eq:Nsn}) in Tab.~\ref{tab:Nbs}, and (ii) the values of $\mJT$ at which each discrete state appears for different $\Phi_b$ in Tab.~\ref{tab:Nth}. We find that there exist
discrete states at finite $\Phi_b$, and that they approach (\ref{eq:Nsn}) as $\Phi_b$ increases, with deviations below $10^{-3}$ at $\Phi_b=10$. The thresholds approach $-\frac14-\frac n2$, with the lowest one deviating by $\mathcal O(\Phi_b^{-2})$. For small $\Phi_b$ there is an additional threshold at positive tension in the regime $\mJT>\Phi_b$ where $U_{\rm m}(L)$ is negative near $L=0$, which has no
analog in the asymptotic boundary limit. Moreover, the phase of $\mathcal A_+$ in \eqref{eq:asymptotics of psi s} agrees with that of the
Whittaker function \eqref{eq:kimit crucial}, $\arg\Gamma(4{\rm i}s)-\arg\Gamma(\frac12+2{\rm i}s+2\mJT)-2s\log(8\Phi_b)$, up to
$\mathcal O(10^{-3})$ for $s\lesssim0.3\,\Phi_b$ at $\Phi_b=10$, while it departs from it for $s\sim\Phi_b$
(Fig.~\ref{fig:Nheun}). This confirms the selection of the Whittaker $W$-function in \eqref{eq:kimit crucial}, and it agrees with the
fact that for $s\ll\Phi_b$ the region $L\ll\log(8\Phi_b)$ is classically forbidden, so that the boundary condition at
$L=0$ only affects energies $s\sim\Phi_b$.

\begin{table}[t]\centering
\begin{tabular}{c c l l}
\toprule
$\Phi_b$ & $\mJT$ & $s_n$ from (\ref{eq:schr Scarf mod}) & $s_n$ from (\ref{eq:Nsn}) \\
\midrule
2 & $-0.3$ & 0.0527 & 0.05 \\
2 & $-0.5$ & 0.2547 & 0.25 \\
2 & $-1.0$ & 0.7610,\ 0.2641 & 0.75,\ 0.25 \\
2 & $-1.5$ & 1.2682,\ 0.7744,\ 0.2753 & 1.25,\ 0.75,\ 0.25 \\
2 & $-2.0$ & 1.7746,\ 1.2764,\ 0.7632,\ 0.2508 & 1.75,\ 1.25,\ 0.75,\ 0.25 \\
2 & $-3.0$ & 2.7816,\ 2.2545,\ 1.6883,\ 1.1186,\ 0.5695,\ 0.0509 & 2.75,\ 2.25,\ 1.75,\ 1.25,\ 0.75,\ 0.25 \\
\bottomrule
\end{tabular}
\caption{Discrete states $E_{\rm ADM}=-s_n^2$ of \eqref{eq:schr Scarf mod} with the Friedrichs extension at $\Phi_b=2$, compared with the
asymptotic boundary limit (\ref{eq:Nsn}).}
\label{tab:Nbs}
\end{table}

\begin{table}[t]\centering
\begin{tabular}{c cccccc c}
\toprule
$\Phi_b$ & $n=0$ & $n=1$ & $n=2$ & $n=3$ & $n=4$ & $n=5$ & \eqref{eq:boun on mu}\\
\midrule
0.25 & $-0.272$ & $-1.044$ & $-1.940$ & $-2.875$ & & & $-0.177$\\
0.5  & $-0.2325$ & $-0.838$ & $-1.614$ & $-2.473$ & & & $-0.224$\\
1    & $-0.2393$ & $-0.729$ & $-1.310$ & $-2.011$ & $-2.791$ & & $-0.243$\\
2    & $-0.2478$ & $-0.741$ & $-1.229$ & $-1.730$ & $-2.292$ & $-2.931$ & $-0.248$\\
5    & $-0.2497$ & $-0.749$ & $-1.248$ & $-1.747$ & $-2.246$ & $-2.744$ & $-0.2497$\\
10   & $-0.2499$ & $-0.7498$ & $-1.2496$ & $-1.7494$ & $-2.2492$ & $-2.749$ & $-0.2499$\\
$\infty$ & $-0.25$ & $-0.75$ & $-1.25$ & $-1.75$ & $-2.25$ & $-2.75$ & $-0.25$\\
\bottomrule
\end{tabular}
\caption{Values of $\mJT$ below which the $n$-th discrete state of \eqref{eq:schr Scarf mod} exists. The last column is the first
threshold \eqref{eq:boun on mu} for the minimal operator ordering \eqref{eq:schrodinger ETW}, which is close to the $n=0$ threshold for $\Phi_b\gtrsim1$, but
differs at small $\Phi_b$.}
\label{tab:Nth}
\end{table}

\bibliographystyle{JHEP}
\bibliography{references.bib}
\end{document}